%% file: Thesis.tex
\RequirePackage{fix-cm}

\documentclass[12pt,dvipsnames]{book}
\usepackage[Sonny]{fncychap}
\usepackage{enumerate}
\usepackage{dsfont}
\usepackage{amsfonts}
\usepackage{amssymb}
\usepackage{textcomp}
\usepackage[paperheight=240mm, paperwidth = 170mm,
    inner = 1.80cm,  
    outer = 1.80cm,  
    top=2.4cm,
    bottom=2.7cm
    ]{geometry}
\usepackage[utf8]{inputenc}
\usepackage[T1]{fontenc}
\usepackage{graphicx}
\usepackage{mathtools}
\usepackage{soul}
\usepackage{bbm}
\usepackage{subfiles}
\usepackage{subcaption}
\usepackage{fancyhdr}
\usepackage{pdfpages}
\usepackage{bm}
\usepackage{minitoc} 
\usepackage[newparttoc,explicit, clearempty]{titlesec}%
\usepackage{titletoc}
\usepackage{chngcntr}
\usepackage{fmtcount}
\usepackage{etoolbox}
\usepackage{setspace}
\usepackage[fontsize=11pt]{scrextend} 
\usepackage{amsthm}
\usepackage{latexcolors}

\usepackage{lmodern} 

\makeatletter
\newcommand*\showfontsize{\f@size{} point}
\makeatother

\usepackage{tikz,tcolorbox}
\usetikzlibrary{shapes,calc}
\tcbuselibrary{skins}
\definecolor{myblue}{rgb}{0.15,0.15,0.53}

\makeatletter
\newtcbox{\titlebox}{
    enhanced,
    overlay={
        \draw[myblue,fill=myblue](frame.south east)--+(0,.2)to[bend right]+(.2,-0)--cycle;},
        colback=myblue,
        top=-1pt,bottom=-2pt,left=2pt,right=2pt,
        boxrule=1pt,
        colframe=myblue,
        sharp corners=south,
        colupper=white,
        fontupper=\bfseries
}
\newtcolorbox{myblock}[1][]{
    enhanced,
    left=2pt,
    right=2pt,
    colframe=myblue,
    boxrule=1pt,
    colback=blue!10,
    overlay={
        \def\myblock@tempa{#1}
        \ifx\myblock@tempa\@empty
        \else
        \draw[myblue,fill=myblue]($(frame.north west)+(.2pt,-.2pt)$)--+(.1,0)to[bend right]+(-0,-.1)--cycle;
        \node [
            anchor=south west,
            inner sep=0pt,
            outer sep=0pt
        ]at(frame.north west){\titlebox{#1}};
    \fi
    },
}
\makeatother

\def\partname{Part}

\titlecontents{part}[0em]{\large\bfseries\protect\addvspace{25pt}\titlerule\addvspace{1.5ex}}%
{\contentslabel[\rlap{\partname~\thecontentslabel}]{0em}\hphantom{\partname~\thecontentslabel\quad}}{}%
{\hfill\contentspage}[\addvspace{1ex}\titlerule\addvspace{1ex}]%

\makeatletter
\newcommand*{\hyperlinkcite}[1]{\hyper@link{cite}{cite.#1}}
\newcommand{\citeref}[2]{\hyperref[{#1}]{\color{black}{[}\color{darkpastelgreen}{{#2}}\color{black}{]}}}
\makeatother

\usepackage[square,numbers]{natbib}

\usepackage[resetlabels,labeled]{multibib}
\usepackage[colorlinks=true,urlcolor=darkred,unicode]{hyperref}
\usepackage[
        noabbrev,
        capitalise,
    ]{cleveref}
\usepackage{textalpha}

\definecolor{darkred}{RGB}{153,0,0}
\hypersetup{
    colorlinks = true,
    linkcolor = brandeisblue, 
    citecolor = 
    darkpastelgreen 
    }

\DeclareCaptionSubType*{figure}

\usepackage[toc, sort=use]{glossaries}

\newcommand{\R}{\mathbb R}

\newcommand{\D}{\mathrm{d}}

\newcommand{\db}{\bar\partial}
\newcommand{\sgn}{\text{sgn}}

\newcommand{\psR}{pseudo-Riemannian }

\newcommand{\ab}{$(\alpha,\beta)$}
\newcommand{\mkrop}{$m$-Kropina }
\newcommand{\nablad}{\bm{\nabla}_{\textit{\hspace{-2.5pt}\tiny D}}\hspace{1pt}} 

\newcommand\YUGEE{\fontsize{41}{48}\selectfont} 
\newcommand\yuge{\fontsize{19}{23}\selectfont} 

\newtheorem{prop}{Proposition}
\AfterEndEnvironment{prop}{\noindent\ignorespaces} 
\numberwithin{prop}{section}

\newtheorem{cor}[prop]{Corollary}
\AfterEndEnvironment{cor}{\noindent\ignorespaces} 

\newtheorem{lem}[prop]{Lemma}
\AfterEndEnvironment{lem}{\noindent\ignorespaces} 

\newtheorem{theor}[prop]{Theorem}
\AfterEndEnvironment{theor}{\noindent\ignorespaces} 
\newtheorem{rem}[prop]{Remark}
\AfterEndEnvironment{rem}{\noindent\ignorespaces}
\newtheorem{defi}[prop]{Definition}
\AfterEndEnvironment{defi}{\noindent\ignorespaces}
\newtheorem{ex}[prop]{Example}
\AfterEndEnvironment{ex}{\noindent\ignorespaces}
\newtheorem{defthm}[prop]{Definition and Theorem}
\AfterEndEnvironment{defthm}{\noindent\ignorespaces}

\AfterEndEnvironment{lem2}{\noindent\ignorespaces}

\AfterEndEnvironment{q}{\noindent\ignorespaces}

\AfterEndEnvironment{proof}{\noindent\ignorespaces} 
\AfterEndEnvironment{itemize}{\noindent\ignorespaces} 
\AfterEndEnvironment{enumerate}{\noindent\ignorespaces} 

\usepackage[framemethod=TikZ]{mdframed}
\newcounter{classbox}[chapter]

\newcites{H}{Publications on which this Dissertation is Based}
\newcites{trash}{foo} 

\makeglossaries
\input{glossary.tex}

 \AtBeginDocument{\addtocontents{toc}{\protect\setlength{\parskip}{2.5
 pt}}} 

\begin{document}

\frontmatter

\setstretch{1} 

\includepdf[pages=-]{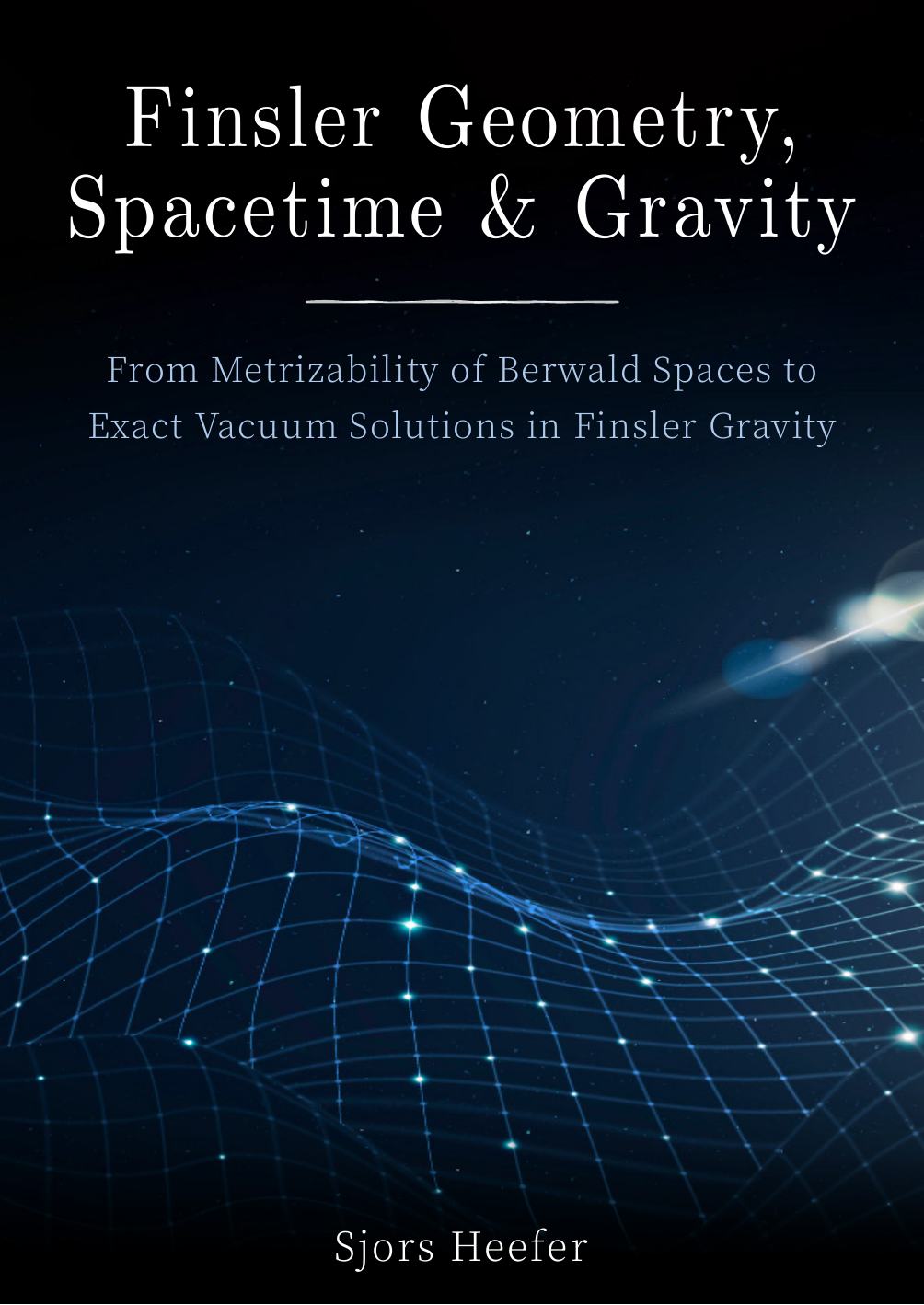} 
\setcounter{page}{1} 

\subfile{TitlepageOfficial}

\title{Metrizability of Berwald Spaces and Exact Vacuum Solutions in Finsler Gravity}
\author{Sjors Heefer}

\subfile{Summary} 

{\hypersetup{linkcolor=black}
\setcounter{tocdepth}{1} 
\doparttoc\tableofcontents}

\subfile{ch_Glossary}

\mainmatter
\subfile{ch_introduction}


\part{Preliminaries}\label{part:preliminaries}

{\hypersetup{linkcolor=black}
\mtcsettitle{parttoc}{}
}

\subfile{intr1}

\subfile{ch_Fundamentals_of_Finsler_geometry}

\subfile{ch_Connections}

\subfile{ch_Berwald_Landsberg}

\part[Berwald Spaces and Metrizability]{Berwald Spaces and Pseudo-Riemann Metrizability}\label{part:Berwald_metrizability}
{\hypersetup{linkcolor=black}
\mtcsettitle{parttoc}{}
}

\subfile{intr2}

\subfile{ch_New_characterization_of_Berwald}
\subfile{ch_Metrizability}

\part{Vacuum Solutions in Finsler Gravity}\label{part:vacuum_sols}
{\hypersetup{linkcolor=black}
\mtcsettitle{parttoc}{}
}

\subfile{intr3}
\subfile{ch_Finsler_gravity}

\subfile{ch_exact_Berwald_solutions}

\subfile{ch_lin_grav_waves}

\subfile{ch_Unicorn_cosmology}

\subfile{Discussion}

\part*{Appendices}

\addcontentsline{toc}{part}{Appendices}
\subfile{Appendix}


\backmatter

\cleardoublepage
\phantomsection



\subfile{ListOfPublications}

\cleardoublepage
\phantomsection

\bibliographystyle{utphys}
\addcontentsline{toc}{chapter}{Bibliography}
\bibliography{GeneralBib}

\subfile{Acknowledgments}



\end{document}

%% file: glossary.tex
\newglossaryentry{M}
{
    name=$M$,
    description={Smooth manifold}
}
\newglossaryentry{x}
{
    name=$x^i$,
    description={Coordinates on the manifold $M$}
}
\newglossaryentry{n}
{
    name=$n$,
    description={Dimension of the manifold $M$}
}

\newglossaryentry{TM}
{
    name=$TM$,
    description={Tangent bundle of $M$}
}
\newglossaryentry{TM0}
{
    name=$TM_0$,
    description={Slit tangent bundle $TM_0 = TM\setminus\{0\}$}
}
\newglossaryentry{conicA}
{
    name=$\mathcal A$,
    description={Conic subbundle of $TM$}
}
\newglossaryentry{conicA_tilde}
{
    name=$\tilde{\mathcal A}$,
    description={$ = \mathcal A\setminus \{\phi'=0\}$ if $F = \alpha\phi(\beta/\alpha)$ is an \ab-metric}
}

\newglossaryentry{pullback_bundle}
{
    name=$\pi^*TM$,
    description={Pull-back bundle of the tangent bundle $TM$ along the canonical projection $\pi:TM\to M$ or its restriction $\pi:\mathcal A\to M$ if $\mathcal A\subset TM$ is a conic subbundle}
}

\newglossaryentry{vector_fields}
{
    name={$\mathfrak X(M)$},
    description={Set of (smooth) vector fields on $M$}
}

\newglossaryentry{1forms}
{
    name={$\Omega^1(M)$},
    description={Set of 1-forms on $M$}
}
\newglossaryentry{E}
{
    name={$E$},
    description={(Total space of a) vector bundle}
}

\newglossaryentry{1forms_with_values_in}
{
    name={$\Omega^1(M,E)$},
    description={Set of 1-forms on $M$ with values in a vector bundle $E$}
}

\newglossaryentry{sections}
{
    name={$\Gamma(E)$},
    description={Sections of a vector bundle $E$}
}
\newglossaryentry{Finsler_vector_fields}
{
    name={$\Gamma(\pi^*TM)$},
    description={Set of all Finsler vector fields on $M$}
}

\newglossaryentry{Finsler_tensor_fields}
{
    name={$\mathcal T^r_s(\pi^*TM)$},
    description={Set of all Finsler $(k,l)$-tensor fields on $M$}
}

\newglossaryentry{Finsler_metric}
{
    name={$F$},
    description={Finsler metric}
}
\newglossaryentry{L}
{
    name={$L$},
    description={Finsler Lagrangian}
}
\newglossaryentry{fundamental_tensor}
{
    name={$g_{ij}$},
    description={Fundamental tensor of a Finsler Lagrangian}
}
\newglossaryentry{cartan}
{
    name={$C_{ijk}$},
    description={Cartan tensor}
}
\newglossaryentry{Nonlin_conn}
{
    name={$N^k_i$},
    description={Nonlinear connection coefficients}
}
\newglossaryentry{partial_der}
{
    name={$\partial_i$},
    description={$=\partial/\partial x^i$, partial derivative}
}
\newglossaryentry{ver_der}
{
    name={$\bar\partial_i$},
    description={$=\partial/\partial y^i$, vertical derivative}
}
\newglossaryentry{hor_der}
{
    name={$\delta_i$},
    description={$=\partial_i - N^k_i\bar\partial_k$, horizontal derivative}
}
\newglossaryentry{Nonlin_curv}
{
    name={$R^k{}_{ij}$},
    description={Nonlinear curvature tensor}
}
\newglossaryentry{Lin_curv}
{
    name={$\bar R^k{}_{lij}$},
    description={Curvature tensor of the canonical linear connection on $TM$ (for Berwald spaces)}
}
\newglossaryentry{Finsler-Ricci-scalar}
{
    name={Ric},
    description={Finsler-Ricci scalar Ric$=R^i_{ij}y^j$}
}

\newglossaryentry{Finsler-Ricci-tensor}
{
    name={$R_{ij}$},
    description={Finsler-Ricci tensor $R_{ij} = \tfrac{1}{2}\bar\partial_i\bar\partial_j\text{Ric}$}
}

\newglossaryentry{Affine Ricci}
{
    name={$\bar R_{ij}$},
    description={Affine Ricci tensor $\bar R_{ij} = \bar R^k{}_{ikj}$ (for Berwald spaces)}
}

\newglossaryentry{Lin_connection}
{
    name={$\nabla$},
    description={Linear connection}
}

\newglossaryentry{DynCovDer}
{
    name={$\nablad$},
    description={Dynamical covariant derivative}
}

\newglossaryentry{Christoffel_symbs_hor}
{
    name={$\Gamma^k_{ij}$},
    description={Christoffel symbols if $E = TM$, or \newline \textit{Horizontal} Christoffel symbols if $E=\pi^*TM$}
}

\newglossaryentry{Christoffel_symbs_ver}
{
    name={$\bar \Gamma^k_{ij}$},
    description={\textit{Vertical} Christoffel symbols of a linear connection on $E=\pi^*TM$}
}

\newglossaryentry{Christoffel_symbs_LC}
{
    name={$\mathring \Gamma^k_{ij}$},
    description={Christoffel symbols of the Levi-Civita connection}
}

\newglossaryentry{symmetrized_tensor_prod}
{
    name={$\D x^i \D x^j$},
    description={$=\tfrac{1}{2}(\D x^i\otimes \D x^j + \D x^j \otimes \D x^i)$, symmetrized tensor product}
}

\newglossaryentry{symmetrized_tensor_comps}
{
    name={$T_{(ij)}$},
    description={$=\tfrac{1}{2}(T_{ij} + T_{ij})$, symmetrized tensor components}
}

\newglossaryentry{anti_symmetrized_tensor_comps}
{
    name={$T_{[ij]}$},
    description={$=\tfrac{1}{2}(T_{ij} - T_{ij})$, antisymmetrized tensor components}
}

\newglossaryentry{a}
{
    name={$a_{ij}$},
    description={Components of a pseudo-Riemannian metric}
}

\newglossaryentry{LCconn}
{
    name={$\mathring\nabla$},
    description={Levi-Civita connection of $a_{ij}$}
}
\newglossaryentry{y}
{
    name={$y^i$},
    description={Induced coordinates on the fibers of $TM$}
}

\newglossaryentry{A}
{
    name={A},
    description={$=a_{ij}y^i y^j$}
}
\newglossaryentry{epsilon}
{
    name={$\epsilon$},
    description={$\epsilon=\text{sgn}(A)$}
}

\newglossaryentry{alpha}
{
    name={$\alpha$},
    description={$=\sqrt{|A|} = \sqrt{|a_{ij}y^i y^j|}$}
}
\newglossaryentry{b}
{
    name={$b_i$},
    description={Components of a 1-form}
}
\newglossaryentry{beta}
{
    name={$\beta$},
    description={$=b_i y^i$}
}
\newglossaryentry{bsquared}
{
    name={$|b|^2$},
    description={$=a_{ij}b^i b^j$}
}
\newglossaryentry{s}
{
    name={$s$},
    description={$=\beta/\alpha$}
}

\newglossaryentry{int}
{
    name={$A^0$},
    description={Topological interior of the set $A$}
}
\newglossaryentry{clo}
{
    name={$\overline{A}$},
    description={Topological closure of the set $A$}
}

\newglossaryentry{GR}
{
    name={GR},
    description={\textit{General Relativity}}
}
\newglossaryentry{FLRW}
{
    name={FLRW},
    description={\textit{Friedmann-Lema\^itre-Robertson-Walker}}
}
\newglossaryentry{iff}
{
    name={iff},
    description={\textit{If and only if}}
}
\newglossaryentry{LHS}
{
    name={LHS},
    description={\textit{Left-hand side}}
}
\newglossaryentry{RHS}
{
    name={RHS},
    description={\textit{Right-hand side}}
}

\newglossaryentry{TFAE}
{
    name={TFAE},
    description={\textit{The following are equivalent}}
}
\newglossaryentry{WLOG}
{
    name={WLOG},
    description={\textit{Without loss of generality}}
}

%% file: TitlepageOfficial.tex
\thispagestyle{empty}

\noindent
\begin{center}
\YUGEE
\Huge
\vphantom{x}
\vspace{140px}

Finsler Geometry,\\ Spacetime \& Gravity\\[20px]

\vspace{5px}


\vfill


\end{center}
 
\newpage
\thispagestyle{empty}
\noindent

\vphantom{x}
\vfill

\noindent Copyright {\copyright} 2024 by Sjors Heefer\\

\noindent\textit{All rights reserved. This dissertation, or parts thereof, may not be reproduced in any form or by any means, electronic or mechanical, including photocopying, recording or any information storage and retrieval system now known or to be invented, without written permission from the author.}\\[10px]

\noindent The research documented in this dissertation has been carried out at the Centre for Analysis, Scientific Computing and Applications (CASA) within the Department of Mathematics and Computer Science, Eindhoven University of Technology.\\

\noindent A catalogue record is available from the Eindhoven University of Technology library\\

\noindent ISBN: 978-90-386-6026-4


\newpage
\thispagestyle{empty}
\noindent

\begin{center}


\YUGEE

\vspace{5em}

Finsler Geometry, \\Spacetime \& Gravity\\[20px]

\vspace{10px}

\yuge From Metrizability of Berwald Spaces to \\
Exact Vacuum Solutions in Finsler Gravity





\vspace{3em}

\Large
PROEFSCHRIFT

\vspace{3em}

ter verkrijging van de graad van doctor aan de Technische Universiteit Eindhoven, op gezag van de rector magnificus prof.dr. S.K. Lenaerts, voor een commissie aangewezen door het College voor Promoties, in het openbaar te verdedigen op vrijdag 24 mei 2024 om 13:30 uur

\vspace{3em}

door

\vspace{3em}

Sjors Jacco Heefer

\vspace{3em}

geboren te 's-Hertogenbosch

\end{center}

\newpage
\thispagestyle{empty}
\normalsize

\noindent
Dit proefschrift is goedgekeurd door de promotoren en de samenstelling van de promotiecommissie is als volgt:

\vspace{1em}
\noindent
\begin{tabular}{lp{9cm}}
voorzitter:     & prof.dr. E.R. van den Heuvel  \\
promotor:       & prof.dr. L.M.J. Florack \\
co-promotor:    & dr. A. Fuster \\
leden:          & prof.dr. F. Toschi\\
                & prof.dr. G.R. Cavalcanti (Universiteit Utrecht)\\
                & prof.dr. M. S\'anchez Caja (Universidad de Granada) \\
                & dr. N. Voicu (Transilvania University of Brasov) \\
adviseur(s):    & dr. C. Pfeifer (University of Bremen)
\end{tabular}

\vfill

\noindent\textit{Het onderzoek dat in dit proefschrift wordt beschreven is uitgevoerd in overeenstemming met de TU/e Gedragscode Wetenschapsbeoefening.}

\newpage

\newlength\longest
\clearpage

\thispagestyle{empty}
\null\vfill

\vspace{55px}

\settowidth\longest{\huge\itshape just as his inclination leads him;}
\centering
\parbox{\longest}{%
  \raggedright{\large\itshape%
  The most beautiful thing we can experience is the mysterious. It is the source of all true art and science.\par\bigskip
  }   
  \raggedleft\normalsize\MakeUppercase{- A. Einstein}\par%
}

\vfill\vfill

\clearpage







%% file: Summary.tex
\cleardoublepage
\phantomsection
\addcontentsline{toc}{chapter}{Abstract}
\chapter*{Abstract}
\fancyhead{}
\fancyhead[LE]{\textit{Abstract}}
\fancyhead[RO]{\textit{Abstract}}

This dissertation has been organized into three parts. \cref{part:preliminaries} introduces the mathematical preliminaries. \cref{part:Berwald_metrizability} focuses on Berwald spaces, their characterization, and (pseudo-)Riemann metrizability. Finally, \cref{part:vacuum_sols} deals with exact vacuum solutions to Finsler gravity. \\

In \cref{part:Berwald_metrizability} we first develop a new characterization of Berwald spaces and discuss several applications. One of these is the Berwald condition for \ab-metrics, which we apply to several specific cases of interest: Randers metrics, (generalized) $m$-Kropina metrics, and exponential metrics. 

Next, we show, by means of a general argument as well as a simple counterexample, that Szab\'o's metrization theorem does not generalize to Finsler spaces of indefinite signature. This important theorem states that the affine connection on a positive definite Berwald space (smooth on the slit tangent bundle) can always be understood as the Levi-Civita connection of some Riemannian metric. In short: every such positive definite Berwald space is Riemann metrizable. In other signatures, however, or more generally in situations with less smoothness, we show that the situation is much more complex.

We investigate the class of $m$-Kropina metrics in detail and obtain several necessary and sufficient conditions for (local) metrizability. Interestingly, local metrizability turns out to be equivalent to the property of having a symmetric (affine) Ricci tensor. The same equivalence holds, trivially, for several other types of metrics, such as Randers metrics. This observation naturally leads us to hypothesize that such an equivalence may hold more generally. 

Then, using our characterization of local metrizability, we classify all Ricci-flat, locally metrizable $m$-Kropina metrics in (1+3)D whose 1-forms have a constant causal character. The latter class of metrics is of interest in physics since any such metric yields an exact vacuum solution to Pfeifer and Wohlfarth's field equation in Finsler gravity. Some of our results also extend to generalized $m$-Kropina metrics.\\

In \cref{part:vacuum_sols}, 
we turn our focus completely to exact solutions of the aforementioned field equations in Finsler gravity. 
We prove that any Finsler metric constructed solely (but arbitrarily) from a vacuum solution in general relativity, $\alpha$, and a covariantly constant 1-form $\beta$, is a vacuum solution in Finsler gravity. This leads to a large class of exact solutions of \ab-type. For specific types of \ab-metrics, we have stronger results. For Randers metrics, for instance, we classify all solutions of Berwald type and show that they can all be understood as Finslerian pp-wave spacetimes. 

We then discuss the physical interpretation and the observational signature of such Finslerian gravitational waves by investigating their effect on interferometric gravitational wave detectors. Remarkably, we come to the conclusion that this effect is indistinguishable from that of a standard gravitational wave in general relativity with the same waveform. Indeed, we compute the expression for the radar distance---the main observable in interferometer experiments---and find that it is identical to its general relativistic counterpart.

Finally, we obtain a solution of unicorn (i.e. Landsberg but not Berwald) type, which is of interest in cosmology. This solution has cosmological symmetry, i.e. is spatially homogeneous and isotropic, and it is additionally conformally flat, with conformal factor depending only on the timelike coordinate. It turns out that, just as in classical Friedmann-Lema\^itre-Robertson-Walker (FLRW) cosmology, this conformal factor can be interpreted as the scale factor of the universe. Our solution describes a linearly expanding (or contracting) Finslerian universe.

\newpage

%% file: ch_Glossary.tex
\clearpage
\glsaddall 
\renewcommand*{\arraystretch}{1.15}
\printglossary[title=List of Symbols and Abbreviations, toctitle = List of Symbols and Abbreviations, type=main,style=long,nonumberlist]

%% file: ch_introduction.tex
\cleardoublepage
\phantomsection
\addcontentsline{toc}{chapter}{Introduction}
\chapter*{Introduction}\label{introduction}

\fancyhead{}
\fancyhead[LE]{\small\textit{Introduction}}
\fancyhead[RO]{\small\textit{Introduction}}


Since its finalization in 1915, Einstein's general theory of relativity (GR) has proven itself as our most successful theory of gravity yet and today it is a cornerstone of modern physics. It has also been clear for a long time, however, that there is no hope of it being the ultimate answer. Indeed, as illustrated by the famous Penrose-Hawking singularity theorems, the theory even predicts its own demise. Besides, while GR is a \textit{classical} theory, the other forces of nature and the known types of matter in the universe all behave according to the laws of quantum mechanics as described in the standard model of particle physics. In order to have a consistent picture of nature it would thus seem that gravity ought to be described quantum mechanically as well. 

The `quantization' of gravity, often dubbed the holy grail of fundamental physics, has proven to be an extraordinary challenge. Even though approaches such as string theory, loop quantum gravity and others have made considerable progress in this direction, it is fair to say that we still do not have a clue which of these approaches, if any, is on the right track. 

Nevertheless, many of the fundamental approaches to quantum gravity seem to converge on the idea that local Lorentz invariance may not be fundamental but rather a low-energy approximation (see \cite{Addazi_2022} for a recent review and references therein). 
Experimental searches for Lorentz invariance violation are actively ongoing as well \cite{Addazi_2022,batista2023whitepaper,Mattingly_2005}. 
Departure from Lorentz invariance has as a direct consequence that Lorentzian geometry---the geometric foundation of GR---is no longer adequate as a description of spacetime in certain regimes. This is where Finsler geometry comes in. 




As concisely put by Chern \cite{chernQuote}, Finsler geometry is just pseudo-Riemannian geometry without the quadratic restriction. In other words, it is the natural extension of pseudo-Riemannian geometry in which the squared line element $\D s^2$ 
is not restricted to be quadratic in the coordinate 1-forms $\D x^i$. Historically, the possibility of considering this type of geometry was already discussed by Riemann in his famous habilitation lecture in 1854 \cite{Riemann1,Riemann2}. But the first systematic study of such spaces appeared only in Finsler's  1918 PhD thesis \cite{Finsler}. After this, the theory was developed further by mathematicians such as Cartan, Chern, Rund, Matsumoto, and many others.

As the initial development of GR took place between 1905-1915, it is only natural that the theory was formulated on the basis of pseudo-Riemannian geometry, which was already well developed at that time. And when it was found in the decades that followed that the theory matched experiment to extraordinary precision, the obvious conclusion was that the theory and in particular its geometric foundation were simply correct. With the knowledge that we have today, however, and specifically in light of what has been said above, 
it is natural to consider the possibility that these foundations may need to be revisited. There are compelling reasons to consider Finsler geometry rather than the more restrictive Lorentzian geometry as the fundamental geometric framework underlying the structure of spacetime and gravity.


First of all, while Lorentzian geometry is obviously inadequate in scenarios with a departure from local Lorentz invariance, it turns out that at the classical level, virtually any such scenario 
can be described in terms of a Finsler geometry on the spacetime manifold \cite{Girelli:2006fw,Raetzel:2010je,Rodrigues:2022mfj}; see e.g. \cite{Amelino-Camelia:2014rga,Lobo_2017, Letizia:2016lew} for applications to specific phenomenological quantum gravity models. 

Second, as suggested already in 1985 by Tavakol and Van den Bergh \cite{TAVAKOL198523, Tavakol_1986, Tavakol2009}, the axiomatic approach to GR by Ehlers, Pirani, and Schild (`EPS axiomatics') \cite{Ehlers2012} does not single out Lorentzian geometry---as was widely believed for a long time---but is compatible with Finsler geometry as well. This was originally overlooked due to artificially restrictive differentiability assumptions, as recently (2018) pointed out in \cite{Lammerzahl:2018lhw} and then worked out in detail in \cite{Bernal_2020}. Other axiomatic approaches also allow for more general types of geometry, see e.g. \cite{Bubuianu2018}. 

And finally, Finsler geometry provides the most general geometric framework that is compatible with the standard formulation of the clock postulate\footnote{We remark that Weyl geometry, another generalization of Lorentzian geometry, is also compatible with the clock postulate, but in that case the definition of proper time has to be revised \cite{Perlick1987}.}, which states that the time measured by a clock between two events is given by the length of the clock's spacetime trajectory connecting these events; in this case the Finslerian length rather than the pseudo-Riemannian length. 

The idea that Finsler geometry might play a role in fundamental physics has been around at least since 1926 when first attempts were made to unify gravity with electromagnetism \cite{Goenner2004,Goenner2014,tonnelat2014einstein}. Several decades later the possibility of extending GR into the realm of Finsler geometry started being considered \cite{Beem,Asanov,Rutz,Pfeifer:2011xi,Horvath1950,Horvath1952, Ikeda1981, Chang:2009pa,Kouretsis:2008ha,Stavrinos2014,Voicu:2009wi,Minguzzi:2014fxa}. But it was only in 2012 that the action-based approach to Finsler gravity was developed by Pfeifer and Wohlfarth \cite{Pfeifer:2011xi,Pfeifer:2013gha}. Structurally, this theory is completely analogous to GR, with the only essential difference being that the space of allowed metrics is enlarged from Lorentzian metrics to Finsler metrics. Einstein's field equation is extended into the realm of Finsler geometry by means of Pfeifer and Wohlfarth's field equation, which is derived from the natural Finsler generalization of the Einstein-Hilbert action and reduces to Einstein's field equation in the Lorentzian limit. In the spirit of Chern, we could say that \textit{Finsler gravity is just general relativity without the quadratic restriction}.

Recently (2019), it was discovered \cite{Hohmann_2019} that Pfeifer and Wohlfarth's field equation is in fact the variational completion \cite{Voicu_2015} of an earlier proposal for a field equation by Rutz \cite{Rutz}, the latter being given simply by the vanishing of the Finsler-Ricci tensor and derived from---or rather motivated by---the geodesic deviation equation. That is to say that, while $R_{\mu\nu}=0$ cannot be obtained as the Euler-Lagrange expression of an action functional, Pfeifer and Wohlfarth's equation is the unique Euler-Lagrange equation that is \textit{as close to it as possible}, in a well-defined sense \cite{Voicu_2015}. For reference, in the \psR setting, the variational completion algorithm transforms Einstein's early proposal for the left-hand side of the field equation---the Ricci tensor $R_{\mu\nu}$---into his final and correct expression---the Einstein tensor $R_{\mu\nu}-\tfrac{1}{2}Rg_{\mu\nu}$---%
that yields not only the correct vacuum equation (which is equivalent to $R_{\mu\nu}=0$) but also a consistent matter coupling \cite{Voicu_2015}. In 2020 the field equations corresponding to the natural Finsler generalization of the Einstein-Hilbert-Palatini action were investigated as well \cite{javaloyes2023einsteinhilbertpalatini}, and for large and important classes of Finsler spaces (Berwald, Landsberg, and weakly Landsberg spaces) these were found to be equivalent to Pfeifer and Wohlfarth's equation. All this puts the latter on firm ground.


Roughly half of this dissertation, \cref{part:vacuum_sols}, 
is devoted to the study of exact solutions to Pfeifer and Wohlfarth's field equation for Finsler gravity, to which we will come back momentarily. The other half, \cref{part:Berwald_metrizability}, 
is more fundamental in nature and concerns the characterization and properties of an important class of Finsler spaces, namely those of Berwald type. Apart from their intrinsic mathematical interest, such spaces are very relevant also in the context of Finsler gravity. Indeed, almost all of the solutions to the Finsler gravity field equations discussed in \cref{part:vacuum_sols} are of this type.
%

A Berwald space is a Finsler space that admits a (necessarily unique) torsion-free, metric-compatible linear connection on the tangent bundle, akin to the Levi-Civita connection of a \psR metric. 
Such spaces may be thought of as being just slightly more general than \psR manifolds. 
While various characterizations of Berwald spaces are known \cite{Szilasi2011}, we present here a novel characterization in terms of an arbitrary auxiliary \psR metric and use it to obtain, in particular, a very useful necessary and sufficient Berwald condition for \ab-metrics---Finsler metrics constructed from a \psR metric $\alpha$ and a 1-form $\beta$ (\cref{ch:Characterization_of_Berwald}). We then apply this result to several specific \ab-metrics of interest such as Randers metrics, exponential metrics, and (generalized) \mkrop metrics, obtaining novel results as well as neatly reproducing some well-known ones.


For any Berwald space, one can ask the natural question of whether its canonical torsion-free metric-compatible connection can be understood as the Levi-Civita connection of some auxiliary \psR metric. This will be referred to as the question of (pseudo-)Riemann metrizability or just metrizability. Simply put: is any Berwald space metrizable? For positive definite Finsler spaces that are smooth on the entire slit tangent bundle the answer, due to Szab\'o \cite{Szabo}, is well-known and affirmative: the affine connection on any such Berwald space is the Levi-Civita connection of some Riemannian metric. However, in alternative signatures like Lorentzian signature, this question remains almost completely unexplored. We demonstrate here that Szab\'o's metrization theorem does not extend to the general setting (\cref{ch:Metrizability}). In particular, it does not hold in Lorentzian signature. For the class of \mkrop metrics, we analyze the question of metrizability in detail and we obtain precise necessary and sufficient conditions for local metrizability. We also classify all Ricci-flat, locally metrizable \mkrop spaces. Since any Ricci-flat Berwald space is automatically an exact solution to Pfeifer and Wohlfarth's field equation, this classification has important implications for Finsler gravity. 




And that leads us naturally into \cref{part:vacuum_sols}, which is devoted to the study of exact solutions to Pfeifer and Wohlfarth's field equation for Finsler gravity. Any solution to Einstein's field equation is trivially also a solution in Finsler gravity, but not many exact, properly Finslerian solutions are known as of yet. Prior to the contributions on which this dissertation is based, the only ones known in the literature were the ($m$-Kropina type) Finsler pp-waves \cite{Fuster:2015tua} and their generalization as \textit{very general relativity} (VGR)  spacetimes \cite{Fuster:2018djw}. Here we will extend the list considerably.

Most of the solutions that we present and investigate are of Berwald type (\cref{ch:Berwald_solutions}). First, we introduce a large class of general Berwald \ab-type solutions; we prove that if the \psR metric $\alpha$ is chosen to be a general relativistic pp-wave and $\beta$ the corresponding canonical covariantly constant null 1-form, then 
\textit{any} Finsler metric constructed from these two building blocks will be an exact vacuum solution. This yields a wide range of Finsler metrics that generalize the well-known pp-waves from GR as well as the (m-Kropina type) Finsler pp-waves obtained in \cite{Fuster:2015tua}.

\sloppy 

For Randers metrics---the most abundant type of \ab-metric---we prove that Pfeifer and Wohlfarth's field equation is in fact equivalent to Rutz's equation, i.e. the vanishing of the Finsler-Ricci tensor. Using this fact, we (locally) classify all exact vacuum solutions of Berwald-Randers type. 
Completely analogous results hold for what we call the \textit{modified} Randers metric---a small modification of the standard Randers metric that we introduce because of its satisfactory causal properties. 

A natural question that arises is how such spacetimes should be interpreted physically, and in particular, whether and how they can be physically distinguished from their general relativistic counterparts, given by the pseudo-Riemannian metric $\alpha$. To answer this question, we apply a two-fold linearization scheme to our class of \ab-metric solutions. We find that the linearized solutions may be interpreted as Finslerian gravitational waves, and we study their observational signature (\cref{ch:lin}). More precisely, we ask the question of what would be observed in a gravitational wave interferometer when such a Finslerian gravitational wave passes the earth, and what would be the difference with a classical general relativistic gravitational wave. To this end, we compute the Finslerian radar distance---the main observable measured by interferometers. Remarkably, when interpreted correctly, the result turns out to be completely equivalent to its GR counterpart \cite{Rakhmanov_2009}. In other words, gravitational wave interferometers simply do not have the ability to distinguish this type of Finsler gravitational waves from the standard ones in GR. We discuss the implications of this result.




Finally, in the last chapter, we present a class of exact solutions of unicorn type (\cref{ch:unicorn_cosm}). Above, we already introduced Berwald spaces, roughly speaking, as those Finsler spaces that are only `slightly' more general than \psR manifolds. One might say that by going up one level in generality, one arrives at the class of Landsberg spaces. Every Berwald space is also Landsberg, but whether or not the opposite is true has been a long-standing open question in the field, provided one adheres to the most strict definition of a Finsler space.  Bao has called these non-Berwaldian Landsberg spaces \textit{`[\dots] unicorns, by analogy with those mythical single-horned horse-like creatures for which no confirmed sighting is available,'} \cite{Bao_unicorns} and Matsumoto stated in 2003 that they represent the next frontier of Finsler geometry \cite{Bao_unicorns}. In the last two decades, some examples of unicorns have been obtained by Asanov \cite{asanov_unicorns}, Shen \cite{shen_unicorns}, and  Elgendi \cite{Elgendi2021a} by slightly relaxing the definition of a Finsler space. Here we present a new exact vacuum solution to Pfeifer and Wohlfarth's field equation which is a unicorn that falls into one of the classes introduced by Elgendi. This solution has cosmological symmetry, i.e. is spatially homogeneous and isotropic, and it is additionally conformally flat, with conformal factor depending only on the timelike coordinate. We show that, just as in classical Friedmann-Lema\^itre-Robertson-Walker (FLRW) cosmology, this conformal factor can be interpreted as the scale factor of the universe; we compute this scale factor as a function of cosmological time, and find that it corresponds to a linearly expanding (or contracting) Finslerian universe.





\newpage

\section*{Organization of this dissertation}

This dissertation is organized into three parts. The following is a concise summary of their contents.

\begin{itemize}
    \item In \cref{part:preliminaries} we introduce the necessary mathematical preliminaries (\cref{ch:finsler_fund}-\ref{ch:BerwaldLandsbergUnicorn}). 
    \item In \cref{part:Berwald_metrizability} we present our results pertaining to the characterization of Berwald spaces and their metrizability. 
    \begin{itemize}
        \item \cref{ch:Characterization_of_Berwald} covers the characterization of Berwald spaces. 
        \item \cref{ch:Metrizability} covers metrizability. 
    \end{itemize}
    \item In \cref{part:vacuum_sols} we present our results concerning exact solutions to Pfeifer and Wohlfarth's field equation in vacuum. 
    \begin{itemize}
        \item \cref{ch:Finsler_gravity} is an introduction to Finsler gravity and reviews the field equation, some of its properties, and its motivation and derivation.
        \item \cref{ch:Berwald_solutions} contains our exact solutions that are of Berwald type.
        \item \cref{ch:lin} is concerned with the physical interpretation and observational signature of the solutions obtained in \cref{ch:Berwald_solutions}. 
        \item \cref{ch:unicorn_cosm} exhibits our cosmological unicorn solutions.
    \end{itemize}
\end{itemize}

\noindent Moreover, each part has an accompanying introduction that is somewhat more specific than the general introduction given above.


\newpage

%% file: intr1.tex
\chapter*{Introduction to \cref{part:preliminaries}: Preliminaries}

The purpose of \cref{part:preliminaries} is to develop the mathematical prerequisites that are necessary to properly understand \cref{part:Berwald_metrizability} and \cref{part:vacuum_sols}. We assume familiarity with \psR geometry (see e.g. \cite{lee2018introduction,o1983semi,sasane2021mathematical}) and standard concepts in differential geometry such as connections on vector bundles (see e.g. \cite{tu2017differential,LoringWTu_Manifolds,lee2012smoothmanifolds}), but we provide a reasonably self-contained exposition of Finsler geometry, starting with the fundamentals in \cref{ch:finsler_fund}. Here we introduce the definition of a Finsler space, of geodesics, and we introduce the pullback bundle $\pi^*TM$ and its associated tensor bundles, leading to the notion of a Finsler tensor field. 

We then go on to review the topic of nonlinear (Ehresmann) connections in \cref{ch:nonlin_conns}, we introduce the associated notions of curvature, torsion, and metric-compatibility, and we prove the existence and uniqueness of a torsion-free metric-compatible homogeneous nonlinear connection associated with a Finsler metric, generalizing the Levi-Civita connection. We then introduce the Finsler-Ricci curvature and the dynamical covariant derivative, which both appear in the Finsler gravity field equations. 

In \cref{ch:lin_conns} we review the theory of linear connections on the pullback bundle $\pi^*TM$ and in particular their torsion and metric-compatibility, we introduce the Chern-Rund and Berwald connections associated with a Finsler metric, and we discuss geodesic deviation. The latter will be used as motivation for the field equations. 

Finally, \cref{ch:BerwaldLandsbergUnicorn} discusses Berwald spaces, Landsberg spaces, and unicorns, i.e. non-Berwaldian Landsberg spaces. These notions are very relevant as \cref{part:Berwald_metrizability} is centered around Berwald spaces, and many of the solutions to the Finsler gravity field equations discussed in \cref{part:vacuum_sols} are of this type. The only exception is our cosmological unicorn solution which, as the name suggests, is of unicorn type.

Our exposition of the material is based on a wide range of literature. For further reference, we refer to \cite{Bao, Szilasi, Bucataru, shen2001lectures, Theory_Of_Sprays,handbook_Finsler_vol1,handbook_Finsler_vol2,shen2013differential,shen2016introduction,BaoRobles2004,Minguzzi:2014fxa,Anastasiei2004,Anastasiei1996,Szilasi2011}.

%% file: ch_Fundamentals_of_Finsler_geometry.tex
\chapter{Fundamentals of Finsler Geometry}\label{ch:finsler_fund}

After briefly outlining some of our notational conventions, we start by introducing Finsler metrics, Finsler Lagrangians, Finsler spaces, and notions such as the Finslerian length of a curve, the fundamental tensor, the Cartan tensor, and Euler's theorem and some of its immediate consequences. We then introduce geodesics as critical points of some `energy' functional. Although an alternative equivalent definition can be given in terms of autoparallel (`straight') curves of some connection, which is perhaps more intuitive\footnote{Mathematically, this is merely a matter of taste. In the context of relativity, on the other hand, the definition in terms of autoparallel curves is the physically relevant one. This is because gravity is understood \textit{not} to be a force and hence the trajectories of freely falling test particles should be the straightest possible lines through the curved spacetime geometry, i.e. autoparallels, i.e. geodesics.}, the one given here can be stated without any reference to a connection and hence it will be our starting point. The equivalence with autoparallel curves will be discussed in detail in the next two chapters. Finally, we introduce the pullback bundle $\pi^*TM$, its associated tensor bundles, and the notion of a Finsler tensor field.

\section{Notation and conventions}
Not only in this chapter, but throughout the dissertation we will often work in local coordinates. Given a smooth manifold $M$ of dimension $n$ we will always assume that some chart $\varphi:U\subset M\to \R^n$ is provided, and we will effectively identify any $p\in U$ with its image $x = (x^1,\dots,x^n)  \coloneqq\varphi(p)\in\R^n$ under $\varphi$. If $p\in U$ then each $X_p\in T_pM$ can be written as $X_p = y^i\partial_i\big|_p \eqqcolon y^i\partial_i$ (we will often be sloppy and suppress the dependence on the point $p$), where the tangent vectors $\partial_i \coloneqq \frac{\partial}{\partial x_i}$  make up the coordinate basis of $T_pM$. This decomposition provides natural local coordinates on the tangent bundle $TM$ via the chart
\begin{align}
\tilde\varphi: TU \to \R^n\times\R^n,\qquad \tilde\varphi(p,Y) = (\varphi(p),y^1,\dots,y^n)\eqqcolon (x,y),
\end{align}
where $TU$ is given by
\begin{align}
    TU = \bigcup_{p\in U} \left\{p\right\}\times T_p M\subset TM.
\end{align}
These local coordinates on $TM$ in turn provide a natural basis $\{\partial_i,\bar\partial_i\}_{i=1,\dots,n}$ of its tangent space $T_{(x,y)}TM$ at $(x,y)$, where we define
\begin{align}
\partial_i \coloneqq \frac{\partial}{\partial x^i}, \qquad\quad  \bar{\partial}_i\coloneqq \frac{\partial}{\partial y^i}.
\end{align}
The space of smooth functions on $M$ will be denoted by $C^\infty(M)$ and the space of smooth vector fields on $M$ by $\mathfrak{X}(M)$. Given a vector bundle $\pi:E\to M$ over $M$, we denote its space of smooth sections by $\Gamma(E)$.

Finally, we will use the notation $\D x^i\D x^j$ for the symmetrized tensor product of 1-forms, i.e. $\D x^i\D x^j \equiv \tfrac{1}{2}(\D x^j\otimes \D x^j + \D x^j \otimes \D x^j)$, and whenever we work in Lorentzian signature we will adhere to the sign convention $(-,+,\dots,+)$ unless otherwise specified.

\section{Finsler spaces}
\label{sec:FinslerSpaces}

Let $M$ be a smooth manifold. A conic subbundle of $TM$ is an open subset $\mathcal A\subset TM$ which is conic in the sense that $(x,\lambda y)\in\mathcal A$ for any $(x,y)\in\mathcal A$ and $\lambda>0$, and which satisfies $\pi(\mathcal A) = M$, where $\pi:TM\to M$ is the canonical projection of the tangent bundle. The latter property says that the fibers $\mathcal A_{x}$ of $\mathcal A$ are nonempty. 
\begin{defi}\label{def:Finsler_Lagrangian}
A Finsler Lagrangian on a conic subbundle $\mathcal A$ is a smooth map $L:\mathcal A\to \R$ such that
\begin{itemize}
	\item $F$ is positively homogeneous of degree two:
	\begin{align}
	L(x,\lambda y) =\lambda^2 L(x, y)\,,\quad \forall \lambda>0\,;
	\end{align}
	\item The \textit{fundamental tensor}, with components $g_{ij} = \db_i\db_j \left(\frac{1}{2}L\right)$, is nondegenerate.
\end{itemize}
\end{defi} 
In \cref{sec:Finsler_tensor_fields} we will see that $g_{ij}$ is indeed a (Finsler) tensor field.

Any \psR metric $a = a_{ij}\D x^i \D x^j$ on $M$ induces a quadratic Finsler Lagrangian $L= a_{ij}(x)y^i y^j$ on $\mathcal A = TM$ with fundamental tensor $g_{ij}=a_{ij}$. We say that such a Finsler Lagrangian $L$ is pseudo-Riemannian. In fact, it is easy to see that a Finsler Lagrangian is \psR if and only if (iff) it is quadratic (in $y$) iff its fundamental tensor has no directional dependence, $g_{ij} = g_{ij}(x)$. In such a situation the theory reduces to pseudo-Riemannian geometry. By extension, we will also say that $L$ is \psR if there exists a finite cover $\{\mathcal A_i\}_i$ of $\mathcal A$ such that $L$ is quadratic on each $\mathcal A_i$. For example, a Lagrangian such as $L = |a_{ij}y^iy ^j|$ on $\mathcal A = \{(x,y)\in TM\,:\, a_{ij}(x)y^iy ^j\neq 0\}$ will also be called pseudo-Riemannian. If no such cover exists, then we say that $L$ is \textit{properly Finslerian}. It turns out that any Finsler Lagrangian defined on $\mathcal A = TM$ must be pseudo-Riemannian. In the properly Finslerian case, the largest possible domain one can hope for is given by $\mathcal A = TM_0$, the slit tangent bundle,
\begin{align}
    TM_0 \coloneqq \{(x,y)\in TM\,:\, y\neq 0\}.
\end{align}
Even though it is the 2-homogeneous Finsler Lagrangian that enters in most (but not all) formulas in Finsler geometry, it is useful to independently define a similar 1-homogeneous object, the Finsler metric. 
\begin{defi}\label{def:Finsler_metric}
A Finsler metric on a conic subbundle $\mathcal A$ is a smooth map $F:\mathcal A\to \R$ such that
\begin{itemize}
	\item $F$ is positively homogeneous of degree one:
	\begin{align}\label{eq:F_homogeneity}
	F(x,\lambda y) =\lambda F(x, y)\,,\quad \forall \lambda>0\,;
	\end{align}
	\item The \textit{fundamental tensor}, with components $g_{ij} = \db_i\db_j \left(\frac{1}{2}F^2\right)$, is nondegenerate.
\end{itemize}
\end{defi} 
The length of a curve $\gamma: (a,b) \to M$ is then defined as\footnote{Strictly speaking the length of $\gamma$ is defined only when $\dot \gamma(\lambda)\in\mathcal A$ for all $a<\lambda<b$.}
\begin{align}\label{eq:lengt_functional}
\ell(\gamma)=\int_a^b  \left|F(\gamma(\lambda),\dot{\gamma}(\lambda))\right|\,\D \lambda, \qquad \qquad\dot{\gamma}=\frac{\D\gamma}{\D\lambda}.
\end{align}
The homogeneity condition \eqref{eq:F_homogeneity} ensures that this length is invariant under (orientation-preserving) reparameterization. 

Note that the two definitions \ref{def:Finsler_Lagrangian} and \ref{def:Finsler_metric} are not equivalent. Any Finsler metric has a canonical Finsler Lagrangian $L=F^2$ associated with it and, conversely, from any Finsler Lagrangian on $\mathcal A$ one may define a canonical Finsler metric $F = \sqrt{|L|}$ on $\mathcal A\cap\{L\neq 0\}$. However, both of these correspondences $F\leftrightarrow L$ are neither injective nor surjective. That said, it is fair to say that \Cref{def:Finsler_Lagrangian} is the more general definition of the two, as in most applications one only cares about $L$ and not (the sign of) $F$. If so, one could choose to restrict the analysis to Finsler metrics that satisfy $F\geq 0$, which is typically done. In that case the correspondence $F\mapsto L=F^2$ \textit{is} injective and with this identification, nonnegative Finsler metrics form a subclass of Finsler Lagrangians. Nevertheless, we will see several times throughout this dissertation that it can be useful to allow $F$ to vary in sign, which is why we incorporate \Cref{def:Finsler_metric} as well.
\begin{defi}\label{def:Finsler_manifold}
    A Finsler space or Finsler manifold is a triple $(M,\mathcal A,L)$ or $(M,\mathcal A,F)$.
\end{defi}
We will sometimes omit the specification of $\mathcal A$ and say, loosely speaking, that $F$ is a Finsler metric on $M$ and $L$ a Finsler Lagrangian on $M$. We say that $F$ or $L$ is positive definite if $g_{ij}$ is positive definite, and more generally that $F$ or $L$ has a certain signature if $g_{ij}$ has that signature. 
We remark that in the positive definite setting, a Finsler space is most often defined as a triple $(M,\mathcal A, F)$, where $\mathcal A = TM_0$ and $F$ is a Finsler metric in the sense of \Cref{def:Finsler_metric} with the additional requirement that $F>0$ and that $g_{ij}$ be positive definite. In other signatures such as Lorentzian signature, however, many important examples have a domain $\mathcal A$ which is strictly smaller and cannot be extended to $TM_0$.

Next, we recall a classical result, known as Euler's theorem for homogeneous functions (stated in a somewhat atypical way). A smooth function $f:U\to\R$ on an open set $U\subset \R^n$ is said to be (positively) homogeneous of degree $k$ or simply $k$-homogeneous if $f(\lambda y) =\lambda^k f(y)$ for all $\lambda>0$. If $f$ is a function on a conic subbundle $\mathcal A\subset TM$ then, as a matter of speaking, we will say that $f$ is $k$-homogeneous if it is $k$-homogeneous in its $y$-dependence.
\begin{theor}[Euler's Theorem \cite{Bao}]
    Let $f:\mathcal A\to \R$ be a smooth, $k$-homogeneous function on a conic subbundle $\mathcal A\subset TM$. Then
    \begin{align}
        y^i\bar\partial_i f(x,y) = k f(x,y).
    \end{align}
\end{theor}
This result has a central importance in Finsler geometry. For example, since the Finsler Lagrangian $L$ is 2-homogeneous, Euler's theorem implies that it can be written as
\begin{align}\label{eq:Fsquared_in_terms_of_g}
L=g_{ij}y^i y^j.
\end{align}
The Cartan tensor (or Cartan torsion) is defined by its components
\begin{align}\label{eq:cartan_tensor}
    C_{ijk} &= \tfrac{1}{2}\bar\partial_i g_{jk} = \tfrac{1}{4}\bar\partial_i\bar\partial_j\bar\partial_k L.
\end{align}
Note that $C_{ijk}$ is completely symmetric. As a second application of Euler's theorem, the fact that $g_{ij}$ is 0-homogeneous implies that
\begin{align}
    y^iC_{ijk} = y^jC_{ijk} = y^kC_{ijk} = 0. \label{eq:Cartan_contraction_identity}
\end{align}    
The mean Cartan tensor (or torsion) is the trace of the Cartan tensor, 
\begin{align}
    C_{k} = g^{ij}C_{ijk}.
\end{align}
Rather than taking the Finsler metric $F$ as the fundamental object, one could equivalently start from the fundamental tensor, but then one needs an additional axiom to guarantee that $g_{ij}$ is the Hessian of some scalar function, namely the constraint that the Cartan tensor associated with $g_{ij}$ be fully symmetric. More precisely (see e.g. \cite{BaoRobles2004}) we have the following. \\

\textit{A symmetric and nondegenerate collection of smooth, $0$-homogeneous,  functions $g_{ij}:\mathcal A\to \R$ on a conic subbundle $\mathcal A$ makes up the fundamental tensor of a Finsler Lagrangian if and only if the associated Cartan tensor is completely symmetric.}

\section{Geodesics}\label{sec:geodesics_0}

The simplest way to define the notion of geodesics directly from the Finsler function or Lagrangian is via a variational approach. Let $\gamma:[a,b]\to M$ be a curve in a Finsler space such that $\dot\gamma(\lambda)\in \mathcal A$ for all $\lambda$. Then we say that $\gamma$ is a geodesic if it is a critical point of the `energy' functional 
\begin{align}\label{eq:energy_functional}
E[\gamma]=\int_a^b  L(\gamma(\lambda),\dot{\gamma}(\lambda))\,\D \lambda,
\end{align}
which is the case iff it is a solution to the Euler-Lagrange equations
\begin{align}\label{eq:E.L.}
    \frac{\partial L}{\partial x^i} - \frac{\D}{\D t}\frac{\partial L}{\partial y^i}=0,
\end{align}
evaluated along $(x,y) = (\gamma(t), \dot\gamma(t))$. %
%
%
%
Using \eqref{eq:Cartan_contraction_identity} it is straightforward to show that Equation \eqref{eq:E.L.} is equivalent to the \textit{geodesic equation},
\begin{align}\label{eq:geodesic_equation}
    \ddot \gamma^k + G^k(\gamma,\dot\gamma) =0,
\end{align}
where the \textit{geodesic spray coefficients}
\begin{align}\label{eq:spray_original_def_from_formal_christ}
    G^k(x,y) = \gamma^k_{ij}(x,y)y^i y^j
\end{align}
are defined in terms of the so-called \textit{formal Christoffel symbols}\footnote{It is a matter of coincidence that both the formal Christoffel symbols and the curve under consideration are represented by the symbol $\gamma$ here. The formal Christoffel symbols are defined independently of any curve.}
\begin{align}
    \gamma^k_{ij} = \tfrac{1}{2}g^{k\ell}\left(\partial_i g_{\ell j} + \partial_j g_{i\ell} - \partial_\ell g_{ij}\right).
\end{align}
We will also refer to $G^k$ simply as the geodesic spray. Note the resemblance to the Christoffel symbols of the Levi-Civita connection in Riemannian geometry. The term `formal' refers to the fact that the $\gamma^k_{ij}$ are not really the Christoffel symbols of a well-defined connection on $TM$, except when $L$ is pseudo-Riemannian, in which case they are just the Christoffel symbols of the Levi-Civita connection. In \Cref{ch:nonlin_conns} we will see that geodesics can also be characterized as autoparallel curves of a canonical connection that generalizes the Levi-Civita connection. We also note that geodesics are critical points of the length functional \eqref{eq:lengt_functional} as well.


\section{Finsler tensor fields}\label{sec:Finsler_tensor_fields}

Apart from tensor fields on $M$ and tensor fields on $\mathcal A\subset TM$ in the standard sense, many of the geometric objects of interest in Finsler geometry are tensor fields on $M$ `with dependence on both position and direction'. Examples we have seen so far are the fundamental tensor $g_{ij}$ and the Cartan tensor $C_{ijk}$. For a careful treatment of such tensors, it is useful to introduce the so-called pullback bundle $\pi^*TM$. A direction-dependent vector field can then be defined as a section of this bundle, and a direction-dependent tensor field as a section of an associated tensor bundle. We will call such objects \textit{Finsler tensor fields}. This formulation in terms of the pullback bundle will also allow us to easily define covariant derivatives of such tensor fields; both linear covariant derivatives (\Cref{ch:lin_conns}) as well as the so-called dynamical covariant derivative arising from the canonical nonlinear connection on a Finsler space (\Cref{ch:nonlin_conns}). 

\subsection{The pullback bundle \texorpdfstring{$\pi^*TM$}{pi*TM}}\label{sec:pullback_bundle}

Given any vector bundle $\pi:E\to M$ over a smooth manifold $M$ and a smooth map $f:N\to M$, where $N$ is another smooth manifold, one can define the pullback bundle $f^*E$, which is a smooth vector bundle over $N$. More precisely, the pullback bundle is the vector bundle $\pi' :f^{*}E\to N$, where the total space is defined as 
\begin{align}
    f^{*}E = \{(n,e) \in N \times E \mid f(n) = \pi(e)\}\subset N\times E,
\end{align}
and the projection is defined as $\pi'(n,e) = n$. 

Consider now the special case where $E$ is the tangent bundle, $\pi:TM\to M$ the canonical projection, and  $f$ its restriction $f=\pi|_{\mathcal A}:\mathcal{A}\to M$ to some conic subbundle $\mathcal A\subset TM$. Then the construction yields a vector bundle on $\mathcal A$, denoted by $\pi^*TM$, and to which we will refer as \textit{the} pullback bundle.

\subsection{Vector fields and tensor fields}

As already alluded to above, we shall adopt the following definition.
\begin{defi}
    A Finsler vector field (or anisotropic vector field) is a section $X\in \Gamma(\pi^*TM)$ of the pullback bundle.
\end{defi}
Let us check that this definition agrees with our intuitive notion of a Finsler vector field as a vector field with a dependence on both position and direction. By construction, the fiber $\pi^*TM_{(x,y)}$ at $(x,y)$ is canonically isomorphic to the vector space $T_xM$, so we may think of $\pi^*TM$ as the vector bundle constructed by erecting a copy of $T_xM$, the fiber, at each point $(x,y)\in \mathcal A$. Given a vector field $X$ on $M$, one can define a corresponding Finsler vector field $\hat X$ by setting $\hat X_{(x,y)} = X_x$, using the identification discussed above. If we take for $X$ the local coordinate basis vectors $\partial_i$, then the corresponding Finsler vector fields $\hat \partial_i$ form a basis for the fibers $\pi^*TM_{(x,y)}$. By some abuse of notation, we will drop the hat and denote these simply by $\partial_i$ as well. Finsler vector fields $X\in \Gamma(\pi^*TM)$ can therefore be represented locally as $X = X^i(x,y)\partial_i$, justifying their definition as vector fields with a dependence on both position and direction. As usual, $X$ is smooth if and only if its component functions $X^i(x,y)$ are smooth. 

From here the route to defining general Finsler tensor fields is straightforward.
Skipping some of the technical details (which can be found in e.g. \cite{Szilasi}), the idea is to define the dual bundle $(\pi^*TM)^*$---which is in fact isomorphic to the pullback 
of the cotangent bundle $T^*M$ along the same map $f=\pi|_{\mathcal A}:\mathcal A\to M$ that we used earlier---and consider tensor products of $\pi^*TM$ and $(\pi^*TM)^*$. 
\begin{defi}
    A Finsler tensor field (or anisotropic tensor field) of type $(k,l)$ is a section of the tensor bundle
\begin{align}
    \underbrace{\pi^*TM\otimes\dots\otimes \pi^*TM}_{k \text{ times}} \otimes \underbrace{(\pi^*TM)^*\otimes\dots\otimes (\pi^*TM)^*}_{l \text{ times}}.
\end{align}
    The space of smooth Finsler $(k,l)$-tensor fields will be denoted by $\mathcal T^k_l(\pi^*TM)$.
\end{defi}
Equivalently, a Finsler $(k,l)$-tensor field $T$ may be regarded as a $C^\infty(\mathcal A)$-multilinear map 
\begin{align}
    T:\underbrace{(\pi^*TM)^*\times\dots\times(\pi^*TM)^*}_{k \text{ times}} \times \underbrace{\pi^*TM\times\dots\times \pi^*TM}_{l \text{ times}}\to C^\infty(\mathcal A),
\end{align}
 and locally it can be expressed as
\begin{align}
    T = T^{i_1\dots i_ k}{}_{j_1\dots j_l} \partial_{i_1}\otimes\dots \otimes\partial_{i_k}\otimes\D x^{j_1}\otimes\dots\otimes \D x^{j_l},
\end{align}
where the Finsler 1-forms $\D x^i$ are defined in a way analogous to the Finsler vector fields $\partial_i$ and where the components $T^{i_1\dots i_ k}{}_{j_1\dots j_l}(x,y)$ are functions on $\mathcal A$ that behave in the standard way under coordinate transformations, namely
\begin{align}
    \tilde T^{i_1\dots i_ k}{}_{j_1\dots j_l} = \frac{\partial \tilde x^{i_1}}{\partial x^{m_1}}\dots \frac{\partial \tilde x^{i_k}}{\partial  x^{m_k}} \frac{\partial  x^{n_1}}{\partial \tilde x^{j_1}}\dots \frac{\partial  x^{n_l}}{\partial \tilde x^{j_l}} T^{m_1\dots m_ k}{}_{n_1\dots n_l}.
\end{align}
Finally, by convention,  we define a Finsler $(0,0)$-tensor field to be just a function $f:\mathcal A\to \R$. Since we will be dealing with Finsler tensor fields a lot, we will often refer to them simply as tensor fields, and similarly for Finsler vector fields, etc. To distinguish Finsler tensor fields from conventional tensor fields on $M$, we will sometimes refer to the latter as \textit{classical} tensor fields in case the distinction is important. Under the relevant identifications, a classical tensor field may be regarded as a Finsler tensor field whose components depend only on $x$ and not on $y$.



%% file: ch_Connections.tex
\chapter{Nonlinear Connections on \texorpdfstring{$\mathcal A\subset TM$}{A in TM}}\label{ch:nonlin_conns}

Nonlinear connections are an essential ingredient in the study of Finsler geometry, as they allow for a generalization of the Levi-Civita connection to Finsler spaces. The name is somewhat misleading, though, as it is perfectly possible for a nonlinear connection to be linear. It would be more accurate to call them \textit{not-necessarily-linear connections}, but we'll stick to the standard terminology.  Below we start by briefly recalling the definition of a nonlinear connection in terms of horizontal distributions and the corresponding notion of parallel transport. Then we discuss the notions of curvature, torsion, and compatibility with a Finsler metric, and we prove that, given any Finsler metric there exists a unique homogeneous, torsion-free, and metric-compatible nonlinear connection. We end the chapter by introducing the dynamical covariant derivative on Finsler tensor fields.

\section{Horizontal distributions and parallel transport}
\subsection{Horizontal distributions}
A nonlinear (or Ehresmann) connection on a conic subbundle $\mathcal A \subset TM$, later to be identified as the domain of a given Finsler metric, is a smooth decomposition of the tangent bundle $T\mathcal A$ into a horizontal and a vertical subbundle\footnote{For alternative equivalent definitions, see \cite{javaloyes2023einsteinhilbertpalatini} and references therein.},
\begin{align}
T\mathcal A = H\mathcal A \oplus V\mathcal A,
\end{align}
where $\oplus$ denotes the Whitney sum of vector bundles. 
The vertical subbundle $V\mathcal A$ is canonically defined on any smooth manifold as
\begin{align}
V\mathcal A = \ker (\D\pi) =\text{span}\left\{\bar\partial_i\right\},
\end{align}
in terms of the vertical coordinate vector fields (or vertical derivatives) $\bar\partial_i$, but in general, there is no preferred choice of the horizontal subbundle. A  nonlinear connection is thus simply a specification of this horizontal subbundle $H\mathcal A$. In a given chart, the latter can be represented as
\begin{align}
H\mathcal A = \text{span}\left\{\delta_i\right\},\qquad \delta_i\equiv \partial_i - N^j{}_i\db_j,
\end{align}
for a collection of smooth functions $N^i_j(x,y)$, $\,i,j=1,\dots,n=\dim M$ on $\mathcal A$ (or, strictly speaking, rather on the induced chart on $\mathcal A$), known as the \textit{connection coefficients}. The $\delta_i$ are called horizontal coordinate vector fields\footnote{Note that, strictly speaking, the $\delta_i$ are not themselves coordinate vector fields on $TM$ in the usual sense of the word (this is prohibited by the Frobenius theorem unless the curvature of the connection vanishes). The terminology is nevertheless natural here, though, as the $\delta_i$ are the horizontal lifts of the coordinate vector fields $\partial_i$ on $M$.} or horizontal derivatives. The connection coefficients have the following transformation behavior under coordinate transformations on $M$:
\begin{align}\label{eq:nonlin_con_transf}
N^k{}_j = \tilde N^l{}_i\,\frac{\partial x^k}{\partial\tilde x^l}\,\frac{\partial \tilde x^i}{\partial x^j} + \frac{\partial x^k}{\partial\tilde x^l}  \, \frac{\partial^2 \tilde x^l}{\partial x^i \partial x^j}\, \tilde y^i.
\end{align}
Conversely, any choice of smooth local functions $N^i_j(x,y)$ satisfying \eqref{eq:nonlin_con_transf} defines a nonlinear connection.

Once a connection has been specified one can introduce the notions of horizontal and vertical vector fields. These are special types of vector fields on $\mathcal A$. We say that a smooth vector field $X\in \mathfrak X(\mathcal A)$ is \textit{horizontal} and write $X\in \mathfrak X^h(\mathcal A)$ if $X_{(x,y)}\in H_{(x,y)}\mathcal A$ for all $(x,y)\in \mathcal A$. Similarly, we say that a smooth vector field $X\in \mathfrak X(\mathcal A)$ is \textit{vertical} and write $X\in \mathfrak X^v(\mathcal A)$ if $X_{(x,y)}\in V_{(x,y)}\mathcal A$ for all $(x,y)\in \mathcal A$.

\subsection{Parallel transport}

A nonlinear connection provides a very general means of defining parallel transport of vectors along curves and it leads in particular to the notion of autoparallel (i.e. `straight') curves. Parallel transport of a vector field $V$ along a curve $\gamma$ on $M$ is characterized\footnote{Here we follow e.g. \cite{Szilasi}. We note that some authors (see e.g. \cite{Bucataru}) choose to define parallel transport in a different way, namely by requiring \textit{a priori} that parallel transport should be linear, which then leads to the alternative parallel transport equation $\dot V^i + N^i_j(\gamma,\dot\gamma) V^j=0$, where $\dot \gamma$ and $V$ are interchanged in the second term with respect to \eqref{eq:nonlinear.parallel.transport.eq}. %
In certain specific situations, this description has its advantages, but in general, it seems less natural to us.} by the requirement that the rate of change $\dot V$ along $\gamma$ be horizontal, i.e. $\dot V\in H\mathcal A$. Here we identify $V$ with the curve $t\mapsto(\gamma(t),V(\gamma(t)))$ in $\mathcal A\subset TM$ and hence in general we have
\begin{align}
\dot V \equiv \tfrac{\D}{\D t} V(\gamma(t)) = (\dot\gamma^1,\dots \dot\gamma^n, \dot V^1,\dots,\dot V^n) = \dot\gamma^i\partial_i\big|_{V} + \dot V^i\bar\partial_i\big|_{V}\in TTM.
\end{align} 
We can decompose this as
\begin{align}
\dot V(t) 
= \dot\gamma^i\delta_i\big|_{V} + \left(N^i_j(\gamma,V) \dot\gamma^j +\dot V^i\right)\bar\partial_i\big|_{V},
\end{align}
in terms of horizontal and vertical coordinate vector fields and hence $V$ is parallel transported along $\gamma$ iff it satisfies the \textit{parallel transport equation},
\begin{align}
\label{eq:nonlinear.parallel.transport.eq}
\dot V^i + N^i_j(\gamma,V)\dot \gamma^j = 0\,.
\end{align}
Notice that parallel transport can be defined only as long as $V(t)$ stays in $\mathcal A$.
%
%
An autoparallel curve, or just autoparallel, is a curve whose tangent vector field is parallel transported along the curve itself. Hence, setting $V = \dot\gamma$, a curve $\gamma$ is an autoparallel if and only if it satisfies the autoparallel equation,
\begin{align}
\label{eq:nonlinear.geodesic.eq}
\ddot \gamma^i + N^i_j(\gamma,\dot\gamma)\dot \gamma^j = 0\,.
\end{align}
%
Assuming for the moment that $\mathcal A = TM$ it is easy to see from \eqref{eq:nonlinear.parallel.transport.eq} that parallel transport is linear in $V$ if and only if $N^i_j$ is linear in $V$. Therefore we say that a connection is linear if its connection components are linear in the tangent space coordinates, i.e. $N^k_i(x,y)=\Gamma^k_{ij}(x)y^j$. Even if $\mathcal A$ is not a linear space, the condition $N^k_i(x,y)=\Gamma^k_{ij}(x)y^j$ for all $(x,y)\in\mathcal A$ allows one to canonically extend the nonlinear connection to $\mathcal A = TM$, so it still makes sense in this case to say the connection is linear, hence we will do so. If the connection is linear then, as a result of \eqref{eq:nonlin_con_transf}, the functions $\Gamma^k_{ij}$ are the Christoffel symbols\footnote{Sometimes the term \textit{Christoffel symbols} is reserved for the connection coefficients of the Levi-Civita connection. For us, the term will apply to the connection coefficients of any linear connection, to distinguish it from the \textit{connection coefficients} of a nonlinear connection.} of a linear (Koszul) connection on $TM$, i.e. an affine connection on $M$, in the standard sense. Thus, essentially, a linear nonlinear connection (excuse the terminology) is nothing but a linear Koszul connection and we will indeed identify the two whenever no confusion is possible. In such a case, \eqref{eq:nonlinear.geodesic.eq} reduces to the familiar autoparallel equation of the corresponding Koszul connection,
\begin{align}
\label{eq:linear.geodesic.eq}
\ddot \gamma^i + \Gamma^i_{jk}(\gamma)\dot \gamma^j\dot \gamma^k = 0\,.
\end{align}
Similarly, we say that a nonlinear connection is positively homogeneous of degree $1$, or simply \textit{homogeneous}, if $N^k_i(x,\lambda y) = \lambda N^k_i(x,y)$ for all $\lambda>0$. This definition makes sense since $\mathcal A$ is conic, by definition.

\section{Curvature, torsion and metric compatibility}

It is the goal of this section to introduce the notions of curvature, torsion, and metric-compatibility for nonlinear connections on $\mathcal A\subset TM$. To this end, we first recall the definitions of the horizontal and vertical lift of a vector field. To any (classical) vector field on the base manifold $M$, we can assign both a horizontal and a vertical (classical) vector field on $\mathcal A$. The horizontal lift $X^h$ of a vector field $X\in\mathfrak X(M)$ is the horizontal vector field $X^h\in \mathfrak X(\mathcal A)$ defined by linear extension of $\partial_i\mapsto \delta_i$, i.e. if $X = X^i\partial_i$ then
\begin{align}
     X^h = X^i\delta_i,
\end{align}
where the components $X^i$ are now interpreted as functions of $x$ and $y$; technically we should write $X^h = (X^i\circ\pi)\delta_i$ but we will usually omit the projection $\pi$.
Similarly, the vertical lift $X^v\in \mathfrak X(\mathcal A)$ of $X$ is defined by linear extension of $\partial_i\mapsto \bar\partial_i$, i.e.
\begin{align}
     X^v = X^i\bar\partial_i.
\end{align}
%

\subsection{Curvature and torsion}\label{}

The horizontal and vertical lifts of vector fields on $M$ provide an intuitive way of understanding the curvature and torsion of a nonlinear connection. The idea is that the curvature measures the failure of two horizontal (coordinate) vector fields to commute, while the torsion measures the failure of a horizontal and a vertical (coordinate) vector field to commute. In other words, we are interested in $[\delta_i,\delta_j]$ and $[\delta_i,\bar\partial_j]$, or more generally, $[X^h,Y^h]$ and $[X^h,Y^v]$ for any $X,Y\in\mathfrak X(M)$. The latter two commutators are not tensorial, however. In order to make them tensorial, that is, $C^\infty(M)$-multilinear, we need to modify the expressions in the following way, 
\begin{align}
    R(X,Y) &= -\left([X^h,Y^h] - [X,Y]^h\right), \\
    T(X,Y) &= [X^h,Y^v] - [Y^h,X^v] - [X,Y]^v.
\end{align}
The tensor fields $R$ and $T$ are called the \textit{curvature} and the \textit{torsion} of the nonlinear connection, respectively\footnote{The overall minus sign in the definition of the curvature tensor is chosen so as to guarantee that the curvature tensor reduces to the standard curvature tensor for linear connections in the special case that the connection is linear.} (see e.g. \cite{Szilasi}). They are smooth $C^\infty(M)$-bilinear maps $\mathfrak X(M)\times \mathfrak X(M)\to \mathfrak X^v(\mathcal A)$. It is easily deduced that their coordinate expressions are given by
\begin{align}
    R(\partial_i,\partial_j) &= -[\delta_i,\delta_j]  \eqqcolon R^k_{ij}\bar\partial_k, \ &R^k_{ij} &= \delta_i N^k_j - \delta_j N^k_i ,\label{eq:def_nonlinear_curvature}\\
    T(\partial_i,\partial_j) &= [\delta_i,\bar\partial_j] - [\delta_j,\bar\partial_k] 
 \eqqcolon T^k_{ij}\bar\partial_k,  &T^k_{ij} &= \bar\partial_jN^k_ i - \bar\partial_i N^k_j.\label{eq:def_nonlinear_torsion}
\end{align}
%
%
If the connection is linear, $N^k_i = \Gamma^k_{ij}(x)y^j$, then a straightforward calculation reveals that the expressions above reduce to
\begin{align}
\label{eq:symm_ricci}
R^k_{ij} &= \bar  R^k{}_{\ell ij} (x)y^\ell, \qquad\quad &T^k_{ij} &= \Gamma^k_{ij}(x) - \Gamma^k_{ji}(x),
\end{align}
in terms of the curvature tensor of the associated affine connection and its Ricci tensor, 
\begin{align}\label{eq:class_ricci_and_riemann}
    \bar R^k{}_{\ell ij} = \partial_i\Gamma^k_{j\ell} - \partial_j\Gamma^k_{i\ell} + \Gamma^k_{im}\Gamma^m_{j\ell} - \Gamma^k_{jm}\Gamma^m_{i\ell}, \qquad \bar R_{lk} = \bar R^i{}_l{}_{ik},
\end{align}
respectively. In other words, we recover the standard expressions for the components of the curvature tensor and torsion tensor of a linear connection.

\subsection{Metric-compatibility}

We say that a nonlinear connection is compatible with a Finsler metric, or simply metric-compatible, if the Finsler Lagrangian $L$ is invariant under parallel transport. Or in other words, if the function $t\mapsto L(\gamma(t),v(t))$ is constant whenever $v$ is a vector field that is parallel along a curve $\gamma:t\mapsto \gamma(t)$. Sometimes this property is also called \textit{weak} metric-compatibility\footnote{In contrast to strong-metric compatibility, which is the property that the fundamental tensor is horizontally constant with respect to the Berwald derivative $\nabla^B$ induced by the nonlinear connection, see e.g. \cite{Anastasiei2004}.}.
\begin{prop}
    A nonlinear connection is metric-compatible if and only if the Finsler Lagrangian $L$ is horizontally constant, i.e. $\delta_i L =0$.
\end{prop}
\begin{proof}
    For parallel transport of a vector $v$ along a curve $\gamma(t)$ in $M$ we have $\dot v^i + N^i_j(\gamma,v)\dot \gamma^j = 0$, by \eqref{eq:nonlinear.parallel.transport.eq}, and hence
    \begin{align}
        \tfrac{\D}{\D t} L(\gamma(t),v(t)) &= \dot\gamma^i \partial_i L + \dot v^i\bar\partial_iL \\
        &= \dot\gamma^i \partial_i L - N^i_j(\gamma,v)\dot \gamma^j\bar\partial_iL \\
        &= \dot\gamma^j\left( \partial_j L - N^i_j(\gamma,v)\bar\partial_iL\right)\\
        &= \dot\gamma^j\delta_j L\big|_{(\gamma(t),v(t))}.
    \end{align}
    This vanishes for all $\gamma,v$ if and only if $\delta_jL=0$.
\end{proof}
In some of the literature on Finsler geometry, an alternative definition of metric-compatibility is used, namely that $\nablad g=0$ in terms of the dynamical covariant derivative $\nablad$ to be introduced shortly in \cref{sec:dyn_cov_der}. The following result shows that this definition is equivalent to ours, provided that the nonlinear connection is assumed to be torsion-free, which it always is in the context of Finsler geometry (cf. \cref{sec:canonical_nonlin_con}).
\begin{prop}\label{prop:metric_compatible_equiv_defs}
Let $F$ be a Finsler metric and $N$ a torsion-free, homogeneous nonlinear connection on $M$. Then 
\begin{align}
    \delta_i L =0\qquad\Longleftrightarrow\qquad \nablad g = 0.
\end{align}
\end{prop}
Since the proof is somewhat lengthy and the result will not be used anywhere else in this dissertation, the proof can be found in \cref{app:metric_compatibility}.
\section{The canonical nonlinear connection}\label{sec:canonical_nonlin_con}

Just as one can assign a canonical linear connection to any pseudo-Riemannian manifold, the Levi-Civita connection, one can similarly assign a canonical connection to any Finsler manifold, if one is willing to give up linearity. Geodesics, as we defined them in \cref{sec:geodesics_0}, can then be understood as autoparallels of this canonical connection, and the notions of nonlinear curvature and Ricci curvature of a Finsler manifold can be defined.

\subsection{The fundamental lemma of Finsler geometry}\label{sec:fundamental_lemma}

The nonlinear connection established by the following theorem is called the \textit{canonical nonlinear connection} or the \textit{Cartan nonlinear connection}. We give a simple proof in local coordinates. The first intrinsic, coordinate-free proof was given by Grifone \cite{Grifone1972}, see also \cite{Szilasi}.

\begin{theor}[Fundamental lemma of Finsler geometry]
    Given a Finsler Lagrangian $L$ defined on $\mathcal A\subset TM$, there is a unique homogeneous nonlinear connection on $\mathcal A$ that is torsion-free and compatible with $L$. Its connection coefficients are given by
\begin{align}\label{eq:nonlinear_connection_explicit}
N_i^j = \tfrac{1}{4}\bar\partial_i\left[g^{jk}\left(y^m\partial_m\bar\partial_k L - \partial_k L\right)\right].
\end{align}
\end{theor}
\begin{proof}
The three desired properties can be stated as follows:
\begin{enumerate}[(i)]
\item $\partial_i L - N_i^j\bar\partial_j L = 0$ (metric-compatibility);
\item $\bar\partial_j N_i^k = \bar\partial_i N_j^k$ (torsion-freeness);
\item $N_i^j(x,\lambda y) = \lambda N_i^j(x,y)$, for all $\lambda>0$ (homogeneity).
\end{enumerate}
To prove uniqueness, we have to show that (i)-(iii) imply \eqref{eq:nonlinear_connection_explicit}. We begin by differentiating (i) with respect to $y^k$ and multiplying by $y^i$ and then in the middle term we apply (ii) followed by (iii) in the form of Euler's theorem:
\begin{align}
0 &= y^i\left(\partial_i\bar\partial_k L - \bar\partial_k N_i^j \bar\partial_j L - N_i^j \bar\partial_j \bar\partial_k L\right)\\
&= y^i\partial_i\bar\partial_k L - y^i\bar\partial_i N_k^j \bar\partial_j L - y^iN_i^j \bar\partial_j \bar\partial_k L \\
&= y^i\partial_i\bar\partial_k L - N_k^j \bar\partial_j L -y^i  N_i^j \bar\partial_j \bar\partial_k L.
\end{align}
Next, we bring the rightmost term to the left-hand side (LHS), which we rewrite in terms of the fundamental tensor, and we apply (i) to the right-hand side (RHS). This yields
\begin{align}
2 y^i N_i^j g_{jk} = y^i\partial_i\bar\partial_k L - \partial_k L
\end{align}
Because $g_{ij}$ is nondegenerate, this is equivalent to 
\begin{align}
2 y^i N_i^j = g^{jk}\left(y^i\partial_i\bar\partial_k L - \partial_k L\right).
\end{align}
The last step is to differentiate both sides with respect to $y^l$ and apply (ii) and Euler's theorem once more. The result, after renaming indices, is \eqref{eq:nonlinear_connection_explicit}. This proves uniqueness, and also existence in a local chart, since \eqref{eq:nonlinear_connection_explicit} satisfies all three properties (i)-(iii). Global existence can be verified by checking that the obtained formula $N_i^j$ has the appropriate transformation behavior, namely
\begin{align}
\tilde N^k{}_j = N^l{}_i\,\frac{\partial \tilde x^k}{\partial x^l}\,\frac{\partial  x^i}{\partial \tilde x^j} +  \frac{ \partial \tilde  x^k}{\partial x^l}  \, \frac{\partial^2  x^l}{\partial \tilde x^i \partial \tilde x^j}\,\tilde y^i.
\end{align}
This is a little tedious but in principle straightforward.
\end{proof}
One can easily check that if $L = g_{ij}(x)y^iy^j$ arises from a pseudo-Riemannian metric $g_{ij}$ then the canonical connection reduces to the Levi-Civita connection (having Christoffel symbols $\Gamma^k_{ij}$) of $g_{ij}$, in the sense that
\begin{align}
N^j_i =  \Gamma^j_{im}y^m , \qquad \Gamma^j_{mi} = \tfrac{1}{2}g^{jk}\left(\partial_ig_{km} + \partial_mg_{ki} - \partial_kg_{mi}\right).
\end{align}

\subsection{Geodesics and autoparallels}

For a general Finsler space, by writing out the term in square brackets in the expression \eqref{eq:nonlinear_connection_explicit} for the nonlinear connection coefficients and using that $L = g_{ij}(x,y)y^i y^j$, we find that 
\begin{align}\label{eq:poeptemp}
    g^{jk}\left(y^m\partial_m\bar\partial_k L \!-\! \partial_k L\right) = g^{jk}\left(2\partial_\ell g_{ik} \!-\! \partial_k g_{i\ell}\right)y^iy^\ell = 2\gamma^j_{i\ell}y^iy^\ell = 2 G^j,
\end{align}
in terms of the geodesic spray coefficients defined in \eqref{eq:spray_original_def_from_formal_christ}. The canonical nonlinear connection and the geodesic spray of a Finsler metric are thus related in the following way:
\begin{align}\label{eq:nonlinear_connection_and_spray_relations}
    N^j_i = \tfrac{1}{2}\bar\partial_i G^j,\qquad G^j = N^j_i y^i,
\end{align}
where the latter equation follows by multiplying the former by $y^i$ and applying Euler's theorem. 
Moreover, \eqref{eq:poeptemp} yields an expression for the spray directly in terms of the Finsler Lagrangian and fundamental tensor
\begin{align}\label{eq:spray_directly_from_L}
G^j  = \tfrac{1}{2}g^{jk}\left(y^m\partial_m\bar\partial_k L - \partial_k L\right).
\end{align}
Computationally, \eqref{eq:spray_directly_from_L} often provides the most efficient way of computing the spray coefficients directly from an explicitly given Finsler metric. Finally, by combining \eqref{eq:nonlinear_connection_and_spray_relations}, \eqref{eq:geodesic_equation}, and \eqref{eq:nonlinear.geodesic.eq} we obtain the following important result.
\begin{prop}\label{prop:geodesic_equals_autoparallel}
    The geodesics of a Finsler metric coincide with the autoparallels of its canonical nonlinear connection. 
\end{prop}
\subsection{Curvatures on a Finsler space}\label{sec:cuvatures_on_Finsler_space}

Given a Finsler metric or Lagrangian, we define its \textit{nonlinear curvature} as the curvature \eqref{eq:def_nonlinear_curvature} of its canonical nonlinear connection. We further define
\begin{align}\label{eq:Ricci_defs}
    \text{Ric} = R^i{}_{ij}y^j,\qquad\quad R_{ij} = \tfrac{1}{2}\db_i \db_j\text{Ric}.
\end{align}
In the literature, $R_{ij}$ is often just referred to as the Ricci tensor, but we will give it the name \textit{Finsler-Ricci tensor} to distinguish it from the \textit{affine Ricci tensor} that will be introduced later for Berwald spaces. Similarly, we will refer to $\mathrm{Ric}$ as the \textit{Finsler-Ricci scalar}. From homogeneity and Euler's theorem, it follows that the second relation in \eqref{eq:Ricci_defs} can be inverted as
\begin{align}
    \text{Ric} = R_{ij} y^i y^j.
\end{align}
Ric and $R_{ij}$ thus contain the same information, and in particular, we have \mbox{$\text{Ric}=0$} if and only if $R_{ij}=0$. 



\section{The dynamical covariant derivative}\label{sec:dyn_cov_der}

A nonlinear connection induces a so-called \textit{dynamical covariant derivative} $\nablad$ that acts on functions $f\in C^\infty(\mathcal A)$ and Finsler vector fields $X \in \Gamma\left(\pi^*TM\right)$ as  
\begin{align}
\nablad f &= y^i\delta_i f, \\
\nablad X &= \left(\nablad X\right)^j\partial_i, \qquad \left(\nablad X\right)^j = y^i\delta_iX^j + N^j_i X^i, \label{eq:bracketsss}
\end{align}
respectively, where $X = X^i\partial_i$ in local coordinates. With this definition, $\nablad$ maps functions to functions and Finsler vector fields to Finsler vector fields. As per standard convention, we will usually omit the brackets in \eqref{eq:bracketsss} and write
\begin{align}\label{eq:lhssss}
    \nablad X^j \equiv \left(\nablad X\right)^j.
\end{align}
We stress that the LHS of \eqref{eq:lhssss} is not to be interpreted as the dynamical covariant derivative of the component function $X^j$, but always as the $j^\text{th}$ component of the vector field $\nablad X$. We will adhere to this convention for all expressions of this type, so that no confusion should be possible as to what is meant. 

By requiring certain natural properties such as the Leibniz rule, $\nablad$ extends uniquely to a dynamical covariant derivative acting on general Finsler tensor fields in the usual way, e.g.
\begin{align}
\nablad g_{ij} \equiv \left(\nablad g\right)_{ij} = y^k\delta_k g_{ij} - N^k_i g_{kj} - N^k_j g_{ik}.
\end{align}
The generalization to higher-order tensor fields should be clear from this example. Given a tensor field $T\in \mathcal T^r_s(\pi^*TM)$, it follows from the transformation behavior of the components of $\nablad T$ that the latter is again a $(k,l)$-tensor field. In other words, $\nablad$ may be thought of as a map 
\begin{align}
    \nablad: \mathcal T^r_s(\pi^*TM)\to \mathcal T^r_s(\pi^*TM).
\end{align}
\begin{lem}\label{lem:cov_der_along_geodesics}
For any $f\in C^\infty(TM)$ we have
\begin{align}
    \nablad f= \dot\gamma^k\delta_k f = \tfrac{\D }{\D t}f
\end{align}
along any geodesic $\gamma(t)$.
\end{lem}
\begin{proof}
    Along an arbitrary curve $\gamma$ we have
    \begin{align}
        \left(\tfrac{\D }{\D t}f\right)(\gamma,\dot\gamma) &= \dot\gamma^i\partial_i f + \ddot\gamma^i \bar\partial_i f \\
        &= \dot\gamma^i\left(\delta_i f + N_i^k\bar\partial_k f\right) + \ddot\gamma^i \bar\partial_i f \\
        &= \dot\gamma^i\delta_i f + \left(\ddot\gamma^k + \dot\gamma^iN_i^k\right)\bar\partial_k f\\
        &= \nablad f + \left(\ddot\gamma^k + \dot\gamma^iN_i^k\right)\bar\partial_k f.
    \end{align}
    If $\gamma$ is a geodesic then the term in parentheses vanishes and the result follows.
\end{proof}
We will see in the \Cref{ch:lin_conns} that this result can be extended when $N^k_i$ is the canonical nonlinear connection of a Finsler space, such that even for tensors in general, the dynamical covariant can be interpreted as the linear covariant derivative along geodesics with respect to either of the linear Berwald or Chern-Rund connection.

\chapter[Linear Connections on the Pullback Bundle \texorpdfstring{$\pi^*TM$}{pi*TM}]{Linear Connections on the Pullback Bundle \texorpdfstring{$\pi^*TM$}{pi*TM}}\label{ch:lin_conns}

The canonical nonlinear connection corresponding to a Finsler metric, discussed in \Cref{ch:nonlin_conns}, leads to a characterization of geodesics as autoparallel curves of this connection, and it allows us to differentiate Finsler tensor fields by means of the dynamical covariant derivative. Despite its name, the latter is not a covariant derivative in the usual sense, essentially because it does not allow for the differentiation along different directions. For various purposes, it is useful to also introduce a covariant derivative which \textit{can} be used to do just that; and the way to do this is to specify a linear (Koszul) connection on the pullback bundle $\pi^*TM$ (cf. \cref{sec:pullback_bundle}). In contrast to the situation for nonlinear connections, the choice of such a linear connection---even in the presence of a Finsler metric---is in general not unique\footnote{We remark, however, that in the alternative framework of \cite{Anisotropic_Conn}, one can single out a unique so-called (linear) `anisotropic connection', which serves essentially the same purpose.}. We will focus here mostly on the Berwald and Chern-Rund connections.

\section{Notation and conventions
}

Recall that a linear (Koszul) connection on a smooth vector bundle $E$ over $M$ is an $\R$-linear map
\begin{align}
    \nabla : \mathfrak{X}(M)\times \Gamma(E)\to \Gamma(E), \qquad (X,s)\mapsto \nabla_X s,
\end{align}
such that $\nabla_{fX} s = f \nabla_X s$ and $\nabla_X (fs) = (Xf)s + f \nabla_X s$ for any $f\in C^\infty(M)$. Consider now the case where $E = \pi^*TM$. Since the base manifold is the conic subbundle $\mathcal A\subset TM$, a linear connection on $\pi^*TM$ is a map 
\begin{align}
    \nabla : \mathfrak{X}(\mathcal A)\times \Gamma(\pi^*TM)\to \Gamma(\pi^*TM),
\end{align}
so in a local coordinate basis $\partial_i\in \mathfrak{X}(M)$ of $TM$, with corresponding basis $\partial_i\in\Gamma(\pi^*TM)$ of $\pi^*TM$ and induced coordinate basis $\partial_i,\bar\partial_j\in \mathfrak X(\mathcal A)$ of $T\mathcal A$, we have, by linearity,
\begin{align}
    \nabla_{\partial_i}\partial_j = \tilde \Gamma_{ij}^k\partial_k, \qquad \nabla_{\bar\partial_i}\partial_j = \bar\Gamma_{ij}^k\partial_k,
\end{align}
for local functions $\tilde \Gamma_{ij}^k,\bar\Gamma_{ij}^k\in C^\infty(\mathcal A)$, called the Christoffel symbols. We have written a tilde above the first set of Christoffel symbols because these are not the ones that we will use. Indeed, if there is a nonlinear connection defined on $\mathcal A$ then we will usually employ the corresponding horizontal and vertical basis vectors $\delta_i,\bar\partial_i$ of $T\mathcal A$, thus defining the Christoffel symbols
\begin{align}\label{eq:def_Christoffel_symbs}
    \nabla_{\delta_i}\partial_j =  \Gamma_{ij}^k\partial_k, \qquad \nabla_{\bar\partial_i}\partial_j = \bar\Gamma_{ij}^k\partial_k.
\end{align}
From these definitions, it follows that
\begin{align}
    \nabla_{\delta_i}X^j &\equiv \left(\nabla_{\delta_i}X\right)^j = \delta_i X^j + \Gamma^j_{ik}X^k \\
    %
     \nabla_{\bar\partial_i}X^j &\equiv \left(\nabla_{\bar\partial_i}X\right)^j = \bar\partial_i X^j + \bar\Gamma^j_{ik}X^k.
\end{align}
As usual, a Koszul connection on $\pi^*TM$ as defined above extends uniquely to a connection on all tensor bundles of $\pi^*TM$, and one obtains analogous formulas for the horizontal and vertical covariant derivatives of arbitrary Finsler tensors.

\section{Torsion and metric-compatibility}

\subsection{Torsion}
Roughly following \cite{Anastasiei1996} (and see also \cite{Minguzzi:2014fxa} for a slightly different but equivalent approach), given a morphism $\rho: T\mathcal A\to\pi^*TM$ of vector bundles over $\mathcal A$, or equivalently a (generalized) soldering form\footnote{Sometimes a soldering form on a vector bundle $E$ over $M$ is defined as an $E$-valued 1-form $\omega\in\Omega(M,E)$ with the additional property that for each $x$, $\omega_x: T_xM\to E_x$ is a linear isomorphism. For our purposes the latter property is not necessary, hence the word \textit{generalized}.} $\omega \in \Omega(\mathcal A,\pi^*TM)$, one can define a torsion tensor $T^{(\rho)}:\mathfrak X(\mathcal A)\times \mathfrak X(\mathcal A)\to \Gamma(\pi^*TM)$ of a linear connection $\nabla$ on $\pi^*TM$ by slightly modifying the formula for the torsion of an affine connection in the following natural way:
\begin{align}
    T^{(\rho)}(\tilde X, \tilde Y) = \nabla_{\tilde X}\rho(\tilde Y) - \nabla_{\tilde Y}\rho(\tilde X) - \rho([\tilde X,\tilde Y]),
\end{align}
where $\rho(\tilde Y)$ is defined as the Finsler vector field such that $\rho(\tilde Y)_{(x,y)} = \rho(\tilde Y_{(x,y)})$. Note, however, that for different morphisms $\rho$, this leads to different notions of torsion. In our case two morphisms are natural. First, let $\rho$ be given by linear extension of $\rho(\delta_i) = \partial_i$ and $\rho(\bar\partial_i)=0$. The corresponding torsion will be called the \textit{horizontal torsion} and denoted by $T_\text{hor}$. It decomposes as
\begin{align}
    T_\text{hor}(\delta_i,\delta_j) &= \left(\Gamma^k_{ij} - \Gamma^k_{ji}\right)\partial_k, \label{eq:torsion1}\\
    T_\text{hor}(\delta_i,\bar\partial_j) &= - \bar\Gamma^k_{ji}\partial_k, \\
    T_\text{hor}(\bar\partial_i,\bar\partial_j) &= 0.
\end{align}
Next, let $\rho$ be given by linear extension of $\rho(\delta_i) = 0$ and $\rho(\bar\partial_i)=\partial_i$. The corresponding torsion will be called the \textit{vertical torsion} and denoted by $T_\text{ver}$. It decomposes as
\begin{align}
    T_\text{ver}(\delta_i,\delta_j) &= R^k_{ij}\partial_k, \\
    T_\text{ver}(\delta_i,\bar\partial_j) &= \left(\Gamma^k_{ij} - \bar\partial_j N^k_ i\right)\partial_k, \\
    T_\text{ver}(\bar\partial_i,\bar\partial_j) &= \left(\bar\Gamma^k_{ij} - \bar\Gamma^k_{ji}\right)\partial_k,\label{eq:torsion6}
\end{align}
where $R^k_{ij}$ is the curvature of the nonlinear connection. To avoid confusion we remark that some authors consider the horizontal $T_\text{hor}$ to be \textit{the} torsion of the linear connection and do not mention the vertical torsion $T_\text{ver}$ at all. Under that terminology the well-known Chern-Rund and Berwald connections in Finsler geometry, to be defined below, would be torsion-free. Their vertical torsions do in general not all vanish, though.


\subsection{Metric compatibility}

A linear connection $\nabla$ on $\pi^*TM$ is said to be compatible with a Finsler metric $F$, or just \textit{metric-compatible}, if the fundamental tensor is covariantly constant with respect to $\nabla$, i.e. $\nabla g=0$, that is $\nabla_{\delta_i}g_{jk} = \nabla_{\bar\partial_i}g_{jk}= 0$. One might hope that it were possible to associate a torsion-free (or at least horizontally torsion-free), metric-compatible linear connection on $\pi^*TM$ with any Finsler metric, in analogy with the Levi-Civita connection. This is not possible, however, unless $F$ is pseudo-Riemannian. A slightly weaker notion of metric-compatibility, to which we will refer as \textit{horizontal} metric-compatibility, is the property that $\nabla_{\delta_i}g_{jk}=0$. In this case, the fundamental tensor is covariantly constant in all horizontal directions, but not necessarily in the vertical directions. Similarly, we may define vertical metric compatibility by the property that $\nabla_{\bar\partial_i}g_{jk}= 0$.

\section[The Berwald and Chern-Rund connections]{Berwald and Chern-Rund connections}
\label{sec:Berwald_and_ChernRund}
We are now ready to introduce two of the much-used linear (Koszul) connections in Finsler geometry, the \textit{Berwald connection} and the \textit{Chern-Rund connection}. While the latter is always defined in the context of a given Finsler metric, the Berwald connection can be defined for any nonlinear connection $N^k_i$, without even specifying a Finsler metric. But by taking $N^k_i$ to be the canonical torsion-free, metric-compatible, homogeneous nonlinear connection associated with a Finsler metric $F$, we obtain the Berwald connection associated with $F$. Other important linear connections in Finsler geometry are the linear Cartan connection and the Hashiguchi connection. We will mention these only briefly  below.

\subsection{The Berwald connection}

Any homogeneous nonlinear connection on $\mathcal A\subset TM$ induces a canonical linear connection $\nabla^B$ on $\pi^*TM$, known as the \textit{Berwald connection}, whose Christoffel symbols are given by
\begin{align}\label{eq:def_Berwald_conn}
    {}^B\Gamma_{ij}^k = \bar\partial_j N^k_i,\qquad {}^B\bar\Gamma_{ij}^k=0.
\end{align}
By Euler's theorem,  this implies that the nonlinear connection coefficients can be obtained from the Christoffel symbols of the Berwald connection as 
\begin{align}\label{eq:Berwald_christoffel_symbs_to_nonlin_conn}
   N^k_i={}^B\Gamma^k_{ij} y^j.
\end{align}
Taking into account \eqref{eq:nonlinear_connection_and_spray_relations} it follows that the Berwald Christoffel symbols can be expressed as
\begin{align}\label{eq:Berwald_Christoffel_symbols_from_spray}
    {}^B\Gamma^k_{ij} = \tfrac{1}{2}\bar\partial_i\bar\partial_j G^k
\end{align}
in terms of the geodesic spray, and conversely, that
\begin{align}\label{eq:Spray_in_terms_of_Berwald_conn}
    G^k =  {}^B\Gamma^k_{ij}y^i y^j.
\end{align}
From the expressions \eqref{eq:torsion1}--\eqref{eq:torsion6} for the torsions and the definition of the Berwald connection \eqref{eq:def_Berwald_conn} it is clear that one can characterize the Berwald connection as the unique linear connection for which both mixed torsions vanish, i.e. $T_\text{hor}(\delta_i,\bar\partial_j)=T_\text{ver}(\delta_i,\bar\partial_j)=0$. Note that these requirements imply that the remaining torsions of the Berwald connection vanish identically as well, except for the parts $T_\text{hor}(\delta_i,\delta_j)=T^k_{ij}\partial_k$ and  $T_\text{ver}(\delta_i,\delta_j)=R^k_{ij}\partial_k$, which are given by the torsion and the curvature of the nonlinear connection, respectively. It is therefore fair to say that the Berwald connection is 
the closest to a completely torsion-free linear connection on $\pi^*TM$ as one can find in the presence of a given nonlinear connection.

If $N^k_i$ is the canonical nonlinear connection on a Finsler space then the corresponding Berwald connection is called the Berwald connection associated with the Finsler space, or simply \textit{the} Berwald connection. In this case (and more generally in any situation where the nonlinear connection is torsion-free) the Berwald covariant derivative is related to the dynamical covariant derivative $\nablad$ in the following way.
\begin{prop}\label{prop:dyn_cov_der_VS_Berwald}
    The dynamical covariant derivative $\nablad$ and the Berwald connection $\nabla^B$, both corresponding to the same homogeneous, torsion-free nonlinear connection, are related by
    \begin{align}
        y^k\nabla^B_{\delta_k} = \nablad.
    \end{align}
\end{prop}
\begin{proof}
    We prove the result for the operators acting on a Finsler vector field. The generalization to general Finsler tensor fields is then straightforward. We have
    \begin{align}
        y^k\nabla^B_{\delta_k}X^i &= y^k\left(\delta_kX^i + \bar\partial_j N^i_k X^j\right)  \\
        &= y^k\delta_kX^i + y^k\bar\partial_k N^i_j X^j \\
        &= y^k\delta_kX^i + N^i_j X^j = \nablad X^i,
    \end{align}
    where we have first used torsion-freeness of the nonlinear connection to switch the indices $j$ and $k$, and then Euler's theorem, using homogeneity of $N^i_j$.
\end{proof}

\subsection{The Chern-Rund connection}

Given a Finsler space with fundamental tensor $g_{ij}$, the Chern-Rund connection $\nabla^C$ is the unique linear connection on $\pi^*TM$ that is horizontally torsion-free, and horizontally metric compatible, i.e.
\begin{align}
    T_\text{hor}=0,\qquad \nabla^C_{\delta_i}g_{jk} = 0,
\end{align}
where the horizontal torsion is defined using the canonical nonlinear connection associated with $F$. One can show that the Christoffel symbols of the Chern-Rund connection are given by
\begin{align}
    {}^C\Gamma^k_{ij}  = \tfrac{1}{2}g^{k\ell}\left(\delta_i g_{\ell j} + \delta_jg_{i\ell} - \delta_\ell g_{ij}\right),\qquad {}^C\bar\Gamma^k_{ij}=0.
\end{align}
Note that the formula for the horizontal Christoffel symbols is formally identical to the formula for the Levi-Civita Christoffel symbols with partial derivatives replaced by horizontal derivatives. Indeed, the proof is also completely analogous to the proof of the formula for the Levi-Civita Christoffel symbols, so we will not repeat it here. From this expression together with the definition of the horizontal derivatives and the property \eqref{eq:Cartan_contraction_identity} of the Cartan tensor, it is straightforward to show that, just as for the Berwald Christoffel symbols, the geodesic spray coefficients \eqref{eq:spray_original_def_from_formal_christ} can be expressed in terms of the Chern-Rund Christoffel symbols as
\begin{align}\label{eq:Spray_in_terms_of_Chern_conn}
    G^k = {}^C\Gamma^k_{ij}y^i y^j = {}^B\Gamma^k_{ij}y^i y^j.
\end{align}
In fact, for both the Chern-Rund and Berwald connection of a Finsler metric, we have the stronger result that
\begin{align}\label{eq:Chern_and_Berwald_christoffel_symbs_to_nonlin_conn}
     {}^C\Gamma^i_{k\ell}y^k={}^B\Gamma^i_{k\ell}y^k= N^i_\ell.
\end{align}
For the Berwald connection, this follows immediately from \eqref{eq:Berwald_christoffel_symbs_to_nonlin_conn} since the Christoffel symbols are symmetric. The justification of the equality with the expression involving the Chern-Rund connection will be postponed to \Cref{cor:Chern_and_Berwald_christoffel_symbs_to_nonlin_conn} (or see \cite{Bao}). It follows from \eqref{eq:Chern_and_Berwald_christoffel_symbs_to_nonlin_conn} that we can extend the result of \Cref{prop:dyn_cov_der_VS_Berwald} for the Berwald connection to the Chern-Rund connection as well.
\begin{prop}\label{prop:dyn_cov_der_VS_Chern_VS_Berwald}
The dynamical covariant derivative $\nablad$, the Berwald connection $\nabla^B$, and the  Chern-Rund connection $\nabla^C$, all corresponding to the same Finsler metric, are related by
    \begin{align}
        y^k\nabla^C_{\delta_k} = y^k\nabla^B_{\delta_k} = \nablad.
    \end{align}
\end{prop}
\begin{proof}
    This follows immediately from the definitions and \eqref{eq:Chern_and_Berwald_christoffel_symbs_to_nonlin_conn}.
\end{proof}
%
%
Both the Berwald and the Chern-Rund connection will be useful when formulating the field equations of Finsler gravity in \Cref{ch:Finsler_gravity}. There are two other notable linear connections in Finsler geometry and although we will not use these in this dissertation, we will mention them for completeness in the following summary of the important linear connections on a Finsler space. From this summary, we see that in order to have better metric compatibility one must sacrifice some torsion-freeness, and vice versa:
\begin{itemize}
    \item the \textbf{Berwald connection} is defined uniquely by:
    \begin{itemize}
        \item $T_\text{hor}(\delta_i,\bar\partial_j)=T_\text{ver}(\delta_i,\bar\partial_j)=0$. \textit{(`as torsion-free as possible')}
    \end{itemize}
    \item the \textbf{Chern-Rund connection} is defined uniquely by:
    \begin{itemize}
        \item $\nabla_{\delta_i}g_{jk}= 0$. \textit{(horizontal metric-compatibility)}
        \item $T_\text{hor}=0$ \textit{(horizontally torsion-free)}
    \end{itemize}
    \item the \textbf{Hashiguchi connection} is defined uniquely by:
    \begin{itemize}
        \item $\nabla_{\bar\partial_i}g_{jk} = 0$.  \textit{(vertical metric-compatibility)}
        \item $T_\text{ver}(\delta_i,\bar\partial_j)=T_\text{ver}(\bar\partial_i,\bar\partial_j) =0$. \textit{(vertically `as torsion-free as possible')}
    \end{itemize}
    \item the (linear) \textbf{Cartan connection} is defined uniquely by:
    \begin{itemize}
        \item $\nabla_{\delta_i}g_{jk} = \nabla_{\bar\partial_i}g_{jk}= 0$  \textit{(full metric-compatibility)}
        \item \sloppy $T_\text{hor}(\delta_i,\delta_j) = T_\text{ver}(\bar\partial_i,\bar\partial_j)=0$  \textit{(minimal torsion-freeness contraint)}.
    \end{itemize}
\end{itemize}


\subsection{Covariant derivatives along curves}\label{sec:cov_der_along_curves}


Given a connection $\nabla$ on $\pi^*TM$ and a Finsler vector field $X\in \Gamma(\pi^*TM)$, the covariant derivative of $X$ along a curve $\eta:I\to \mathcal A, t\mapsto \eta(t) = (\gamma(t),v(t))$ can be defined in the standard way as
\begin{align}\label{eq:def:cov_der_along_curve}
    \tfrac{D}{D t} X^k \equiv \nabla_{\dot\eta}X^k = \nabla_{\dot\gamma^i\partial_i + \dot v^i \bar\partial_i}X^k = \dot\gamma^i\nabla_{\partial_i}X^k + \dot v^i \nabla_{\bar\partial_i}X^k,
\end{align}
where it is understood that the expressions are evaluated along the curve, i.e. at $(\gamma(t),v(t))\in \mathcal A\subset TM$. 

Now let us assume that the covariant derivative is given by either the Chern-Rund or the Berwald connection on a Finsler space. Expressing the partial derivatives in terms of the horizontal and vertical derivatives, and using the definition \eqref{eq:def_Christoffel_symbs} of the Christoffel symbols together with the fact that $\bar\Gamma^k_{ij}=0$ for both connections, we obtain
\begin{align}
    \tfrac{D}{D t} X^k = \dot\gamma^i \delta_i X^k(\gamma,v) + \dot\gamma^i\Gamma^k_{ij}(\gamma, v)X^j + \left(\dot v^i + N^i_j(\gamma,v)\dot\gamma^j\right)\bar\partial_jX^k(\gamma,v)
\end{align}
If we further assume that the curve $\eta$ is the lift to $TM$ of a curve in $M$ (necessarily $\gamma:I\to M$), in the sense that $v = \dot\gamma$, then we find that
\begin{align}
    \tfrac{D}{D t} X^k &= \dot\gamma^i \delta_i X^k(\gamma,\dot\gamma) + \dot\gamma^i\Gamma^k_{ij}(\gamma,\dot\gamma)X^j + \left(\ddot \gamma^i + N^i_j(\gamma,\dot\gamma)\dot\gamma^j\right)\bar\partial_jX^k(\gamma,\dot\gamma) \\
    &= \dot\gamma^i \delta_i X^k(\gamma,\dot\gamma) + N^k_j(\gamma,\dot\gamma)X^j + \left(\ddot \gamma^i + N^i_j(\gamma,\dot\gamma)\dot\gamma^j\right)\bar\partial_jX^k(\gamma,\dot\gamma) \\
    &= \nablad X^k + \left(\ddot \gamma^i + N^i_j(\gamma,\dot\gamma)\dot\gamma^j\right)\bar\partial_jX^k(\gamma,\dot\gamma) 
\end{align}
in terms of the dynamical covariant derivative $\nablad$ introduced in \cref{sec:dyn_cov_der}. Here we have used that $y^i\Gamma^k_{ij} = N^k_j$ for the Chern-Rund as well as the Berwald connection, i.e. \eqref{eq:Chern_and_Berwald_christoffel_symbs_to_nonlin_conn}. Hence if $\gamma$ is a geodesic, the last term vanishes by the geodesic equation, and we have $\tfrac{D}{D t} X^k = \nablad X^k$. %
%
%
%
In fact, the calculation done here for vector fields generalizes to any type of tensor field. Hence we obtain the following result, which generalizes \Cref{lem:cov_der_along_geodesics}.
\begin{prop}\label{prop:dynamical_cov_der_is_cov_der_along_geodesics}
The Berwald and Chern-Rund covariant derivatives along a geodesic $\gamma$ are both given by the dynamical covariant derivative
    \begin{align}
    \tfrac{\text{D}}{\text{D} t} = \nablad.
\end{align}
\end{prop}
We remark, just to be clear, that by the covariant derivative along $\gamma$ we mean, strictly speaking, the covariant derivative along its lift $\eta$. We immediately have the following corollary, where covariant constancy refers to covariant constancy with respect to either (or equivalently, both) of the Chern-Rund and Berwald connections.
\begin{cor}\label{cor:cov_constancy_along_geodesics}
    A tensor on the pullback bundle is covariantly constant along geodesics if and only if its dynamical covariant derivative vanishes.
\end{cor}
This result will lead to a geometric interpretation of Landsberg spaces in \Cref{ch:BerwaldLandsbergUnicorn}.

\section{Geodesic deviation}

Using the properties of the Berwald and Chern-Rund connection developed in \cref{sec:Berwald_and_ChernRund} we are now in a position to formulate the geodesic deviation equation for geodesics of a Finsler metric, which plays an important role in the `derivation of'---or at least the motivation for---the field equations of Finsler gravity that we will encounter in \Cref{ch:Finsler_gravity}.

Consider a smooth 1-parameter family of geodesics along a curve in a Finsler space $M$, in the sense that there exist real numbers $a<b, c<d$ and a smooth 
map $H: (a,b)\times (c,d)\to M$, $(s,t)\mapsto H(s,t)$ such that for each fixed value of $t$ the curve $\gamma_t: s\mapsto H(s,t)$ is a geodesic. Let $u=u(s,t)$ be the vector field on $M$ that is tangent to the geodesics, i.e. $u = H_* (\partial_s) = (\partial H^i/\partial s)\partial_i$ is the push-forward of $\partial_s$, and let $v$ be the deviation vector field $v = H_* (\partial_t) = (\partial H^i/\partial t)\partial_i$. Then it can be shown that the following differential equation is satisfied, known as the \textit{geodesic deviation} equation
\begin{align}\label{eq:geodesic_deviation_ds}
%
\frac{D^2 v^k}{D s^2}  + R^k{}_i (u)v^i = 0,
\end{align}
where we have suppressed the $x$-arguments for clarity (as we will continue to do below). Here the \textit{geodesic deviation operator} $R^k{}_i$ is given by $R^k{}_i = R^k_{ij}y^j$ in terms of the nonlinear curvature $R^k_{ij}$, see \eqref{eq:def_nonlinear_curvature}, and $D/Ds$ is the (Chern-Rund or equivalenly, Berwald) covariant derivative along the geodesic, introduced in \cref{sec:cov_der_along_curves}. By application of \cref{prop:dynamical_cov_der_is_cov_der_along_geodesics} and \Cref{lem:cov_der_along_geodesics}, this covariant derivative may be written as
\begin{align}
    \frac{D v^k}{D s} = \nablad v^k = \frac{\D v^k}{\D s} + N^k_j(u) v^j.
\end{align}
Hence an alternative but equivalent form of the geodesic deviation equation \eqref{eq:geodesic_deviation_ds} is given by
\begin{align}\label{eq:geodesic_deviation_nabla}
    \nablad^{\hspace{-3.5pt}2}v^k + R^k{}_i v^i = 0,
\end{align}
with the understanding that the expression is evaluated at $y=u$. For a proof of the geodesic deviation equation, we refer e.g. to \cite[\S5.7]{Bucataru} or \cite[\S6.1]{shen2001lectures}.


%% file: ch_Berwald_Landsberg.tex
\chapter{Berwald, Landsberg and Unicorn Metrics
}\label{ch:BerwaldLandsbergUnicorn}


In this last chapter of the preliminaries we discuss two important classes of Finsler metrics: Berwald metrics and Landsberg metrics. Berwald metrics are central to \Cref{part:Berwald_metrizability} of this dissertation, where we investigate the \psR metrizability of such metrics, and also to \Cref{ch:Berwald_solutions} and \cref{ch:lin} in \Cref{part:vacuum_sols}, which deal with solutions in Finsler gravity of Berwald type. Landsberg metrics are only slightly more general than Berwald metrics, but the exact meaning of the word `slightly' here has not completely been settled and is the domain of the so-called unicorn problem, which we briefly discuss as well. In \cref{ch:unicorn_cosm} we will employ Landsberg metrics of unicorn type in the context of cosmology to model a Finslerian expanding universe.

\section{Berwald metrics}\label{sec:Berwald}
A Finsler space is said to be of Berwald type, or simply Berwald, if the canonical nonlinear connection \eqref{eq:nonlinear_connection_explicit} reduces to a (necessarily smooth) linear connection on $TM$, or in other words, an affine connection on the base manifold $M$, meaning that the connection coefficients are of the form $N^k_i = \Gamma^k_{ij}(x)y^j$ for certain coefficients $\Gamma^k_{ij}(x)$. As already alluded to in \Cref{ch:nonlin_conns} it can be inferred from the transformation behavior of $N^i_j$ that the functions $\Gamma^i_{jk}$ have the correct transformation behavior to be the Christoffel symbols of a (torsion-free) affine connection on $M$. We will refer to this affine connection as the associated affine connection, or simply \textit{the affine connection} on the Berwald space. 
Below we specialize some of the important results obtained in \Cref{ch:nonlin_conns} to the setting of Berwald spaces. 

First, the parallel transport \eqref{eq:nonlinear.parallel.transport.eq} and autoparallel equations \eqref{eq:nonlinear.geodesic.eq} reduce in this case to the familiar equations 
\begin{align}
\dot V^i + \Gamma^i_{jk}(\gamma)\dot \gamma^j V^k = 0, \qquad \ddot \gamma^i + \Gamma^i_{jk}(\gamma)\dot \gamma^j \dot \gamma^k = 0,
\end{align}
and indeed the relation between the affine Christoffel symbols and the spray coefficients is given by (cf. \eqref{eq:Spray_in_terms_of_Chern_conn})
\begin{align}
    G^i = \Gamma^i_{jk}(x)y^j y^k.
\end{align}
The curvature tensors of a Berwald space can be written as
\begin{align}
\label{eq:symm_ricci_Berwald}
R^k_{ij} = \bar  R^k{}_{\ell ij} (x)y^\ell, \,\,\,\,\,\, \text{Ric} = \bar R_{ij}(x)y^i y^j, \,\,\,\,\,\, R_{ij}= \tfrac{1}{2}\left(\bar R_{ij}(x) + \bar R_{ji}(x)\right)
\end{align}
in terms of the curvature tensor and Ricci tensor, respectively, of the affine connection, defined by \eqref{eq:class_ricci_and_riemann}, i.e.
\begin{align}\label{eq:affine_curvatures}
    \bar R^k{}_{\ell ij} = \partial_i\Gamma^k_{j\ell} - \partial_j\Gamma^k_{i\ell} + \Gamma^k_{im}\Gamma^m_{j\ell} - \Gamma^k_{jm}\Gamma^m_{i\ell}, \qquad \bar R_{lk} = \bar R^i{}_l{}_{ik}.
\end{align}
We will also refer to $\bar R^k{}_{\ell ij}$ and $\bar R_{lk}$ as the \textit{affine curvature tensor} and \textit{affine Ricci tensor}, respectively. Since the last identity in \eqref{eq:symm_ricci_Berwald} will prove to be of particular interest later, we restate it as a lemma.
\begin{lem}
\label{lem:RicciTensors}
The Finsler-Ricci tensor of a Berwald space is the symmetrization of the affine Ricci tensor.
\end{lem}
It is worth pointing out that in the classical positive definite case with $\mathcal A = TM_0$, it is not necessary to symmetrize since in that case the affine Ricci tensor is always symmetric. This is a consequence, for instance, of Szab\'o's metrization theorem. In other signatures or, more accurately, in cases where $\mathcal A \neq TM_0$, the symmetrization is really necessary. We will come back to these matters in detail in \Cref{ch:Metrizability}.

Berwald manifolds can be characterized in many equivalent ways, some of which we state below. We will only sketch the proof of their equivalences. In \Cref{ch:Characterization_of_Berwald} we will encounter an additional equivalent characterization of Berwald spaces, introduced in \citeref{ref1}{H1}, based on an auxiliary pseudo-Riemannian metric. 
\begin{theor}\label{thm:Berwald_equiv_definitions}
    The following are all equivalent characterizations of Berwald manifolds:
    \begin{enumerate}[(i)]
        \item The canonical nonlinear connection of $F$ is linear;
        \item $F$ admits a (necessarily unique) smooth, torsion-free, metric-compatible affine connection;
        \item The Christoffel symbols of the Chern-Rund connection do not depend on $y$, i.e. ${}^C\Gamma^k_{ij} = {}^C\Gamma^k_{ij}(x)$;
        \item The Christoffel symbols of the Berwald connection do not depend on $y$, i.e. ${}^B\Gamma^k_{ij} = {}^B\Gamma^k_{ij}(x)$;
        \item The spray is quadratic in $y$;
        \item $\bar\partial_j\bar\partial_k\bar\partial_\ell G^i=0$.
    \end{enumerate}
\end{theor}
\begin{proof}[Proof sketch]
    $(i)\Leftrightarrow(iv)$ follows immediately from the definition of the Berwald Christoffel symbols in terms of the nonlinear connection coefficients, and $(iv)\Leftrightarrow (v)$ follows immediately from \eqref{eq:Berwald_Christoffel_symbols_from_spray} and the 2-homogeneity of $G^k$. $(v)\Leftrightarrow (vi)$ is obvious. $(i)\Leftrightarrow(ii)$ follows from the fact that the canonical connection is the unique torsion-free metric-compatible homogeneous connection on $TM$. The equivalence with $(iii)$ requires some more work and for this, we refer to \cite{Bao}. For more detailed proofs of the other equivalences we also refer to \cite{Szilasi} and \cite{Szilasi2011}.
\end{proof}
From the fact that the Berwald and Chern-Rund Christoffel symbols only depend on $x$ for a Berwald metric, together with the fact that the spray can be expressed in the same way in terms of both, namely as $G^k = \Gamma^k_{ij}(x)y^i y^j$, cf. \eqref{eq:Spray_in_terms_of_Berwald_conn} and \eqref{eq:Spray_in_terms_of_Chern_conn}, and finally the fact that both connections are horizontally torsion-free, it follows that the Berwald and Chern-Rund Christoffel symbols coincide for Berwald metrics. Moreover, their horizontal Christoffel symbols coincide with the Christoffel symbols of the (unique) torsion-free, metric-compatible affine connection.

\section{Landsberg metrics}\label{sec:Landsberg}

We have seen above that for Berwald spaces, the difference between the Berwald and Chern-Rund connection vanishes. In general, this difference defines a tensor, called the \textit{Landsberg tensor}. A Landsberg metric is defined by the property that this tensor vanishes. In other words, a Landsberg metric is one for which the Berwald and Chern-Rund connection coincide. But as for Berwald metrics, there are several equivalent and perhaps more enlightening characterizations of Landsberg spaces.
\begin{defthm}\label{defthm:Landsberg_tensor}
The Landsberg curvature $S_{ijk}$ tensor is given by any and each of the following equivalent definitions:
\begin{align}
S_{jkl} &=  
%
-\tfrac{1}{4}y_i \, \bar\partial_j\bar\partial_k\bar\partial_\ell G^i  \qquad (\text{in terms of the spray}) \label{eq:Landsberg_def_intermsof_Spray}\\ 
&= g_{ij}\left({}^B\Gamma^i_{k\ell}-{}^C\Gamma^i_{k\ell}\right)\qquad (\text{Berwald minus Chern-Rund $\Gamma$-symbols})\\
&= \nablad C_{jkl}  \qquad (\text{dynamical cov. der. of the Cartan tensor})\\
&= -\tfrac{1}{2}\nabla^B_{\delta_j}g_{kl}. \qquad (\text{hor. Berwald cov. der. of $g_{ij}$})
\end{align}
\end{defthm}
We do not attempt to give a complete proof of these equivalences here but refer instead to \cite{Bao} for details. Clearly, it follows that the Landsberg tensor is completely symmetric. The \textit{mean Landsberg curvature} is defined as the trace $S_j = g^{k\ell}S_{jk\ell}$ of the Landsberg tensor and can in particular be expressed as
\begin{align}
S_{j} = \nablad C_j = \nablad \left( \bar\partial_j\ln\sqrt{|\det g|}\right).
\end{align}
With \Cref{defthm:Landsberg_tensor} we can define the notion of a \textit{Landsberg space} as follows:
\begin{defi} A Landsberg space is a Finsler space that satisfies any and hence each of the following equivalent conditions:
\begin{itemize}
\item  $S_{jkl} = -\frac{1}{4}y_i \, \bar\partial_j\bar\partial_k\bar\partial_\ell G^i=0$
\item The Berwald and the Chern-Rund Christoffel symbols coincide
\item $\nablad C_{jkl}=0$, i.e. the Cartan tensor is covariantly constant along geodesics.
\item $\nabla^B_{\delta_k}g_{ij}=0$, i.e. the fundamental tensor is horizontally constant with respect to the Berwald connection.
\end{itemize}
\end{defi}
The interpretation in terms of covariant constancy (with respect to the Berwald as well as Chern-Rund connection) along geodesics in the third equivalent condition is justified by \Cref{cor:cov_constancy_along_geodesics}. As a corollary of \Cref{defthm:Landsberg_tensor} we also obtain the following result that we already used in \cref{sec:Berwald_and_ChernRund} (see \eqref{eq:Chern_and_Berwald_christoffel_symbs_to_nonlin_conn}), but the justification of which we had postponed. 
\begin{cor}\label{cor:Chern_and_Berwald_christoffel_symbs_to_nonlin_conn}
    Both the Chern-Rund and Berwald connection satisfy 
    \begin{align}
        {}^C\Gamma^i_{k\ell}y^k={}^B\Gamma^i_{k\ell}y^k= N^i_\ell.
    \end{align}
\end{cor}
\begin{proof}
    It follows from the Leibniz rule for $\nablad$, the fact that $\nablad y^k =0$ and the fact that $y^kC_{ijk}=0$ that $y^k\nablad C_{ijk}=0$. It thus follows from the theorem that $y^kg_{ij}\left({}^B\Gamma^i_{k\ell}-{}^C\Gamma^i_{k\ell}\right)= 0$ and it then follows by nondegeneracy of $g_{ij}$ that ${}^C\Gamma^i_{k\ell}y^k={}^B\Gamma^i_{k\ell}y^k$, the latter being equal to $N^i_\ell$, by \eqref{eq:Berwald_christoffel_symbs_to_nonlin_conn}.
\end{proof}
Finally, we say that a Finsler space is \textit{weakly Landsberg} if the mean Landsberg curvature vanishes. It thus follows immediately from the definitions that we have the following inclusions:
\begin{align*}
   &\text{pseudo-Riemannian} \subset \text{Berwald} \subset \text{Landsberg} \subset \text{Weakly Landsberg}.
\end{align*}
%
%

\section{The unicorn problem}
It has been a long-standing open question whether the inclusion $\text{Berwald} \subset \text{Landsberg}$ is strict. Do there exist Landsberg spaces that are not Berwald? This question, along with some variations of it, is known as the \textit{unicorn problem}. For the important case where the domain is all of $\mathcal A = TM_0$ the answer is unknown. If the condition $\mathcal A = TM_0$ is relaxed, however, some examples of non-Berwaldian Landsberg spaces have been obtained, but such examples are still exceedingly rare. As such, non-Berwaldian Landsberg spaces are referred to as \textit{unicorns} \cite{Bao_unicorns}. We recommend \cite{Bao_unicorns,UnicornSurvey} for reviews on the unicorn problem. 

The first unicorns were found by Asanov \cite{asanov_unicorns} in 2006 and his results were generalized by Shen \cite{shen_unicorns} a few years later. These were the only known examples of unicorns until Elgendi very recently provided some additional examples \cite{Elgendi2021a}. One of the families of unicorns introduced by Elgendi will be our point of departure in \Cref{ch:unicorn_cosm}. In fact, by modifying Elgendi's metrics slightly in such a way that they retain the unicorn property, we will find that there exist exact solutions to Pfeifer and Wohlfarth's field equation of Finsler gravity, of unicorn type, that can be interpreted in a cosmological setting.

%% file: intr2.tex
\chapter*{Introduction to\\
\cref{part:Berwald_metrizability}: Berwald Spaces and Pseudo-Riemann Metrizability}

In \cref{part:preliminaries} we have introduced the notion of a Berwald space and reviewed some of the standard characterizations of such spaces. Here, in \cref{part:Berwald_metrizability}, we obtain a novel characterization of Berwald spaces and use it, among other things, to study the question of pseudo-Riemann metrizability, i.e. the question of whether the canonical torsion-free metric-compatible connection on a Berwald space can be understood as the Levi-Civita connection of some auxiliary \psR metric. While a well-known theorem due to Szab\'o \cite{Szabo} states that this is always possible in the classical positive definite scenario, the question of metrizability in other signatures, such as Lorentzian signature, and in scenarios with less stringent smoothness constraints, is still almost completely unexplored.  

\cref{part:Berwald_metrizability} starts with \cref{ch:Characterization_of_Berwald}, where we present a novel necessary and sufficient condition for a Finsler metric to be of Berwald type and we explore its consequences. This characterization makes use of an arbitrary auxiliary \psR metric, which renders it especially useful in situations with a preferred 
\psR metric. This is the case, for instance, when dealing with Finsler metrics that are (anisotropically) conformal to a pseudo-Riemannian metric, or when dealing with \ab-metrics. \ab-metrics are simply Finsler metrics constructed from a \psR metric $\alpha$ and a 1-form $\beta$ and such metrics will play a major role throughout both \cref{part:Berwald_metrizability} and \cref{part:vacuum_sols}. Indeed, we obtain as a corollary a very useful necessary and sufficient Berwald condition for \ab -metrics and we apply this result to several specific cases of interest: Randers metrics, exponential metrics, and generalized \mkrop metrics, reproducing some well-known results as well as presenting new ones. Our generalized \mkrop metrics are a generalization of the standard \mkrop metric; they naturally appear as the most general \ab-metrics consistent with a certain form of the affine connection. 

Next, \cref{ch:Metrizability} is devoted to the pseudo-Riemann metrizability of Finsler metrics in arbitrary signatures. We show that Szab\'o's metrization theorem cannot be extended to arbitrary signatures such as Lorentzian signature. We then go on to obtain necessary and sufficient conditions for (generalized) \mkrop spaces to be locally metrizable, resulting in a complete characterization of locally metrizable \mkrop spaces whose 1-forms have a constant causal character, and a partial characterization of locally metrizable generalized \mkrop spaces. We end the chapter with a classification of locally metrizable Ricci-flat \mkrop spaces whose 1-forms have a constant causal character.

%% file: ch_New_characterization_of_Berwald.tex
\chapter{Characterization of Berwald Spaces}\label{ch:Characterization_of_Berwald}

The purpose of this chapter is to provide a novel characterization of Berwald spaces, \cref{thm:Berwald_condition_thm}, and explore its consequences, including new results as well as well-known ones that follow from it particularly easily or that are somehow relevant to later chapters. As an important application, a major part of the chapter will be devoted to \ab-metrics of Berwald type. Many of the results discussed here have been published in \citeref{ref1}{H1}, albeit in many cases in a substantially different form. In particular, the results are presented here with a heightened focus on mathematical rigor. Apart from the applications that we discuss here, \cref{thm:Berwald_condition_thm} has also been used to classify all spatially homogeneous and isotropic Berwald spacetimes, i.e. Berwald spacetimes with cosmological symmetry \cite{Hohmann:2020mgs, VoicuHabilitationThesis}. 

\section{The Berwald condition}
In \citeref{ref1}{H1} we introduced a new necessary and sufficient condition for a Finsler metric to be of Berwald type, a condition that employs an auxiliary pseudo-Riemannian metric. Here we provide an alternative proof to that which is given in the original paper.
\begin{theor}\label{thm:Berwald_condition_thm}
    Let $(M,\mathcal A, L)$ be a Finsler space and let $A$ be an auxiliary pseudo-Riemannian Finsler Lagrangian on $M$ such that $L = \Omega A$ on $\mathcal{A}$ for some $\Omega\in C^\infty(\mathcal A)$. Then:
    \begin{enumerate}[(i)]
        \item $L$ is of Berwald type if and only if there exists a smooth symmetric classical $(1,2)$-tensor field $T^j{}_{ik}=T^j{}_{ki}$ on $M$, such that
        \begin{align}\label{eq:BerwCond_compact}
		A\mathring \delta_i \Omega  = T^j{}_{ik}y^k \bar\partial_j L,
	\end{align}
	where $\mathring \delta_i$ denotes the horizontal derivative with respect to the Levi-Civita connection of $A$.
	\item If so, the Christoffel symbols of the torsion-free $L$-compatible        affine connection are given by
		\begin{align}\label{eq:CsymbstfLcomp}
			\Gamma^j_{ik} =\mathring \Gamma^j{}_{ik} + T^j{}_{ik},
		\end{align}
            where $\mathring \Gamma^j{}_{ik}$ are the Levi-Civita Christoffel symbols of $A$. Moreover the Christoffel symbols  $\Gamma^j_{ik}$ coincide with the horizontal Chern-Rund as well as Berwald Christoffel symbols.
    \end{enumerate}
\end{theor}
Note that given a Finsler Lagrangian $L$, any positive definite $A$ will do. In that case there always exists an $\Omega$ such that $L=\Omega A$, namely $\Omega = L/A$. However, as one might expect, the theorem is most useful in situations where there is some preferred pseudo-Riemannian metric $A$. Typical examples of this are Finsler metrics that are (anisotropically) conformal\footnote{The adverb `anisotropically' refers to the fact that the conformal factor is allowed to depend not only on $x$ but also on $y$.} to a pseudo-Riemannian metric, and $(\alpha,\beta)$-metrics, which we study in detail in the next two sections. We also point out that whenever $A\neq 0$ the condition \eqref{eq:BerwCond_compact} can be written as 
\begin{align}\label{eq:BerwCond_expanded}
            \partial_i\Omega - \mathring \Gamma_{ik}^jy^k\bar\partial_j\Omega  = T^j{}_{ik}y^k \left(\bar\partial_j \Omega + \frac{2 y_j \Omega}{A}\right),
\end{align}
which is how it originally appeared in \citeref{ref1}{H1}. Here indices are lowered with the metric $A$.
\begin{proof}[Proof of \Cref{thm:Berwald_condition_thm}.]
    One of the equivalent characterizations of a Berwald metric (see \cref{thm:Berwald_equiv_definitions}) is that there exists a (necessarily unique) torsion-free linear connection on $TM$, with Christoffel symbols denoted by $\Gamma^k_{ij}$, compatible with $L$, i.e. such that $L$ is horizontally constant with respect to it. Denoting the corresponding (linear) nonlinear connection by $N^j_i$, this means that
    \begin{align}
        \delta_i L = \partial_i L - N_i^j\bar\partial_j L = \partial_i L - \Gamma^j_{ik}y^k\bar\partial_j L = 0.
    \end{align}
    After substituting $L = \Omega A$ this can be written as
    \begin{align}\label{Berw_proof_eq}
        \left(\partial_i \Omega  - \Gamma^j_{ik}y^k\bar\partial_j \Omega\right) A = -\Omega \left(\partial_i A  - \Gamma^j_{ik}y^k\bar\partial_j A\right).
    \end{align}
    If $\mathring \Gamma^j_{ik}$ denotes the Levi-Civita Christoffel symbols of the metric $A$, then we can write $\Gamma^j_{ik} = \mathring \Gamma^j_{ik} + T^j_{ik}$, for some symmetric (in $i\leftrightarrow k$) classical tensor field $T^j_{ik}$, since the difference between two linear connections is necessarily tensorial. Since $\tilde\Gamma^j_{ik}$ is compatible with $A$ it follows that the term in between brackets on the RHS of \eqref{Berw_proof_eq} reduces to 
    \begin{align}
        \partial_i A  - \Gamma^j_{ik}y^k\bar\partial_j A = \underbrace{\partial_i A  -  \mathring \Gamma^j_{ik}y^k\bar\partial_j A}_{=0} - T^j_{ik}y^k\bar\partial_j A  = -T^j_{ik}y^k\bar\partial_j A,
    \end{align}
    and we can rewrite \eqref{Berw_proof_eq} as
    \begin{align}
        \left(\partial_i \Omega  -  \mathring \Gamma^j_{ik}y^k\bar\partial_j \Omega\right) A =  T^j_{ik}y^k  \left(A\bar\partial_j\Omega + \Omega\bar\partial_j A\right) =  T^j_{ik}y^k \bar\partial_j L.
    \end{align}
    Recognizing on the LHS the $\mathring \Gamma$-horizontal derivative of $\Omega$, the first result follows. Since $\Gamma^j_{ik} = \mathring \Gamma^j_{ik} + T^j_{ik}$, the formula for the Christoffel symbols of the L-compatible linear connection follows immediately as well, and as discussed in \cref{sec:Berwald}, these have to coincide with the Chern-Rund as well as Berwald horizontal Christoffel symbols.
\end{proof}
An immediate consequence of the theorem is a characterization of spray invariance under (anisotropically) conformal transformations of a pseudo-Riemannian metric, first obtained by Tavakol and Van den Bergh \cite{Tavakol_1986}. It follows essentially by setting $T^j_{ik}=0$.
%
%
\begin{cor}\label{cor:TvdB}
    Suppose that $L = \Omega A$ is a Finsler Lagrangian on $\mathcal A = TM_0$, where $A$ is a pseudo-Riemannian metric and $\Omega\in C^\infty(\mathcal A)$. Then the geodesic spray of $L$ coincides with that of $A$, i.e. $G^k =\mathring G^k=\mathring\Gamma^k_{ij} y^i y^j$, if and only if $\Omega$ is horizontally constant w.r.t. $A$, i.e.
    \begin{align}\label{eq:TvdB}
	\mathring\delta_i \Omega=\partial_i \Omega - \mathring\Gamma^b{}_{ik}y^k\bar\partial_b \Omega = 0.
    \end{align}
\end{cor}
\begin{proof}
   If the geodesic spray of $L$ coincides with that of $A$ then $T^j_{ik}=0$ and it follows by the theorem that \eqref{eq:TvdB} must hold whenever $A\neq 0$. 
   Since $TM\setminus\{A=0\}$ is dense in $\mathcal A=TM_0$ it follows by continuity of $\mathring\delta_i \Omega$ that $\eqref{eq:TvdB}$ must in fact hold everywhere on $\mathcal A$. Conversely, suppose that \eqref{eq:TvdB} holds. Then condition \eqref{eq:BerwCond_compact} holds with $T^j{}_{ik} =0$. Hence, by the theorem, $L$ is Berwald and the spray of $L$ coincides with that of $A$.
\end{proof}

\section{\texorpdfstring{$(\alpha,\beta)$}{(\textalpha,\textbeta)}-metrics of Berwald type}\label{sec:ab_metrics_of_Berw_type}

\subsection{\texorpdfstring{$(\alpha,\beta)$}{(\textalpha,\textbeta)}-metrics}\label{sec:abMetricsNotation}

One instance in which \Cref{thm:Berwald_condition_thm} is particularly useful is when dealing with $(\alpha,\beta)$-metrics. Here 
$\alpha = \sqrt{|a_{ij}(x) y^iy^j|}$ and $\beta = b_i(x) y^i$ are scalar variables on $TM$ defined in terms of a (smooth) pseudo-Riemannian metric $a=a_{\mu\nu}\D x^\mu \D x^\nu$ on $M$ and a (smooth) 1-form $b=b_\mu\D x^\mu$ on $M$, and an $(\alpha,\beta)$-metric is essentially a Finsler metric that is constructed only from some $\alpha$ and $\beta$, i.e. $F = f(\alpha, \beta)$ for some function $f$, or more precisely, $F(x,y) = f(\alpha(x,y),\beta(x,y))$. Due to homogeneity it follows that whenever $\alpha\neq 0$ on $\mathcal A$ any such $F$ can be written in the standard form $F = \alpha\phi(\beta/\alpha)$ for a smooth function $\phi$. In what follows we will always tacitly assume that $\alpha\neq 0$ on $\mathcal A$, which is the case for many examples of interest. We will therefore work with the following definition.
\begin{defi}\label{def:ab_metric}
    An \ab-metric is a Finsler metric with domain $\mathcal A$ not including $\alpha=0$ that can be written in the form $F = \alpha\phi(\beta/\alpha)$, where $\phi$ is smooth on its domain.
\end{defi}
We will often denote $\beta/\alpha$ by $s$, so that we can view $\phi$ as a function of the real variable $s$.

In \cref{sec:ab_determinant} we derive an expression \eqref{sec:detg} for the fundamental tensor of a not necessarily positive definite \ab-metric and its determinant. 
From this, it follows in particular that $\phi$ never vanishes. In accordance with the notation used above, we will denote by $A = a_{ij}y^iy^j$ the Finsler Lagrangian corresponding to $\alpha$, or rather to $a_{ij}$. 
By some abuse of language, we will sometimes refer to either of $\alpha$, $A$, $a$ and $a_{ij}$ as the \psR metric, and to $\beta$, $b$ and $b_i$ as the 1-form, and sometimes we will even write expressions such as $\beta = \D t$ and
\begin{align}
    \alpha = \sqrt{\left|-\D t^2 + \D x^2 + \D y^2 + \D z^2\right|},
\end{align}
but the interpretation of such expressions should be clear. We also denote 
\begin{align}
    |b|^2 = a_{ij}b^i b^j,\qquad  \text{sgn}(A)=\epsilon.
\end{align}
In the presence of an \ab-metric, indices will be raised and lowered with $a_{ij}$ unless otherwise specified. Well-known examples of $(\alpha,\beta)$-metrics are:
\begin{itemize}
\item pseudo-Riemannian Finsler metrics $F = \alpha$;
\item Randers metrics $F = \alpha + \beta$;
\item Kropina metrics $F = \frac{\alpha^2}{\beta}$;
\item $m$-Kropina metrics $F = \alpha^{1+m}\beta^{-m}$, where $m\in\R$, also known as generalized Kropina metrics, Bogoslovsky metrics or Bogoslovsky-Kropina metrics.
\end{itemize}
%
%
From a physics perspective, $(\alpha,\beta)$-metrics are very relevant as they provide a means of deforming \psR metrics $\alpha$ into properly Finslerian Finsler metrics. And it turns out, as we will see in \Cref{part:vacuum_sols}, that these types of metrics can in fact be used to generalize some of the solutions to Einstein's field equations to properly Finslerian solutions to the field equation in Finsler gravity.

As a quick first application of \Cref{thm:Berwald_condition_thm} we obtain the following well-known result (see e.g. \cite{handbook_Finsler_vol2_matsumoto}), where we denote by $\mathring\nabla$ the Levi-Civita connection of $\alpha$.
\begin{cor}\label{eq:ab_cc_pre}
    Let $F=\alpha\phi(\beta/\alpha)$ and suppose that $\mathring\nabla_i b_j=0$. Then $F$ is Berwald and the affine connection and geodesic spray of $F$ coincide with those of $\alpha$.
\end{cor}
\begin{proof}
    It suffices to show that $\mathring \delta_i \Omega =0$, for in that case condition \eqref{eq:BerwCond_compact} holds with $T^j{}_{ik} =0$, showing $F$ is Berwald with affine connection (and hence geodesic spray) equal to that of $\alpha$. We have $\mathring \delta_i \Omega = \frac{\partial\Omega}{\partial\alpha}\mathring \delta_i\alpha + \frac{\partial\Omega}{\partial\beta}\mathring \delta_i\beta$, where $\mathring \delta_i\alpha=0$ because $A$ is horizontally constant w.r.t. its own Levi-Civita connection, by definition. And for any 1-form on $M$ and any (linear) connection, it can easily be checked that $\mathring\delta_i\beta = (\mathring\nabla_ib_j)y^j$, and hence in the particular case at hand we also have $\mathring  \delta_i\beta=0$, resulting in $\mathring \delta_i \Omega =0$ and completing the proof.
\end{proof}

\subsection{The Berwald condition for \texorpdfstring{$(\alpha,\beta)$}{(\textalpha,\textbeta)}-metrics}


The general Berwald condition for $(\alpha,\beta)$-metrics can be obtained by applying \eqref{eq:BerwCond_compact} directly to Finsler metrics of the form $F = \alpha\phi(\beta/\alpha)$. 
We remind the reader that $\epsilon=\text{sgn}(A)$ and that indices are raised and lowered with $a_{ij}$. Hence in what follows, $y_i = a_{ij}y^j$. 
%
%
%
\begin{theor}\label{theor:ab_Berwald_condition_general}
    An $(\alpha,\beta)$-metric $F = \alpha\phi(\beta/\alpha)$ is of Berwald type if and only if there exists a smooth tensor field $T^j{}_{ik}(x)=T^j{}_{ki}(x)$ on $M$ such that
    \begin{align}\label{ab_Berwald_condition}
        y^j\mathring\nabla_i b_j &= T^j{}_{ik}y^k \left[b_j + \frac{\epsilon}{\alpha}\left(\frac{\phi}{\phi'} - \frac{\beta}{\alpha}\right)y_j\right] 
    \end{align}
    whenever $\phi'\neq 0$ and $T^j{}_{ik}y^ky_j=0$ whenever $\phi'= 0$. In that case, the affine connection is given by 
    \begin{align}\label{eq:afconsumbs}
			\Gamma^j_{ik} = \mathring \Gamma^j{}_{ik} + T^j{}_{ik},
    \end{align}
    where $\mathring \Gamma^j{}_{ik}$ are the Christoffel symbols of $\alpha$.
\end{theor}
%
%
\begin{proof}
    The Lagrangian corresponding to $F$ reads $L = F^2 = \epsilon A \phi(\beta/\alpha)^2$, so we identify $\Omega = \epsilon \phi^2$, which is smooth on $\mathcal A$ since this set does not include $A=0$, by assumption. We start by noting that, since $\mathring \delta_i \alpha = 0$, we have
    \begin{align}\label{eq:temp1}
        \mathring \delta_i \Omega = \Omega'\mathring \delta_i\left(\frac{\beta}{\alpha}\right) = \frac{\Omega'}{\alpha}\mathring \delta_i\beta = \frac{\Omega'}{\alpha}(\mathring\nabla_ib_j)y^j,
    \end{align}
    where $\Omega'= \D \Omega/\D s$. On the other hand, making use of the identities 
     \begin{align}
        \bar\partial_i \alpha = \frac{\epsilon y_i}{\alpha} ,\qquad \bar\partial_i s = \frac{1}{\alpha}\left(b_i - \frac{\beta y_i}{A}\right),
     \end{align}
    a straightforward computation shows that
    \begin{align}
        \bar\partial_j L &= \Omega ' \alpha \left(b_j - \epsilon\frac{\beta}{\alpha^2}y_j\right) + 2\epsilon\Omega y_ j \\
        &= \Omega ' \alpha b_j + \epsilon\left(2\Omega-\frac{\beta}{\alpha}\Omega'\right)y_ j. \label{eq:temp2}
    \end{align}
    Plugging \eqref{eq:temp1} and \eqref{eq:temp2} into \eqref{eq:BerwCond_compact} yields
    \begin{align}
        \alpha\Omega'\left(\mathring\nabla_i b_j\right)y^j = T^j{}_{ik}y^k \left[\Omega'\alpha b_j + \epsilon\left(2\Omega - \frac{\beta}{\alpha}\Omega '\right)y_j\right].
    \end{align}
    It follows from nondegeneracy and \eqref{eq:det_ab} that $\phi\neq 0$, and since we are also assuming that $\alpha\neq 0$, we can divide by $2\epsilon\alpha\phi$. Using $\Omega = \epsilon \phi^2$ and $\Omega' = 2\epsilon\phi\phi'$ the equation can thus be rewritten as 
    \begin{align}\label{eq:tempcond}
        \phi'y^j\mathring\nabla_i b_j = T^j{}_{ik}y^k \left[\phi' b_j + \frac{\epsilon}{\alpha}\left(\phi - \frac{\beta}{\alpha}\phi'\right)y_j\right]
    \end{align}
    and by \cref{thm:Berwald_condition_thm}, $F$ is Berwald iff \eqref{eq:tempcond} is satisfied for all $(x,y)\in\mathcal A$. Now we distinguish the two cases. If $(x,y)$ is such that $\phi'\neq 0$ then we can divide by $\phi'$, leading directly to \eqref{ab_Berwald_condition}. If, on the other hand, $\phi'=0$ then the condition \eqref{eq:tempcond} reduces to $T^j{}_{ik}y^k\frac{\epsilon}{\alpha}\phi y_j=0$, or equivalently, $T^j{}_{ik}y^ky_j = T_{jik}y^ky^j=0$. Note that since the set $\{\phi'=0\}$ is not necessarily open this does not imply that $T^j{}_{ik}=0$. The formula for the Christoffel symbols follows from \eqref{eq:CsymbstfLcomp}.
\end{proof}
%
%
This result is closely related to \cite[Prop. 6.3.1.1]{handbook_Finsler_vol2_matsumoto}. %
%
%
%
%
In many cases of interest, the set $\{\phi'= 0\} \subset \mathcal A$ has empty interior or equivalently, the set $\tilde{\mathcal A} \equiv \mathcal A\setminus \{\phi'=0\}$ is dense in $\mathcal A$. In such a scenario it suffices to consider only the situation for points satisfying $\phi'\neq 0$.
\begin{theor}\label{theor:ab_Berwald_condition_general2}
    Let $F = \alpha\phi(\beta/\alpha)$ be an $(\alpha,\beta)$-metric such that the set $\{\phi'= 0\} \subset \mathcal A$ has empty interior. Then TFAE:
    \begin{enumerate}[(i)]
        \item $F$ is of Berwald type;
        \item  there exists a smooth tensor field $T^j{}_{ik}(x)=T^j{}_{ki}(x)$ on $M$ such that \eqref{ab_Berwald_condition} holds whenever $\phi'\neq 0$.
    \end{enumerate}
    In that case, the affine connection is given by \eqref{eq:afconsumbs}.
\end{theor}
%
%
%
%
\begin{proof}
    The implication $(i)\Rightarrow(ii)$ follows immediately from \Cref{theor:ab_Berwald_condition_general}. For the other implication, we start by noting that, since the subset $\phi'=0$ of $\mathcal A$ has empty interior, the set $\tilde{\mathcal A} \equiv \mathcal A\setminus \{\phi'=0\}$ is dense in $\mathcal A$. If \eqref{ab_Berwald_condition} is satisfied whenever $\phi'\neq 0$, then $F$ is Berwald on $\tilde{\mathcal A}$, interpreted as a conic subbundle of the tangent bundle $T\pi(\tilde{\mathcal A})$ of $\pi(\tilde{\mathcal A})\subset M$. (Note that $\tilde{\mathcal A}$ is a conic subbundle of $TM$ iff $\pi(\tilde{\mathcal A})=M$.) Indeed, $\pi(\tilde{\mathcal A})\subset M$ is open since the canonical projection map $\pi$ is an open map and $\tilde{\mathcal A}$ is open. Thus, $\pi(\tilde{\mathcal A})$ is itself a smooth manifold and it is easily verified that $\tilde{\mathcal A}$ satisfies all axioms for a conic subbundle of $T\pi(\tilde{\mathcal A})$. Moreover, it will be useful to note that $\pi(\tilde{\mathcal A})\subset \pi(\mathcal A)=M$ is dense, as the continuous image of a dense set. 
    
    As we observed above, $F$ is Berwald on $\tilde{\mathcal A}$. Then $\bar\partial_j\bar\partial_k\bar\partial_\ell G^i=0$ holds on $\tilde{\mathcal A}$, by \Cref{thm:Berwald_equiv_definitions} and so by continuity of the LHS, we must in fact have $\bar\partial_j\bar\partial_k\bar\partial_\ell G^i=0$ on the closure of $\tilde{\mathcal A}$ (in $\mathcal A$), which is just $\mathcal A$. But then, by \Cref{thm:Berwald_equiv_definitions}, $F$ is Berwald on $\mathcal A$. This proves $(ii)$.
    
    We now prove the remaining statement in the theorem. $F$ is Berwald on $\mathcal A$, so by (the proof of) \Cref{theor:ab_Berwald_condition_general}, there exists a symmetric smooth (classical) tensor field $\tilde T^j{}_{ik}$ on $M$ that satisfies \eqref{eq:tempcond} everywhere on $\mathcal A$. Moreover, the corresponding affine connection on $M$ is given by $\Gamma^j_{ik} = \mathring \Gamma^j{}_{ik} + \tilde T^j{}_{ik}$. But we also know that the affine connection on $\pi(\tilde {\mathcal A})\subset M$ is given by $\Gamma^j_{ik} = \mathring \Gamma^j{}_{ik} + T^j{}_{ik}$, so by the uniqueness of the affine connection, we must have $\tilde T^j{}_{ik}(x)=T^j{}_{ik}(x)$ for all $x\in\pi(\tilde {\mathcal A})$. Thus $T^j{}_{ik}$ and $\tilde T^j{}_{ik}$ are continuous maps on $M$ that coincide on the dense subset $\pi(\tilde {\mathcal A})\subset M$ and as such they must coincide on $M$. It follows that the affine connection on all of $M$ is given by $\Gamma^j_{ik} = \mathring \Gamma^j{}_{ik} + T^j{}_{ik}$, as desired.
\end{proof}
We can now extend the result of \Cref{eq:ab_cc_pre} in the opposite direction, reproducing the following well-known result.
\begin{cor}\label{cor:abmetric_same_apray_iff_cc}
    A properly Finslerian $(\alpha,\beta)$-metric has the same spray as $\alpha$ if and only if $\mathring\nabla_i b_j=0$.
\end{cor}
\begin{proof}
    One direction is given by \cref{eq:ab_cc_pre}. The other direction follows immediately by setting $T=0$ in the Berwald condition \eqref{ab_Berwald_condition} and performing a $y$-derivative, which is possible since the set $\{\phi'\neq 0\}\subset\mathcal A$ is open and also nonempty, since the latter would imply that $F$ is pseudo-Riemannian.
\end{proof}
\section{Important classes of \texorpdfstring{$(\alpha,\beta)$}{(\textalpha,\textbeta)}-metrics}

\subsection{Essential properties of \texorpdfstring{$\alpha$}{\textalpha}, \texorpdfstring{$\alpha^2$}{\textalpha\textasciicircum 2} and \texorpdfstring{$\beta$}{\textbeta}}

Our next goal is to apply \Cref{theor:ab_Berwald_condition_general} to some specific $(\alpha,\beta)$-metrics, both obtaining new results and reproducing some standard ones. In order to do so, we require several lemmas that concern the irrationality of the function $\alpha = \sqrt{|A|}$ and the irreducibly and divisibility of the polynomials $A$ and $\beta$, viewed as functions of $y$. We say that a polynomial is irreducible (or prime) if it cannot be written as a product of two polynomials both with degree $>0$. We will use, without proof, the well-known result (see e.g. \cite{Algebra}) that the ring of multivariate polynomials over any field, and in particular over $\R$, is a factorial ring, i.e. a unique factorization domain (UFD), meaning that any such polynomial can be expressed uniquely (up to some constants) as a product of irreducible polynomials. In this sense, the irreducible polynomials play a role analogous to that of the prime numbers in number theory. In particular, this implies that if a product $PQ$ of polynomials $P, Q$ is divisible by an irreducible polynomial $R$ then either $P$ or $Q$ must be divisible by $R$. 
\begin{lem}\label{lem:irreducible_A}
    If $a = a_{ij}\D x^i\D x^j$ is a pseudo-Riemannian metric on $M$ with $\dim M>2$ then for any fixed $x\in M$,  $A=a_{ij}y^iy^j$ is an irreducible polynomial in $y$.
\end{lem}
\begin{proof}
    Suppose that $A$ were not irreducible, i.e. that $A = PQ$, with $P,Q$ two polynomials of $\deg >0$. Since $\deg A=2$ this implies that $\deg P = \deg Q = 1$ and we may write $P = p_iy^i$ and $Q = q_iy^i$, since any constant terms in $P,Q$ would render the product $PQ$ inhomogeneous, whereas $A$ is homogeneous of degree $2$. Then differentiating $A=PQ$ twice yields $2a_{ij} = p_iq_j + q_ip_j$. This shows that $a_{ij}$ has at most rank 2, because the image of any vector $v^i$ will always be of the form $a_{ij}v^j = c_1 p_i + c_2q_i$ with $c_1,c_2\in\mathbb R$. Since we're assuming that $\dim M>2$ this implies that $a$ is degenerate and hence not a pseudo-Riemannian metric, which is a contradiction.
\end{proof}
Note that $\beta$ is irreducible as well since any linear polynomial is.
\begin{lem}\label{lem:irrational_alpha}
    Let $a = a_{ij}\D x^i\D x^j$ be a pseudo-Riemannian metric on $M$ with $n=\dim M>1$ and fix $x\in M$. Then the restriction of $\alpha = \sqrt{|a_{ij}(x)y^i y^j|}$ to any open subset in $y$-space $\R^n$ is an irrational function of $y$.
\end{lem}
\begin{proof}
    Fix some $x$ and suppose that there were some open subset $U$ of $y$-space on which $\alpha$ were rational in $y$, i.e. $\alpha = P/Q$, with $P,Q$ polynomials in $y$, where we will assume without of generality (WLOG) that the fraction $P/Q$ is in lowest form, i.e. $P$ and $Q$ have no common polynomial divisors of degree $>0$. Then $\alpha^2 = P^2/Q^2$.
    Note that $\alpha^2$ is polynomial on both of the open sets $U_\pm= U\cap \{\pm a_{ij}(x)y^i y^j >0\}$, and that $U_+\cup U_-$ is necessarily nonempty since otherwise $U$ could not be open. Suppose then WLOG that $U_+$ is nonempty (otherwise switch the labels $U_+$ and $U_-$). 
    
    Since  $\alpha^2$ is polynomial on $U_+$, and since $\alpha^2 = P^2/Q^2$, it follows that $P^2/Q^2$ must be polynomial on $U_+$ and hence (as a rational function) it must be polynomial on $\R^n$. Since $P$ and $Q$ and hence $P^2$ and $Q^2$ have no common polynomial divisors of degree $>0$  this implies that $Q$ must be of degree $0$, i.e. constant, and by a redefinition of $P$ we may then WLOG assume that $Q=1$. Then $\alpha^2 = P^2$, and similar to the proof of \Cref{lem:irreducible_A} we may write $P = p_i y^i$ and differentiate twice to reveal that $a_{ij} = p_ip_j$, which shows that $a_{ij}$ has at most rank 1. Since we're assuming that $\dim M>1$ this implies that $a$ is degenerate and hence not a pseudo-Riemannian metric, which is a contradiction.
\end{proof}
The phrase `restriction of $\alpha$ to any open subset' in \Cref{lem:irrational_alpha} is significant, since on its whole domain, $\alpha = \sqrt{|a_{ij}(x)y^i y^j|}$ would trivially be irrational as a consequence of the absolute value signs below the square root even if $a_{ij}$ were degenerate. For instance, $a_{ij} = c_ic_j$ would lead to $\alpha = |c_iy^i|$, which is irrational on $\R$ but rational on its restriction to the open subset $c_iy^i>0$. Such cases are excluded by the lemma.
\begin{lem}\label{lem:div_by_A_nor_beta}
    Fix $x\in M$ with $\dim M>2$, let $c_1,c_2\in \R$ and consider the polynomial $P=c_1 A+ c_2\beta^2$ in $y$, where $\beta$ is not identically vanishing at $x$. If $c_1\neq 0$ then $P$ is not divisible by $\beta$ and if $c_2\neq 0$ then $P$ is not divisible by $A$.
\end{lem}
\begin{proof}
    Suppose first for contradiction that $c_2\neq 0$ and $P$ \textit{is} divisible by $A$. Then $c_1 A+ c_2\beta^2 = c_3 A$ for some constant $c_3$ and hence $c_2\beta^2 = (c_3-c_1)A$, which implies, since $A$ is not divisible by $\beta$ by \Cref{lem:irreducible_A}, that $c_3 = c_1$, which implies that $c_2=0$, contrary to our assumption. Next, suppose for contradiction that $c_1\neq 0$ and $P$ is divisible by $\beta$. Then $c_1 A+ c_2\beta^2=\beta Q$ for some linear polynomial $Q$, which implies that $A = \tfrac{1}{c_1}(Q-c_2\beta )\beta$ and in particular implies that $A$ can be factorized into linear polynomials. This contradicts \Cref{lem:irreducible_A}, hence completing the proof.
\end{proof}
\subsection{Randers metrics and exponential metrics}

We now turn to some specific \ab-metrics, for which we will deduce, among other things, the precise Berwald condition, starting with the Randers metric \cite{Randers}, for which the result is well-known (see e.g. \cite{handbook_Finsler_vol2_matsumoto}).
\begin{prop}\label{prop:Randers_Berwald_cond}
    A Randers metric $F = \alpha+\beta$ with $\dim M>1$ is Berwald if and only if $\mathring\nabla_ib_j=0$. In that case, the affine connection of $F$ coincides with the Levi-Civita connection of $\alpha$.
\end{prop}
\begin{proof}
    For Randers metrics we have $\phi(s) = 1+s$, $\phi'(s) = 1$, so the Berwald condition can be written as
    \begin{align}
        y^j\mathring\nabla_i b_j - T^j{}_{ik}y^k b_j = T^j{}_{ik}y^k\frac{\epsilon y_j}{\alpha}.
    \end{align}
    If $\mathring\nabla_ib_j=0$ then the condition is clearly satisfied by $T^j{}_{ik}=0$ and hence $F$ is Berwald. On the other hand, suppose that $F$ is Berwald. Then the LHS of the Berwald condition is linear and, in particular, is a rational function in $y$, so the RHS must be as well. However, as the product of an irrational function $1/\alpha$ (by \Cref{lem:irrational_alpha}) with a (quadratic and hence) rational function, the RHS is irrational unless $T^j_{ik}y^ky_j = T_{jik}y^ky^j=0$, which implies, by taking two $y$-derivatives, that $T_{jik}+T_{kij}=0$. (Recall that indices are raised and lowered by $a_{ij}$, so $T_{jik} = a_{jl}T^l{}_{ik}$ depends only on $x$.) Christoffel's trick then shows that
    \begin{align}
        T_{jik} &= \tfrac{1}{2}\left[\left(T_{jik}+T_{kij}\right) + \left(T_{ikj}+T_{jki}\right) - \left(T_{kji}+T_{ijk}\right)\right] = 0,
    \end{align}
    Hence $y^j\mathring\nabla_i b_j=0$ and hence $\mathring\nabla_i b_j=0$. The last statement in the proposition follows immediately from the fact that $T^k_{ij}=0$.
\end{proof}
As a second quick application of \Cref{theor:ab_Berwald_condition_general2} we consider a kind of exponential \ab-metrics given by $\phi(s) = e^{cs^2/2}$ for some constant $c$. To the best of our knowledge, the following is a new result, first obtained in \citeref{ref1}{H1}.
\begin{prop}\label{prop:otherproof}
    If $b_i$ is a nowhere vanishing 1-form on $M$ and $c\neq 0$ then $F = \alpha e^{c (\beta/\alpha)^2/2}$ is of Berwald type if and only if $\mathring\nabla_i b_j = 0$.
\end{prop}
\begin{proof}
     Since $\phi'(s) = cs \phi = 0$ iff $s=0$ iff $\beta = 0$, which cannot hold on an open neighborhood of a point $(x,y)$ unless $b_i(x)=0$, the subset $\{\phi'=0\}\subset \mathcal A$ has empty interior and we may apply \cref{theor:ab_Berwald_condition_general2}. The Berwald condition reduces to
    \begin{align}
        y^j\mathring\nabla_i b_j - T^j{}_{ik}y^k b_j = \epsilon T^j{}_{ik}y^ky_j\left(\frac{\alpha^2 - c\beta^2}{c\,\alpha^2\beta}\right),
    \end{align}
    whenever $\beta \neq 0$. Now the implication to the left is obvious, for if $\mathring\nabla_i b_j = 0$ then the condition is satisfied with $T^j{}_{ik}=0$. Conversely, supposing that $F$ is Berwald, the linear LHS of the condition implies that the RHS must be linear as well, so that both $\alpha^2$ and $\beta$ must be divisors of the polynomial $T^j{}_{ik}y^ky_j\left(\alpha^2 - c\beta^2\right)$. It follows by \Cref{lem:div_by_A_nor_beta}, however, that neither of these is a divisor of $\left(\alpha^2 - c\beta^2\right)$. Hence both $\alpha^2$ and $\beta$ must be divisors of $T^j{}_{ik}y^ky_j$, but if $T^j{}_{ik}y^ky_j\neq 0$ this is impossible since the latter is only quadratic and not cubic or higher. We must therefore have $T^j{}_{ik}y^ky_j=0$. Hence, again taking two derivatives and applying Christoffel's trick as in the proof of \cref{prop:Randers_Berwald_cond}, we conclude that $T^j{}_{ik}=0$. Then it follows by \cref{cor:abmetric_same_apray_iff_cc} that $\mathring\nabla_i b_j = 0$, as desired.
\end{proof}
And as a final quick application before we move on to the more intricate case of generalized \mkrop metrics, we consider another type of exponential \ab-metric, given by $\phi(s) = e^{cs}$. We present it here for the first time.
\begin{prop}
    If $c\neq 0$ then $F = \alpha e^{c \beta/\alpha}$ is of Berwald type if and only if $\mathring\nabla_i b_j = 0$.
\end{prop}
\begin{proof}
     We start by noting that $\phi'(s) = c \phi \neq 0$, since otherwise the metric would be degenerate, by \eqref{eq:det_ab}. The Berwald condition can be written as
    \begin{align}
        (\epsilon c \alpha^2)(y^j\mathring\nabla_i b_j - T^j{}_{ik}y^k b_j) + c\beta T^j{}_{ik}y^ky_j = \alpha T^j{}_{ik}y^ky_j.
    \end{align}
    Again the implication to the left is obvious, so we focus on the implication to the right. For each $x$, the LHS is rational in $y$, whereas the RHS is irrational unless $T^j{}_{ik}y^ky_j=0$, by \cref{lem:irrational_alpha}. It follows that the equation can only be satisfied if $T^j{}_{ik}y^ky_j=0$. As we have seen twice before now, in the proofs of \cref{prop:Randers_Berwald_cond} and \cref{prop:otherproof}, this implies that $T^j{}_{ik}=0$ and hence that $\mathring\nabla_i b_j = 0$, by \cref{cor:abmetric_same_apray_iff_cc}.
\end{proof}

\subsection{Generalized \texorpdfstring{$m$}{m}-Kropina metrics}

Next, we consider a generalization of the $m$-Kropina metric that we will refer to as the \textit{generalized} $m$-Kropina metric, given by
\begin{align}
    F = \alpha\phi(\beta/\alpha), \qquad \phi(s) = \pm s^{-m}(c + d s^2)^{(m+1)/2}, \qquad m,c,d = \text{const.}
\end{align}
Such metrics were first introduced in \citeref{ref1}{H1} as the most general Berwald metrics compatible with a particular form of $T^j_{ik}$, as will be made precise in \Cref{theor:generalized_mKrop} below, and they reduce to the $m$-Kropina class whenever $\pm =+$, $c=1$ and $d=0$ and to the standard Kropina class \cite{Kropina} whenever, additionally, $m=1$. For generalized $m$-Kropina metrics we will always assume 
$\mathcal A$ to be contained in the subset of $TM$ characterized by $s>0$, $c+d s^2>0$, in order for any real power of those expressions to be defined (but note that this is strictly speaking not required if $m$ is, for instance, an integer). This implies, in particular, that we may assume without loss of generality that $b_i$ is nowhere vanishing since this would imply that $s=0$ at some point in $\mathcal A$. We will take this fact for granted in what follows. Our next main goal is to obtain the Berwald condition for generalized $m$-Kropina metrics. To this end, we start with a somewhat technical lemma, where we set $\tilde{\mathcal A} \equiv \mathcal A\setminus \{\phi'=0\}$.
\begin{lem}\label{lem:templem}
For a properly Finslerian generalized $m$-Kropina metric with $\dim M>1$ the following are true:
\begin{itemize}
    \item $cm\neq 0$;
    \item $\{\phi'=0\}\subset\mathcal A$ has empty interior;
    \item $\pi(\tilde{\mathcal A})=M$. In particular, $\tilde{\mathcal A}$ is a conic subbundle of $TM$.
\end{itemize}
\end{lem}
\begin{proof}
    We have $\phi = \pm s^{-m}(c + d s^2)^{(m+1)/2}$ and $\phi' = (d s^2-c m)s^{-1- m} (c + d s^2)^{(-1 + m)/2}$, so $\phi'=0$ iff $ds^2=c m$. Note that $c\neq 0$ (otherwise $F=c_1\beta$ for some constant $c_1$ and the fundamental tensor would be degenerate) and $m\neq 0$ (otherwise $F$ would not be properly Finslerian). Hence $cm\neq 0$. 
    
    We now claim that the set $\{\phi'=0\}\subset\mathcal A$ has empty interior. This is trivially true if $d=0$, so we focus on the case $d\neq 0$. Suppose the claim were false. Then there would be an open $U\subset \mathcal A$ such that $s^2=c m/d$ for all $(x,y)\in U$, or equivalently, $(b_iy^i)^2=(c m/d)|a_{ij}y^iy^j|$ for all $(x,y)\in U$. By differentiating twice this would imply that $b_ib_j=\pm(c m/d) a_{ij}$, which is impossible, since the LHS is degenerate, since $\dim M>1$, while the RHS is nondegenerate, thus proving the claim. 
    
    Finally, it remains to be shown that $\pi(\tilde{\mathcal A})=M$. Suppose this were not true. Then there exists $x_0\in M$ such that $\phi'(s(x_0,y))=0$ for all $(x_0,y)\in\mathcal A$, which would imply that $(b_iy^i)^2=(c m/d)|a_{ij}y^iy^j|$ for all $(x_0,y)\in \mathcal A$. Since the set of $y$ for which $(x_0,y)$ lies in the open set $\mathcal A$ is open, we can perform the same differentiation argument as before, leading to the contradiction $b_i(x_0)b_j(x_0)=\pm
    (c m/d) a_{ij}(x_0)$. Since $\tilde{\mathcal A}$ is also open and conic, this completes the proof.
\end{proof}
\begin{prop}\label{prop:generalized_mKtop_berw}
    A properly Finslerian generalized $m$-Kropina metric $F = \alpha\phi(\beta/\alpha)$, $\phi(s) = \pm s^{-m}(c + d s^2)^{(m+1)/2}$ with $n=\dim M>2$ is Berwald if and only if there exists a smooth vector field $f^i(x)$ on $M$ such that
    \begin{align}\label{eq:generalized_mKrop_Berwald_cond_prop}
       \hspace{-5px} \mathring\nabla_ib_j \!=   cm(f^kb_ k)a_{ij} \!+\! (c\!+\!\epsilon d |b|^2)f_i b_j \!-\! (cm \!-\! \epsilon d |b|^2) b_i f_j \!-\! \epsilon d (f^k b_k)b_i b_j.
    \end{align}
    In that case, the affine connection is given by 
    \begin{align}\label{eq:generalized_mKrop_Berwald_affine_connection}
        \Gamma^k_{ij} = \mathring \Gamma^k_{ij} + a^{k\ell}\left(f_ih_{j\ell} + f_jh_{i\ell} - f_\ell h_{ij}\right), \quad h_{ij} \coloneqq  \epsilon d b_i b_j-c m a_{ij}.
    \end{align}
    Whenever $|b|^2\neq 0$ and $c+\epsilon d |b|^2\neq 0$, the vector field $f^i$ can be expressed explicitly as
    \begin{align}
        f_i = \frac{\partial_i\ln||b|^2|}{2(c+\epsilon d |b|^2)}
    \end{align}
    and moreover, in this case, $\D f = 0$.
\end{prop}
\begin{proof}
    In view of \Cref{lem:templem} we may apply \Cref{theor:ab_Berwald_condition_general2}, which says that $F$ is Berwald if and only if there exists a symmetric smooth tensor field $T^j{}_{ik}(x)$ on $M$ such that the condition
    \begin{align}\label{eq:generalized_Berwald_cond_mKrop_pre}
        y^j\mathring\nabla_i b_j = T^j{}_{ik}y^k \left(b_j + \frac{\epsilon y_j}{\alpha}\frac{c(1+m)s}{d s^2 - c m} \right)
    \end{align}
   is satsfied on $\tilde{\mathcal A}$, i.e. whenever $\phi'\neq 0$, or equivalently,  $d s^2 - c m\neq 0$. So suppose that $F$ is Berwald, i.e. the latter is true. Equation \eqref{eq:generalized_Berwald_cond_mKrop_pre} is equivalent to
    \begin{align}
        (d s^2 - c m) \left( y^j\mathring\nabla_i b_j - T^j{}_{ik}y^k  b_j\right) =  \frac{\epsilon y_j}{\alpha}T^j{}_{ik}y^k c(1+m)s,
    \end{align}
     and substituting $s = \beta/\alpha$ and noting that $\epsilon\alpha^2 = A$, this is, in turn, equivalent to
      \begin{align}\label{eq:generalized_Berwald_cond_mKrop_pre2}
        (\epsilon d \beta^2 - c m A ) \left( y^j\mathring\nabla_i b_j - T^j{}_{ik}y^k  b_j\right) =    c(1+m) y_jT^j{}_{ik}y^k \beta.
    \end{align}     
    Since the RHS is divisible by $\beta$, the LHS must be as well. However, since $cm\neq 0$, it follows from \Cref{lem:div_by_A_nor_beta} that the first factor $(\epsilon d \beta^2 - c m A )$ of the LHS is not divisible by $\beta$, which is an irreducible polynomial. Hence the second factor $ y^j\nabla_i b_j - T^j{}_{ik}y^k  b_j$ must be divisible by $\beta$ for any $x\in \pi(\tilde{\mathcal A})$. But $\pi(\tilde{\mathcal A})=M$, by \Cref{lem:templem}, so there exists a vector field $\ell^i$ on $M$ such that
    \begin{align}
        \left( y^j\mathring\nabla_i b_j - T^j{}_{ik}y^k  b_j\right) = \ell_i \beta.
    \end{align}
    Note that there is \textit{a priori} no reason why $\ell_i$ should be smooth. Indeed, at this stage, all we can infer is that the product $\ell_i\beta$ must be smooth. We will see below that $\ell_i$ does turn out to be smooth, though. The formula above shows on the one hand, by taking a derivative, that
    \begin{align}\label{eq:generalized_mKrop_step}
        \mathring\nabla_i b_j = T^k_{ij}b_k + \ell_ib_j,
    \end{align}
    and on the other hand, by plugging the relation back into \eqref{eq:generalized_Berwald_cond_mKrop_pre2}, that
    \begin{align}
        (\epsilon d \beta^2 - c m A ) \ell_i =    c(1+m) y_jT^j{}_{ik}y^k,
    \end{align}
    or equivalently, 
    \begin{align}\label{eq:fsmo}
       T^j{}_{ik}y^ky_j \equiv T_{jik}y^ky^j =  f_i H,
    \end{align}
    where $\ell_i = c(1+m)f_i$ and $H = \epsilon d \beta^2-c m A$. We claim that this implies that $f^k$ is smooth. To see this, note that for every $x_0\in M$ there is a $y_0\in \R^n$ such that $(x_0,y_0)\in\mathcal A$ and $H(x_0,y_0)\neq 0$. For otherwise we would have $H(x_0,y)=0$ for all $y$ in some open set, so that, by taking two derivatives, we would obtain $\epsilon d b_ib_j=c m a_{ij}$, which is a contradiction as the LHS is degenerate while the RHS is nondegenerate. Since $H$ is continuous, it follows that there is an open neighborhood $U\subset\mathcal A$ of $(x_0,y_0)$ such that $H\neq 0$ on $U$. Then it follows from \eqref{eq:fsmo} that $f_i=(T^j{}_{ik}y^ky_j)/H$, viewed as a function of $x$ \textit{and} $y$, is smooth on $U$. In particular, $f_i$ is smooth at $(x_0,y_0)$, and since $f_i$ only has a dependence on $x$, not $y$, this means that $f_i$ is smooth at $(x_0,y)$ for any $y$, and since $x_0$ was arbitrary, that just means that $f_i$ is smooth as a function of $x$ and $y$, and hence it is smooth as a function on $M$. Hence $f^k$ is smooth, proving the claim. 
    Furthermore, taking two derivatives of \eqref{eq:fsmo} shows that 
    \begin{align}
        \tfrac{1}{2}\left(T_{jik}+T_{kij}\right) = f_ih_{jk},\qquad h_{ij} \coloneqq  \epsilon d b_i b_j-c m a_{ij},
    \end{align}
    and using Christoffel's trick, we find that
    \begin{align}
        T_{jik} &= \tfrac{1}{2}\left[\left(T_{jik}+T_{kij}\right) + \left(T_{ikj}+T_{jki}\right) - \left(T_{kji}+T_{ijk}\right)\right] \\
        &= f_ih_{jk} + f_kh_{ij} - f_jh_{ik}.\label{eq:generalized_mKropTtensor}
    \end{align}
    Plugging this into \eqref{eq:generalized_mKrop_step}, we find that
    \begin{align}
       \mathring\nabla_ib_j =   cm(f^kb_ k)a_{ij} + (c+\epsilon d |b|^2)f_i b_j - (cm \!-\! \epsilon d |b|^2) b_i f_j - \epsilon d (f^k b_k)b_i b_j.
    \end{align}
     Thus what we have shown so far is that if $F$ is Berwald then this condition must be satisfied on $\tilde{\mathcal A}$ for some smooth vector field $f^k$ on $M$. But since the equation does not depend on $y$ anymore and since $\pi(\tilde{\mathcal A})=M$, the condition must in fact hold for all $x\in M$. Conversely, 
    if this condition is satisfied for some smooth vector field $f^k$ on $M$ and all $x\in M$ then it is straightforward (although a little tedious) to check that \eqref{eq:generalized_Berwald_cond_mKrop_pre} is satisfied on $\tilde{\mathcal A}$. This proves the `if and only if' statement of the theorem. The formula for the Christoffel symbols follows directly from \eqref{eq:generalized_mKropTtensor} by raising the first index. \\
    Next, in order to derive the explicit formula for $f_i$, we contract \eqref{eq:generalized_mKrop_Berwald_cond_prop} with $b^j$, leading to
    \begin{align}
        \tfrac{1}{2}\partial_i |b|^2 =  (c+\epsilon d |b|^2)|b|^2f_i,  
    \end{align}
    which leads to the desired formula provided that $|b|^2\neq 0 $ and $c+\epsilon d |b|^2\neq 0$. Finally, it is now easy to check that $\partial_i f_j = \partial_j f_i$, meaning that $f$ is in this case necessarily closed, $\D f = 0$.
\end{proof}
In the special case where the 1-form is known to be closed the Berwald condition simplifies considerably. 
\begin{prop}\label{prop:gen_mKrop_Berwald_closed}
    A properly Finslerian generalized $m$-Kropina metric $F = \alpha\phi(\beta/\alpha)$, $\phi(s) = \pm s^{-m}(c + d s^2)^{(m+1)/2}$ with closed 1-form $\beta$ and with $n=\dim M>2$ is Berwald if and only if there exists a smooth function $p:M\to \R$ such that
    \begin{align}\label{eq:generalized_mKrop_Berwald_cond_prop_closed}
        \mathring\nabla_i b_j = p\left\{cm|b|^2 a_{ij} + \left[c(1-m) + \epsilon d |b|^2\right]b_i b_j\right\}
    \end{align}
    In that case, the affine connection is given by 
    \begin{align}\label{eq:generalized_mKrop_Berwald_affine_connection_closed}
        \Gamma^k_{ij} = {}^\alpha\Gamma^k_{ij} + p\left[\epsilon d b_i b_j b^k - cm\left(b_i\delta^k_j + b_j\delta^k_i - b^ka_{ij}\right)\right].
    \end{align}
    Conversely, if \eqref{eq:generalized_mKrop_Berwald_cond_prop_closed} holds then $\beta$ is closed.
\end{prop}
\begin{proof}
    $F$ is Berwald iff the Berwald condition \eqref{eq:generalized_mKrop_Berwald_cond_prop} holds for some smooth $f^k$. The Berwald condition implies that $(\D b)(\partial_i,\partial_j)= \partial_i b_j -\partial_j b_i = \mathring\nabla_i b_j - \mathring\nabla_j b_i = c(1+m)(f_i b_j - f_j b_i)$, so since $b_i$ is closed
    , this expression vanishes and hence $f_i b_j = f_j b_i$ must hold for all $i,j$, which is only possible if $f_i$ is proportional to $b_i$ in the sense that $f_k = pb_k$ for some function $p$ on $M$ (this can be checked easily at any given point in $M$ by choosing coordinates in which $b_i$ has only one nonvanishing component at that point). In this case \eqref{eq:generalized_mKrop_Berwald_cond_prop} reduces to \eqref{eq:generalized_mKrop_Berwald_cond_prop_closed} and \eqref{eq:generalized_mKrop_Berwald_affine_connection} reduces to \eqref{eq:generalized_mKrop_Berwald_affine_connection_closed}. Moreover, since $f_k$ is smooth, $pb_k$ must be smooth, and this implies that $p$ is smooth whenever $b_k\neq 0$, and since $b_k$ is nowhere vanishing, this implies that $p$ is smooth on $M$. The opposite holds trivially as well: \eqref{eq:generalized_mKrop_Berwald_cond_prop_closed} implies that $\partial_i b_j -\partial_j b_i = \mathring\nabla_i b_j - \mathring\nabla_j b_i=0$ and hence $\beta$ is closed. 
\end{proof}
The Berwald condition for $m$-Kropina spaces follows immediately by setting $\pm = +, c=1, d=0$. Again, under the additional assumption that the 1-form is closed, this condition simplifies considerably.
%
%
\begin{cor}\label{cor:mKrop_berwald_conditions}
Let $F= \alpha^{1+m}\beta^{-m}$ be a properly Finslerian $m$-Kropina metric on a manifold $M$ with dimension greater than two.
    \begin{itemize}
        \item $F$ is of Berwald type if and only if there exists a smooth vector field $f^i$ on $M$ satisfying 
        \begin{align}\label{eq:VGR_Berwald_new2}
            \mathring\nabla_j b_i = m (f_k b^k)a_{ij} + b_i f_j  - m f_i b_j.
        \end{align}
        In this case, the affine connection is given by
        \begin{align}\label{eq:affine_conn_general}
            \Gamma^\ell_{ij} = \mathring\Gamma^\ell_{ij} + m a^{\ell k}\left(a_{ij}f_k - a_{jk}f_i - a_{ki}f_j\right).
        \end{align}
        \item If $\beta$ is closed, then $F$ is Berwald if and only if there exists a smooth function $c:M\to \R$ such that
        \begin{align}\label{eq:VGR_Berwald_old2}
            \mathring\nabla_j b_i = c \left[ m|b|^2 a_{ij} + (1-m)b_i b_j\right],
        \end{align}
        In this case, the affine connection is given by
        \begin{align}\label{eq:affine_conn}
            \Gamma^\ell_{ij} = \mathring\Gamma^\ell_{ij} + m c \left(a_{ij}b^\ell - \delta^\ell_j b_i - \delta^\ell_i b_j\right),
        \end{align}
        \item Conversely, \eqref{eq:VGR_Berwald_old2} implies that $\beta$ must be closed.
    \end{itemize}
\end{cor}
The condition \eqref{eq:VGR_Berwald_new2} has also been obtained in \cite{handbook_Finsler_vol2_matsumoto} under the assumptions $|b|^2\neq 0$ (essentially replacing our assumption that $\dim M>2$) and $A>0$. The less general condition \eqref{eq:VGR_Berwald_old2} was obtained in \cite{Fuster:2018djw}, where it was believed to be the general Berwald condition due to some details that were overlooked. In our paper \citeref{ref2}{H3} we identified the error in the proof in \cite{Fuster:2018djw} and corrected it; we refer to \citeref{ref2}{H3} for the details of that discussion. We remark that our $c$ in \eqref{eq:VGR_Berwald_old2} is related to $C(x)$ in \cite{Fuster:2018djw} by $C(x) = (1+m)c/2$ and that our power $m$ is related to the power $n$ in \cite{Fuster:2018djw} by $n = -2m/(1+m)$.
 

Our next aim is to prove \Cref{theor:generalized_mKrop}, which may be viewed in some sense as a generalization of \Cref{cor:abmetric_same_apray_iff_cc}. It shows that generalized $m$-Kropina metrics are the most general $(\alpha,\beta)$-metrics that are consistent with a certain form of the affine connection. In the version of the theorem we prove below, we will assume that $\phi'\neq 0$ on $\mathcal A$. However, this assumption is superfluous and the theorem can be proven also without it. Since the proof below is already quite lengthy, we ignore such generalizations for the moment and refer instead to \cref{sec:mKropGenerality} for details. In what follows, we will denote by $s$ both the function $s:(x,y)\mapsto s(x,y)=\beta(x,y)/\alpha(x,y)$ as well as its function values $s\in \R$. So we will write, for instance, that $s\in s(U)$ if the \textit{value} $s$ lies in the image of the set $U$ under the \textit{function} $s$. Moreover, if $A\subset X$ is a subset of a topological space $X$, we will denote by $A^0$ and $\overline{A}$ its (topological) interior and closure, respectively. We start with a technical but important lemma.
\begin{lem}\label{lem:top}
    Suppose that $b_i$ is nowhere vanishing. If $U\subset\mathcal A$ is open and connected then the image $s(U)$ is an interval with nonempty interior and hence it satisfies, in particular, $s(U)\subset \overline{s(U)^0\setminus C}$ for any finite set $C$. 
\end{lem}
\begin{proof}
    Since $U$ is connected and $s$ continuous, $s(U)$ is a connected subset of $\R$ and hence it is an interval $s(U)=I\subset \R$. Furthermore, it is an interval containing more than one point. For suppose it contained only one point $s_0$. Then we would have $\beta^2 = s_0\alpha^2 = s_0 |A|$ for all $(x,y)\in U$. Now fix some $x_0$ such that $(x_0,y)\in U$ and define $V_\pm = \{ y\in\R^n\,:\, (x_0,y)\in U, \pm A(x_0,y)>0\}$. Then at least one of $V_+$ and $V_-$ is open and nonempty. Assume WLOG that $V_+$ is open and nonempty. Then $\beta(x_0,y)^2 = s_0 A(x_0,y)$ for all $y\in V_+$ and hence this equation between polynomials must hold for all $y\in\R^n$. Since $0\neq \mathcal A$, by assumption, $s_0\neq 0$. But then $s_0^{-1}\beta(x_0,y)^2 = A(x_0,y)$ and so $A$ is divisible by $\beta$, contradicting \cref{lem:irreducible_A}. %
    %
    %
    Hence $I$ is an interval that contains at least two points and as such, the interior $I^0$ of $I$ must be a nonempty open interval. Removing a finite set of points $C$ from $I^0$ thus yields a union $I^0\setminus C = \cup_{j=1}^k (a_j,b_j)$ of open intervals with
    \begin{align}
        a_1<b_1=a_2<b_2=a_2<\dots <b_{k-1}=a_k<b_k.
    \end{align}
    It is then clear that $\overline{I^0\setminus C} = [a_1,b_k]$ and hence that $I\subset \overline{I^0\setminus C}$, completing the proof.
\end{proof}
As a result, if some ODE for $\phi(s)$ is satisfied for all $s\in s(U)\setminus C$, where $U\subset \mathcal A$ is open and $C$ a finite set, then the ODE may be solved (explicitly) on the open set $s(U)^0\setminus C$ using standard methods, and this uniquely determines $\phi$ on all of $s(U)$ by continuous extension.
 \begin{theor}\label{theor:generalized_mKrop}
     Let $F = \alpha\phi(\beta/\alpha)$ be a properly Finslerian $(\alpha,\beta)$-metric with $\dim M>2$. Suppose furthermore that $\mathcal A$ is connected and that $\phi'\neq 0$ on $\mathcal A$. Then TFAE:
     \begin{enumerate}[(i)]
         \item F is Berwald and there exist nowhere vanishing $\lambda,\rho,\sigma\in C^\infty(M)$ such that
         \begin{align}\label{eq:TtensorTheorem}
             T^k{}_{ij} = \lambda b^k b_i b_j + \rho\left(b_ i\delta^k_j+ b_ j\delta^k_i \right) + \sigma b^k a_{ij};
         \end{align}
         \item $\beta$ is closed and there are nonvanishing constants $c,d,m$ and a nowhere vanishing $p\in C^\infty(M)$ such that         
         \begin{align}
         \phi(s) &= \pm s^{-m}(c + d s^2)^{(m+1)/2}\\
        \mathring\nabla_i b_j &= p\left\{cm|b|^2 a_{ij} + \left[c(1-m) + \epsilon d |b|^2\right]b_i b_j\right\} \label{eq:generalized_mKrop_Berwald_cond_theor_closed}
    \end{align}
     \end{enumerate}
     In that case $\rho = -\sigma = -cmp$ and $\lambda=\epsilon d p$.
 \end{theor}
 \begin{proof}
     Assuming (ii) then, according to \Cref{prop:gen_mKrop_Berwald_closed}, $F$ is Berwald and the affine connection is given by \eqref{eq:generalized_mKrop_Berwald_affine_connection_closed} and from this we infer that $T^k_{ij}$ is clearly of the form \eqref{eq:TtensorTheorem} with $\lambda = \epsilon d p$ and $\rho = -\sigma = -c m p$. Next, suppose that $(i)$ holds. Then the Berwald condition \eqref{ab_Berwald_condition} must be satisfied on all of $\mathcal A$ (since $\phi'$ never vanishes) with $T^k_{ij}$ given by \eqref{eq:TtensorTheorem}, in which case this condition reduces to
     \begin{align}
         y^j\mathring\nabla_i b_j = \lambda |b|^2 \beta b_i + 2\rho \beta b_i + \sigma |b|^2 y_i + \frac{\epsilon \psi}{\alpha}\left( \lambda\beta^2 b_i + \rho(\beta y_i + A b_i) +\sigma \beta y_i \right),
     \end{align}
     where $\psi = \phi/\phi' - s$. We can write this also as
     \begin{align}\label{eq:to_subst_into}
         y^j\mathring\nabla_i b_j - \left( \lambda |b|^2 \beta b_i + 2\rho \beta b_i + \sigma |b|^2 y_i\right) = \frac{\epsilon \psi}{\alpha}\left( \lambda\beta^2 + \rho A \right) b_i +  \frac{\epsilon \psi}{\alpha}\left( \rho + \sigma\right)\beta y_i,
     \end{align}
     where we have collected all manifestly linear terms on the LHS and grouped the RHS by a $b_i$ term and a $y_i$ term. Since the LHS is linear, the RHS must be so as well, and in fact, by \Cref{prop:lemapp} this can only be achieved if both terms on the RHS are linear, i.e. if  $\frac{\epsilon \psi}{\alpha}\left( \rho + \sigma\right)\beta$ is independent of $y$ and if $\frac{\epsilon \psi}{\alpha}\left( \lambda\beta^2 + \rho A \right)$ is linear. Since $A=0$ is excluded, by definition, from $\mathcal A$ and since $\mathcal A$ has only a single connected component, by assumption, $\mathcal A$ must be contained in one of the open sets $A>0$ or $A<0$. And that, in particular, implies that $\epsilon=\sgn(A)$ is just a constant on $\mathcal A$. Thus we need $\frac{\psi}{\alpha}\left( \rho + \sigma\right)\beta$ to be independent of $y$ and $\frac{ \psi}{\alpha}\left( \lambda\beta^2 + \rho A \right)$ to be linear.\\
     
     \noindent\textbf{Step 1) Claim:} $\rho = -\sigma$\\
     First we consider the $y_i$ term, which is linear in $y$ iff $\psi\left( \rho + \sigma\right)\beta/\alpha$ is independent of $y$. The only way this can be true, assuming $F$ is properly Finslerian, is if $\rho = -\sigma$. To see this, suppose that $\rho\neq -\sigma$. Then we must have $ \psi\beta/\alpha = \psi s\eqqcolon\ell$ where $\ell$ is a priori a function of $x$, but since the LHS depends only on $s$, $\ell$ must actually be constant. Hence $\psi = \phi/\phi ' - s = \ell/s$ for all $(x,y)\in \mathcal A$ with $s\neq  0$. As a differential equation in the variable $s$ this must hold for all $s\in s(\mathcal A)\setminus\{0\}$. Since $\mathcal A$ is open and connected, we can apply \Cref{lem:top} and its consequences: we can integrate the differential equation for $\phi(s)$ on the interior of $s(\mathcal A)\setminus\{0\}$ and then extend the solution by continuity to all of $s(\mathcal A)$, i.e. for all relevant values of $s$.
     
     We will now apply this scheme. Assuming WLOG that $\ell\neq 0$ (for otherwise $\phi -s \phi '=0$ and $\det g_{ij}=0$, by \eqref{eq:det_ab}) we integrate to find $\ln |\phi| = \int \left(s + \ell/s\right)^{-1}\D s= \ln(|c(\ell + s^2)|)/2$, where $c$ is an integration constant. Then $\phi = \sqrt{c(\ell + s^2)}$ for $s$ in the interior of $s(\mathcal A)\setminus\{0\}$, and by the argument given above, $\phi$ must in fact have this form for all relevant values of $s$. Hence $F^2 = \alpha^2 \phi^2 = c(\ell\alpha^2 + \beta^2)$ is pseudo-Riemannian. This is excluded in the premise of the theorem and hence it is a contradiction, so we must indeed have $\rho = -\sigma$.\\

     \noindent\textbf{Step 2) Deriving an ODE in $s$}\\
     Next, we consider the $b_i$ term, setting $\sigma=-\rho$. This term is linear iff the coefficient of $b_i$ is linear, which is the case iff 
     \begin{align}\label{eq:second_der_for_linearity}
         \bar\partial_i\bar\partial_j\left[\frac{\psi}{\alpha}\left( \lambda\beta^2 + \rho A \right)\right] =0.
     \end{align}
     For the first derivative, we have
     \begin{align}
         \bar\partial_i\left[\frac{\psi}{\alpha}\left( \lambda\beta^2 + \rho A \right)\right] &= \frac{\psi'\bar\partial_i s}{\alpha}\left( \lambda\beta^2 + \rho A \right) - \frac{\psi}{\alpha^2}\bar\partial_i\alpha\left( \lambda\beta^2 + \rho A \right) \\
         &\qquad+ \frac{\psi}{\alpha}\left( 2\lambda\beta b_i + 2\rho y_i \right) \\
         &\hspace{-60px}=  \frac{\psi'}{\alpha^2}\left(b_i - \frac{\beta y_i}{A}\right)\left( \lambda\beta^2 + \rho A \right) - \frac{\epsilon \psi y_i}{\alpha^3}\left( \lambda\beta^2 + \rho A \right)  \\
         &\qquad+ \frac{\psi}{\alpha}\left( 2\lambda\beta b_i+ 2\rho y_i \right) \\
         &\hspace{-60px}=  \frac{1}{\alpha^2}\left[-\left(\frac{\beta \psi'}{A} + \frac{\epsilon \psi}{\alpha}\right)\left( \lambda\beta^2 + \rho A \right)  + 2\alpha \rho\psi\right] y_i \\
         &\qquad+ \{b_i\text{ terms}\},\label{eq:first_der_for_linearity}
     \end{align}
     where we have used the identities
     \begin{align}
        \bar\partial_i \alpha = \frac{\epsilon y_i}{\alpha} ,\qquad \bar\partial_i s = \frac{1}{\alpha}\left(b_i - \frac{\beta y_i}{A}\right),
     \end{align}
     and separated the $y_i$ and the $b_i$ parts, the latter of which are irrelevant. The reason for this is the following. By inspecting the situation (or working it out exactly) it is clear that the second derivative in \eqref{eq:second_der_for_linearity} will be of the form $f a_{ij} + g h_{ij}$, where $f,g$ are functions and $h_{ij}$ is a linear combination of $b_i b_j, b_i y_j, y_i b_j$ and $y_i y_j$. Hence $h_{ij}$ has at most rank 2, which is strictly less than the rank of $a_{ij}$. That means that \eqref{eq:second_der_for_linearity}, i.e. the equation $f a_{ij} + g h_{ij}=0$, can only hold if $f=0$, for otherwise we could write $a_{ij} =(g/f)h_{ij}$ and $a_{ij}$ and $h_{ij}$ would have the same rank. It turns out that it suffices for what we aim to prove to consider this coefficient $f$, as we will see shortly. Differentiation of the $b_i$ terms in \eqref{eq:first_der_for_linearity}, however, will never yield terms proportional to $a_{ij}$. In fact, the only way we get terms proportional to $a_{ij}$ is by directly differentiating $y_i$ in \eqref{eq:first_der_for_linearity}. No other terms resulting from the product rule are proportional to $a_{ij}$ either. This  simplifies this calculation enormously because it means that in order to get the $a_{ij}$-coefficient $f$ from \eqref{eq:first_der_for_linearity} we effectively need only replace $y_i$ by its derivative $\bar\partial_j y_i = a_{ij}$ to obtain
     \begin{align}
         \bar\partial_i\bar\partial_j\left[\frac{\psi}{\alpha}\left( \lambda\beta^2 + \rho A \right)\right] =\qquad\qquad\qquad\qquad\qquad\qquad\qquad\qquad\qquad\\
          \frac{1}{\alpha^2}\left[-\left(\frac{\beta \psi'}{A} + \frac{\epsilon \psi}{\alpha}\right)\left( \lambda\beta^2 + \rho A \right)  + 2\alpha \rho\psi\right]a_{ij} + g h_{ij}.
     \end{align}
     As we argued above, the vanishing of this implies that we must have
     \begin{align}
          0&=\frac{1}{\alpha}\left[-\left(\frac{\beta \psi'}{A} + \frac{\epsilon \psi}{\alpha}\right)\left( \lambda\beta^2 + \rho A \right)  + 2\alpha \rho\psi\right]\\
          &= -\left(\frac{s \psi'}{\epsilon \alpha^2} + \frac{\epsilon \psi}{\alpha^2}\right)\left( \lambda\beta^2 + \rho A \right)  + 2 \rho\psi\\
          &= -\epsilon\left(s \psi'+ \psi\right)\left( \lambda s^2 + \rho \epsilon \right)  + 2 \rho\psi, 
     \end{align}
     or, after rewriting,
     \begin{align}\label{eq:ODE_for_psi}
         \frac{\psi'}{\psi} = \frac{ \eta- s^2}{ s( s^2 + \eta )}, \qquad \eta \equiv \epsilon\rho/\lambda,
     \end{align}
     whenever $s\neq 0$ and $s^2+\eta\neq 0$. Note that $\psi$ never vanishes for otherwise  $\phi -s \phi '=0$ and $\det g_{ij}=0$, by \eqref{eq:det_ab}. As before, since the LHS depends only on $s$, the RHS should also not depend on $x$ explicitly, which implies that $\eta$, which is a priori a function on $M$, should actually be a constant\footnote{We can invert \eqref{eq:ODE_for_psi} to obtain $\eta$ as a function of $s$, except for values of $s$ that satisfy $\psi+s\psi'=0$, which can only constitute a set with empty interior, for otherwise $F$ would be pseudo-Riemannian. Hence, by continuous extension, the formula for $\eta(s)$ must also hold on this set.}. Hence \eqref{eq:ODE_for_psi} is an ODE for $\psi(s)$ that should be satisfied for all for all $s\in K\equiv s(\mathcal A)\setminus\{\sqrt{-\eta},0,-\sqrt{-\eta}\}$. \\
     
     \noindent\textbf{Step 3) Solving the ODE to obtain $\phi(s)$}\\
     In analogy with how we solved the differential equation for $\phi$ in step 1 of this proof, this one is solved uniquely, for all $s\in K^0$, by 
     \begin{align}\label{eq:psi_unique_sol}
         \psi = \frac{\tilde c s}{s^2 + \eta},
     \end{align}
      where $\tilde c$ is an integration constant. We find $\phi$ on $K^0$ via $\psi = \phi/\phi' - s$. This can now be written as
     \begin{align}
         \frac{\phi'}{\phi} = \left(\frac{\tilde cs}{s^2 + \eta}+s\right)^{-1} = \frac{d s^2 -  c m}{d s^3 + c s},
     \end{align}
     where we have introduced an arbitrary constant $d$ and defined $ c = d(\tilde c+\eta)$ and $m = - d \eta/ c = -\eta/(\eta+\tilde  c)$. This is uniquely solved by
     \begin{align}
         \phi(s) = \tilde d s^{-m}( c+d s^2)^{(m+1)/2},
     \end{align}
     where we may absorb the integration constant $\tilde d$ (up to sign) into $c$ and $d$. Hence $\phi$ attains the desired form
     \begin{align}\label{eq:pghut}
         \phi(s) = \pm s^{-m}( c+d s^2)^{(m+1)/2}
     \end{align}
     for all $s\in K$. Once again, by \Cref{lem:top}, we can extend the solution by continuity to all relevant values of $s$. In other words, $\phi$ has the form \eqref{eq:pghut} for all relevant values of $s$, as desired. In principle, the constants $c,d,m$ and the sign $\pm$ could have different values on different connected components of $s(\mathcal A$), but since $\mathcal A$ is assumed to be connected, $s(\mathcal A)$ is also connected and hence the constants and the sign are actually fixed over all of $s(\mathcal A)$.\\
     %
     %
     %
     
     \noindent\textbf{Step 4) The condition on $\mathring\nabla_i b_j$}\\
     Finally we substitute the form \eqref{eq:psi_unique_sol} of $\psi$ and the relation between $\rho,\sigma$ and $\lambda$ into the Berwald condition \eqref{eq:to_subst_into}, leading to
     \begin{align}\label{eq:consistency_check}
         y^j\mathring\nabla_i b_j &= \left( \lambda |b|^2 \beta b_i + 2\rho \beta b_i + \sigma |b|^2 y_i\right) + \epsilon \lambda \tilde c\beta b_i,
     \end{align}
     from which we can infer by differentiation that for $\mathring\nabla_i b_j$ is symmetric under $i\leftrightarrow j$. This implies that the 1-form $\beta$ is closed and hence, by \Cref{prop:gen_mKrop_Berwald_closed}, that the desired Berwald condition holds. This completes the proof. 
 \end{proof}
 \begin{rem}
      As a consistency check for the last part of the proof, we may also derive the desired Berwald condition directly from \eqref{eq:consistency_check}. We note that the definitions of $c$ and $m$ in terms of $\tilde c$ and $\eta$ above may be inverted as $\tilde c = c(1+m)/d$ and $\eta  = -mc/d$. Using also the relations between $\lambda,\rho,\sigma$, we find that \eqref{eq:to_subst_into} turns into
     \begin{align}
         y^j\mathring\nabla_i b_j &= \left( \lambda |b|^2 \beta b_i + 2\beta b_i + \sigma |b|^2 y_i\right) + \epsilon \lambda \tilde c\beta b_i \\
         &= \left( \lambda |b|^2 \beta b_i + 2\epsilon\eta\lambda \beta b_i - \epsilon\eta\lambda  |b|^2 y_i\right) + \epsilon \lambda \frac{c(1+m)}{d}\beta b_i\\
         &=  \lambda\left[\left( |b|^2   + 2\epsilon\eta    + \epsilon  \frac{c(1+m)}{d} \right) \beta b_i -\epsilon\eta |b|^2 y_i\right]\\
         &=  \lambda\left[\left( |b|^2   - \frac{2\epsilon mc}{d}    +   \frac{\epsilon c(1+m)}{d} \right) \beta b_i +\frac{\epsilon mc}{d} |b|^2 y_i\right]\\
         %
         &=  \lambda\left[\left( |b|^2   + \frac{\epsilon c}{d}\left(1-m\right) \right) \beta b_i -\epsilon\frac{-mc}{d} |b|^2 y_i\right],
     \end{align}
     from which we can infer by differentiation that
     \begin{align}
         \mathring\nabla_i b_j &= \frac{\epsilon \lambda}{d}\left\{\left[ \epsilon d |b|^2   +  c\left(1-m\right) \right]  b_ib_j + mc |b|^2 a_{ij}\right\},
     \end{align}
     which is precisely \eqref{eq:generalized_mKrop_Berwald_cond_theor_closed} provided we identify $p=\epsilon \lambda/d$.
 \end{rem}
%
%
%
%
%
We now turn to an explicit nontrivial example of a class of Berwald spaces that will turn out to be more general than it may seem, as we will see shortly in \Cref{prop:closed_null_gen_mKrop}. We will come back to the special case where $F$ is $m$-Kropina (i.e. $c=1,d=0$) in detail in the next chapter. In what follows we will use the convention that indices $i,j\dots$ run from $1$ to $n$, whereas indices $a,b\dots$ run from $3$ to $n$.
%
%

%
%
%
\begin{ex}[A family of generalized $m$-Kropina spaces of Berwald type]\label{ex:Berwald_spacetime}
    Consider a generalized $m$-Kropina metric 
    \begin{align}
        F = \alpha\phi(\beta/\alpha),\qquad\qquad \phi(s) = \pm s^{-m}(c + d s^2)^{(m+1)/2},
    \end{align}
    with $m\neq 1$,  $\alpha$ given by
	\begin{align}\label{Kundt_ex_Berwald}
	a= -2\D u\D v + H(u,v,x)\; \D u^2 + 2W_a(u,x)\,\D x^a\D u +h_{ab}(u,x) \, \D x^a \D x^b,
	\end{align}
	and $1$-form $\beta=\D u$, expressed in coordinates $(u,v,x)\coloneqq (u,v,x^3,\dots, x^n)$, where $H$, $W_a$, $a=3,\dots n$ are arbitrary smooth functions and $h_{ab}$ is a pseudo-Riemannian metric of dimension $n-2$. 
    In the special case that $h_{ab}$ and hence $a_{ij}$ is Lorentzian, metrics of the form \eqref{Kundt_ex_Berwald} are known as Kundt metrics \cite{Kundt1961,Kundt:1962svr,stephani_kramer_maccallum_hoenselaers_herlt_2003}, see also \cref{sec:Kundt_sort_of}. The 1-form $\beta$ is null w.r.t. $\alpha$ and the only (possibly) nonvanishing component of the covariant derivative of $\beta$ is:
	\begin{align}
	\mathring\nabla_ub_u = -\tfrac{1}{2}\partial_v H\,.
	\end{align}
	Hence condition \eqref{eq:generalized_mKrop_Berwald_cond_prop_closed} is satisfied with $|b|^2 =0$ and $p = -\partial_v H/2c(1-m)$ and hence this geometry is Berwald. The affine connection is determined by the tensor $T^k_{ij}$, which is of the form \eqref{eq:TtensorTheorem} with coefficients given by
    \begin{align}
        \lambda = \frac{-\epsilon d}{2c(1-m)}\partial_v H, \qquad \rho = - \sigma = \frac{m}{2(1-m)}\partial_v H.
    \end{align}
\end{ex}
In fact, as shown by the following proposition, this example exhausts essentially all possible generalized $m$-Kropina spaces of Berwald type with closed null 1-form.
\begin{prop}\label{prop:closed_null_gen_mKrop}
    Let $F = \alpha\phi(\beta/\alpha)$, $\phi(s) = \pm s^{-m}(c + d s^2)^{(m+1)/2}$ be a generalized $m$-Kropina space with $n>2$, $m\neq 1$ and suppose that $\beta$ is a closed null 1-form. TFAE:
    \begin{enumerate}[(i)]
        \item F is Berwald
        \item $\mathring\nabla_i b_j = p \,b_i b_j$ for a smooth function $p$;
        \item $F$ is given, locally, by the Finsler metric of \Cref{ex:Berwald_spacetime};
    \end{enumerate}
    In this case $p =- \tfrac{1}{2}\partial_v H$ and 
    \begin{align}
            \Gamma^\ell_{ij} &= \mathring\Gamma^\ell_{ij} + \frac{m p}{1-m} \left(a_{ij}b^\ell - \delta^\ell_j b_i - \delta^\ell_i b_j\right).
    \end{align}
\end{prop}
\begin{proof}
    Setting $|b|^2=0$ in \Cref{prop:gen_mKrop_Berwald_closed} it follows that $F$ is Berwald iff there exists a smooth function $p$ on $M$ such that $\mathring\nabla_i b_j = pc(1-m) b_i b_j$. Since $c\neq 0$ by \cref{lem:templem}, and since we have assumed that $(1-m)\neq 0$, this is equivalent to condition (ii), proving the equivalence between (i) and (ii). The implication (iii)$\Rightarrow$(ii) is provided by \Cref{ex:Berwald_spacetime}, and finally, the implication (ii)$\Rightarrow$(iii) follows by \Cref{lem:coordinates} below.
\end{proof}
In contrast to most of the results in the chapter, \cref{prop:closed_null_gen_mKrop} as well as \cref{lem:coordinates} below are based on \citeref{ref2}{H3} rather than \citeref{ref1}{H1}. 
\begin{lem}\label{lem:coordinates}
    Let $n>2$ and let $a$ be a \psR metric and $b$ a nowhere vanishing 1-form. If $|b|^2=0$ and $\mathring\nabla_i b_j = p \,b_i b_j$ for some function $p$ then around each point in $M$ there exist local coordinates $(u,v,x^3,\dots,x^n)$ such that $b = \D u$ and
    \begin{align}
    {\normalfont a = -2\D u\D v + H(u,v,x)\D u^2 + 2W_a(u,x)\D u\D x^a + h_{ab}(u,x)\D x^a\D x^b},
    \end{align}
    with $h$ some pseudo-Riemannian metric of dimension $n-2$, and $p$ is then given by $p =- \tfrac{1}{2}\partial_v H$.
\end{lem}
\begin{proof}
    Because $\mathring\nabla_i b_j$ is symmetric, $b$ is closed and hence locally exact, and, since $|b|^2=0$, by assumption, it follows by \Cref{prop:metric_w_closed_null_1form} that the metric $a$ can be written in the form
    \begin{align}
        a = -2\D u\D v + H\D u^2 + 2W_b\D u\D x^b + h_{bc}\D x^b\D x^c.
    \end{align}
    It remains to be shown that the functions $W_a$ and $h_{ab}$ do not depend on coordinate $v$. To this end, we employ on the one hand the condition $\mathring\nabla_i b_j = p\,b_ib_j = p\, \delta_i^u \delta_j^u$ expressed in our preferred coordinates, and on the other hand the expression for $\mathring\nabla_i b_j$ computed explicitly in the preferred coordinates. Using the fact that $b_i = \delta^u_i$ and $a^{ui} = -\delta^i_v$ and $a_{iv} = 0$, we find after a straightforward computation that the latter is given by $\mathring\nabla_ib_j  = -\frac{1}{2}\frac{\partial a_{ij}}{\partial v}$, thus leading to the requirement that 
    \begin{align}
    p\, \delta_i^u \delta_j^u = -\frac{1}{2}\frac{\partial a_{ij}}{\partial v}.
    \end{align}
    Plugging in the explicit form of $a_{ij}$ in our preferred coordinates, this
    leads to
    \begin{align}
    p =- \tfrac{1}{2}\partial_v H, \qquad \partial_v W_a = \partial_vh_{ab} = 0,
    \end{align}
    as desired.
\end{proof}
As a final remark, we note that some of the results obtained in this chapter for Berwald spaces can easily be generalized to so-called \textit{generalized} Berwald spaces, defined by the property that they admit a (not necessarily torsion-free) metric-compatible affine connection. The only difference is that in that case, $T^k_{ij}$ need not be symmetric in $i\leftrightarrow j$, as it has to be for Berwald spaces. We note, however, that we have applied \textit{Christoffel's trick} in some of the proofs, which is based precisely on this symmetry, so these parts will not necessarily generalize immediately.

%% file: ch_Metrizability.tex
\chapter{Pseudo-Riemann Metrizability}\label{ch:Metrizability}

Given a Finsler space of Berwald type, the Cartan nonlinear connection defines a linear connection on $TM$ that is, by definition, torsion-free and metric-compatible. The natural question thus arises whether there exists a pseudo-Riemannian metric that has this connection as its Levi-Civita connection. Simply put, is (the affine connection on) every Berwald space metrizable? For positive definite Finsler spaces defined on $\mathcal A = TM_0$ this is indeed the case, as was first shown by Szab\'o.
\sloppy
\begin{theor}[Szab\'o's Metrization Theorem \cite{Szabo}]
Any positive definite Berwald space with $\mathcal A = TM_0$ is metrizable by a Riemannian metric.
\end{theor}
The proofs of this theorem \cite{Szabo, Vincze2005} rely on procedures such as averaging \cite{CrampinAveraging} over the indicatrix $S_x = \{(x,y)\in T_xM\,:\,F(x,y)=1\}$ for which it is essential that the Finsler metric $F$ is sufficiently smooth and defined everywhere on $\mathcal A = TM_0$. In the case of Finsler metrics of indefinite signature, however, the domain where $F$ is defined is typically only a proper subset of $TM_0$ and hence the proofs do not extend to this case. And even if $\mathcal A = TM_0$, the fact that the indicatrix is not compact in indefinite signatures poses problems.

It was shown in \citeref{ref3}{H2} that Szab\'o's metrization theorem is indeed not valid for Finsler spaces in the general sense of our \cref{def:Finsler_manifold}, and in \citeref{ref2}{H3} the situation was further investigated and necessary and sufficient conditions for (local) metrizability were obtained in the specific case of $m$-Kropina spaces with closed null 1-form. In this chapter we establish these results and expand on them considerably, covering not only $m$-Kropina metrics but also generalized $m$-Kropina metrics, and arbitrary 1-forms. 

The culprit behind all known counterexamples to Szab\'o's theorem is the fact that the affine Ricci tensor is in general not symmetric. This property, symmetry of the affine Ricci tensor, is clearly a necessary condition for metrizability, and we will see that for all specific metrics investigated here, this is in fact also a sufficient condition, at least locally. In other words, if the affine Ricci tensor is symmetric then any point in $M$ has a neighborhood on which the geometry is metrizable. This property is called \textit{local metrizability}, and it will be the main focus of this chapter. In what follows we will thus use the following definition.
\begin{defi}
    A Berwald space with underlying manifold $M$ is said to be locally metrizable if each point in $M$ has a neighborhood that admits a pseudo-Riemannian metric whose Levi-Civita connection coincides with the affine connection restricted to that neighborhood.
\end{defi}
After establishing the necessary and sufficient conditions for local metrizability, we classify all locally metrizable and Ricci-flat $m$-Kropina metrics in $(1+3)$D whose 1-forms have a constant causal character, and we discuss the implications for vacuum solutions to the field equations in Finsler gravity. But before we turn to these specific types of spaces, we give the general argument for nonmetrizability, together with a simple explicit counterexample.

\section{Nonmetrizability of Berwald spaces}

We start with an adaptation of the results developed in \citeref{ref3}{H2}. A simple way to see why not all Berwald spacetimes are metrizable is to consider the lack of symmetry of the affine Ricci tensor $\bar R_{ij}$. Recall that the affine Ricci tensor is given by \eqref{eq:affine_curvatures} in terms of the affine connection induced by the canonical nonlinear connection. Recall further that the curvature tensor of the affine connection of a Berwald space is related to the nonlinear curvature via \eqref{eq:symm_ricci_Berwald}. Finally, define the function $f = \ln\sqrt{|\det g|}$ and note that 
\begin{align}
    \delta_i f &=   \frac{1}{2 |\det g|} \delta_i |\det g| = \frac{1}{2}g^{kl}\delta_i g_{kl} = \Gamma^k{}_{ik}, \\
	\bar{\partial}_i f &= \frac{1}{2 |\det g|} \bar{\partial}_i |\det g| = \frac{1}{2}g^{kl}\bar{\partial}_i g_{kl} = g^{kl}C_{ikl} = C_i\,.
\end{align}
Then, using the definition of the nonlinear curvature \eqref{eq:def_nonlinear_curvature} in terms of commutators of horizontal vector fields and the fact that $\delta_i\Gamma^k_{jl}=\partial_i\Gamma^k_{jl}$, since $\bar\partial_i\Gamma^k_{jl}=0$, we find that
\begin{align}
    \bar R_{ij} - \bar R_{ji} &= \delta_i \Gamma^{ k}{}_{jk} - \delta_j \Gamma^{ k}{}_{ik} =\delta_i \delta_j f - \delta_j \delta_i f\\
    & = [\delta_i, \delta_j]f = R^k{}_{ij}\bar{\partial}_kf = \bar R^k{}_{lij}y^l\bar{\partial}_kf \\
    &= \bar R^k{}_{lij} y^lC_k.
\end{align}
Hence the skew-symmetric part of the affine Ricci tensor is given by
\begin{align}\label{eq:Ricci_asymmetry_general}
    \bar R_{[ij]} = \tfrac{1}{2}\bar R^k{}_{lij} y^lC_k. 
\end{align}
If the Finsler Lagrangian is pseudo-Riemannian then $C_k=0$ and $\bar R_{ij} = \bar R_{ji}$, but for general Finsler metrics of Berwald type the RHS of \eqref{eq:Ricci_asymmetry_general} need not vanish. We will give an explicit example below. In such a case it is immediately clear that the affine connection cannot be obtained as the Levi-Civita connection of a pseudo-Riemannian metric, because the Levi-Civita connection always has a symmetric Ricci tensor. This shows that not all Berwald spaces are pseudo-Riemann metrizable, as was demonstrated originally in \citeref{ref3}{H2}.
\begin{ex}[Counterexample to Szab\'o's theorem]
    Let the Lorentzian metric $\alpha$ and the $1$-form $\beta$ be given by 
    \begin{align}
    a = -2\, \D u \D v +  v \,\psi(x,y)\,\D u^2 + \D x^2 + \D y^2,\qquad \beta = \D u,
    \end{align}
    where $\psi$ is any scalar function that is not constant, and let $F$ be any generalized $m$-Kropina metric $F = \alpha\phi(\beta/\alpha)$, $\phi(s) = \pm s^{-m}(c + d s^2)^{(m+1)/2}$ constructed from these building blocks. Note that these form a subclass of the family of Berwald metrics described in \Cref{ex:Berwald_spacetime}. A straightforward calculation (or application of \Cref{prop:ricci} below) shows that
    \begin{align}
    \tfrac{1}{2}\left(\bar R_{ux}-\bar R_{xu}\right) = \tfrac{m}{m-1}\partial_x\psi, \qquad \tfrac{1}{2}\left(\bar R_{uy}-\bar R_{yu}\right) = \tfrac{m}{m-1}\partial_y\psi.
    \end{align}
    Since $\psi$ is not constant, it is clear that unless $m=0$, the affine Ricci tensor is not symmetric and hence the affine connection of $F$ does not coincide with the Levi-Civita connection of any pseudo-Riemannian metric.
\end{ex}
The culprit behind the fact that $\bar R_{ij}$ fails to be symmetric in general, and indeed the culprit behind all currently known counterexamples to Szab\'o's theorem, seems to be not the signature, but the fact that $\mathcal A \subsetneq TM_0$ is a proper subset. This is shown by the following theorem, the proof of which uses the properties of the (positively) projective tangent space at a point $x\in M$, defined as the quotient $PT_xM^+\equiv \left(T_xM\setminus 0\right)/\!\!\sim$ where $y\sim y'$ iff $y'= \lambda y$ for some $\lambda>0$. This construction is similar to that of standard projective spaces, see e.g. \cite{LoringWTu_Manifolds, chern1999lectures}; for details we refer to \cite{Hohmann_2019} and references therein. The two essential properties that we will use in the proof below are that (i) $PT_xM^+$ is compact, and that (ii) any smooth $0$-homogeneous function on $T_x M\setminus 0$ can be identified with a smooth function $\hat f$ on $PT_xM^+$ via $\hat f([y]) = f(y)$, where $[y]$ is the equivalence class of $y\in T_x M\setminus 0$ under the equivalence relation $\sim$.
\begin{theor}\label{thm:Symmetric_Ricci_on_TM0}
	For a Berwald space 
    with $\mathcal{A} = TM_0$, the affine Ricci tensor is symmetric, $R_{ij} = R_{ji}$.
\end{theor}
\begin{proof}
    We employ the expression $R_{ij} - R_{ji}  =  R^k{}_{ij}\bar{\partial}_kf$ derived above for the skew-symmetric part of the Ricci tensor and we start by noting that the LHS only depends on $x$, not $y$, for Berwald spaces. It thus suffices to show that for each $x\in M$ there exists at least one $y\in\mathcal{A} = TM_0$ such that $R^k{}_{ij}\bar{\partial}_kf=0$; this then implies that the expression vanishes on all of $TM_0$. So let us fix some $x\in M$. Since $y\mapsto f(x,y)$ is smooth and homogeneous of degree $0$ it can be interpreted as a smooth function $\hat f$ on $PT_xM^+$. Since $\hat f$ is in particular continuous and $PT_xM^+$ is compact, $\hat f$ attains its minimum on $PT_xM^+$. Suppose this minimum is attained at the equivalence class $[y]\in PT_xM^+$. That means that the original function $y\mapsto f(x,y)$ attains its minimum at $y \in T_xM\setminus 0$, so that $\bar\partial_i f|_{(x,y)} = 0$ and hence $R^k{}_{ij}\bar{\partial}_kf|_{(x,y)} =0$. Since $x$ was arbitrary this completes the proof.
\end{proof}

\section[Generalized \texorpdfstring{$m$}{m}-Kropina spaces with closed null 1-form]{Metrizability of generalized \texorpdfstring{$m$}{m}-Kropina spaces with closed null 1-form}
\label{sec:mKrop_closed_null}

This section focuses on generalized $m$-Kropina metrics with closed null 1-form (i.e. $\D b = 0$ and $|b|^2 = a_{ij}b^ib^j=0$) of Berwald type. We will also assume throughout this section that $n\coloneqq\dim M>2$ and that $c\neq 0$ and $m\neq 1$ and we remind the reader that, since $\mathcal A$ is assumed to be contained in the subset of $TM$ characterized by $s>0$, $c+d s^2>0$, in order for any real power of those expressions to be defined, we may assume without loss of generality that $b_i$ is nowhere vanishing. Furthermore, we will stick to our convention that indices $a,b,c,\dots$ run from $3$ to $n$, whereas indices $i,j,k,\dots$ run from $1$ to $n$. Many results in this section have been published in \citeref{ref2}{H3}, but only with regard to \textit{standard} $m$-Kropina spaces. It turns out that exactly the same analysis holds up for \textit{generalized} \mkrop spaces, and we present this extended analysis here for the first time.

Assuming our generalized $m$-Kropina space is Berwald it follows from \Cref{prop:closed_null_gen_mKrop} that in suitable coordinates we have $\beta=du$ and 
\begin{align}
	a=-2\D u \D v + H(u,v,x)\; \D u^2 + 2W_a(u,x)\,\D x^a\D u + h_{ab}(u,x) \, \D x^a \D x^b,
\end{align}
and that
\begin{align}\label{eq:mKrop_closed_null_affine_conn_extra}
    \Gamma^\ell_{ij} &= \mathring \Gamma^\ell_{ij} + \frac{m p}{1-m} \left(a_{ij}b^\ell - \delta^\ell_j b_i - \delta^\ell_i b_j\right)
\end{align}
with $p =- \tfrac{1}{2}\partial_v H$. Substituting the form of $p$ into \eqref{eq:mKrop_closed_null_affine_conn_extra} and using that $b_i = \delta^u_ i$ and hence that $b^\ell = a^{\ell k}b_k =a^{\ell k}\delta^u_k = a^{\ell u} = -\delta^\ell_v$, we obtain the following.
\begin{lem}\label{lem:connection_coeff}
In the coordinates $(u,v,x^3,\dots,x^n)$, the affine connection coefficients can be expressed in terms of the Levi-Civita Christoffel symbols $\mathring\Gamma^k_{ij}$ of the pseudo-Riemannian metric $\alpha$ as
\begin{align}
\Gamma^k_{ij}  = \mathring\Gamma^k_{ij}  + \Delta\Gamma^k_{ij} , \qquad \Delta\Gamma^k_{ij} =  \frac{m}{2(1-m)}\partial_v H \left(a_{ij}\delta^k_v + \delta^k_j \delta^u_ i + \delta^k_i \delta^u_ j\right). 
\end{align}
\end{lem}
We can use the preceding results to analyze the (deviation from the) symmetry of the affine Ricci tensor, which has a very simple expression in these coordinates, as the following result shows.
\begin{prop}\label{prop:ricci}
In the coordinates $(u,v,x^3,\dots,x^n)$, the skew-symmetric part of the affine Ricci tensor is given by
\begin{align}
\bar R_{[ij]} = -\frac{mn}{4(1-m)}(\delta^u_i\partial_j\partial_vH-\delta^u_j\partial_i\partial_vH).
\end{align}
\end{prop}
\begin{proof}
From the definition \eqref{eq:affine_curvatures} of the affine Ricci tensor of a Berwald space and the fact that $\Gamma^k_{ij} = \Gamma^k_{ji}$ it follows that the skew-symmetric part of the affine Ricci tensor can be written as
\begin{align}
\bar R_{[ij]} \equiv \tfrac{1}{2}\left(\bar R_{ij} - \bar R_{ji}\right)  = \partial_{[i}\Gamma^k_{j]k}.
\end{align}
Since the Levi-Civita connection $\mathring\Gamma^k_{ij}$ has a symmetric Ricci tensor, it follows that $\partial_{[i}\mathring\Gamma^k_{j]k}=0$ and hence we have 
\begin{align}\label{eq:add_eq_}
    \bar R_{[ij]} &= \partial_{[i}\Gamma^k_{j]k} = \partial_{[i}\Delta \Gamma^k_{j]k}=\tfrac{1}{2}\left(\partial_{i}\Delta\Gamma^k_{kj}-\partial_{j}\Delta\Gamma^k_{ki}\right).
\end{align}
Using \Cref{lem:connection_coeff} we find that 
\begin{align}
\Delta\Gamma^k_{kj} = \frac{m}{2(1-m)}\partial_v H\left(a_{vj} + \delta^u_ j+ n \delta^u_j\right) = \frac{mn}{2(1-m)}\partial_v H \delta^u_j,
\end{align}
where we have used that $a_{vj} = -\delta^u_j$, and hence we obtain
\begin{align}
\bar R_{[ij]} = \frac{mn}{4(1-m)}\left(\delta^u_j\partial_{i}\partial_v H-\delta^u_i\partial_{j}\partial_v H  \right).
\end{align}
\end{proof}
%
%
Now we will prove the main result of this section, characterizing the local metrizability of generalized $m$-Kropina spaces of Berwald type with closed null 1-form.
\begin{theor}\label{theorem:main}
Let $F$ be a generalized $m$-Kropina metric of Berwald type with closed null 1-form and with $m\neq 0$ and $\dim M>2$. The following are equivalent:
\begin{enumerate}[(i)]
\item $F$ is locally metrizable;
\item The affine Ricci tensor is symmetric, $\bar R_{ij} = \bar R_{ji}$;
\item Around each point in $M$ there exist coordinates $(u,v,x^3,\dots,x^n)$ 
such that {\normalfont$b = \D u$} and
\begin{align}
a = -2\D u\D v + \left[\tilde H(u,x) + \rho(u)v\right]\D u^2 + 2&W_a(u,x)\D u\D x^a \nonumber\\
&+ h_{ab}(u,x)\D x^a\D x^b,
\end{align}
with $\tilde H, \rho, W_3,\dots W_{n}$ smooth functions and $h$ a pseudo-Riemannian metric of dimension $n-2$.
\end{enumerate}
In this case, the affine connection restricted to the chart corresponding to the coordinates $(u,v,x^3,\dots x^n)$ is metrizable by the following pseudo-Riemannian metric:
\begin{align}\label{eq:metrizing_metric}
{\normalfont \tilde a = e^{\frac{m}{1-m}\int^u \rho(\tilde u)\D \tilde u } a}
\end{align}
\end{theor}
Before presenting the proof, we note that if $\rho = 0$ then $\tilde a = a$, i.e. the affine connection is metrizable by the defining pseudo-Riemannian metric $\alpha$. This was to be expected, since in that case the 1-form $\beta$ is parallel with respect to $\alpha$, and hence the result follows by \Cref{cor:abmetric_same_apray_iff_cc}. 
%
%
\begin{proof}
(i) trivially implies (ii). For (ii)$\Rightarrow$(iii) we use the preferred coordinates $(u,v,x^3,\dots,x^n)$. By \Cref{prop:ricci}, the only nonvanishing components of the skew-symmetric part of the affine Ricci tensor are
\begin{align}
\bar R_{[uj]} = -\frac{mn}{4(1-m)}\partial_j\partial_vH,\qquad j = 2,\dots, n.
\end{align}
By assumption, the Ricci tensor is symmetric, so all of these must vanish. The $uv$ component yields  $\partial_v^2 H = 0$ and the remaining components yield $\partial_v\partial_{a} H=0$, $a = 3,\dots n$. %
%
In other words, $H$ must be linear in $v$ and the corresponding linear coefficient can depend only on the coordinate $u$.  That is,
\begin{align}
a = -2\D u\D v + \left[\tilde H(u,x) + \rho(u)v\right]\D u^2 + 2W_a(u,x)\D u\D x^a + h_{ab}(u,x)\D x^a\D x^b.
\end{align}
This proves $(ii)\Rightarrow(iii)$. For the final implication $(iii)\Rightarrow (i)$, recall from \Cref{lem:connection_coeff} that the affine connection coefficients can be expressed as 
\begin{align}
\Gamma^k_{ij}  = \mathring\Gamma^k_{ij}  + \frac{m}{2(1-m)}\partial_v H \left(a_{ij}\delta^k_v + \delta^k_j \delta^u_ i + \delta^k_i \delta^u_ j\right).\label{eq:temp:afC1}
\end{align}
On the other hand, an elementary calculation shows that the Levi-Civita Christoffel symbols of the pseudo-Riemannian metric $\tilde a = e^{\psi(u)}a$ can be expressed in terms of the original Levi-Civita Christoffel symbols corresponding to $a$ as
\begin{align}
{}^{\tilde a}\Gamma^k_{ij} = \mathring\Gamma^k_{ij} +\tfrac{1}{2}\psi'(u) \left(a_{ij}\delta^k_v + \delta^k_j \delta^u_ i + \delta^k_i \delta^u_ j\right). \label{eq:temp:afC2}
\end{align}
In this case, $\psi(u) = \frac{m}{1-m}\int^u \rho(\tilde u)\D \tilde u$, so  $\psi'(u)  = \frac{m}{1-m}\rho(u) = \frac{m}{1-m}\partial_v H$, and hence it is clear that \eqref{eq:temp:afC1} and \eqref{eq:temp:afC2} coincide. This shows that 
the connection coefficients of $\tilde a$ coincide with the affine connection coefficients of $F$, and hence $F$ is locally metrizable, i.e. (i) holds. This completes the proof of the theorem.
\end{proof}
\Cref{theorem:main} provides necessary and sufficient conditions for a generalized m-Kropina space with closed null 1-form to be locally metrizable. Next, we apply the theorem to an explicit example from the physics literature.
\begin{ex}\label{ex:FinslerVSI}
Consider the Finsler VSI spacetimes presented in \cite{Fuster:2018djw}, with the 4-dimensional Finsler metric
\begin{align}
F = \left|-2\D u \D v +(\tilde H + \rho\, v)\D u^2 + 2W_a\D u\D x^a + \delta_{ab}\D x^a \D x^b\right|^{\frac{1+m}{2}}\left(\D u\right)^{-m}, \label{eq:Gyr_ppW_F_ex}
\end{align}
where the metric functions $\tilde H, \rho, W_a$ depend only on $u$ and $x^a$, not on $v$. The way to interpret the RHS is to view all 1-forms as real-valued functions on $TM$, products, sums, and powers of which are defined pointwise. The name `Finsler VSI spacetimes' stems from the fact that the \psR part $\alpha$ is a VSI spacetime, i.e. it has vanishing scalar curvature invariants \cite{VSI-4D}. $F$ is an $m$-Kropina metric and by \Cref{prop:closed_null_gen_mKrop}, it is of Berwald type. By \Cref{theorem:main} it is locally metrizable if and only if $\partial_a\rho= 0$. The case $\rho=0$ provides a Finsler version of the gyratonic pp-wave metric \cite{gyraton,Maluf2018}, which is trivially metrizable by the Lorentzian gyratonic pp-wave metric itself.
\end{ex}
As a special case of \Cref{ex:FinslerVSI} we obtain a particularly simple, nontrivial example of a locally metrizable Finsler metric.
\begin{ex}
Let $F$ be given by \eqref{eq:Gyr_ppW_F_ex} with $\tilde H(u,x) = 0$, $\rho(u,x)=u$ and $W_a(u,x) =0$. Relabeling the two coordinates $x^a$ as $x$ and $y$, this leads to the Finsler metric
\begin{align}
F = \left|-2\D u \D v + u\,v\,\D u^2 + \D x^2 + \D y^2\right|^{\frac{1+m}{2}}\left(\D u\right)^{-m},
\end{align}
which has an affine connection given by the following nonvanishing Christoffel symbols:
\begin{align}
\Gamma^u_{uu} &= \frac{1+m}{1-m}\frac{u}{2}, \qquad  \Gamma^v_{uu} = \left(\frac{u^2}{1-m} - 1\right)\frac{v}{2}, \\ 
\Gamma^v_{uv} &= -\frac{u}{2}, \qquad \Gamma^v_{xx} = \Gamma^v_{yy} = \Gamma^x_{ux} = \Gamma^y_{uy} = \frac{m}{1-m}\frac{u}{2}.
\end{align}
As indicated by \eqref{eq:metrizing_metric} in \Cref{theorem:main} this connection is metrizable by the Lorentzian metric
\begin{align}
\tilde g = e^{\frac{m u^2}{2(1-m)} }\left(-2\D u \D v + u\,v\,\D u^2 + \D x^2 + \D y^2\right).
\end{align}
\end{ex}

\section{Metrizability of \texorpdfstring{$m$}{m}-Kropina spaces}

The results of the previous section have been obtained using the specific coordinate system introduced in \Cref{ex:Berwald_spacetime} and \Cref{prop:closed_null_gen_mKrop}, which presupposes that the 1-form $\beta$ is closed and null. In this section we consider general 1-forms $\beta$ and so the methods used in the previous section cannot be generalized straightforwardly. However, using different methods it is possible to obtain some interesting results on the metrizability of (standard) $m$-Kropina spaces with arbitrary 1-form, without the need for introducing a special coordinate system. As a special case, this will reproduce the results obtained above for \mkrop spaces with a closed null 1-form, providing an alternative proof of those results. The results in this section were obtained only recently and have not been published yet; we present them here for the first time. In contrast to the results obtained in \cref{sec:mKrop_closed_null} they do not generalize in an obvious way to \textit{generalized} $m$-Kropina spaces.

Hence throughout this section, we will assume $F$ is an $m$-Kropina metric. Again we restrict our attention to the case $n=\dim M>2$. Recall from \Cref{cor:mKrop_berwald_conditions} that an $m$-Kropina metric is of Berwald type if and only if there exists a smooth vector field $f^i$ on $M$ such that 
\begin{align}\label{eq:C-mtr_mKrop_betrald_cond}
    \mathring\nabla_j b_i = m (f_k b^k)a_{ij} + b_i f_j  - m f_i b_j,
\end{align}
and if so, the affine connection is given by
\begin{align}\label{eq:delta_gamma_General}
    \Gamma^\ell_{ij} = \mathring\Gamma^\ell_{ij} + \Delta \Gamma^k_{ij},\qquad \Delta \Gamma^k_{ij} = m a^{\ell k}\left(a_{ij}f_k - a_{jk}f_i - a_{ki}f_j\right)
\end{align}
in terms of the Levi-Civita connection $\mathring\Gamma^\ell_{ij}$ of $\alpha$. We start, in analogy with the previous section, by deriving a precise formula for the skew-symmetric part of the affine Ricci tensor, which we will denote by $\mathcal A(\bar R)$, i.e. $\mathcal A(\bar R)_{ij} = \bar R_{[ij]}$. Moreover, at the risk of stating the obvious, we remark that the expression $\D f$ below denotes the exterior derivative of the 1-form $f$ with components $f_i = a_{ij}f^j$ appearing in the Berwald condition \eqref{eq:C-mtr_mKrop_betrald_cond}.
\begin{lem}\label{prop:ricci_general}
The skew-symmetric part of the affine Ricci tensor is given by 
\begin{align}\label{eq:lem_skew_R_formula}
    \mathcal A(\bar R) = -\frac{mn}{2} \D f
\end{align}
\end{lem}
\begin{proof}
In general we have $\bar R_{[ij]} =\tfrac{1}{2}(\partial_{i}\Delta\Gamma^k_{kj}-\partial_{j}\Delta\Gamma^k_{ki})$ as shown in the proof of \Cref{prop:ricci} (see \eqref{eq:add_eq_} in particular). Employing the expression \eqref{eq:delta_gamma_General} for $\Delta \Gamma^k_{ij}$, we find that
\begin{align}\label{eq:fsmooth}
    \Delta\Gamma^k_{kj} =  - m n f_j
\end{align}
and this yields
\begin{align}
\bar R_{[ij]} = \frac{mn}{2}(\partial_jf_i-\partial_if_j),
\end{align}
which is precisely the coordinate expression of \eqref{eq:lem_skew_R_formula}, as desired.
\end{proof}
We then have the following characterization of local metrizability. 
\begin{theor}\label{theor:metrizability_main}
For an $m$-Kropina metric  $F = \alpha^{1+m}\beta^{-m}$ of Berwald type with $\dim M>2$, the following are equivalent:
\begin{enumerate}[(i)]
    \item $F$ is locally metrizable;
    \item The affine Ricci tensor is symmetric, $\bar R_{ij} = \bar R_{ji}$;
    \item $f_i$ is a closed 1-form.
\end{enumerate} 
In that case we can locally write  $f_i =\partial_i\psi$ for some $\psi\in C^\infty(M)$ and we may define $\tilde a_{ij} = e^{-2m\psi}a_{ij}$ and $\tilde b_i = e^{-(1+m)\psi}b_i$ and define $\tilde \alpha$ and $\tilde \beta$ accordingly. Then $F$ can be written as $F = \tilde\alpha^{1+m}\tilde\beta^{-m}$, $\tilde \beta$ is covariantly constant with respect to $\tilde \alpha$, and $\tilde a_{ij}$ provides a local metrization of $F$.
\end{theor}
\begin{proof}
    The first two implications are rather straightforward. (i)$\Rightarrow$(ii): If $F$ is locally metrizable then it follows trivially that the Ricci tensor is symmetric. (ii)$\Rightarrow$(iii): If the Ricci tensor is symmetric then \Cref{prop:ricci_general} shows that $\D f =0$. The remaining nontrivial part of the proof is showing that (iii) implies (i). (iii)$\Rightarrow$(i): If $f_i$ is closed then $f$ is locally exact, i.e. $f_i = \partial_i\psi$ for some (locally defined) function $\psi$ on $M$. Consider the following transformed metric and 1-form, $\tilde a_{ij} = e^{-2m\psi} a_{ij}$ and $\tilde b_i = e^{-(1+m)\psi}b_i$ and define $\tilde \alpha$ and $\tilde\beta$ accordingly. Then $F = \alpha^{1+m}\beta^{-m}=\tilde\alpha^{1+m}\tilde\beta^{-m}$ and it is straightforward to show that the Christoffel symbols of $\tilde a_{ij}$ are given by
    \begin{align}
        \tilde\Gamma^k_{ij} = \mathring\Gamma^k_{ij} - m\left(f_i\delta^k_j+f_j\delta^k_i - f^ka_{ij}\right)
    \end{align}
    in terms of those of $a_{ij}$. Hence if we denote by $\tilde\nabla$ the Levi-Civita connection of $\tilde \alpha$ then we find that
    \begin{align}
        \tilde\nabla_j\tilde b_i = e^{-(1+m)\psi}\left[\mathring\nabla_j b_i -( m (f_k b^k)a_{ij} + b_i f_j  - m f_i b_j)\right] = 0,
    \end{align}
    which vanishes, by \eqref{eq:C-mtr_mKrop_betrald_cond}. In other words, $\tilde b_i$ is covariantly constant w.r.t. $\tilde a_{ij}$, and since we can write $F$ as $F = \tilde\alpha^{1+m}\tilde\beta^{-m}$, it follows from \Cref{eq:ab_cc_pre} that the affine connection on $F$ is just the Levi-Civita connection of $\tilde a_{ij}$. Hence $F$ is locally metrizable, namely by $\tilde a_{ij}$, completing the proof. 
\end{proof}
The mapping $(\alpha,\beta)\mapsto (\tilde\alpha,\tilde\beta)$ is a generalization of what is called the $f$\textit{-change} in \cite[Chapter 6]{handbook_Finsler_vol2_matsumoto}. 

We can now distinguish several cases.
\begin{enumerate}
    \item $b$ is nowhere null. 
    In this case, $f_i$ is necessarily closed, which follows immediately from \Cref{prop:generalized_mKtop_berw} and which was also observed in \cite{handbook_Finsler_vol2_matsumoto}. Hence we have the following.
\end{enumerate}
\begin{cor}\label{prop:nonnullmetrizable}
    An $m$-Kropina space with nowhere null 1-form is locally metrizable.
\end{cor}
\begin{enumerate}
    \item[2.] If $b$ is null and closed then \cref{sec:mKrop_closed_null} completely characterizes local metrizability in terms of local coordinates.
    \item[3.] If $b$ is null but not necessarily closed then the local metrizability of $F$ can also be characterized in local coordinates, as will be discussed next. 
\end{enumerate}
\begin{prop}\label{prop:null_metrization_classification}
If $|b|^2=0$ then $F$ is locally metrizable if and only if around each point in $M$ there exist coordinates $(u, v, x^3, \dots, x^n)$ such that
\begin{align}
    a &= e^{2m\psi}\left(-2\D u\D v + H\D u^2 + 2W_a\D u\D  x^a + h_{ab}\D x^a \D x^b\right), \label{eq:pra}\\
    b &= e^{(1+m)\psi}\D u, \label{eq:prb}
\end{align}
where $\psi$ is any smooth function on $M$ and the metric functions $H, W_a, h_{ab}$ depend only on $u$ and $x^a$. In that case, $F$ can be written as
\begin{align}\label{eq:standard_pp_wave_m_Kropina}
    F = \left|-2\D u\D v + H\D u^2 + 2W_a\D u\D  x^a + h_{ab}\D x^a \D x^b\right|^{(1+m)/2}\left(\D u\right)^{-m}
\end{align}
and the metric $-2\D u\D v + H\D u^2 + 2W_a\D u\D  x^a + h_{ab}\D x^a \D x^b$ provides a local metrization of $F$.
\end{prop}
\begin{proof}
We first prove the `only if' part. By \Cref{theor:metrizability_main}, if $F$ is locally metrizable then there exist $\tilde a_{ij} = e^{-2m\psi}a_{ij}$ and $\tilde b_i = e^{-(1+m)\psi}b_i$ with corresponding $\tilde \alpha$ and $\tilde \beta$ such that $F = \tilde\alpha^{1+m}\tilde\beta^{-m}$ and such that $\tilde \beta$ is covariantly constant with respect to $\tilde a_{ij}$. By \Cref{prop:coords_cc_null_1form} (recall that we may assume WLOG that $b_i$ does not vanish by our definition of $\mathcal A$) it thus follows that locally, we may write
\begin{align}
    \tilde a &= -2\D u\D v + H\D u^2 + 2W_a\D u\D  x^a + h_{ab}\D x^a \D x^b, \\
    \tilde b &= \D u,
\end{align}
where the functions $H, W_a, h_{ab}$ depend only on $u$ and $x^a$. It is immediately verified that $a,b$ and $F$ now attain the desired form. Conversely, if $a$ and $b$ have the stated form, then $F$ is given by \eqref{eq:standard_pp_wave_m_Kropina} in terms of $\tilde a$ and $\tilde b$, and since $\tilde b=\D u$ is covariantly constant with respect to $\tilde \alpha$, the latter provides a local metrization of $F$, by \Cref{eq:ab_cc_pre}.
\end{proof}
As a special case of \Cref{prop:null_metrization_classification} we recover the main result of the previous section, \Cref{theorem:main}, as shown by the following corollary.
\begin{cor}
If $b$ is null and closed then $F$ is locally metrizable if and only if around each point in $M$ there exist coordinates $(u,v,x^3,\dots,x^n)$ such that
\begin{align}
    F &= \left|-2\D u\D v + H\D u^2 + 2W_a\D u\D  x^a + h_{ab}\D x^a \D x^b\right|^{(1+m)/2}\left(\D u\right)^{-m},
\end{align}
where
\begin{align}
    H =  \tilde H(u,x^a) + \rho(u)v, \qquad W_a = W_a(u,x^a), \qquad h_{ab} =h_{ab}(u,x^a).
    \end{align}
\end{cor}
\begin{proof}
    Suppose that $F$ is locally metrizable, then $a$ and $b$ are given by \eqref{eq:pra} and \eqref{eq:prb}, where the metric functions do not depend on the coordinate $v$. Furthermore, the fact that the 1-form $b = e^{(1+m)\psi}\D u$ is closed implies that $\psi = \psi(u)$. In this case, it can be verified that the coordinate transformation given by $\bar u = \int^u e^{(1+m)\psi(u)}\D u$ and $\bar v = e^{(m-1)\psi(u)}v$ brings the 1-form into the form $b = \D\bar u$ and the metric into the form
    \begin{align}
        a = -2\D \bar u\D \bar v + \bar H\D \bar u^2 + 2\bar W_a\D \bar u\D  x^a + \bar h_{ab}\D x^a \D x^b,
    \end{align}
    where
    \begin{align}
        \bar H = e^{-2\psi}\left[H + 2(m-1) \psi'v\right],\quad \bar W_a =  e^{(m-1)\psi} W_a, \quad \bar h_{ab} &= e^{2m\psi} h_{ab}.
    \end{align}
    Identifying $\tilde H(\bar u,x^a) = e^{-2\psi(u(\bar u))}H$ and $\rho(\bar u) = 2e^{-(1+m)\psi(u(\bar u))}(m-1)\psi'(u(\bar u))$ it can be verified that also $a$ and hence $F$ attains the desired form in the new coordinates.
    
    Conversely, if $F$ has the stated form, then we define $\psi(u)$ such that $\rho(u) = 2(m-1)\psi'e^{-(1+m)\psi} = 2(m-1)d\psi/d\bar u$, by setting $\psi(u) = \tfrac{1}{2(m-1)}\int^{\bar u(u)}\rho(u)\D u$, and then we can simply perform the same coordinate transformation backward. It is clear that this results in a metric and 1 form of the form \eqref{eq:pra} and \eqref{eq:prb} and hence, by \Cref{prop:null_metrization_classification}, $F$ is locally metrizables.
\end{proof}

\section{Ricci-flatness of \texorpdfstring{$m$}{m}-Kropina spaces}\label{sec:classification_ricci_flat}

We continue our investigation of $m$-Kropina metrics and turn to Ricci-flatness. Recall that a Finsler space is said to be Ricci-flat if the Finsler-Ricci tensor vanishes, $R_{ij}=0$, or equivalently, if the Ricci scalar vanishes, $\text{Ric}=0$. We restrict ourselves in this section to the case where $M$ is four-dimensional, which is the case that is relevant to Finsler gravity. The results in this section have not been published yet; we present them here for the first time.

\subsection{Ricci-flat nonnull \texorpdfstring{$m$}{m}-Kropina metrics in \mbox{(1+3)D}}

First, we investigate Ricci-flat $m$-Kropina metrics for which $b$ is nowhere null in $(1+3)$D, by which we mean that the underlying manifold $M$ has dimension $4$ and that $a_{ij}$ has Lorentzian signature. 
\begin{prop}\label{prop:Ricci_flat_F}

    A $(1+3)$D $m$-Kropina space with nowhere null 1-form is Ricci-flat if and only if it is locally of the form
    \begin{align}\label{eq:unique_nonnull_ricci_flat_mKrop}
        F = \left|\eta_{ij}\D x^i \D x^j\right|^{(1+m)/2}\left(c_i\D x^i\right)^{-m}
    \end{align}
    with $\eta_{ij}=$ diag$(-1,1,1,1)$ and $c_i=$ constant.
\end{prop}
\begin{proof}
We first prove the `if' direction. Since $c_i\D x^i$ is covariantly constant with respect to $\eta_{ij}$, it follows by \Cref{eq:ab_cc_pre} that any $F$ defined by \eqref{eq:unique_nonnull_ricci_flat_mKrop} is Ricci-flat, since the Ricci tensor of $\eta_{ij}$ (which vanishes) must coincide with the affine Ricci tensor of $F$, which therefore vanishes, implying that also the Finsler-Ricci tensor vanishes, by \cref{lem:RicciTensors}. 

Conversely, to prove the `only if' direction, suppose that $F$ is Ricci-flat. Then, defining $\tilde \alpha,\tilde \beta$ as in \Cref{theor:metrizability_main}, we may write $F = \tilde\alpha^{1+m}\tilde\beta^{-m}$ and $F$ has the same affine connection as $\tilde \alpha$, hence if $F$ is Ricci-flat then $\tilde \alpha$ is Ricci-flat. But $\tilde \alpha$ admits a nonnull covariantly constant 1-form, namely $\tilde b$, so if $\tilde \alpha$ is Ricci-flat it must actually be flat, by \Cref{cor:EFE_sol_with_parallel_field}, and hence, in suitable coordinates it can be written as $\tilde a = \eta_{ij}\D x^i \D x^j$. Since $\tilde \beta$ is covariantly constant with respect to this metric, it must have constant coefficients in these coordinates, and the conclusion of the theorem follows.
\end{proof}
We remark that the assumption about the dimensionality is not void, because the conclusion of \Cref{cor:EFE_sol_with_parallel_field} is not necessarily true if $\dim M \neq 4$. We also point out that the \textit{representation} of $F$ in terms of a specific choice of $\alpha$ and $\beta$ is, of course, not unique. However, since any $\alpha$ is related to the $\tilde\alpha$ from the proposition by a conformal transformation, we have the following two corollaries.
%
%
%
\begin{cor}
    If $F$ is Ricci-flat and $b$ nowhere null then $\alpha$ is conformally flat.
\end{cor}
\begin{cor}
    If $\alpha$ is Ricci-flat and $b$ nowhere null then $F$ is Ricci-flat only if $\alpha$ is flat and $\beta$ has constant components in flat coordinates.
\end{cor}
\begin{proof}
        If $F$ is Ricci-flat and $b$ nowhere null then, by the previous corollary, $\alpha$ is conformally flat as well as Ricci-flat. It is thus a simple consequence of the Ricci-decomposition of the Riemann tensor that $\alpha$ must be flat. From \Cref{prop:Ricci_flat_F} it then follows that $\beta$ must have constant coefficients in flat coordinates.
\end{proof}
These results have important consequences for the construction of solutions to the field equations in Finsler gravity, the topic of \cref{part:vacuum_sols}. Most (properly Finslerian) solutions currently known in the literature are obtained through constructions where a vacuum solution $\alpha$ in general relativity is deformed into a Finslerian solution $F$ that has the form of an $(\alpha,\beta)$-metric. Then the Ricci-flatness of $F$ (which is a sufficient condition to be a vacuum solution in Finsler gravity) is usually derived somehow from the Ricci-flatness of $\alpha$ (the exact condition to be a vacuum solution to Einstein's field equation). For $m$-Kropina metrics of Berwald type, our results above show that if $b$ nowhere null, this procedure can only lead to a single possible Finsler gravity vacuum solution\footnote{Strictly speaking, this is only true locally. Our results do not say anything about the global topology of the manifold.}, namely the metric \eqref{eq:unique_nonnull_ricci_flat_mKrop}. It follows that, apart from this trivial solution, $m$-Kropina spacetimes only lend themselves to constructing nontrivial Finslerian vacuum solutions in the above way for \textit{null} 1-forms $\beta$ (assuming the causal character of the 1-form does not change). We consider those next.
%
%
%

\subsection{Ricci-flat null \texorpdfstring{$m$}{m}-Kropina metrics in \mbox{(1+3)D}}

While an $m$-Kropina space with nowhere null 1-form is always locally metrizable, as we have seen in \Cref{prop:nonnullmetrizable}, this is not always the case when the 1-form is null. With regard to Ricci-flatness, it turns out that in the $|b|^2=0$ case we have to add the additional assumption that $F$ is locally metrizable in order to obtain nice results. Alternatively, one may choose to replace everywhere the two conditions \textit{locally metrizable and Ricci-flat} by the single condition \textit{affinely Ricci-flat}, the latter meaning that the \textit{affine} Ricci tensor vanishes (rather than the Finsler-Ricci tensor, which is its symmetrization, by \cref{lem:RicciTensors}). Affine Ricci-flatness is not a very common notion in the literature, though, so we will stick to locally metrizable and Ricci-flat.
\begin{prop}
    Let $F$ be an $m$-Kropina space. Then 
    \begin{align}
        \text{F is affinely Ricci-flat}\quad\Leftrightarrow\quad \text{F is locally metrizable and Ricci-flat} \nonumber
    \end{align}
\end{prop}
\begin{proof}
    If $F$ is affinely Ricci-flat then it is trivially Ricci-flat, since the Finsler-Ricci tensor is the symmetrization of the affine Ricci tensor. It is also locally metrizable, by \Cref{theor:metrizability_main}, since the (vanishing) affine Ricci tensor is, in particular, symmetric. For the converse implication, if $F$ is locally metrizable then, by \Cref{theor:metrizability_main}, its Ricci tensor is symmetric and hence the affine Ricci tensor coincides with the Finsler-Ricci tensor. The vanishing of the Finsler-Ricci tensor thus implies, in this case, the vanishing of the affine Ricci tensor.
\end{proof}
The following lemma will allow us to obtain the $|b|^2=0$ analog of \Cref{prop:Ricci_flat_F}, and it will prove to be useful more generally in \cref{part:vacuum_sols}.
\begin{lem}\label{lem:coordchange}
    Let a Lorentzian metric $a$ and 1-form $b$ on a $4$-dimensional manifold be given by
    \begin{align}
        a &= -2\D u \D v + H(u,x)\, \D u^2 + 2W_a(u,x)\,\D x^a\D u +h_{ab}(u,x) \D x^a \D x^b, \label{eq:original_pp_wave_metricl} \\
        b &= \D u.\label{eq:original_1forml}
    \end{align}
    If $a$ is Ricci-flat then $a$ and $b$ can be expressed, in suitable coordinates, as
    \begin{align}\label{eq:pp_waves_final_form_body2s}
        a &= -2\D \bar u\D \bar v + \bar H(\bar u,\bar x,\bar y)\D \bar u^2 + \D \bar x^2 + \D \bar y^2,\\
        b &= \D \bar u,
\end{align}
such that $\bar H$ satifies $\partial^2_{\bar x} \bar  H + \partial^2_{\bar y} \bar H=0$.
\end{lem}
\begin{proof}
We first consider only the metric $a$. Since $b$ is null and covariantly constant, it follows from \cref{eq:metric_cc_null_1form_vac_sol} that $a$ can be expressed in suitable coordinates as
\begin{align}\label{eq:pp_waves_final_form_bodyl}
    a = -2\D u\D v + H(u,x,y)\D u^2 + \D x^2 + \D y^2
\end{align}
such that $\delta^{ab}\partial_a \partial_b H=0$.
We are, however, not only interested in the transformation behavior of $a$ alone but also in that of $b$. In \cite{ehlers1962exact}, on which \cref{eq:metric_cc_null_1form_vac_sol} is based, no explicit coordinate transformation is given, so to see why we may assume without loss of generality that $b = \D u$ remains invariant, we will use the properties of the general set of coordinate transformations
\begin{align}\label{eq:coordTraffol}
(u,v,x^1,x^2)\mapsto (\bar u,\bar v,\bar x^1,\bar x^2)
\end{align}
that leave the form of the metric \eqref{eq:original_pp_wave_metricl} invariant (but that generally change the expressions for the metric functions $H, W_a, h_{ab}\mapsto \bar H, \bar W_a, \bar h_{ab}$). Such transformations have been classified \cite[Section 31.2]{stephani_kramer_maccallum_hoenselaers_herlt_2003} and they all have the specific property that $u = \phi(\bar u)$ for some function $\phi$ depending on $\bar u$ alone. Hence this applies in particular to the desired transformation that relates \eqref{eq:original_pp_wave_metricl} to \eqref{eq:pp_waves_final_form_bodyl}. Let's be a little bit more specific with the notation. What we have argued so far is that there must exist a coordinate transformation \eqref{eq:coordTraffol} such that \eqref{eq:original_pp_wave_metricl} turns into
\begin{align}\label{eq:pp_waves_final_form_body2l}
    a = -2\D \bar u\D \bar v + \bar H(\bar u,\bar x,\bar y)\D \bar u^2 + \D \bar x^2 + \D \bar y^2
\end{align}
and such that $u = \phi(\bar u)$. The latter implies we may express the 1-form as $b = \D  u = \phi'(\bar u)\D \bar u$, or equivalently $\bar b_\mu = \phi'(\bar u)\delta_\mu^u$. The trick is now to realize that since $b$ is covariantly constant (cf. \cref{prop:coords_cc_null_1form}) with respect to $a$, we must have $\bar \nabla_\mu\bar b_\nu=0$ also in the new coordinates.  All  Christoffel symbols $\bar \Gamma^u_{\mu\nu}$ of the metric \eqref{eq:pp_waves_final_form_body2l} with upper index $u$ vanish identically, however (see \eqref{eq:christu}), and hence $\phi$ must satisfy
\begin{align}
0\stackrel{!}{=}\bar \nabla_{\bar u} \bar b_{\bar u} = \partial_{\bar u} b_{\bar u} - \bar \Gamma^u_{uu}\phi'(\bar u) = \phi''(\bar u).
\end{align}
It follows that $\phi'(\bar u)= C =$ constant, i.e. $b = C \D \bar u$. In this case, it is easily seen that scaling $\bar u$ by $C$ and scaling $\bar v$ by $1/C$ leaves the form of the metric \eqref{eq:pp_waves_final_form_body2l} invariant (while changing $\bar H$) and brings the 1-form back into its original form $b = \D\bar u$. This shows that we may assume without loss of generality that $C=1$ and hence that the 1-form remains invariant under the coordinate transformation relating \eqref{eq:original_pp_wave_metricl} to \eqref{eq:pp_waves_final_form_bodyl}. This completes the proof.
\end{proof}
\begin{prop}
    Any locally metrizable $(1+3)$D $m$-Kropina space with $|b|^2=0$ is Ricci-flat if and only if it is locally of the form
\begin{align}\label{eq:nullmKroptrivialRicciFlat}
    F = \left|-2\D u\D v + H(u,x)\D u^2 + \delta_{ab}\D x^a \D x^b\right|^{(1+m)/2}\left(\D u\right)^{-m}
\end{align}
such that $H$ satifies $\delta^{ab}\partial_a \partial_b H=0$.
\end{prop}
\begin{proof}
First of all, since $\D u$ is covariantly constant, it follows by \Cref{eq:ab_cc_pre} that any $F$ defined by \eqref{eq:nullmKroptrivialRicciFlat} is Ricci-flat, since the Ricci tensor of the Lorentzian metric (which vanishes, by \cref{eq:metric_cc_null_1form_vac_sol}) must coincide with the affine Ricci tensor of $F$, which therefore vanishes, implying that also the Finsler-Ricci tensor vanishes, by \cref{lem:RicciTensors}. 

Conversely, suppose that $F$ is Ricci-flat. Since $F$ is locally metrizable it can be written in the standard form \eqref{eq:standard_pp_wave_m_Kropina} with $a$ and $b$ given by \eqref{eq:original_pp_wave_metricl} and \eqref{eq:original_1forml}, respectively. Since the 1-form $b = \D u$ is covariantly constant with respect to the corresponding metric $a$, it follows that the affine connection on $F$ coincides with the affine connection on $a$. So since $F$ is Ricci-flat, $a$ is Ricci-flat as well. Hence it follows from \cref{lem:coordchange} that $F$ can be written locally in the form \eqref{eq:nullmKroptrivialRicciFlat}, as desired.
\end{proof}
%
%
%
With this, we have classified (locally, at least) all locally metrizable, Ricci-flat $m$-Kropina metrics in $(1+3)$D whose 1-forms have a constant causal character. We end this chapter with a final remark about the implications of our findings for the study of vacuum solutions in Finsler gravity. The results show that any Berwald $m$-Kropina solution to the Finsler field equations in vacuum that is not of one of the trivial forms \eqref{eq:unique_nonnull_ricci_flat_mKrop} or \eqref{eq:nullmKroptrivialRicciFlat}, must be one for which at least one of the following three conditions is satisfied:
\begin{enumerate}[(i)]
    \item the 1-form is null and $F$ is not locally metrizable;
    \item $F$ is not Ricci-flat;
    \item the causal character of $b$ is not constant.
\end{enumerate}
In light of the fact (cf. \cref{part:vacuum_sols}) that for Berwald $m$-Kropina spaces with null 1-form $\beta$ the vacuum field equations are equivalent to the vanishing of the Finsler-Ricci tensor \cite{Fuster:2018djw}, it follows that we can equivalently write these conditions as:
\begin{enumerate}[(i)]
    \item the 1-form is null and $F$ is not locally metrizable;
    \item the 1-form is not null and $F$ is not Ricci-flat.
    \item the causal character of $b$ is not constant.
\end{enumerate}
To complete the classification of exact vacuum solutions to the field equations within the class of Berwald $m$-Kropina spaces, a better understanding of these two scenarios is required. While it is currently not known whether vacuum solutions satisfying (ii) or (iii) exist at all, some examples of Ricci-flat spaces satisfying (i) have been obtained, as shown by the example below. Nevertheless, a thorough investigation of each of the cases (i)-(iii) has yet to be carried out. 
\begin{ex}\label{ex:cond_i_RicciF}
Consider the Finsler metric from \cref{ex:FinslerVSI},
\begin{align}
F = \left|-2\D u \D v +(\tilde H + \rho\, v)\D u^2 + 2W_a\D u\D x^a + \delta_{ab}\D x^a \D x^b\right|^{\frac{1+m}{2}}\left(\D u\right)^{-m},
\end{align}
with $m\neq 1$, denote  $x^1 = x$ and $x^2 = y$ and set 
\begin{align}
    \tilde H(u,x,y) = k x^4 + \tfrac{1}{12} y^4, \qquad k = \frac{-1+3m-m^2}{12(m-1)^2},\\
    \rho(u,x,y) = x, \qquad  W_1(u,x,y)=0, \qquad W_2(u,x,y) = xy.
\end{align}
Then $\beta=\D u$ is null and it follows from the proof of \Cref{theorem:main} that $F$ is not locally metrizable, hence $F$ satisfies condition (i). On the other hand, it follows from the results in \cite{Fuster:2018djw} that $F$ is Ricci-flat. This shows that Ricci-flat $m$-Kropina spaces satisfying (i) do exist.
\end{ex}

%% file: intr3.tex
\chapter*{Introduction to \\
\cref{part:vacuum_sols}: Vacuum Solutions in Finsler
Gravity}

In \cref{part:Berwald_metrizability} we already briefly touched upon some aspects of the Finsler gravity field equation when we (locally) classified all Ricci-flat metrizable $m$-Kropina metrics. Such metrics are automatically vacuum solutions. \cref{part:vacuum_sols} focuses completely on the field equation---which we properly introduce in \cref{ch:Finsler_gravity}---and exact vacuum solutions to it. 

Most of the solutions that we present and investigate are of Berwald type. These are the topic of \cref{ch:Berwald_solutions}. Here, we first introduce a large class of general \ab-type solutions; we prove that when $\alpha$ is chosen to be a classical pp-wave metric and $\beta$ is its defining covariantly constant null 1-form, then any \ab-metric constructed from these two building blocks will be an exact vacuum solution. In the specific case of Randers metrics of Berwald type, we also show that \textit{any} Berwald vacuum solution must be of this type, resulting in a (local) classification of Berwald-Randers vacuum solutions. 
Completely analogous results hold for so-called \textit{modified} Randers metrics, obtained as a small modification of the standard Randers metric and introduced because of their preferable causal properties. 

An important question that arises is how such Finslerian spacetimes should be interpreted physically, and in particular, whether and how they can be physically distinguished from their general relativistic counterparts, given by the \psR metric $\alpha$. To answer this question we apply, in \cref{ch:lin}, a two-fold linearization scheme to the \ab-metric solutions obtained in \cref{ch:Berwald_solutions} and conclude that the linearized solutions may be interpreted as Finslerian gravitational waves. We then study the observational signature of these gravitational waves by investigating the question of what would be observed in an interferometer experiment when such a Finslerian gravitational wave would pass the earth, and what would be the difference with the effect of a classical general relativistic gravitational wave. To this end, we compute the Finslerian radar distance, the typical observable measured in interferometer experiments. Remarkably, when interpreted correctly, the expression for the Finslerian radar distance turns out to be completely equivalent to its general relativistic analog, obtained in \cite{Rakhmanov_2009}. In other words, we show that gravitational wave interferometers are not able to distinguish between our Finsler gravitational waves and standard general relativistic ones, which is a remarkable conclusion. 




Finally, in \cref{ch:unicorn_cosm} we present a class of exact vacuum solutions of unicorn type. A particularly interesting subfamily of this class has cosmological symmetry, i.e. is spatially homogeneous and isotropic, and is additionally conformally flat, with conformal factor depending only on the timelike coordinate. We show that, just as in classical Friedmann-Lema\^itre-Robertson-Walker (FLRW) cosmology, this conformal factor may be interpreted as the scale factor of the universe. We compute this scale factor as a function of cosmological time, and we show that the solution corresponds to a linearly expanding (or contracting) Finslerian universe.\\

Since \cref{part:vacuum_sols} is more physics-oriented than the preceding parts and generally deals with Finsler spaces of Lorentzian signature, we will (unless otherwise specified) assume that $\dim M = 4$ and use the convention that Greek indices $\mu,\nu,\rho,\sigma,\dots$ run from $0$ to $3 =\dim M-1$, while Latin ones $i,j,k,l,\dots$ run from $1$ to $3$. As in \cref{part:Berwald_metrizability}, we will also encounter Latin indices $a,b,c\dots$ from the beginning of the alphabet and these will run from $1$ to $2=\dim M-2$, consistent with our earlier conventions. 

%% file: ch_Finsler_gravity.tex
\chapter{Finsler Gravity}\label{ch:Finsler_gravity}


The whole field of Finsler gravity can be concisely summarized by the statement that \textit{Finsler gravity is just general relativity without the quadratic restriction}\footnote{This sentence is an adaptation of Chern's famous phrase saying that `Finsler geometry is just Riemannian geometry without the quadratic restriction.'}. The `quadratic restriction' refers to the fact that general relativity (GR) is based on \psR geometry, in which the squared line-element $\D s^2$ is, by definition, quadratic in the coordinate displacements. The essential idea is that Finsler gravity generalizes GR by lifting this quadratic restriction---leading to Finsler geometry as the relevant underlying mathematical framework---while retaining other fundamental aspects of the theory as much as possible. As an example, the time measured by a clock between two events is still given, as it is in GR, by the length of the clock's spacetime trajectory connecting these events; in this case the Finslerian length rather than the pseudo-Riemannian length. In fact, Finsler geometry is the most general geometric framework that is compatible with this formulation of the clock postulate\footnote{We remark that Weyl geometry, another generalization of Lorentzian geometry, is also compatible with the clock postulate, but in that case, the definition of proper time has to be revised \cite{Perlick1987}.}.

As discussed in some detail in the introduction there are various compelling arguments that motivate the study of Finsler gravity. While we will not repeat that discussion in detail here, we recall that much of this motivation comes from 
\begin{enumerate}[(i)]
    \item quantum gravity phenomenology \citeH{Addazi_2022} and specifically the idea of Lorentz invariance violation (or deformed Lorentz symmetry), which has been linked to Finsler geometry \cite{Girelli:2006fw,Raetzel:2010je,Rodrigues:2022mfj,Amelino-Camelia:2014rga,Lobo_2017, Letizia:2016lew};
    \item axiomatic approaches to spacetime and relativity \cite{TAVAKOL198523, Tavakol_1986, Tavakol2009,Ehlers2012,Lammerzahl:2018lhw,Bernal_2020,Bubuianu2018,Pfeifer_2019} that have been found to be compatible with geometric frameworks more general than pseudo-Riemannian geometry. And by the clock postulate argument given above, it seems that Finsler geometry is the most general framework one can aim for.
\end{enumerate}
There are also hints that Finsler gravity could provide a more accurate description of the gravitational field of a kinetic gas, by taking into account the individual motion of each gas particle \cite{Hohmann:2019sni,Hohmann:2020yia}, which is usually averaged over in the standard Einstein-Vlasov treatment \cite{Andreasson:2011ng,10.1063/1.4817035}. And there are hints that it might play a role in the understanding of dark matter and dark energy \cite{Hama2022}. Furthermore, some teleparallel gravity models can be described within the framework of Finsler geometry as well \cite{YehengFinslerTeleparallel}. 

Finally, apart from all of this, the simple fact that a theory as beautiful and successful as GR \textit{can} be generalized in a natural geometric way, without breaking it, elicits a natural curiosity as to what such an extension may bring us. I would even argue that this on its own provides ample motivation for taking the idea seriously.

\section{Finsler spacetimes}

\subsection{The basic mathematical definition}
As the starting point for our mathematical definition of a Finsler spacetime, we will simply stick to \Cref{def:Finsler_manifold} with the additional requirement that the signature of $g_{\mu\nu}$ be Lorentzian in some subset of the domain $\mathcal A$ of the Finsler metric. We purposely remain vague about this subset for the moment but we will come back to it later. The homogeneity condition \eqref{eq:F_homogeneity} ensures that the length is invariant under (orientation-preserving) parameterization and hence the time measured by a clock between two events can consistently be defined as the Finslerian length of the clock's spacetime trajectory connecting these events\footnote{We remark that time-orientation needs to be taken into account as well in case the Finsler function is not symmetric, i.e. if $F(x,y)\neq F(x,-y)$.}. 

As a definition of a Finsler spacetime, this is a relatively weak one in the sense that there are many other definitions appearing in the literature that are more restrictive.  In fact, various nonequivalent definitions exist (see e.g. \cite{Beem, Asanov, Pfeifer:2011tk, Pfeifer:2011xi,Lammerzahl:2012kw, Javaloyes2014-1, Javaloyes2019}) and a consensus on what is the `best' definition---if such a thing exists at all---has yet to be reached. It is not our current aim to significantly advance this endeavor, so the definition employed here has been chosen so as to be as general as possible. Indeed, most of the results we will discuss in what follows can be proven without further restrictions. It should be understood, however, that in order to guarantee that a viable physical interpretation is possible, the geometry should be subjected to more stringent requirements. Some of these are outlined below, but we will mostly discuss such considerations on a case-by-case basis throughout the chapters that follow. 


\subsection{Some remarks on causal structure and signature}

Given a Finsler spacetime geometry, it is natural to postulate, in analogy with GR, that matter travels along timelike geodesics and light travels on null geodesics. The notion of a \textit{null} direction is generalized in a mathematically straightforward way: a vector $y^u$ at a point $x^\mu$ is said to be null (or lightlike) if $L(x,y) = g_{\mu\nu}(x,y)y^\mu y^\nu=0$. However, the structure of the light cone, composed of such null vectors, may be nontrivial. In GR it is always the case that the light cone separates the tangent space at each point into three connected components, which we may interpret as consisting of forward-pointing timelike vectors, backward-pointing timelike vectors, and spacelike vectors, respectively. It is then consistent to define a timelike vector as one that has negative (or positive, depending on the sign convention) norm; that precisely singles out what we would intuitively call the inside or the interior\footnote{We're not talking about the topological interior here.} of the light cone, sometimes called the timelike cone. For a generic Finsler spacetime metric, however, these properties of the light cone structure are by no means guaranteed, and as such it is not obvious in general how to even define what one means by timelike vectors. It certainly is not obvious that these should be defined as vectors with negative norm. We will not consider this issue any further in its full generality here but rather discuss the details on a case-by-case basis. In many of the explicit examples of Finsler spacetimes that we will encounter, the causal structure (i.e. light cone structure) has exactly the desirable properties mentioned above, just as in the case of GR, allowing for a straightforward physical interpretation.

In a scenario where a natural (forward and/or backward) timelike cone can be identified,  this cone should ideally be contained in the subbundle $\mathcal A$. Intuitively, this condition essentially states that the geometry is well defined for all timelike infinitesimal spacetime separations. Moreover, in the ideal case, the signature is Lorentzian throughout the entire timelike cone. That is, $g_{\mu\nu}(x,y)$ has Lorentzian signature whenever $y$ is a timelike vector at $x$. It can be argued that it is not strictly necessary for the signature of $g_{\mu\nu}$ to be Lorentzian at spacelike vectors $y$, as it would not be possible, even in principle, to perform any physical experiment that probes such spacelike directions. This is true even in GR, but in that case $g_{\mu\nu}$ does not depend on $y$ and so if the metric is Lorentzian in any direction it is automatically Lorentzian in all directions. In Finsler geometry this is not the case and hence it makes sense to make this distinction. For further arguments along these lines, we refer to \cite{Bernal_2020} and references therein. Whether the light cone should be contained in $\mathcal A$ (or its topological closure) and whether the signature should be Lorentzian at the light cone is a more delicate question, which we will not further explore here. Mathematically such a property is very convenient (see e.g. \cite{Javaloyes2019,javaloyes2023einsteinhilbertpalatini}), but many examples of interest simply do not seem to have this property.


\section{The field equations in vacuum}

We now turn to the field equation in Finsler gravity, which generalizes Einstein's field equation to the more general setting of Finsler spacetimes. While several earlier proposals exist in the literature \cite{Horvath1950,Horvath1952, Ikeda1981, Asanov1983, Chang:2009pa,Kouretsis:2008ha,Stavrinos2014,Voicu:2009wi,Minguzzi:2014fxa}, we consider here Pfeifer and Wohlfarth's field equation \cite{Pfeifer:2011xi,Pfeifer:2013gha,Hohmann_2019}, which we deem to have the firmest basis.

\subsection{Motivation for the field equations}

Pfeifer and Wohlfarth's field equation can be derived, or rather motivated, by the following simple argument. In Newtonian gravity, the separation vector $\vec v$ between two nearby freely falling particles satisfies, in Cartesian coordinates, the deviation equation
\begin{align}\label{eqq;1}
\frac{\D^2 v^k}{\D t^2}  + H^k{}_i v^i = 0, \qquad H^k{}_i = \delta^{kj}\frac{\partial^2\phi}{\partial x^i \partial x^j},
\end{align}
where $\phi(x)$ is the gravitational potential. In vacuum, the equation that determines the gravitational potential is the Laplace equation $\Delta \phi = \delta^{ij}\frac{\partial^2\phi}{\partial x^i \partial x^j}=0$, which can be written simply as $H^i{}_i = 0$, i.e. the \textit{vanishing of the trace of the deviation tensor}. Of course, this principle extends to GR, where the gravitational field equation in vacuum is again equivalent to the vanishing of the trace of the geodesic deviation tensor, $R_{\mu\nu}=0$. 

We can immediately generalize this to the Finsler geometric setting. The analogous situation is then governed by the Finslerian geodesic deviation equation \eqref{eq:geodesic_deviation_ds} along a geodesic $\gamma$,
\begin{align}\label{eqq;2}
\frac{D^2 v^\rho}{D s^2}  + R^\rho{}_\mu (\gamma,\dot\gamma)v^\mu = 0,
\end{align}
where the geodesic deviation operator is given by $R^\rho{}_\mu(x,y) = R^\rho{}_{\mu\nu}(x,y)y^\nu$ in terms of the nonlinear curvature $R^\rho{}_{\mu\nu}$ of the Finsler metric. 
In analogy with Newtonian gravity and GR, it is natural to postulate that in the Finslerian case, the trace of the deviation tensor must vanish as well, leading to the candidate field equation $R^\mu{}_\mu = 0$. Since $R^\mu{}_\mu = \text{Ric}$, the Finsler-Ricci scalar (see \cref{sec:cuvatures_on_Finsler_space}), this leads to the equation $\text{Ric}=0$ that was proposed by Rutz \cite{Rutz}. When restricted to Lorentzian metrics, Rutz's equation reduces to Einstein's vacuum equation $R_{\mu\nu}=0$, since in that case $\text{Ric} = R_{\mu\nu}(x) y^\mu y^\nu$. Hence the equation yields the correct general relativistic limit.

It turns out, however, that Rutz's equation is not variational; there does not exist an action functional that yields the Rutz equation as its Euler-Lagrange equation. Although this may not necessarily be problematic, it is not obvious in such a case how to couple the theory to matter, let alone how to eventually quantize the theory. The variational completion of Rutz's equation, i.e. the unique variational equation that is \textit{as close to it as possible}, in a well defined sense \cite{Voicu_2015}, turns out to be the field equation that was proposed by Pfeifer and Wohlfarth in \cite{Pfeifer:2011xi} using a Finsler extension of the Einstein-Hilbert action \cite{Hohmann_2019} (details below). 

This is completely analogous to the situation in GR, where the correct\footnote{Correct in the sense that it also leads to the correct matter coupling when the energy-momentum tensor is inserted on the RHS.} form of the vacuum field equation $R_{\mu\nu} - \frac{1}{2}g_{\mu\nu}R = 0$ is also precisely the variational completion of Einstein's earlier proposal for a vacuum equation, $R_{\mu\nu}=0$, obtained from the geodesic deviation argument above\footnote{We remark that Einstein himself arrived at both his early proposal $R_{\mu\nu}=0$ as well as his final expression $R_{\mu\nu} - \frac{1}{2}g_{\mu\nu}R = 0$ for the vacuum equation by alternative means \cite{Landsman2021}.} \cite{Voicu_2015}. While in the GR case, the variationally completed equation happens to be equivalent to the original one, this is not true any longer in the Finsler setting.

\subsection{Equivalent formulations of the field equation}

In four spacetime dimensions, Pfeifer and Wohlfarth's field equation in vacuum reads\footnote{We note that in \cite{Pfeifer:2011xi} the Landsberg tensor is defined with a relative minus sign with respect to our definition and in both references \cite{Pfeifer:2011xi,Hohmann_2019} the Ricci curvatures are defined with a relative minus sign with respect to our definition.} \cite{Pfeifer:2011xi,Hohmann_2019}
\begin{align}\label{eq:Pfeifer_Wohlfarth_eq}
-2\text{Ric} + \frac{2 L}{3}g^{\mu\nu}R_{\mu\nu}+\frac{2 L}{3}g^{\mu\nu}\left(\bar\partial_\mu\left(\nablad S_\nu\right) - S_\mu S_\nu  + \nabla^C_{\delta_\mu}S_\nu \right) = 0,
\end{align}
in terms of (the horizontal part of) the Chern-Rund connection $\nabla^C_{\delta_\mu}$ and the dynamical covariant derivative $\nablad$. This is essentially the Euler-Lagrange equation in four dimensions corresponding to the action functional
\begin{align}\label{eq:Finsler_Einstein_Hilbert_action}
    S = \int_{SM}\text{Ric}\,\D \mu_{SM},
\end{align}
where $SM = \{(x,y)\in TM\,:\,F(x,y)=1\}$ is the unit tangent bundle, or \textit{indicatrix}, with volume element $\D \mu_{SM}$. In order to make rigorous sense of this action functional and its critical points, however, one needs to use a construction in terms of compact subsets of the (positively) projective tangent bundle. When $F$ is pseudo-Riemannian, \eqref{eq:Finsler_Einstein_Hilbert_action} reduces to the Einstein-Hilbert action, \eqref{eq:Einstein_Hilbert_action} below, up to a boundary term. For details, we refer to \cite{Hohmann_2019}. 

Equivalently, the equation \eqref{eq:Pfeifer_Wohlfarth_eq} can be expressed in terms of the Berwald connection rather than the Chern-Rund connection. Using the fact that the difference between the horizontal part of the Berwald and Chern-Rund connection is given by the Landsberg tensor (see \cref{sec:Landsberg}), the field equation \eqref{eq:Pfeifer_Wohlfarth_eq} can be rewritten as\footnote{Since $g^{ij}\nabla^B_{\delta_i}S_j= g^{ij}\left(\nabla^C_{\delta_i}S_j  - S_iS_j \right)$.}
\begin{align}
-2\text{Ric} + \frac{2 L}{3}g^{\mu\nu}R_{\mu\nu}+\frac{2 L}{3}g^{\mu\nu}\left(\bar\partial_\mu\left(\nablad S_\nu\right) + \nabla^B_{\delta_\mu}S_\nu \right) = 0.
\end{align}
This formulation\footnote{We note that also in this reference both the Landsberg tensor and the Ricci tensor(s) are defined with a relative minus sign with respect to our definitions.} appears for instance in \cite{Fuster:2018djw}.

It is worth pointing out that a very similar equation has been derived in the case of positive definite Finsler metrics. In \cite{Chen-Shen}, Chen and Shen consider the action functional
\begin{align}\label{eq:normalized_Finsler_Einstein_Hilbert_action}
    S = \frac{1}{\text{Vol}(SM)^{1-2/n}}\int_{SM}\text{Ric}\,\D \mu_{SM},
\end{align}
where $n = \dim M$. This is the normalized version of \eqref{eq:Finsler_Einstein_Hilbert_action}. In terms of the Chern-Rund connection, the Euler-Lagrange equations in four dimensions read
\begin{align}\label{eq:Chen_Shen_field_eq}
-2\text{Ric} + \frac{2 L}{3}g^{\mu\nu}R_{\mu\nu}+\frac{2 L}{3}g^{\mu\nu}\left(\bar\partial_\mu\left(\nablad S_\nu\right) - S_\mu S_\nu + \nabla^C_{\delta_\mu}S_\nu \right) = \frac{2L}{3}r,
\end{align}
where 
\begin{align}
    r = \frac{1}{\text{Vol}(SM)}\int_{SM}\text{Ric}\,\D \mu_{SM}
\end{align}
is the average value of Ric on $SM$. Clearly the only difference between  \eqref{eq:Pfeifer_Wohlfarth_eq} and \eqref{eq:Chen_Shen_field_eq} is the presence of a possibly nonvanishing (constant) value of $r$ on the RHS, playing the role of a kind of cosmological constant. Indeed, in the case that the geometry is pseudo-Riemannian, \eqref{eq:Chen_Shen_field_eq} reduces to the equation for an Einstein space,
\begin{align}
    R_{\mu\nu} = r g_{\mu\nu},
\end{align}
or, in other words, the vacuum field equation for GR in the presence of a possibly nonvanishing cosmological constant $\Lambda = r$. Similarly, the Pfeifer-Wohlfarth equation \eqref{eq:Pfeifer_Wohlfarth_eq} reduces to Einstein's equation in vacuum without cosmological constant, $R_{\mu\nu}=0$.

For reference, it is worth briefly comparing the results discussed above with the classical results for Riemannian metrics. It is well-known (see e.g. \cite{RiemannFunctionals}) that for Riemannian metrics on a compact manifold $M$ the critical points of the Einstein-Hilbert action functional 
\begin{align}\label{eq:Einstein_Hilbert_action}
    S_\text{EH} =\int_{M}R\,\D \mu_{M},
\end{align}
are precisely the Ricci-flat metrics, $R_{\mu\nu}=0$, while the critical points of the normalized Einstein-Hilbert action
\begin{align}\label{eq:normalized_Einstein_Hilbert_action}
    S = \frac{1}{\text{Vol}(M)^{1-2/n}}\int_{M}R\,\D \mu_{M},
\end{align}
are precisely the Einstein metrics, $R_{\mu\nu}=\Lambda g_{\mu\nu}$.\\

\textit{We will refer to Pfeifer and Wohlfarth's field equation, \eqref{eq:Pfeifer_Wohlfarth_eq} simply as \textit{the} field equation in Finsler gravity, and solutions to it will be referred to simply as solutions in Finsler gravity.}

\subsection{Weakly Landsberg and Berwald field equation}

The field equations simplify considerably if one restricts attention to the class of weakly Landsberg metrics (see \cref{sec:Landsberg}), defined by the condition that $S_\mu =0$, which holds in particular for all Berwald spaces as well as all Landsberg spaces. In that case, all terms involving the mean Landsberg curvature vanish identically and \eqref{eq:Pfeifer_Wohlfarth_eq} reduces to
\begin{align}\label{eq:Pfeifer_Wohlfarth_eq_Landsberg_pre}
\text{Ric} - \frac{ L}{3}g^{\mu\nu}R_{\mu\nu} = 0,
\end{align}
or equivalently, 
\begin{align}\label{eq:Pfeifer_Wohlfarth_eq_Landsberg}
\left(L g^{\mu\nu}-3 y^\mu y^\nu \right)R_{\mu\nu} = 0.
\end{align}
From this, it follows that within the class of weakly Landsberg spaces, the vanishing of the Finsler-Ricci tensor, $R_{\mu\nu}=0$, is a sufficient condition for being a vacuum solution to the field equations. It turns out that in many cases of interest (but not in general) it is a necessary condition as well. For instance, for Randers metrics of Berwald type, the vacuum field equation is equivalent to  $R_{\mu\nu}=0$, as we will prove in \cref{sec:RandersSolutions}, and similar results have been obtained in the case of null $m$-Kropina metrics \cite{Fuster:2018djw} and Finsler spacetimes satisfying strict smoothness requirements \cite{javaloyes2023einsteinhilbertpalatini}. This equivalence also holds for spherically symmetric and asymptotically flat Berwald spacetimes; details will be provided in a forthcoming article. By a recent generalization of Birkhoff's theorem to spherically symmetric, Ricci-flat Berwald spaces \cite{voicu2023birkhoff}, this implies in particular, that any vacuum solution of Berwald type that is spherically symmetric and asymptotically flat is either flat (in the sense that the nonlinear curvature tensor vanishes identically) or given by the Schwarzschild metric. In general, the vacuum equation is not equivalent to Ricci-flatness, though: non--Ricci-flat vacuum solutions do exist. An explicit example illustrating this will also be provided in a forthcoming article.

Since any Berwald space is weakly Landsberg,  \eqref{eq:Pfeifer_Wohlfarth_eq_Landsberg} applies in particular to Berwald spaces. In this case $R_{\mu\nu} = R_{\mu\nu}(x)$ is independent of $y$, which is an incredible convenience, computationally. When $L$ is pseudo-Riemannian, and hence, in particular, Berwald, the inverse fundamental tensor $g^{\mu\nu}=g^{\mu\nu}(x)$ is also independent of $y$ and hence both terms in the field equation \eqref{eq:Pfeifer_Wohlfarth_eq_Landsberg} are quadratic in $y$. Differentiating twice with respect to the $y$-coordinates implies that $3 R_{\mu\nu} - g_{\mu\nu}R=0$ and taking the trace then reveals that $R=0$ and hence $R_{\mu\nu}=0$. Hence for pseudo-Riemannian metrics, the field equation in vacuum indeed reduces to Einstein's field equations in vacuum.

\subsection{Exact solutions}

The list of known exact solutions in Finsler gravity is very short. Indeed, prior to the contributions on which this dissertation is based, the only\footnote{Specifically in the realm of positive definite Finsler geometry, an additional class of solutions has been obtained \cite{Chen-Shen}.} ones known in the literature were the
\begin{enumerate}[(i)]
    \item ($m$-Kropina type) Finsler pp-waves \cite{Fuster:2015tua} and their generalizations as
    \item \textit{very general relativity} (VGR) spacetimes \cite{Fuster:2018djw}.
\end{enumerate}
%
%
%
Over the last couple of years, this list has been extended by the
\begin{enumerate}[(i)]
    \setcounter{enumi}{2}
    \item Randers pp-waves \citeref{ref4}{H4} with their extension to
    \item generic \ab-metric pp-waves \citeref{ref5}{H5}, and finally
    \item the unicorn solutions \citeref{ref6}{H6}.
\end{enumerate}
Additionally, vacuum solutions of Berwald type with cosmological symmetry have been classified and properly Finslerian spherically symmetric vacuum solutions have been obtained, the details of which will appear in future publications. It is the aim of the chapters that follow to properly introduce and discuss the solutions $(iii)$--$(v)$ in detail.

%% file: ch_exact_Berwald_solutions.tex
\chapter{Exact Vacuum Solutions of Berwald Type}\label{ch:Berwald_solutions}

In this chapter, we present and investigate those of our solutions that are of Berwald type. We begin in \cref{sec:ab_metric_solutions} with generic \ab-metric solutions, then specialize to Randers metrics in \cref{sec:RandersSolutions}, and to modified Randers metrics in \cref{sec:Modified_Randers_Metrics}. Regarding \ab-metrics, we stick to the notation introduced in \cref{sec:abMetricsNotation} and we recall, in particular, that by some abuse of language, we will refer to either of $\alpha$, $A$, $a$ and $a_{ij}$ as the \psR metric, and to $\beta$, $b$ and $b_i$ as the 1-form, and sometimes we will write expressions such as $\beta = \D t$ and $\alpha = \sqrt{\left|-\D t^2 + \D x^2 + \D y^2 + \D z^2\right|}$,
but the interpretation of such expressions should be clear. In the presence of an \ab-metric, indices will be raised and lowered with $a_{ij}$ unless otherwise specified. Moreover, when discussing properties of the 1-form $\beta$ such as being `covariantly constant' (or `parallel', which has the same meaning) or `null', we will always mean that $\beta$ is covariantly constant \textit{with respect to $\alpha$} or null \textit{with respect to $\alpha$}, respectively, unless otherwise specified.

\section{\texorpdfstring{$(\alpha,\beta)$}{(\textalpha,\textbeta)}-metrics}\label{sec:ab_metric_solutions}

The results in this section are based mostly on \citeref{ref5}{H5}.

\subsection{Construction of exact solutions}

From a physical perspective, $(\alpha,\beta)$-metrics (see \cref{sec:abMetricsNotation}) allow us to deform a general relativistic (i.e. Lorentzian) spacetime $\alpha$ into a Finsler spacetime, given a 1-form $\beta$. And it turns out, as we will prove below, that these types of metrics can be used to generalize some of the vacuum solutions to Einstein's field equations to properly Finslerian vacuum solutions to Pfeifer and Wohlfarth's field equation for Finsler gravity. This procedure is possible whenever such a classical solution admits a covariantly constant vector field, or equivalently, 1-form. More precisely, if the Lorentzian metric $\alpha$ solves the classical Einstein equations in vacuum and the 1-form $\beta$ is covariantly constant with respect to $\alpha$ then any $(\alpha,\beta)$-metric constructed from the given $\alpha$ and $\beta$ is a solution to the Finslerian field equations. 

To see why this is true, we first recall \Cref{eq:ab_cc_pre}, which states that if $\beta$ is covariantly constant with respect to $\alpha$ then any \ab-metric $F$ constructed from these building blocks is of Berwald type and the affine connection of $F$ coincides with the (Levi-Civita) affine connection of $\alpha$.

If the affine connections coincide, the associated curvature tensors and in particular the (affine) Ricci tensors must also coincide. So if $\alpha$ happens to be a vacuum solution to Einstein gravity, i.e. its Ricci tensor vanishes, then it follows that the affine Ricci tensor of $F$ must vanish as well. And since the Finsler-Ricci tensor is the symmetrization of the affine Ricci tensor, by \cref{lem:RicciTensors}, the Finsler-Ricci tensor vanishes as well. This implies that \eqref{eq:Pfeifer_Wohlfarth_eq_Landsberg} is satisfied, i.e. $F$ is a vacuum solution in Finsler gravity. We may summarize this result with the following theorem.
\begin{theor}\label{theor:(alphabeta)solutions}
Let $F$ be any $(\alpha,\beta)$-metric such that:
\begin{enumerate}[(i)]
    \item $\alpha$ solves the classical Einstein equations in vacuum;
    \item $\beta$ is covariantly constant.
\end{enumerate}
Then $F$ is a Ricci-flat vacuum solution to the field equation in Finsler gravity. Moreover, the affine connection of $F$ coincides with the Levi-Civita connection of $\alpha$. 
\end{theor}
%
In this way, $(\alpha,\beta)$-metrics provide a mechanism to \textit{Finslerize} any vacuum solution to Einstein's field equations, as long as the solution admits a covariantly 1-form or equivalently, a covariantly constant vector field. The theorem generalizes some of the results obtained in \citeref{ref4}{H4} for Randers metrics and in \cite{Fuster:2015tua,Fuster:2018djw} for $m$-Kropina metrics (or VGR spacetimes) to arbitrary Finsler spacetimes with $(\alpha,\beta)$-metric. In particular, all pp-wave type solutions in Finsler gravity currently known in the literature are of this type.

\subsection{pp-Wave solutions}\label{sec:ab_pp_wave_solutions}

Let us investigate the solutions of the type of \cref{theor:(alphabeta)solutions} in some more detail. It turns out that if a vacuum solution $\alpha$ to Einstein's field equations admits a covariantly constant 1-form $\beta$ then at least one of the following conditions must hold:
\begin{enumerate}
    \item  $\alpha$ is flat;
    \item $\beta$ is null.
\end{enumerate}
This was shown for the first time in \cite{EhlersKundt1960} (see also \cite{Hall_2000,Batista_2014}) and in \cref{app:metrics_w_parallel_vector_field} we provide a self-contained proof, starting from the Frobenius theorem and culminating in \Cref{cor:EFE_sol_with_parallel_field}. We remark that the result assumes that the dimension of the pseudo-Riemannian manifold is $4$; it is generally not true in higher dimensions. Hence \cref{theor:(alphabeta)solutions} leads to two classes of solutions.

\subsubsection{Class 1: flat solutions}
If $\alpha$ (and hence $F$) is flat then in coordinates 
in which all Christoffel symbols vanish, $\beta$ must have constant components. Hence the resulting class of solutions is in some sense trivial---yet not necessarily uninteresting. Provided $\alpha$ is of Lorentzian signature\footnote{The result can easily be generalized to other signatures as well.}, it can always be written in suitable coordinates in the following way, where we stick to the notation introduced in \cref{sec:abMetricsNotation}.\\


\begin{myblock}[Class 1: flat solutions]
       Any \ab-metric constructed from 
        \begin{align}\label{eq:class1_ab_sols}
            a = -\D t^2 + \D x^2 + \D y^2 + \D z^2 , \qquad b = b_\mu \D x^\mu,
        \end{align}
        where $b_\mu=$ constant, is a vacuum solution to the field equations in Finsler gravity. The resulting geometry is of Berwald type with all affine connection coefficients vanishing identically in these coordinates.
   \end{myblock}



\subsubsection{Class 2: pp-wave solutions}
The second possibility, the one where $\beta$ is null, leads to a richer and more interesting class of solutions. In this case, $\alpha$ is a CCNV metric, meaning that it admits a covariantly constant null vector (CCNV), namely $\beta$, or rather its vector equivalent via the isomorphism induced by $\alpha$. In Lorentzian signature, CCNV metrics are also known as \textit{pp-waves} (plane-fronted gravitational waves with parallel rays) and these have been studied in detail in \cite{EhlersKundt1960,ehlers1962exact} (see \cite[Section 24.5]{stephani_kramer_maccallum_hoenselaers_herlt_2003} for a summary). 

In \cref{part:Berwald_metrizability} we have already encountered such metrics. In particular, we proved the important \cref{lem:coordchange}. Combined with \cref{prop:coords_cc_null_1form}, and using the fact that $\alpha$ is a vacuum solution and hence Ricci-flat, this lemma says that coordinates $(u,v,x,y)$ can always be chosen such that $\alpha$ and $\beta$ attain the form
    \begin{align}\label{eq:pp_waves_final_form_body}
    a = -2\D u\D v + H(u,x,y)\D u^2 + \D x^2 + \D y^2, \qquad b = \D u,
\end{align}
and such that $\Delta_{(x,y)}H = (\partial_{x}^2+\partial_{y}^2)H = 0$. The last statement also follows easily from the fact that the Ricci tensor of the metric $a$ is given (up to a nonzero constant) by the $(x,y)$-Laplacian $\Delta_{(x,y)}H$ of $H$. We may therefore characterize the second class of solutions in the following way---again assuming for concreteness that $\alpha$ is a Lorentzian metric.\\



\begin{myblock}[Class 2: pp-wave solutions]
    Any \ab-metric constructed from
    \begin{align}
        a &= -2\D u\D v + H(u,x,y)\, \D u^2+\D x^2 + \D y^2, \label{eq:reduced_pp_wave_metric2} \\
        b &= \D u \label{eq:reduced_pp_wave_1_form2},
    \end{align}
    such that $\Delta_{(x,y)}H = 0$, is a vacuum solution to the field equations in Finsler gravity. The resulting geometry is of Berwald type with affine connection identical to the Levi-Civita connection of $a$.
   \end{myblock}

\noindent Note that when $H=0$ the geometries in \textit{Class 2} are also contained in \textit{Class 1}. It is not the case, however, that \textit{Class 1} is a subset of \textit{Class 2} because in \textit{Class 1} the 1-form $\beta$ need not be null,  necessarily. The preceding line of argument shows that these two classes of solutions in fact exhaust all possibilities, which we encapsulate in the following theorem.
\begin{theor}\label{theor:(alphabeta)solutions2}
Any $(\alpha,\beta)$-metric such that
\begin{enumerate}[(i)]
    \item $\alpha$ solves the classical Einstein equations in vacuum;
    \item $\beta$ is covariantly constant,
\end{enumerate}
(i.e. \textit{any} vacuum solution of the type of \Cref{theor:(alphabeta)solutions}) must belong to either \textit{Class 1} or \textit{Class 2}, provided that $\alpha$ has Lorentzian signature.
\end{theor}
Before we move on to solutions of plane wave type, we end this section with some remarks about the relation between our \ab-type solutions and other solutions in the literature.
\begin{itemize}
\item The ($m$-Kropina type) Finsler pp-waves \cite{Fuster:2015tua} and the Randers pp-waves obtained in \citeref{ref4}{H4} are special cases of the solutions introduced here.
\item For Randers metrics of Berwald type \textit{any} vacuum solution must be of the type described in \cref{theor:(alphabeta)solutions}, that is, $\alpha$ is necessarily a vacuum solution in Einstein gravity and $\beta$ is necessarily covariantly constant. Any such solution is therefore either in \textit{Class 1} or \textit{Class 2}. This was first shown in \citeref{ref4}{H4} and we will prove it in \cref{sec:RandersSolutions}.
\item For $m$-Kropina metrics, some of the vacuum solutions obtained \cite{Fuster:2018djw} are of a more general type than the ones introduced here, as illustrated by \cref{ex:cond_i_RicciF}; our solutions are all trivially metrizable (by $\alpha$), whereas the metric in \cref{ex:cond_i_RicciF} is not metrizable. In fact, our results from \cref{sec:classification_ricci_flat} show that any of the solutions obtained in \cite{Fuster:2018djw} belongs to either \textit{Class 1} or \textit{Class 2} if and only if the geometry is metrizable, and \cref{ex:FinslerVSI} provides the precise
 necessary and sufficient condition for this to be the case.
\end{itemize}
This list comprises all exact solutions to Pfeifer and Wohlfarth's equation for Finsler gravity currently known in the literature, apart from:
\begin{itemize}
    \item the unicorn solutions \citeref{ref6}{H6} that are the topic of \cref{ch:unicorn_cosm};
    \item \psR metrics that solve Einstein's vacuum field equation, which are trivially vacuum solutions in Finsler gravity as well.
\end{itemize}

\subsection[Plane wave solutions]{Plane wave solutions in Brinkman and Rosen coordinates}

Equation \eqref{eq:reduced_pp_wave_metric2} expresses the pp-wave metric in Brinkmann form \cite{Brinkmann:1925fr}. In this form the metric reads
\begin{align}\label{eq:Brinkmann_2}
    a &= -2\D u\D v + H(u,x^a)\, \D u^2+\delta_{ab}\, \D x^a \D x^b,
\end{align} 
where we have used the notation $x = x^1$, $y=y^1$, which will be somewhat more convenient for our present purposes, and where indices $a,b,c\dots$ run from $1$ to $2$, consistent with the conventions we have stuck to all along. For the description of the physical effects of (plane) gravitational waves in general relativity, it is sometimes more convenient to use a different coordinate system, known as Rosen coordinates \cite{rosen1937plane}. In our experience, this remains true in the Finsler setting. When we compute the effect on the radar distance of a passing (linearized) Randers gravitational wave in \cref{ch:lin}, our starting point will be the expression for the (exact) wave in Rosen coordinates. Therefore we briefly review the relation between the two coordinate systems here.

Rosen coordinates can be introduced for the subclass of pp-waves known as \textit{plane waves}. These are characterized by the property that the curvature tensor does not change (i.e. is covariantly constant) along the Euclidean `wave surfaces' given in Brinkmann coordinates by $\D u = \D v = 0$, i.e.
\begin{align}\label{eq:invariance_of_Riemann_curvature}
\nabla_{\partial_{x^1}} R^\rho{}_{\sigma\mu\nu} = \nabla_{\partial_{x^2}} R^\rho{}_{\sigma\mu\nu} = 0.
\end{align}
We remark as a side note that $\nabla_{\partial_v} R^\rho{}_{\sigma\mu\nu} = 0$ always holds, identically, so invariance along $\D u = \D v = 0$ is actually equivalent to invariance along the hypersurfaces $\D u = 0$. The conditions \eqref{eq:invariance_of_Riemann_curvature} are equivalent to the statement that $\partial_a\partial_b\partial_c H = 0$ in Brinkmann coordinates \eqref{eq:Brinkmann_2}, i.e. that $H(u,x,y)$ is a quadratic polynomial in $x^a$ (for each fixed value of $u$). In that case there always exists a coordinate transformation that removes the linear and constant terms  
\cite[Section 24.5]{stephani_kramer_maccallum_hoenselaers_herlt_2003} and brings the metric into the form
\begin{align}\label{eq:plane_wave_brinkmann}
a = -2\D u \D v + A_{ab}(u)x^ax^b\, \D u^2+\delta_{ab}\, \D x^a \D x^b
\end{align}
This is the standard expression for a plane-wave metric in Brinkmann form. Moreover, an argument completely analogous to the one given in the proof of \cref{lem:coordchange} shows that we may assume without loss of generality that the 1-form $b = \D u$ remains unchanged under the transformation to these new coordinates. 

On the other hand, any such plane wave metric can also be written in Rosen form
\begin{align}\label{eq:Rosen_form}
a = -2\D U\D V + h_{ij}(U)\D y^i \D y^j,
\end{align}
where $h_{ij}$ is a two-dimensional Riemannian metric. And conversely, any metric of Rosen type \eqref{eq:Rosen_form} can be cast in the form \eqref{eq:plane_wave_brinkmann}. The two coordinate systems are related via
\begin{align}
U = u,\quad V = v - \dfrac{1}{2}\dot E_{ai} E^i{}_b x^a x^b, \quad x^a = E^a{}_iy^i,
\end{align}
where $A_{ab} = \ddot E_{ai}E^i{}_b$ and $E^a{}_i$ is a vielbein for $h_{ij}$ in the sense that $h_{ij} = E^a{}_i E^b{}_j \delta_{ab}$, satisfying the additional symmetry condition $\dot E_{ai} E^i{}_b = \dot E_{bi} E^i{}_a$. Such a vielbein can always be chosen. For details, we recommend the lecture notes \cite{Blau2011} and references therein (see also the Appendix of \cite{Blau_2003}). Note that we have momentarily labeled the $y$-coordinates by indices $i,j,k,\dots$ so as to distinguish them from indices $a,b,c,\dots$ in order that we may employ the usual notation with regard to the vielbein indices: $E^i{}_a$ represents the (matrix) inverse of $E^a{}_i$ and indices $a,b,c\dots$ are raised and lowered with $\delta_{ab}$, whereas indices $i,j,k,\dots$ are raised and lowered with $h_{ij}$. In this case $i,j,k,\dots$ thus take the values 1 and 2, contrary to our standard convention. The dot that sometimes appears above the vielbein represents a $U$-derivative. Since the vielbein depends only on $U=u$, this derivative is equivalent to a $u$-derivative, and moreover, the act of raising and lowering the $a,b,c,\dots$ indices commutes with taking such a derivative of the vielbein.

It is again the case that, after relabeling $U,V\mapsto u,v$, the form of the 1-form $b = \D u = \D U$ remains unchanged under this transformation, which in this case follows trivially, since $u=U$. After also relabelling $y\mapsto x$, we conclude that we can express any \textit{Class 2} solution of plane-wave type in Rosen coordinates as follows,
\begin{align}
F = \alpha\, \phi(\beta/\alpha), \qquad a = -2\D u\D v + h_{ij}(u)\D x^i \D x^j, \qquad b = \D u \label{eq:ab_plane_wave},
\end{align}
Conversely, for any choice of $\phi$ and $h_{ij}(u)$, this is a vacuum solution to the field equations in Finsler gravity if $a$ is a vacuum solution to Einstein's field equation. The resulting geometry is of Berwald type with affine connection identical to the Levi-Civita connection of $a$. The latter follows from the line of argument above but is also easy to check explicitly since all Christoffel symbols of $a$ in Rosen coordinates with an upper index $u$ vanish identically, and as a consequence, $b = \D u$ is covariantly constant (as we already knew, of course), so that the result follows by \Cref{theor:(alphabeta)solutions}.

\section{Randers metrics}
\label{sec:RandersSolutions}

Next, we turn to Randers metrics \cite{Randers}, i.e. \ab-metrics of the specific form $F =  \alpha + \beta$. We already encountered such metrics in \cref{ch:Characterization_of_Berwald}. Apart from their intrinsic value as relatively simple and computable, yet nontrivial examples of Finsler metrics, Randers metrics have also found a wealth of applications in a wide variety of fields such as biology, psychology, and physics. Most of the results developed in this section are based on \citeref{ref4}{H4}. We point out, though, that \citeref{ref4}{H4} uses a signature convention that is opposite to the one employed here. Before turning to the field equations, we briefly discuss the domain of definition of the standard Randers metric.

\subsection{Domain of definition}\label{sec:RandersDomain_of_def}


 In the case where $\alpha$ is positive definite, it is well-known (see e.g. \cite{ChernShen_RiemannFinsler}) that $F=\alpha+\beta$ is a positive definite Finsler metric if and only if $|b|<1$. A crucial ingredient that leads to this conclusion is the Cauchy-Schwarz inequality $|\beta|\leq |b| \alpha $, which does not extend in this simple form to indefinite signatures. In the latter case, the situation therefore becomes somewhat more complex. It can be shown using the general formula \eqref{eq:det_ab} for the determinant of the fundamental tensor of an \ab-metric that 
\begin{align}\label{eq:determinant_Randers}
\det g =  \sgn(A)^n\left(\frac{\alpha+\beta}{\alpha}\right)^{n+1}\det a
\end{align}
for Randers metrics, where $n=\dim M$. This expression has in general no well-defined limit as $\alpha\to 0$, so from this, it is clear that in Lorentzian signature, the subbundle $\mathcal A$ can only include points $(x,y)$ for which $\alpha\neq 0$. In order to have a connected subbundle, one might propose to work on the subbundle consisting of $\alpha$-timelike vectors, i.e. the (forward and backward) timelike cone of $\alpha$. However, this does not work in general; one has to restrict $\mathcal A$ just a little bit further. The following result was proven in \citeref{ref4}{H4} (cf. \cref{sec:Modified_Randers_Metrics} and \cite{Voicu23_Finsler_ab_spacetime_condition} for related results).
\begin{prop}\label{prop:Randers_well_defined_on_future_timecone}
Given a Lorentzian metric $\alpha = \sqrt{|a_{\mu\nu}y^\mu y^\nu|}$ on a manifold $M$ with time-orientation $T$ and a past-pointing 1-form $\beta$ that is either null or timelike with respect to $a$, the Randers metric $F=\alpha+\beta$ defines a Finsler spacetime with Lorentzian signature on 
\begin{align}
\mathcal A = \left\{(x,y)\in TM\,: a_{\mu\nu}(x)y^\mu y^\nu<0,\,\, a_{\mu\nu}(x)T^\mu(x)y^\nu<0\right\},
\end{align}
i.e. the forward timelike cone of $\alpha$.
\end{prop}
%
%
%
In order to prove this, we start with some lemmas. Note that the result of \cref{prop:Randers_well_defined_on_future_timecone}, as well as that of \cref{lem:Randers_well_defined_on_future_timecone}, is dependent on the signature convention, which we recall has been chosen here as $(-,+,+,+)$. The first lemma is a standard result.
\begin{lem}\label{lem:constant_signature}
    If $\{M_\lambda\}_{\lambda\in [0,1]}$ is a continuous family of $n\times n$-matrices such that $\det M_{\lambda}\neq 0$ for all $\lambda$, then the signature of $M_{\lambda}$ is constant for $\lambda\in [0,1]$.
\end{lem}
\begin{proof}
    We use the well-known fact that the $n$ eigenvalues of $M_\lambda$ each depend continuously on $\lambda$ (see e.g. \cite[Cor. VI.1.6]{MatrixAnalysisBhatia}). We need to show that the respective signs of these $n$ eigenvalue functions are constant. Suppose for contradiction that the $k$-th eigenvalue is positive for some $\lambda_1$ and negative for some $\lambda_2$. By continuity, it follows from the intermediate value theorem that there must exist some $\lambda$ between $\lambda_1$ and $\lambda_2$ for which the eigenvalue vanishes. In that case, the determinant of $M_\lambda$ vanishes for that value of $\lambda$, which is a contradiction. Hence the result follows.
\end{proof}
\begin{lem}\label{lem:Randers_well_defined_on_future_timecone}
Let $y$ be a timelike and future-oriented vector at $x\in M$, and $b$ a timelike or null, and past-oriented vector at $x$, all with respect to some time-oriented Lorentzian metric $g_{\mu\nu}$. Then $g_{\mu\nu}y^\mu b^\nu>0$.
\end{lem}
\begin{proof}
Let the time orientation be defined in terms of a nowhere-vanishing timelike vector field $v$. Then the future-orientation of $y$ and past-orientation of $b$ imply that $g_{\mu\nu}y^\mu v^\nu<0$ and $g_{\mu\nu}b^\mu v^\nu> 0$. Now notice that we can always pick a basis of $T_xM$ such that $v$ has coordinates $(v^0,0,\dots,0)$ with $v^0>0$ and such that $g_{\mu\nu}=\eta_{\mu\nu}=$ diag$(-1,1,1,1)$ is the Minkowski metric (first pick a basis such that the metric has the form of the standard Minkowski metric, then apply a suitable spatial rotation so that all spatial components of $v^\mu$ except $v^1$ are mapped to $0$, and finally apply an appropriate one-dimensional Lorentz boost to make $v^1$ vanish as well). In this basis we write $\vec y = (y^1, y^2, \dots , y^{n-1})$. With this notation, the future-- and past-pointing properties of $y$ and $b$, respectively, translate to $y^0>0$ and $b^0<0$ and the remaining properties can be stated as $|y^0|>|\vec y|$ and $ |b^0|\geq |\vec b|$. Combined they read $|\vec y|< y^0$ and $ |\vec b|\leq -b^0$. Then we have $|\vec y||\vec b| < -y^0 b^0$ and hence, in terms of the standard dot product on $\R^{n-1}$, we have 
\begin{align}
    \vec y\cdot\vec b \geq -|\vec y||\vec b| > y^0 b^0,
\end{align}
by the Cauchy-Schwarz inequality, $-\vec y\cdot\vec b\leq |\vec y\cdot\vec b|\leq |\vec y||\vec b|$. It thus follows that
\begin{align}
g_{\mu\nu}y^\mu b^\nu=\eta_{\mu\nu}y^\mu b^\nu &= -y^0b^0 + \vec y\cdot\vec b  > -y^0b^0 + y^0 b^0 = 0.
\end{align}
\end{proof}
\Cref{lem:Randers_well_defined_on_future_timecone} thus says that under the stated conditions, $\beta>0$. The proof of \cref{prop:Randers_well_defined_on_future_timecone} is now quite simple.
\begin{proof}[Proof of \cref{prop:Randers_well_defined_on_future_timecone}]
   \Cref{lem:Randers_well_defined_on_future_timecone} shows that $\beta>0$ and hence that $F = \alpha + \beta>0$ on the forward timelike cone of $\alpha$, so that $\det g$ is strictly negative, by \eqref{eq:determinant_Randers}. The remainder of the proof is similar to its counterpart in the positive definite scenario. Introduce a parameter $\lambda$ and consider the family of functions $F_\lambda = \alpha+\lambda\beta$ for $\lambda\in [0,1]$. For each value of $\lambda$, $\det g_\lambda$ is strictly negative and so, by \Cref{lem:constant_signature}, $g_\lambda$ has constant signature for $\lambda\in [0,1]$. In particular $g_{\lambda=1}$ has the same signature as $g_{\lambda=0}$, which is just the statement that $g$ is Lorentzian. Since the homogeneity of $F$ is clear, this completes the proof that $F$ defines a Finsler spacetime on the forward timelike cone of $\alpha$ with Lorentzian signature on all of $\mathcal A$.
\end{proof}


\subsection{The field equations}

In general, as discussed, the vanishing of the Finsler-Ricci tensor is a sufficient condition for a Berwald spacetime to be a solution to Pfeifer and Wohlfarth's field equation \eqref{eq:Pfeifer_Wohlfarth_eq_Landsberg}. For Berwald-Randers metrics it is a sufficient condition as well.
\begin{prop}\label{prop:Randers_EFEs}
For Berwald-Randers spacetimes, the field equation is equivalent to Ricci-flatness, i.e. 
\begin{align}
    \left(L g^{\mu\nu}-3 y^\mu y^\nu\right)R_{\mu\nu}=0\quad\Leftrightarrow\quad R_{\mu\nu}=0\quad\Leftrightarrow\quad \mathrm{Ric}=0.\nonumber
\end{align}
\end{prop}
\begin{proof}
The equivalence between the $\mathrm{Ric}=0$ and $R_{\mu\nu}=0$ holds in general, and each of these trivially implies the field equation on the left. The nontrivial part of the proof consists of showing that the field equation implies Ricci-flatness. We first compute the fundamental tensor for a Randers spacetime $F = \alpha+\beta$. A straightforward computation using that $\partial_\mu \alpha = \epsilon y_\mu/\alpha$ and \mbox{$\partial_\mu\partial_\nu \alpha = \tfrac{1}{\alpha}(\epsilon a_{\mu\nu} - y_\mu y_\nu/\alpha^2)$}, where $\epsilon = \sgn(a_{\mu\nu}y^\mu y^\nu)$, shows that
\begin{align}\label{eq:Randers_fund_tensor}
    g_{\mu\nu} = \frac{\epsilon F}{\alpha}a_{\mu\nu} + b_\mu b_\nu + (b_\mu y_\nu + b_\nu y_\mu) \frac{\epsilon}{\alpha} -  y_\mu y_\nu \frac{\beta}{\alpha^3}.
\end{align}
Although a little tedious, it's also straightforward to check that its inverse is given by
\begin{align}
    g^{\mu\nu} = \frac{\epsilon\,\alpha}{F}a^{\mu\nu} - (b^\nu y^\mu + b^\mu y^\nu)\frac{\epsilon\,\alpha}{F^2}+ y^\mu y^\nu \frac{(\beta  + \epsilon |b|^2 \alpha)}{F^3}\,.
\end{align}
Substituting this into the equation $\left(L g^{\mu\nu}-3 y^\mu y^\nu\right)R_{\mu\nu}=0$ and using the fact that $R_{\mu\nu}$ is symmetric leads to
\begin{align}
     \epsilon R \alpha (\alpha + \beta)^2  - 2\epsilon \alpha (\alpha + \beta) R_{\mu\nu} b^\mu y^\nu  + (&\beta + \epsilon |b|^2 \alpha ){R}_{\mu\nu}y^\mu y^\nu \nonumber\\
     &- 3 ( \alpha + \beta){R}_{\mu\nu} y^\mu y^\nu = 0,
\end{align}
where we have denoted $R \coloneqq a^{\mu\nu}R_{\mu\nu}$. This is a polynomial equation in $\alpha$ and we can separate it into a rational part (even powers of $\alpha$) and an irrational part (odd powers of $\alpha$) as
\begin{align}\label{eq:irr_rat}
   \epsilon R \alpha^3 +  (\epsilon R \beta^2 - 2\epsilon  \beta R_{\mu\nu}y^\mu b^\nu &+ (\epsilon |b|^2-3) R_{\mu\nu} y^\mu y^\nu )\alpha \nonumber\\
   &=  2 \epsilon ( R_{\mu\nu} b^\mu y^\nu - R \beta )\alpha^2 + 2 \beta{R}_{\mu\nu} y^\mu y^\nu \,.
\end{align}
On either of the open sets $\pm A>0$ (i.e. $\epsilon=\pm 1=$ const), \eqref{eq:irr_rat} is of the form $P\alpha = Q$, where $P$ and $Q$ are both polynomial in $y$. We claim that this implies that $P=Q=0$. To see this, suppose for contradiction that $P\neq 0$. Then we would be able to write $\alpha = Q/P$ and hence $\alpha$ would be a rational function of $y$, which contradicts \Cref{lem:irrational_alpha}. Hence we must indeed have $P = 0$, which implies that $Q=0$ as well. It thus follows that both sides of \eqref{eq:irr_rat} must vanish separately. We focus on the vanishing of the rational part, 
\begin{align}\label{eq:feqrat}
    2 ( R_{\mu\nu} b^\mu y^\nu - R \beta )A+ 2 \beta{R}_{\mu\nu} y^\mu y^\nu = 0\,,
\end{align}
where we have used that $A = \epsilon \alpha^2$. Differentiating twice with respect to $y$ (while keeping in mind that $R_{\mu\nu}$ depends only on $x$, by the Berwald property) and contracting with $a^{\mu\nu}$ leads to
\begin{align}
   4(8 R_{\mu\nu}b^\mu y^\nu-5 R \beta) = 0,
\end{align}
and substituting this back into equation \eqref{eq:feqrat} yields
\begin{align}
    \label{eq:last_eq_proof}
    \beta\left(2 {R}_{\mu\nu} y^\mu y^\nu- \tfrac{3}{4}R A\right)=0,
\end{align}
which implies that the expression in parentheses must vanish. This is obviously true whenever $\beta\neq 0$, but the remaining points where $\beta=0$ constitute a set with empty topological interior, and hence the conclusion that the expression in parentheses must vanish can be extended by continuity to this set as well. Now we perform the same trick once again; differentiating this expression twice with respect to $y$ and contracting with $a^{\mu\nu}$ results in $-2R=0$. Substituting this back into \eqref{eq:last_eq_proof} yields $\mathrm{Ric}=R_{\mu\nu}y^\mu y^\nu =0$, as desired.
\end{proof}
In other words, in the context of Berwald-Randers spacetimes, the vacuum field equations of Finsler gravity are formally identical to the vacuum field equations of general relativity and they also coincide with Rutz's equation \cite{Rutz}, $\text{Ric}=0$. It turns out that these equivalences hold more generally under strict additional smoothness requirements on the Finsler Lagrangian \cite{javaloyes2023einsteinhilbertpalatini}. Without such additional constraints, however, the equivalence does not extend even to general $(\alpha,\beta)$-metrics. This will be demonstrated in a forthcoming article. 

\begin{rem}
    In the final stages of writing this dissertation it came to our attention (through discussions with Nicoleta Voicu)  that the proof of \cref{prop:Randers_EFEs} can be simplified by noticing the following. Since $b^\mu$ is necessarily covariantly constant for a Berwald-Randers metric by \cref{prop:Randers_Berwald_cond}, it follows directly from the definition of the Riemann curvature tensor $\bar R^\rho{}_{\sigma\mu\nu}$ of $a_{\mu\nu}$ that $\bar R^\rho{}_{\sigma\mu\nu}b^\sigma=0$, and hence the Ricci-tensor of $a_{\mu\nu}$ satisfies $\bar R_{\mu\nu}b^\nu=0$. Since $\bar R_{\mu\nu} = R_{\mu\nu}$ it follows that all occurrences of $R_{\mu\nu}b^\nu$ in the proof may be replaced by zero immediately.
\end{rem}

\subsection{Randers pp-waves}

\cref{prop:Randers_EFEs} allows us to classify all Berwald-Randers vacuum solutions to Pfeifer and Wohlfarth's field equation. 
The first step toward this end is to reduce the problem to a classification problem in general relativity rather than Finsler gravity.
\begin{theor}
\label{theor:main}
A Randers metric of Berwald type, $F = \alpha + \beta$, is a vacuum solution in Finsler gravity if and only if $\alpha$ is a vacuum solution to Einstein's field equation.
\end{theor}
\begin{proof}
Since $F$ is Berwald-Randers, \Cref{prop:Randers_Berwald_cond} implies that its affine connection coincides with the Levi-Civita connection of $\alpha$. It follows that the associated curvature tensors and in particular the (affine) Ricci tensors must also coincide. In particular, the affine Ricci tensor is symmetric and hence it is equal to its symmetrization, the Finsler-Ricci tensor. Hence the Finsler-Ricci tensor vanishes if and only if the Ricci tensor of $\alpha$ vanishes. By \Cref{prop:Randers_EFEs} this implies the result.
\end{proof}
Taking into account \Cref{prop:Randers_Berwald_cond}, it is clear that \cref{theor:main} can also be formulated in the following alternative way.
\begin{theor}
\label{theor:main2}
For a Randers metric $F = \alpha + \beta$, the following are equivalent:
\begin{enumerate}[(i)]
\item $F$ is a Berwald vacuum solution in Finsler gravity;
\item $\alpha$ solves Einstein's equation in vacuum and $F$ is Berwald;
\item $\alpha$ solves Einstein's equation in vacuum and $\beta$ is covariantly constant.
\end{enumerate}
\end{theor}
Hence any vacuum solution of Berwald-Randers type is of the type of \cref{theor:(alphabeta)solutions}. Combining this with \Cref{theor:(alphabeta)solutions2} leads to the conclusion that any Berwald-Randers metric that solves the vacuum field equation in Finsler gravity must belong to either \textit{Class 1} or \textit{Class 2} in the terminology introduced in \cref{sec:ab_pp_wave_solutions}---provided $\alpha$ has Lorentzian signature. We thus arrive at the following (local) classification.
\begin{cor}\label{cor:Randers_classification_sols}
A Randers metric $F = \alpha + \beta$ of Berwald type, where $\alpha$ is a Lorentzian metric and $\beta$ a nonvanishing 1-form, is a vacuum solution in Finsler gravity if and only if one of the following statements is true:
\begin{itemize}
\item There exist local coordinates $(x^\mu)=(u,v,x,y)$ such that
\begin{align}
F = \alpha + \beta = \sqrt{\left|-2\D u \D v + H(u,x,y)\, \D u^2 +\D x^2 + \D y^2 \right|}+ \D u,
\end{align}
with $\Delta_{(x,y)}H=0$;
\item There exist local coordinates  $(x^\mu)=(t,x,y,z)$ such that
\begin{align}
F = \alpha + \beta = \sqrt{\left|-\D t^2+ \D x^2+ \D y^2+ \D z^2\right|}+ b_\mu \D x^\mu,
\end{align}
with $b_\mu =$ constant.
\end{itemize}
\end{cor}
The second class of solutions in the theorem may be viewed as a Randers analog of the Bogoslovsky/\textit{very special relativity} line element \cite{Cohen:2006ky,Gibbons:2007iu,Bogoslovsky1973,Bogoslovsky}. The first class, on the other hand, may be viewed as a Randers analog of the Finsler pp-waves of $m$-Kropina type \cite{Fuster:2015tua}, and to a large extent also as an analog of the \textit{very general relativity} spacetimes introduced in \cite{Fuster:2018djw}.

\section{Modified Randers metrics}\label{sec:Modified_Randers_Metrics}

In the preceding section, we discussed solutions of Randers type, with Finsler metric $F = \alpha+\beta$. We will argue below, however, that in order to obtain a physically natural causal structure (including, for instance, both a forward and a backward light cone), the conventional definition of the Randers metric must be modified slightly. As we will see, this will not impact the characterization of exact solutions at all. The content in this section is based mostly on \citeref{ref5}{H5}.

After a heuristic argument that motivates the desired modification, we show that the modified Randers metric has, in particular, a very satisfactory---even pseudo-Riemannian---causal structure. As a result of this, a clear (future and past) timelike cone can be identified and within these timelike cones the signature of the Fundamental tensor is Lorentzian everywhere. The only constraint is that $|b|^2\equiv a^{\mu\nu}b_\mu b_\nu>-1$, which, interestingly, is in some sense the opposite\footnote{If one were to adopt the opposite signature convention to ours, however, the constraint in the Lorentzian case would also turn out to be $|b|^2<1$, matching the positive definite case.} of the condition $|b|^2<1$ that appears in the well-known positive definite case, see e.g. \cite{ChernShen_RiemannFinsler}. Note that in our case, $|b|^2$ is allowed to be positive as well as negative, which is one of the properties---others will be discussed below---that distinguishes the modified Randers metrics from standard Randers metrics, for which the constraint is given by\footnote{This constraint applies to the situation where the signature is required to be Lorentzian within the whole timelike cone. It should not be confused with the situation corresponding e.g. to \cref{prop:Randers_well_defined_on_future_timecone} where the region with Lorentzian signature might be strictly smaller.} $-1<|b|^2\leq 0$ \cite{Voicu23_Finsler_ab_spacetime_condition}.

In contrast to the preceding sections and the rest of \cref{part:vacuum_sols}, we will momentarily set aside our assumption that $\mathrm{dim}M=n=4$ and allow our manifold to have any dimension.



\subsection{Motivation and definition}
\label{sec:Randers_modified1}


First of all, let us review why it is not that obvious in Lorentzian signature what the `most natural' definition of the Randers metric should be, starting at the very beginning. The original definition of a Randers metric, in positive definite Finsler geometry, is that it is just a Finsler metric of the form $F = \alpha + \beta$, where $\alpha = \sqrt{a_{ij}y^i y^j}$ is a Riemannian metric and $\beta = b_i y^i$ any 1-form. This is well-defined as long as $a_{ij}$ is positive-definite because in that case $A \equiv a_{ij}y^iy^j$ is always positive for nonvanishing vectors $y$. If we allow $a_{\mu\nu}$ to be a Lorentzian metric, however, the quantity $A$ can become negative, in which case $\sqrt{A}$ is ill-defined, as we want $F$ to be a real function. One way to remedy this, at least at a technical level, is to simply restrict the domain $\mathcal A\subset TM_0$ to those vectors for which $a_{\mu\nu}y^\mu y^\nu >0$ (or alternatively, $a_{\mu\nu}y^\mu y^\nu<0$). 
This is the approach that was taken in \citeref{ref4}{H4} and \Cref{prop:Randers_well_defined_on_future_timecone} can be interpreted in that context\footnote{Since the signature convention in \citeref{ref4}{H4} is opposite to the one employed here, the condition $a_{\mu\nu}y^\mu y^\nu >0$ in that case precisely select the timelike vectors. With our current conventions, however, the relevant condition is $a_{\mu\nu}y^\mu y^\nu <0$ as in \Cref{prop:Randers_well_defined_on_future_timecone}.}. %
%
%
However, restricting $\mathcal A$ in this way leads to issues when it comes to the physical interpretation, so here we take a different approach.

The obvious first alternative to restricting $\mathcal A$ to vectors with positive norm is to simply replace $A$ by $|A|$ and define $\alpha = \sqrt{|A|}$, as we have done throughout this dissertation. Such a definition still has the side effect that $F$ is not smooth on vectors satisfying $A=0$, but the upside is that there's no need \textit{a priori} to restrict $\mathcal A$ to the timelike cone of $\alpha$ anymore. This modification leads to a Randers metric of the form $F = \sqrt{|A|} +\beta$, which is just the standard Randers metric as we have defined it earlier. Mathematically it clearly is the most natural extension of the original definition. 

From a physics perspective, however, this definition has the undesirable consequence that light rays can only propagate into one half of the tangent space, namely the half given by $\beta\leq 0$. This follows immediately from the null condition $F=0$. Already in the simplest case---the one in which the 1-form is timelike---this leads to a situation where the light cone separates the tangent space into only two rather than three connected components\footnote{This can be checked easily in suitable coordinates adapted to $\beta$.}, resulting in only one timelike cone, not both a forward \textit{and} a backward one. Consequently, there is not a straightforward interpretation in terms of (future \textit{and} past) timelike, spacelike, and lightlike directions, at least not in the conventional way\footnote{We note that in some approaches 
a single, future pointing (by definition) cone is deemed to be sufficient \cite{Javaloyes2019,Bernal_2020}.}. 

We therefore take the viewpoint that outside of the half-space $\beta\leq 0$ in each tangent space, this version of the Randers metric is not valid, but needs to be modified. 
It turns out that it is possible to remove the constraint $\beta\leq 0$---extending the lightcone to the other half-space $\beta>0$---by changing $F$ into $F=\sgn(A)\sqrt{|A|}+\sgn(\beta)\beta = \sgn(A)\sqrt{|A|}+|\beta|$. The result of this is, effectively, that under some mild assumptions (details will follow below), the single light cone (from the $\beta\leq 0$ half space) is mirrored to the complementary $\beta\geq 0$ half space, whereas in the cone consisting of $\alpha$-timelike vectors---the domain of the standard Randers metric (cf. \Cref{prop:Randers_well_defined_on_future_timecone})--- $F$ reduces to the standard Randers metric with an overall minus sign, $F = -(\alpha+\beta)$, by \Cref{prop:Randers_well_defined_on_future_timecone}. The overall minus sign is not of any relevance and may simply be removed as a matter of convention. In particular, $F$ is now reversible, i.e. invariant under $y\to-y$. 

The preceding heuristic argument motivates the following definition.
\begin{defi}
A \textit{modified Randers metric} is a Finsler metric of the form,
\begin{align}
F = \sgn(A)\alpha + |\beta|,\label{eq:modified_randers_metric}
\end{align}
where we recall for completeness that $\alpha = \sqrt{|A|}$,  $A = a_{\mu\nu}y^\mu y^\nu$, $\beta = b_\mu y^\nu$. 
\end{defi}
It is worth pointing out that we could have chosen a minus sign instead of a plus sign in the definition of the modified Randers metric, which would have been just as natural. But it turns out that in that case the resulting Finsler metric would not be guaranteed to have Lorentzian signature everywhere inside of the timelike cones\footnote{In case one employs the opposite signature convention $(+,-,-,-)$ the converse would be true. In that case the preferable choice would be $F=\sgn(A)\alpha-|\beta|$ rather than $F=\sgn(A)\alpha+|\beta|$.}. The plus variant does have this property (as long as $|b|^2>-1$), as we will discuss in detail below.\\

\subsection{Causal structure}
\label{sec:Randers_causality}

Our task is now to show that the modified Randers metric \eqref{eq:modified_randers_metric} indeed has very nice properties, starting with its causal structure, i.e. the structure of its light cone. By definition, the light cone is given by
\begin{align}
F = 0 \qquad \Leftrightarrow \qquad  \sgn(A)\leq 0 \,\land\, |A| = |\beta|^2 \qquad \Leftrightarrow \qquad  A =  -\beta^2,
\end{align}
and thus it follows that
\begin{align}
F=0 \qquad \Leftrightarrow\qquad (a_{\mu\nu}+ b_\mu b_\nu)\D x^\mu \D x^\nu = 0,
\end{align}
meaning that the light cone of $F$ is given precisely by the light cone of $\tilde a_{\mu\nu} = a_{\mu\nu}+b_\mu b_\nu$, which is itself a Lorentzian metric provided that $|b|^2>-1$, as shown by the following Lemma.
\begin{lem}
    Let $a_{\mu\nu}$ be a Lorentzian metric and $b_\mu$ a 1-form such that $|b|^2= a_{\mu\nu} b^\mu b^\nu >-1$. Then $\tilde a_{\mu\nu} = a_{\mu\nu}+b_\mu b_\nu$ is also a Lorentzian metric.
\end{lem}
\begin{proof}
    We have to show that $\tilde a_{\mu\nu}$ has Lorentzian signature. First, the matrix determinant lemma for rank-one updates (see e.g. \cite{harville2008matrix}) says that 
\begin{align}
\det \tilde a= (1+|b|^2)\det a.
\end{align}
Since $|b|^2>-1$ this implies that $\det \tilde a$ has the same sign as $\det a$, and since $a_{\mu\nu}$ is Lorentzian, this implies $\tilde a$ has negative determinant everywhere. In 4D this immediately implies that $\tilde a$ is Lorentzian (although the signs of the eigenvalues might be flipped with respect to $a_{\mu\nu}$), but we'll allow the dimensionality to be arbitrary here. Consider the continuous family of 1-forms $b^{(\eta)}_\mu= \eta b_\mu$, where $\eta\in[0,1]$, and define $a^{(\eta)}_{\mu\nu}=a_{\mu\nu} + b^{(\eta)}_\mu b^{(\eta)}_\nu$. For any $\eta$ we have
\begin{align}
\det  {a^{(\eta)}} = \left[1+|b^{(\eta)}|^2\right]\det a = \left[1+\eta^2 |b|^2\right]\det a,
\end{align}
so $\det  {a^{(\eta)}}\neq 0$ and hence it follows by \Cref{lem:constant_signature} that $ {a^{(\eta)}}$ has the same signature for all values of $\eta$. In particular, $\tilde a =  {a^{(1)}}$ has the same signature as $a =  {a^{(0)}}$, which is Lorentzian signature.
\end{proof}
The fact that, as long as $|b|^2>-1$, the light cone of $F$ coincides with the light cone of a \psR metric, implies that this light cone separates the tangent space at each point into three connected components, which we can naturally interpret as the forward timelike cone, backward timelike cone, and the remainder consisting of spacelike vectors. Coincidentally we note that
\begin{align}
F<0 \qquad \Leftrightarrow\qquad (a_{\mu\nu}+ b_\mu b_\nu) y^\mu y^\nu < 0,
\end{align}
from which it also follows that
\begin{align}
F>0 \qquad \Leftrightarrow\qquad (a_{\mu\nu}+ b_\mu b_\nu)y^\mu y^\nu > 0.
\end{align}
This leads to the additional convenience that $F$-timelike vectors are precisely given by $F<0$, and $F$-spacelike vectors by $F>0$, in addition to the null vectors that are given, by definition, by $F=0$. We summarize these results in the following proposition.
%
%
\begin{prop}\label{prop:randers_causal_structure}
    As long as $|b|^2>-1$, the causal structure of the modified Randers metric $F = \sgn(A)\alpha + |\beta|$ is identical to the causal structure of the \textit{Lorentzian} metric $a_{\mu\nu}+ b_\mu b_\nu$, with null vectors given by $F=0$, timelike vectors given by $F<0$, and spacelike vectors by $F>0$.
\end{prop}
As a result of these nice features of the causal structure of the modified Randers metric, it is possible to define time orientations in a way similar to how they are defined in general relativity, namely by means of a nowhere vanishing timelike vector field $T$. Such $T$ selects one of the two timelike cones as the `forward' one, namely the one that contains $T$. Then another timelike vector $y$ is future-oriented (i.e. lies in the forward timelike cone) if and only if $(a_{\mu\nu}+ b_\mu b_\nu) T^\mu y^\nu  <0$. We note that similar characterizations of time orientations in Finsler spacetimes (although not in terms of an auxiliary pseudo-Riemannian metric, like $a_{\mu\nu}+ b_\mu b_\nu$) have been discussed in \cite{Javaloyes2019}. 

In the special case that $\beta$ is covariantly constant with respect to $\alpha$ (equivalently, $F$ is Berwald)---a case that is interesting from the point of view of exact solutions---we have even stronger results. In that case, not only the causal structure but also the affine structure (i.e. affine connection) of $F$ can be understood in terms of $a_{\mu\nu}+b_\mu b_\nu$. In order to prove this, we start by characterizing the affine structure of the latter.
\begin{lem}\label{lem:a_plus_bb_Christoffel_symbs}
Assuming that $|b|^2>-1$, the Christoffel symbols of $\tilde a_{\mu\nu}=a_{\mu\nu}+b_\mu b_\nu$ can be expressed as
\begin{align*}
    \widetilde\Gamma^\rho_{\mu\nu} &= \mathring\Gamma^\rho_{\mu\nu} +\frac{1}{1+|b|^2}b^\rho \mathring\nabla_{(\mu}b_{\nu)}  - \left(a^{\rho\lambda} - \frac{1}{1+|b|^2}b^\rho b^\lambda\right)\left( b_\mu\mathring\nabla_{[\lambda} b_{\nu]} + b_\nu\mathring\nabla_{[\lambda} b_{\mu]}\right),
\end{align*}
in terms of the Christoffel symbols $\mathring\Gamma^\rho_{\mu\nu}$ and Levi-Civita connection $\mathring\nabla$ of $a_{\mu\nu}$.
\end{lem}
Before proving this, we point out the following immediate corollary\footnote{This corollary can also be proven without reference to \cref{lem:a_plus_bb_Christoffel_symbs}, by simply noticing that $\mathring\nabla_\rho \tilde a_{\mu\nu} =0$, which implies that $\mathring\nabla$ must also be the Levi-Civita connection of $\tilde a_{\mu\nu}$.}.
\begin{cor}\label{cor:a_plus_bb_Christoffel_symbs_cc}
If $\mathring\nabla_{\mu}b_{\nu}=0$ then $\widetilde\Gamma^\rho_{\mu\nu} = \mathring\Gamma^\rho_{\mu\nu}$.
\end{cor}
\begin{proof}[Proof of \Cref{lem:a_plus_bb_Christoffel_symbs}.]
It can easily be checked that the inverse of $\tilde a_{\mu\nu}$ is given by 
\begin{align}
    \tilde a^{\mu\nu} =a^{\mu\nu} - \frac{1}{1+|b|^2}b^\mu b^\nu.
\end{align}
Unless otherwise specified (as in the case of $\widetilde\Gamma$ below!) indices are raises and lowered with $a_{\mu\nu}$. Writing $\mathring\Gamma_{\lambda\mu\nu}=a_{\lambda\rho}\mathring\Gamma^\rho_{\mu\nu}$ and $\widetilde\Gamma_{\lambda\mu\nu}=\tilde a_{\lambda\rho}\widetilde\Gamma^\rho_{\mu\nu}$ we first note that we can express the latter as
\begin{align}
    \widetilde\Gamma_{\lambda\mu\nu} &=\tfrac{1}{2} \left(
    \partial_\mu \tilde a_{\lambda\nu} + \partial_\nu \tilde a_{\mu\lambda} - \partial_\lambda \tilde a_{\mu\nu}\right) \\
    &= 
\mathring\Gamma_{\lambda\mu\nu}  + b_\lambda\partial_{(\mu} b_{\nu)} - b_\mu\partial_{[\lambda} b_{\nu]} - b_\nu\partial_{[\lambda} b_{\mu]},
\end{align}
where $(\mu,\nu)$ denotes symmetrization and $[\mu,\nu]$ denotes anti-symmetrization. It thus follows that
\begin{align}
\widetilde\Gamma^\rho_{\mu\nu} &= \tilde a^{\rho\lambda}\widetilde\Gamma_{\lambda\mu\nu} \\
&= \left(a^{\rho\lambda} - \frac{1}{1+|b|^2}b^\rho b^\lambda\right)\left(\mathring\Gamma_{\lambda\mu\nu}  + b_\lambda\partial_{(\mu} b_{\nu)} -b_\mu\partial_{[\lambda} b_{\nu]} -b_\nu\partial_{[\lambda} b_{\mu]}\right) \\
&=\mathring\Gamma^\rho_{\mu\nu} - \frac{1}{1+|b|^2}b^\rho b_\lambda\mathring\Gamma^\lambda_{\mu\nu} 
+\left(a^{\rho\lambda} - \frac{1}{1+|b|^2}b^\rho b^\lambda\right)b_\lambda\partial_{(\mu} b_{\nu)} \\
&\qquad\qquad\qquad\quad- \left(a^{\rho\lambda} - \frac{1}{1+|b|^2}b^\rho b^\lambda\right)\left( b_\mu\partial_{[\lambda} b_{\nu]} + b_\nu\partial_{[\lambda} b_{\mu]}\right).
\end{align}
The second and third term add up to
\begin{align}
-\frac{1}{1+|b|^2}b^\rho & b_\lambda\mathring\Gamma^\lambda_{\mu\nu} +\left(a^{\rho\lambda} - \frac{1}{1+|b|^2}b^\rho b^\lambda\right)b_\lambda\partial_{(\mu} b_{\nu)} \\
&= -\frac{1}{1+|b|^2}b^\rho b_\lambda\mathring\Gamma^\lambda_{\mu\nu} +b^\rho\partial_{(\mu} b_{\nu)} - \frac{|b|^2}{1+|b|^2} b^\rho \partial_{(\mu} b_{\nu)} \\
&=-\frac{1}{1+|b|^2}b^\rho b_\lambda\mathring\Gamma^\lambda_{\mu\nu} \ + \frac{1}{1+|b|^2} b^\rho \partial_{(\mu} b_{\nu)}\\
&=\frac{1}{1+|b|^2}b^\rho \left(\partial_{(\mu}b_{\nu)}-b_\lambda\mathring\Gamma^\lambda_{\mu\nu} 
\right)\\
&=\frac{1}{1+|b|^2}b^\rho \nabla_{(\mu} b_{\nu)},
\end{align}
which results in
\begin{align}
    \widetilde\Gamma^\rho_{\mu\nu} &= \mathring\Gamma^\rho_{\mu\nu} +\frac{1}{1+|b|^2}b^\rho \nabla_{(\mu}b_{\nu)}  - \left(a^{\rho\lambda} - \frac{1}{1+|b|^2}b^\rho b^\lambda\right)\left( b_\mu\partial_{[\lambda} b_{\nu]} + b_\nu\partial_{[\lambda} b_{\mu]}\right).
\end{align}
Finally, we may replace all partial derivatives with covariant ones because
\begin{align}
    2\nabla_{[\lambda} b_{\nu]}= \nabla_\lambda b_\nu - \nabla_\nu b_\lambda = 
    \partial_\lambda b_\nu- \partial_\nu b_\lambda  = 2\partial_{[\lambda} b_{\nu]}.
\end{align}
That yields the desired formula.
\end{proof}
\begin{theor}\label{theor:mod_Randers_final_result}
If $\beta$ is covariantly constant with respect to $\alpha$ and satisfies $|b|^2>-1$ then the causal structure and the affine structure of the modified Randers metric $F = \sgn(A)\alpha + |\beta|$ are identical to those of the Lorentzian metric $\tilde a_{\mu\nu} = a_{\mu\nu} + b_\mu b_\nu$. In other words, the timelike, spacelike, and null geodesics of $F$ coincide with the timelike, spacelike, and null geodesics, respectively, of $\tilde a_{\mu\nu}$.
\end{theor}
\begin{proof}
\cref{prop:randers_causal_structure} guarantees that the causal structures coincide. By \cref{prop:Randers_Berwald_cond} it follows that the affine structure of $F$ coincides with that of $\alpha$ and \cref{cor:a_plus_bb_Christoffel_symbs_cc} shows that this affine structure is the same as that of $\tilde a_{\mu\nu}$.
\end{proof}
We note that in the proof of \cref{theor:mod_Randers_final_result}, we have applied \cref{prop:Randers_Berwald_cond}, which is a result for standard Randers metrics. Strictly speaking, however, the modified Randers metric $F$ is not even an \ab-metric, let alone a Randers metric. This is because $F$ depends not only on $\alpha$ and $\beta$ but also on the sign of $A$. The following argument shows that \cref{prop:Randers_Berwald_cond} is nevertheless valid also in this case. 

If we denote by $\mathcal A$ the maximal conic subbundle of $TM$ on which $F$ satisfies all axioms of a Finsler metric then we have $\mathcal A \subset \mathcal A_{++}\cup \mathcal A_{+-}\cup \mathcal A_{-+}\cup \mathcal A_{--}$ in terms of the conic subbundles
\begin{align}
    \mathcal A_{\pm\pm} = \{(x,y)\in \mathcal A\,:\, \pm A >0, \pm \beta>0\},
\end{align}
where the first (second) $\pm$ on the LHS refers to the first (second) $\pm$ on the RHS. This decomposition holds because $F$ is differentiable neither at $A=0$ nor at $\beta=0$, so that these zero sets are not contained in $\mathcal A$. On each of the subdomains $\mathcal A_{\pm\pm}$, $F$ is just a standard Randers metric (perhaps up to an irrelevant overall minus sign) and this is why many of the results for standard Randers metrics carry over to modified Randers metrics. This is true in particular for \cref{prop:Randers_Berwald_cond} and hence the proof of \cref{theor:mod_Randers_final_result} is perfectly valid. We will come back to this in the context of exact solutions later.

Apart from the intrinsic significance of \Cref{theor:mod_Randers_final_result}, one of its nice consequences is the existence of so-called radar neighborhoods \cite{Perlick2008}. In words, this means that, given any observer and any spacetime point sufficiently close to the observer's worldline, there is (at least locally) exactly one future-pointing light ray and one past-pointing light ray that connect that point to the worldline of the observer. The precise statement is as follows.
\begin{prop}[Existence of radar neighborhoods]\label{prop:radar_nbh_Randers}
    Consider a Finsler spacetime with modified Randers metric of Berwald type and with $|b|^2>-1$, let $\gamma$ be a timelike curve 
    and let $p=\gamma(\lambda_0)$ be some point on $\gamma$. Then there are open subsets $U$ and $V$ with $p\in U \subset V$ such that every point $q\in U \setminus \mathrm{image}(\gamma)$ can be connected to $\gamma$ by precisely one future-pointing and precisely one past-pointing null geodesic that stays within $V$.
\end{prop}
\begin{proof}
    This result has been proven in \cite{Perlick2008} for spacetimes with a Lorentzian metric rather than a modified Randers metric. However, the statement relies only on the properties of null geodesics and timelike curves. By \Cref{theor:mod_Randers_final_result}, null geodesics and timelike curves of the modified Randers metric can be understood as null geodesics and timelike curves, respectively, of some Lorentzian metric (note that $F$ being Berwald implies that $\mathring\nabla_\mu b_\nu =0$, by \cref{prop:Randers_Berwald_cond}). Hence the result extends immediately to our case.
\end{proof}
The set $U$ is called a radar neighborhood of $p$ with respect to $\gamma$. The existence of radar neighborhoods will play an important role in \Cref{ch:lin} when we calculate the radar distance to a given spacetime point during the passing of a Finslerian gravitational wave of modified Randers type and compare it to the result for a classical general relativistic gravitational wave. \cref{prop:radar_nbh_Randers} essentially says that this radar distance is a well-defined notion to begin with. 

\subsection{Signature and regularity}
Our next objective is to determine the signature and regularity (i.e. degree of smoothness) of $F$. Provided $\alpha$ is Lorentzian, we will find the signature of $F$ to be Lorentzian everywhere within the timelike cone, except (trivially) at those points where $F$ is degenerate or not smooth, i.e. at the irregularities of $F$. This holds irrespective of the properties of the 1-form. Therefore we will be interested in the subset of the timelike cone where $F$ is not regular. Ideally, this subset is empty, and we will see that when $\beta$ is timelike or null this is indeed the case.

\subsubsection{Signature of the fundamental tensor}

The determinant of the fundamental tensor of an \ab-metric $F = \alpha \phi(s)$, with $s = \beta/\alpha$ is given by
\begin{align}\label{eq:ab_det_formula_main_text}
\det g = \det(A)^n\phi^{n+1}(\phi-s\phi')^{n-2}(\phi-s\phi' +  (\sgn(A) |b|^2-s^2)\phi'')\det a,
\end{align}
where $n=\dim M$ and the prime denotes differentiation with respect to $s$. This formula, which appeared (with a typo) originally in \citeref{ref5}{H5}, is a generalization of a well-known result for positive definite \ab-metrics, and its proof can be found in \cref{sec:ab_determinant}. Because of the appearance of $\sgn(A)$ the expression is slightly different from its positive definite counterpart, to which it reduces when $A>0$, i.e. $\sgn(A)=1$. For a modified Randers metric of the form $F = \text{sgn}(A)\alpha + |\beta|$ the function $\phi$ is given by $\phi(s) = \text{sgn}(A) + |s|$, so \eqref{eq:ab_det_formula_main_text} reduces to 
\begin{align}\label{eq:det_mod_Randers}
\frac{\det g}{\det a} =  \sgn(A)^{2n-1}\left(\sgn(A)+|s|\right)^{n+1} = \sgn(A)^{n}\left(\sgn(A)\frac{F}{\alpha}\right)^{n+1}.
\end{align}
%
%
We denote by $\mathcal A$ the maximal conic subbundle of $TM$ on which $F$ satisfies all axioms of a Finsler metric. The precise characteristics of $\mathcal A$ are yet to be determined, but since $F$ is differentiable neither at $\alpha=0$ nor at $\beta=0$, $\mathcal A$ can clearly only contain points $(x,y)\in TM_0$ that satisfy $\alpha\neq 0$ and $\beta\neq 0$. We will tacitly use this fact in what follows.
\begin{lem}\label{lem:mod_Randers_Lorentzian_sign}
    In even dimension $n$, the subset of $\mathcal A$ on which $g$ has Lorentzian signature is given precisely by those points in $\mathcal A$ that satisfy $\sgn(A)F>0$.
\end{lem}
\begin{proof}
     Consider the family of functions $F_\lambda = (1-\lambda)\sgn(A)\alpha+ \lambda F = \sgn(A) \alpha + \lambda|\beta|$ for $\lambda\in [0,1]$ and suppose first that $\sgn(A)F>0$. Then both $\sgn(A)F_0 = \alpha>0$ and $\sgn(A)F_1=\sgn(A)F>0$, and since $F_\lambda = (1-\lambda)F_0 + \lambda F_1$ is a convex sum, this implies that $\sgn(A)F_\lambda>0$ for each $\lambda$. Since we also know that $\det a <0$, \eqref{eq:det_mod_Randers} then implies that $\det g_\lambda$ has the same sign for each $\lambda$ and hence it follows from \Cref{lem:constant_signature} that the signature of $g=g_1$ is the same as that of $\epsilon a = g_0$, i.e. Lorentzian.\\
     Conversely, if $\sgn(A)F\leq 0$ then \eqref{eq:det_mod_Randers} shows that $\det g \geq 0$, as $n+1$ is assumed to be odd. Hence in this case $g$ is certainly not Lorentzian.
\end{proof}
Let us investigate what this set $\sgn(A)F>0$ looks like. First note that $F<0$ trivially implies $A<0$, meaning that the $F$-timelike cone (recall \cref{prop:randers_causal_structure}) lies completely within the $A$-timelike cone. It follows that if $F<0$ then $\sgn(A)F>0$ and hence the region $F<0$ has Lorentzian signature. In other words, within the entire timelike cone of $F$ (intersected with $\mathcal A$), the signature of the fundamental tensor is Lorentzian. Similarly, $A>0$ implies $F>0$ and hence $\sgn(A)F>0$, so the region $A>0$ also has Lorentzian signature. What remains is the region where simultaneously $A\leq 0$ and $F\geq 0$, and in this case, it is clear that $\sgn(A)F\leq 0$. Hence, apart from possible irregularities, this is precisely the subset of $TM$ where the signature is \textit{not} Lorentzian. The inequalities describing this region can be interpreted as saying that $y$ lies within the $A$-timelike cone but outside of the $F$-timelike cone (or possibly on the boundary of either). In other words, this is precisely the region in between the light cone of $A$ and that of $F$.  
We thus conclude that $F$ has Lorentzian signature everywhere on $\mathcal A$ except in the region in between the two light cones (including their boundaries\footnote{On the light cone of $A$ the Finsler metric is not even differentiable, so it certainly does not have Lorentzian signature there. On the light cone of $F$, either $A=0$, in which case the previous comment applies, or $A\neq 0$, in which case $\det g = 0$, by \eqref{eq:det_mod_Randers}, so that the point does not lie in $\mathcal A$.}).

\cref{fig:lightcones_and_signatures} shows a $(1+2)$D projection of the light cone of the modified Randers metric, and it also shows the region (in green) where the signature of the fundamental tensor is Lorentzian, 
for each possible causal character of the 1-form. In each subfigure, the inner light cone is that of $F$ and the outer light cone that of $A$.
\begin{figure}
	\begin{center}
	\begin{subfigure}[b]{0.47\textwidth}
    	\includegraphics[width=\textwidth]{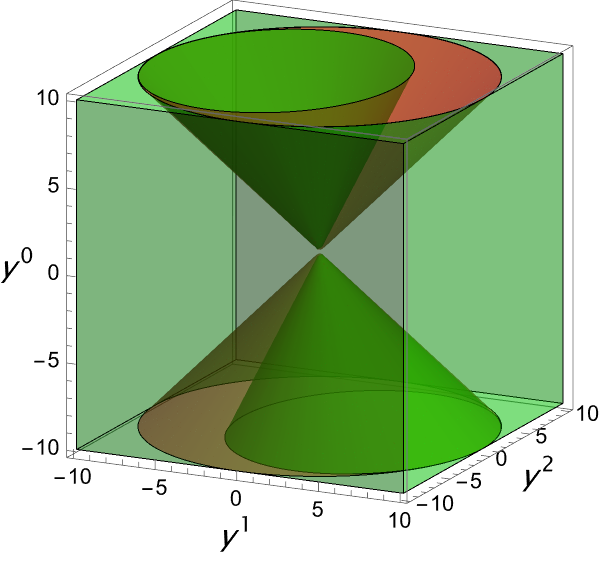}
    	\caption{Null 1-form, $\rho=0.6$}
    	\label{fig:L1}
    \end{subfigure}
    \hspace{10px}
    \begin{subfigure}[b]{0.47\textwidth}
    	\includegraphics[width=\textwidth]{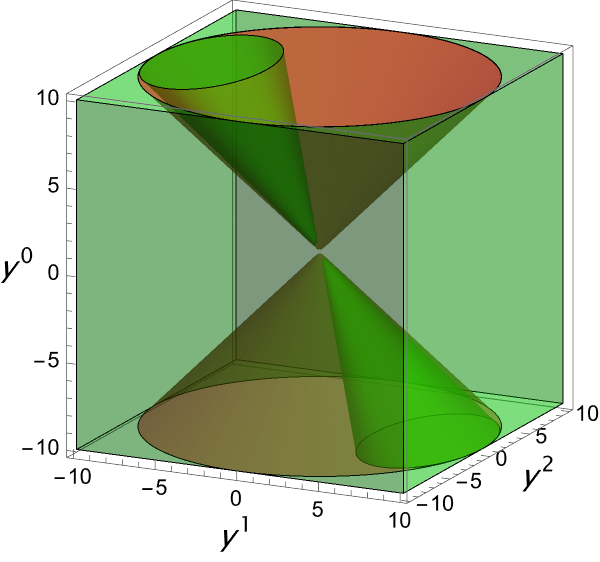}
    	\caption{Null 1-form, $\rho=1.4$}
    	\label{fig:L3}
    \end{subfigure}
    \end{center}
    \caption{The figures show, for several representative values of $\rho$, a $3$D projection of the light cone and the signature of the fundamental tensor of the (even dimensional) modified Randers metric $F = \sgn(A)\alpha + |\beta|$, in coordinates such that at $x\in M$ one has $A = -(y^0)^2+(y^1)^2+\dots +(y^2)^{n-1}$ and $\beta = \rho \,y^0$ (in the timelike case) or $\beta = \rho(y^0+y^1)$ (in the null case) or $\beta = \rho \,y^1$ (in the spacelike case). Such coordinates can always be chosen. Green regions correspond to Lorentzian signature and red regions to non-Lorentzian signature. Figures  \ref{fig:L1} - \ref{fig:S3} show the physically reasonable scenarios, where $|b|^2>-1$. In that case, two cones can be observed. The inner cone is the true light cone of $F$ (i.e. the set $F=0$), and the outer cone is the light cone of $a_{ij}$ (i.e. the set $A=0$). The only region with non-Lorentzian signature (disregarding irregularities) is precisely the gap in between the two cones, including their boundaries. If on the other hand $|b|^2=-1$ (\cref{fig:T5}) then the light `cone' of $F$ is the line $y^1=y^2=y^3=0$. And if $|b|^2<-1$ (\cref{fig:T4}) then the light `cone' of $F$ consists only of the origin. Therefore we deem the latter two cases not physically interesting.}
    \label{fig:lightcones_and_signatures}
\end{figure}
\begin{figure}
    \addtocounter{figure}{-1}
    \begin{center}
    \begin{subfigure}[b]{0.47\textwidth}
        \addtocounter{subfigure}{2}
    	\includegraphics[width=\textwidth]{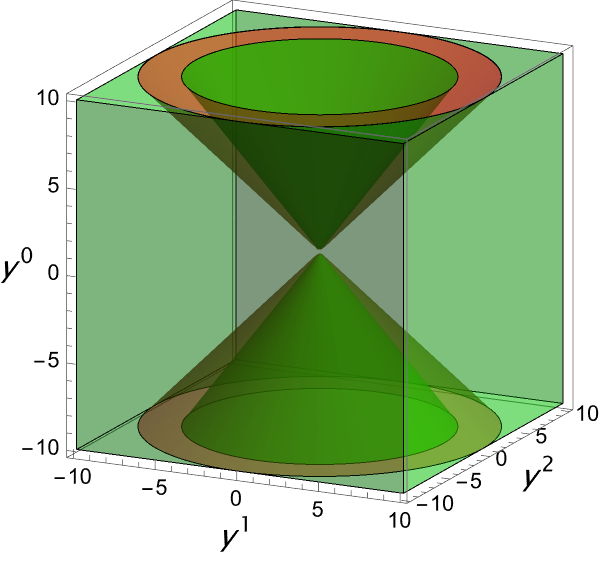}
    	\caption{Timelike 1-form, $\rho=0.65$}
    	\label{fig:T1}
    \end{subfigure}
    \hspace{10px}
    \begin{subfigure}[b]{0.47\textwidth}
    	\includegraphics[width=\textwidth]{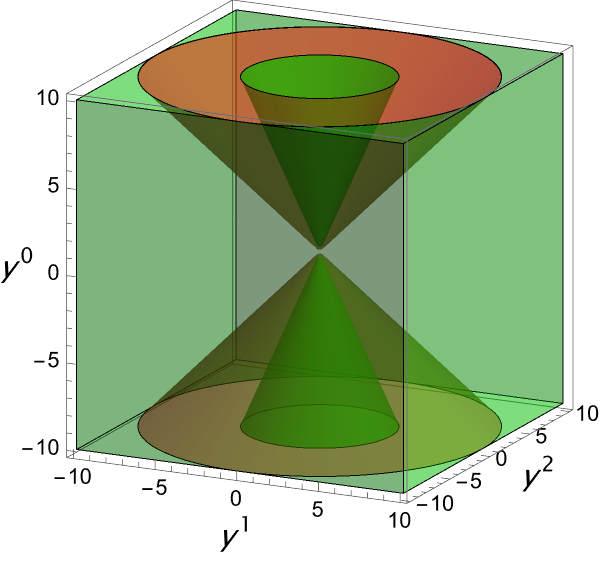}
    	\caption{Timelike 1-form, $\rho=0.9$}
    	\label{fig:T3}
    \end{subfigure}
    \end{center}
\end{figure}
\begin{figure}
    \addtocounter{figure}{-1}
    \begin{center}
    \begin{subfigure}[b]{0.47\textwidth}
        \addtocounter{subfigure}{4}
    	\includegraphics[width=\textwidth]{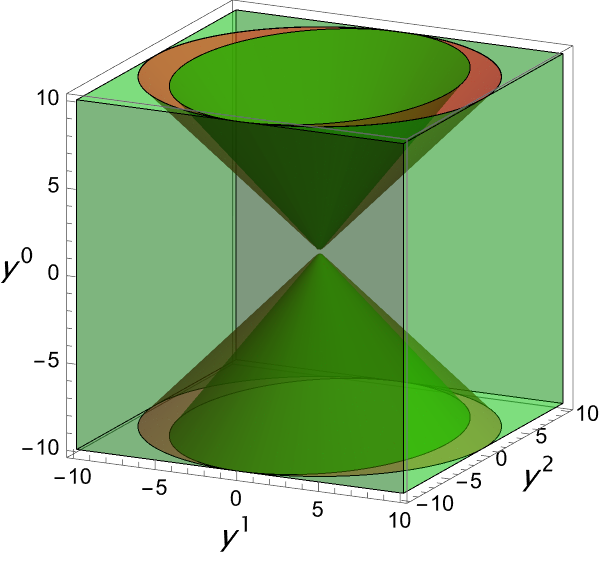}
    	\caption{Spacelike 1-form, $\rho=0.8$}
    	\label{fig:S1}
    \end{subfigure}
    \hspace{10px}
    \begin{subfigure}[b]{0.47\textwidth}
    	\includegraphics[width=\textwidth]{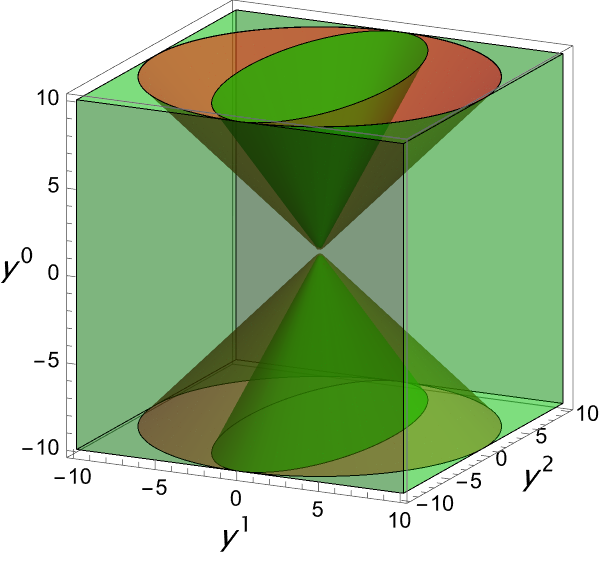}
    	\caption{Spacelike 1-form, $\rho=2$}
    	\label{fig:S3}
    \end{subfigure}
    \end{center}
\end{figure}
\begin{figure}
    \addtocounter{figure}{-1}
    \begin{center}
    \begin{subfigure}[b]{0.47\textwidth}
        \addtocounter{subfigure}{6}
    	\includegraphics[width=\textwidth]{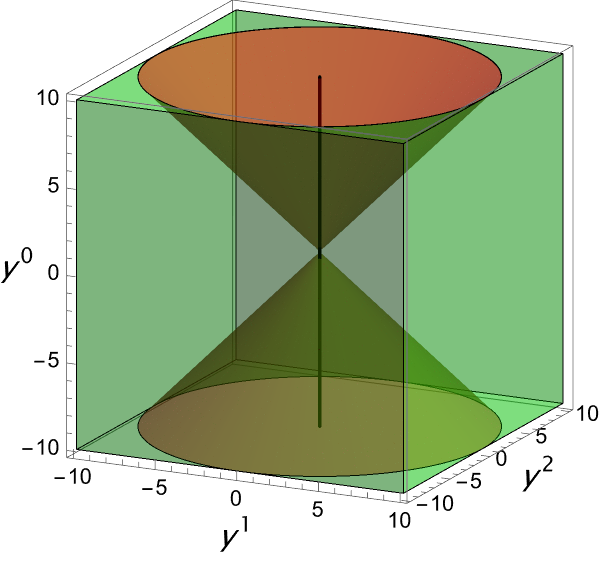}
    	\caption{Timelike 1-form, $|b|^2=-1$.}
    	\label{fig:T5}
    \end{subfigure}
    \hspace{10px}
    \begin{subfigure}[b]{0.47\textwidth}
    	\includegraphics[width=\textwidth]{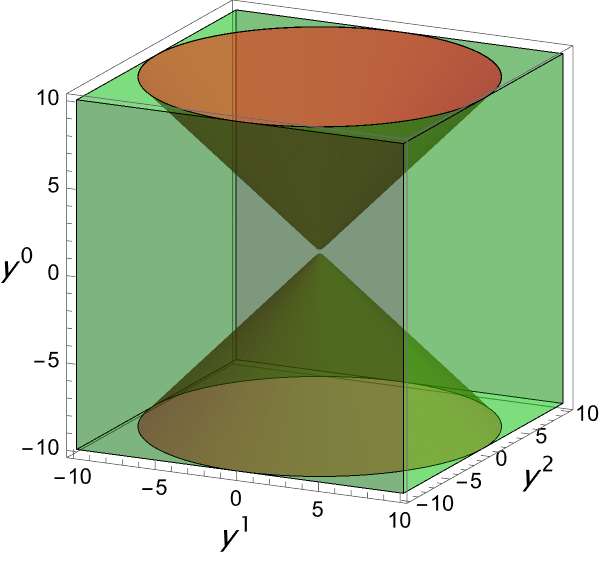}
    	\caption{Timelike 1-form, $|b|^2<-1$.}
    	\label{fig:T4}
    \end{subfigure}
    \end{center}
\end{figure}
\subsubsection{Regularity}

From the definition of the modified Randers metric, it is clear that the set of points where $F$ (or more relevantly, $F^2$) is \textit{not} smooth is given by $A=0$ or $\beta=0$. We would like to know which of these points, if any, lie within the timelike cone of $F$, or perhaps on its boundary, the light cone. First, we focus on the possibility that $A=0$. As observed above and confirmed visually in \cref{fig:lightcones_and_signatures}, the $F$-timelike cone lies completely within the $A$-timelike cone, which implies that none of the $A=0$ irregularities lie within the $F$-timelike cone. But they might lie on the light cone, namely if the light cone of $A$ and the light cone of $F$ have a nonempty intersection. The specifics of this are straightforward to derive, but depend on the (pointwise) causal character of the 1-form $\beta$; we list the results below and leave the details to the reader. The easiest way to check these properties is to use coordinates in which
\begin{align}\label{eq:coords_adapted_to_beta}
A = -(\D x^0)^2+\delta_{ij}\D x^i\D x^j,\quad \beta =\left\{\begin{array}{ll}
\rho \,\D x^0 & \text{\textit{timelike}}\\ 
\rho\,(\D x^0+\D x^1) &\text{\textit{null}}\\
\rho \,\D x^1 & \text{\textit{spacelike}}
\end{array}\right. 
\end{align}
at a given point $x\in M$. Such coordinates always exist\footnote{We recall that this can be seen as follows. First, since $a_{ij}$ is Lorentzian, it is always possible to choose coordinates such that $A$ is just the Minkowski metric at a given point $x\in M$. Writing $b^\mu = (b^0,b^1,\dots b^{n-1})$ in these coordinates, we may apply a spatial rotation to the coordinates $b^1,\dots,b^{n-1}$, such that they are transformed into $(b^1,0,\dots,0)$, leaving the metric at $x$ unchanged. Then $b^\mu = (b^0,b^1,0,\dots,0)$. Now we separate the three cases. If $|b|^2=0$, it follows that $b^1=\pm b^0$ and by applying---if necessary---a spatial reflection in the $x^1$ direction we may choose either sign. If $|b|^2<0$ then we may go to the local rest frame by a Lorentz transformation, resulting in $b^1=0$. If on the other hand $|b|^2>0$ we may perform a Lorentz transformation resulting in $b^0=0$.}. In each of the following cases, we consider the light cone(s) at the fixed point $x$.
\begin{itemize}
\item If $\beta$ is timelike at $x$ and $|b|^2>-1$ then the light cones do not intersect (apart from the trivial intersection in the origin). Hence not only the $F$-timelike cone but also its boundary, the light cone, lies completely within the (interior of the) light cone of $A$.
\item If $\beta$ is null at $x$ then the intersection of the two lightcones is given by those $y^\mu$ that are multiples of $b^\mu$. This intersection thus spans a single line in the tangent space. 
\item If $\beta$ is spacelike at $x$ (and assuming $\dim M>2$) then $a_{ij}$ induces an $(n-1)$-dimensional Lorentzian metric 
on the hypersurface $\beta=0$. The two light cones intersect along the light cone of this induced Lorentzian metric. In \cref{fig:S1} and \cref{fig:S3}, which are $3$D projections, the latter is given by $y^0=\pm y^2$; hence the intersection appears as a pair of lines.
\item If $|b|^2 = -1$ then the light cone of $F$ is given by a single line, namely the line consisting of all multiples of $b^\mu$. In \cref{fig:T5} this is just the $y^0$-axis.  
\item If $|b|^2 < -1$ then the light cone of $F$ is given by the single point $y=0$. 
\end{itemize}
This sums up the $A=0$ irregularities on the light cone of $F$. Regarding the $\beta=0$ irregularities, note that $\beta = a_{\mu\nu} b^\mu y^\mu=0$ if and only if $y^\mu$ is orthogonal to $b^\mu$. The orthogonal complement to a timelike vector is spacelike, so if $b_\mu$ is timelike, $\beta$ cannot vanish on $A$-timelike vectors and hence in particular $\beta\neq 0$ in the $F$-timelike cone. If $b_\mu$ is null, on the other hand, then the only nonspacelike vectors orthogonal to it are multiples of $b^\mu$ itself. These coincide with the $A=0$ irregularities in the second bullet point above. Only in the case that $b_\mu$ is spacelike will there be $\beta=0$ irregularities inside the $F$-timelike cone, namely all $A$-timelike vectors orthogonal to $b_\mu$.

Now let us combine everything we have worked out in this section so far. A modified Randers metric with $|b|^2>-1$ has Lorentzian signature everywhere within its timelike cone, with the exception of possible irregularities in the cone. The latter are given either by $\beta = 0$, characterized in the preceding paragraph, or by $A=0$, corresponding to one of the intersections listed in the bullet list above, depending on the causal character of $\beta$. We may thus summarize the results as follows, focusing for clarity on the case where the 1-form has a global causal character, i.e. is everywhere timelike, spacelike or null.\\

%

\begin{myblock}[Regularity and Signature - Summary]
    Let $F$ be a modified Randers metric with $|b|^2>-1$ and $n$ even. 
    \begin{itemize}
        \item If $\beta$ is \textbf{timelike} then $F$ is smooth and has Lorentzian signature within its timelike cone. On the light cone, $F$ is smooth but degenerate.
        \item If $\beta$ is \textbf{null} then $F$ is smooth and has Lorentzian signature within its timelike cone. On the light cone, $F$ is degenerate and there is a $1$--dimensional irregularity where $F$ is not smooth, namely the line spanned by $b^\mu$.
         \item If $\beta$ is \textbf{spacelike} then within its timelike cone, $F$ has Lorentzian signature but has an $(n-1)$-dimensional irregularity where $F$ is not smooth, given by the hyperplane $\beta=0$. On the light cone, $F$ is degenerate and there is an $(n-2)$-dimensional irregularity where $F$ is not smooth, given by the light cone of the induced metric on the submanifold $\beta=0$.
    \end{itemize}
    If $|b|^2\leq -1$ the timelike cone is the empty set, suggesting that this case is not physically relevant.
   \end{myblock}

\noindent Clearly, the timelike case is the winner in terms of desirable properties from the point of view of physics. But the null case is a close runner-up, and even the spacelike scenario is acceptable. In particular, the set where $F$ is not smooth has measure zero and empty interior in each of the cases, provided $\dim M>2$. This is true even when the 1-form is spacelike. On the other hand, the \textit{classical} Randers metric can only be considered a physically reasonable Finsler spacetime if the 1-form is either null or timelike  \cite{Voicu23_Finsler_ab_spacetime_condition}, and even in these cases, there is only a future timelike cone (or only a past timelike cone, but not both). This motivates the use of modified Randers metrics rather than standard Randers metrics in Finsler gravity.

%
%



\subsection{Exact vacuum solutions}

As noted already below \cref{theor:mod_Randers_final_result}, a modified Randers metric $F$ is `locally' just a standard Randers metric, in the sense that the domain $\mathcal A$ of $F$ can be decomposed into several parts, on each of which $F$ reduces to a standard Randers metric. That means that on each of the individual subdomains, all results obtained in \cref{sec:RandersSolutions} are applicable. And since the classification results pertaining to exact solutions do not depend on the specifics of the domain, it follows that the results remain true also for modified Randers metrics considered on their entire domain $\mathcal A$. We summarize the most important ones below.
\begin{itemize}
    \item For modified Randers metrics of Berwald type, Pfeifer and Wohlfarth's field equation is equivalent to Rutz's equation, which comes down to Ricci-flatness, $\mathrm{Ric}=0$, or equivalently, $R_{\mu\nu} = 0$. This is the analog of \cref{prop:Randers_EFEs}. 
    \item From this, it follows that a modified Randers metric $F = \sgn(A)\alpha + |\beta|$ of Berwald type is a vacuum solution (to any and each of these equations) if and only if $\alpha$ is a vacuum solution to Einstein's equations. This is the analog of \cref{theor:main}. 
    \item The analog of the final classification result, \cref{cor:Randers_classification_sols}, holds for modified Randers metrics as well.
\end{itemize}
%
%
%
%
%

%% file: ch_lin_grav_waves.tex
\chapter{Finsler Gravitational Waves}\label{ch:lin}

The aim of this chapter, which is based mostly on \citeref{ref5}{H5}, is to attach a physical interpretation to, and investigate the physical consequences of, the (plane) pp-wave solutions of \ab-metric type obtained in \cref{sec:ab_metric_solutions}. We will achieve this by means of a two-fold linearization procedure, after which the whole class of solutions essentially reduces to a single family of Randers gravitational waves parameterized by a $+$ and a $\times$ polarization, for which it is possible to calculate exactly what happens when such a wave passes, say, a gravitational wave interferometer on Earth. 

The relevant observable measured in interferometer experiments is essentially the radar distance, so we compute this radar distance for our Finslerian gravitational waves, reproducing in the appropriate limit the radar distance formula for a standard gravitational wave in general relativity (GR) \cite{Rakhmanov_2009}. Although at first sight, the expression for the Finsler radar length looks quite different from its GR counterpart, we show that this is nothing but a coordinate artifact. Remarkably, when the Finslerian expression is interpreted correctly in terms of observable quantities, it becomes clear that the two expressions are in fact identical, suggesting that there is no observational difference between the Finsler and GR case, at least as far as radar distance measurements are concerned. 

This is the first example of a situation in which an explicit expression for the Finslerian radar distance could be obtained in the case of finite spacetime separations\footnote{In the case of infinitesimal spacetime separations the radar length has been studied in \cite{Pfeifer_2014,Gurlebeck:2018nme}.}, and as such our results may be viewed in part as a proof of concept.


\section{Linearized pp-wave solutions}

There are two steps to the linearization. First, we linearize in the departure from flatness by writing the Lorentzian part $\alpha$ (or rather $A$) as the flat Minkowski metric plus a small deviation, resulting in a standard general relativistic linearized gravitational wave. Second, we linearize in the departure from GR, reducing the class of \ab-metrics to metrics of Randers type. Our starting point will be the expression \eqref{eq:ab_plane_wave} for the plane wave solutions in Rosen coordinates.

\subsection{Linearized gravitational wave solutions}
\label{sec:linearized_Randers_sols}



As anticipated, we start by linearizing the plane wave metric 
\begin{align}
    A = -2\D u\D v + h_{ij}(u)\D x^i \D x^j
\end{align}
appearing in \eqref{eq:ab_plane_wave} in its departure from flatness. In other words, we consider the scenario that $\alpha$ is very close to the flat Minkowski metric. In this case we may write $h_{ij}(u) = \delta_{ij}+\varepsilon f_{ij}(u)$ with $\varepsilon \ll  1$. The linearized field equations (i.e. to first order in $\varepsilon$) for $\alpha$ then read\footnote{The (linearized) vacuum field equation for $F$ is more complicated in general, but as discussed extensively in \cref{sec:ab_metric_solutions}, if the vacuum field equation for $\alpha$ is satisfied then so is the field equation for $F$. In the case of Randers metrics, to which everything will reduce momentarily, the field equation for $F$ is even equivalent to the field equations for $\alpha$. Hence for our present purposes, the field equations for $\alpha$ suffice.}
\begin{align}
f_{11}''(u) + f_{22}''(u)=0.
\end{align}
Hence $f_{11}$ and $-f_{22}$ must be equal up to an affine function of $u$. Here we will focus on the case where $f_{11}=-f_{22}$, which can always be achieved by means of the transverse traceless gauge\footnote{We leave open the question of whether the form of the 1-form $\beta=\D  u$ always remains invariant under such a transformation to the transverse traceless gauge.}. Conventionally one writes the subscripts as $f_{11} = -f_{22} \eqqcolon f_+$ and $f_{12} = f_{21} \eqqcolon f_\times$, denoting the plus-- and cross-polarization of the gravitational wave, so we will stick to that notation from here onwards. 
That brings us to the Finsler metric
\begin{align}
F = &\alpha\, \phi(\beta/\alpha), \label{eq:ab_grav_waves}\\
&\left\{\begin{array}{ll}
 A = -2\D u \D v + (1+\varepsilon f_+(u)) \D x^2 + (1-\varepsilon f_+(u))\D y^2 + 2\varepsilon f_\times (u) \D x\,\D y \\ 
\beta = \D u
\end{array}\right.\nonumber
\end{align} 
Note that if we substitute $u= (t-z)/\sqrt{2}$ and $v = (t+z)/\sqrt{2}$, then $A$ reduces to the standard expression for a gravitational wave metric in GR. In these coordinates, the Finsler metric reads
%
\begin{align}
F &= \alpha\, \phi(\beta/\alpha), \label{eq:ab_grav_waves_2}\\
&\left\{\begin{array}{ll}
 A = -\D t^2+ \D z^2 + (1+\varepsilon f_+(t-z)) \D x^2 + (1-\varepsilon f_+(t-z))\D y^2 \\
 \,\,\,\,\, \qquad \qquad\qquad\qquad\qquad\qquad\qquad\qquad\quad\,\,\,\,\,\,\, + 2\varepsilon f_\times (t-z) \D x\,\D y  \\
\beta = \frac{1}{\sqrt{2}}\left(\D t - \D z\right).
\end{array}\right. \nonumber 
\end{align}
We will thus refer to such Finsler metrics as (Finslerian) \textit{gravitational wave} metrics. 

%

\subsection{Linearized \texorpdfstring{$(\alpha,\beta)$}{(\textalpha,\textbeta)}-metrics are Randers metrics}
\label{sec:lin_ab_metric_is_randers}

It is natural to linearize not only in $\varepsilon$, characterizing the departure from flatness, but to also use a perturbative expansion in the `size' of the 1-form, characterizing the departure from GR and pseudo-Riemannian geometry. In other words, we consider gravitational waves that deviate only slightly from the well-known general relativistic ones. The purpose of this section is to highlight that any \ab-metric is equivalent to a Randers metric, to first order in this expansion, so that from a physics point of view, Randers metrics are quite a bit more general than they might seem at first glance. 

So consider an \ab-metric constructed from a pseudo-Riemannian metric $\alpha$ and a 1-form $\beta$ such that $\beta\ll 1$. To see what happens in such a scenario, we replace $\beta$ with $\lambda\beta$ and expand to first order in $\lambda$. Then we obtain
\begin{align}
F = \alpha \phi\left(\frac{\lambda\beta}{\alpha}\right) \approx \alpha \left(\phi(0)+ \lambda \phi'(0)\frac{\beta}{\alpha}\right) = \alpha \phi(0) + \lambda\phi'(0)\beta = \tilde\alpha + \tilde\beta.
\end{align}
Hence to first order in $\lambda$, any $(\alpha,\beta)$-metric is indeed equivalent to a Randers metric\footnote{Actually, this is not true for \textit{all} $(\alpha,\beta)$-metrics but only those which allow an expansion around $s = \beta/\alpha = 0$. This excludes some of the $m$-Kropina metrics, for instance, because these are not always well-behaved in the limit $\beta\to 0$.}. Consequently, by replacing $\D u$ by $\lambda\D u$ in \eqref{eq:ab_grav_waves}, which technically can be achieved by a coordinate transformation that scales $u$ by $\lambda$ and  $v$ by $1/\lambda$, it follows that to first order in $\lambda$ the Finsler metric of the \ab-type gravitational waves takes the form,
\begin{align}
F &=  \alpha + \beta, \label{eq:metric_naive_Randers_grav_wave0}\\
&\left\{\begin{array}{ll}
 A = -2\D u \D v + (1+\varepsilon f_+(u)) \D x^2 + (1-\varepsilon f_+(u))\D y^2 + 2\varepsilon f_\times (u) \D x\,\D y \\ 
\beta = \lambda\D u
\end{array}\right.\nonumber
\end{align} 
\sloppy The parameter $\lambda$ then characterizes the departure from GR and pseudo-Riemannian geometry. We will assume without loss of generality that $\lambda> 0$. 

%
One more step is necessary to arrive at the Finsler metric that we will take as the starting point for the calculation of the radar distance in \cref{sec:radar_distance}. And this step has to do with the causal structure of the Randers metric. We carefully argued in \cref{sec:Modified_Randers_Metrics} that the physically relevant version of the Randers metric is not the standard one, but that one should prefer the \textit{modified Randers metric}. Accordingly, the physically relevant version of the Finslerian gravitational wave metric is given by
\begin{align}
F &= \sgn(A)\alpha + |\beta|, \label{eq:metric_Randers_grav_wave0}\\
&\left\{\begin{array}{ll}
 A = -2\D u \D v + (1+\varepsilon f_+(u)) \D x^2 + (1-\varepsilon f_+(u))\D y^2 + 2\varepsilon f_\times (u) \D x\,\D y \\ 
\beta = \lambda\D u
\end{array}\right.\nonumber
\end{align} 
For completeness, we also include the form of the Finsler metric in the coordinates $(t,x,y,z)$ defined right below \eqref{eq:ab_grav_waves}:
\begin{align}
F &= \sgn(A)\alpha + |\beta|, \label{eq:metric_Randers_grav_wave}\\
&\left\{\begin{array}{ll}
 A = -\D t^2+ \D z^2 + (1+\varepsilon f_+(t-z)) \D x^2 + (1-\varepsilon f_+(t-z))\D y^2 \\
\,\,\,\,\, \qquad \qquad\qquad\qquad\qquad\qquad\qquad\qquad\quad\,\,\,\,\,\,\,+ 2\varepsilon f_\times (t-z) \D x\,\D y  \\
\beta = \frac{\lambda}{\sqrt{2}}\left(\D t - \D z\right),
\end{array}\right. \nonumber 
\end{align}
Let us briefly explain why this modification of the Randers metric is perfectly allowed, even though it might seem at first glance to be in conflict with the result obtained above that any \ab-metric should reduce to a \textit{conventional} Randers metric. What is important to note is that there is in principle the possibility that different regions of the tangent bundle (subdomains of $\mathcal A$) could house different \ab-metrics, hence leading to different Randers metrics in each subdomain. 

More precisely, we could define one \ab-metric $F_1$ on a conic subbundle $\mathcal A_1\subset TM_0$ and another \ab-metric, $F_2$, on a different conic subbundle $\mathcal A_2\subset TM_0$. If the two subbundles do not overlap then this defines a perfectly valid \ab-type Finsler spacetime on the union $\mathcal A = \mathcal A_1\cup\mathcal A_2$. Admittedly, it would not satisfy the definition of an \ab-metric in the strict sense of \cref{def:ab_metric} on its domain as a whole, but there is no reason at this point in the analysis why we should be so strict. For our present purposes, having a physically viable causal structure is clearly more important than adhering to a mathematically convenient---but arbitrary from the physics perspective---definition. 

To first order in the deviation from pseudo-Riemannian geometry, the so-obtained Finsler metric would reduce to a certain Randers metric on $\mathcal A_1$ and to a different Randers metric on $\mathcal A_2$. Hence on $\mathcal A$ as a whole, the resulting linearized metric might not be expressible as a single standard Randers metric. This is indeed precisely the case for the modified Randers metric. On each of the four subsets of $\mathcal A$ given by $\pm A>0, \pm \beta>0$, the modified Randers metric $F = \sgn(A)\alpha + |\beta|$ reduces to a standard Randers metric (up to a possible overall minus sign, which is irrelevant). Our use of the modified Randers metric is therefore completely consistent with the previous results, and it is preferred here over the standard Randers metric because of its physically natural causal structure and its other desirable properties. 

\section{Observational signature}
\label{sec:radar_distance}

Now we are finally in the position to analyze the physical effects of a passing Finslerian gravitation wave of \ab-type. Our starting point will be the linearized Finsler metric \eqref{eq:metric_Randers_grav_wave}. Since actual gravitational wave measurements are very often done by interferometers, which effectively measure the \textit{radar distance}, the aim of this section is to compute this radar distance during the passing of a gravitational wave of the form \eqref{eq:metric_Randers_grav_wave}. 

The setup is as follows. A light ray is emitted from some spacetime location with coordinates $(t_0, x_0, y_0, z_0)$, travels to another location in spacetime with coordinates $(t_0+\Delta t, x_0+\Delta x, y_0 + \Delta y, z_0+\Delta z)$, where it is reflected and after which it travels back to the original (spatial) location, with spacetime coordinates $(t_0+\Delta t_{\text{tot}}, x_0, y_0, z_0)$, being received there again. We are interested in the amount of proper time that passes between emission and reception of the light ray, as measured by an  `inertial' observer\footnote{In this context, we say that an observer is intertial if it would be considered an intertial observer in the absence of the wave (i.e. for $f_+ = f_\times=0$). In other words, thinking of the gravitational wave as having a finite duration as it passes the Earth, an observer is inertial precisely if it is inertial before and after the wave passes. In the absence of the wave, an observer is said to be inertial if it has constant spatial position $(x,y,z)$ in some coordinate system in which $A = -\D t^2 +\D x^2+\D y^2 +\D z^2$ and $\beta = \D u$.} located at the spatial coordinates $(x_0, y_0, z_0)$. Because light travels forward \textit{and} backward during this time interval, one half of the time interval is usually called the \textit{radar distance} between the two spacetime points (in GR the value is sometimes multiplied by the velocity of light, $c$, which we have set to 1, so as to obtain a quantity with the dimensions of distance). In other words, the radar distance can be expressed as $R = \Delta \tau/2$. Mathematically, to obtain the radar distance (as measured by the interferometer) to the point $(t_0+\Delta t, x_0+\Delta x, y_0 + \Delta y, z_0+\Delta z)$, we have to find the unique (cf. \cref{prop:radar_nbh_Randers}) two null geodesics that connect this point to the worldline of the observer.

The expression for the radar distance of a standard gravitational wave in GR has been obtained in \cite{Rakhmanov_2009} and our calculation below follows essentially the same methods. At each step of the calculation, we will clearly point out the differences with the corresponding situation in GR, so that it is clear where each of the Finslerian effects (three separate effects can be identified) enters precisely. In addition to linearizing in $\varepsilon$, we also use a perturbative expansion in $\lambda$, as argued for in \cref{sec:lin_ab_metric_is_randers}. In fact, rather than working to first order in $\lambda$, we will work to second order in the Finslerian parameter, as certain important Finslerian effects only enter at second order, as we will see. We do neglect terms of combined order $\varepsilon\lambda^2$ and higher.


\subsection{Null geodesics}\label{sec:geodesics}

As explained above, our goal is to find the unique two null geodesics that connect this point to the worldline of the observer. As long as the interferometer is sufficiently small in its spatial extent and the time window of observation is not too large, \cref{prop:radar_nbh_Randers} guarantees that this can be done. The first important observation here is that the geodesics in a modified Randers gravitational wave spacetime with Finsler metric $F = \text{sgn}(A)\alpha + |\beta|$ as in \eqref{eq:metric_Randers_grav_wave} coincide with the geodesics of the GR spacetime with metric $A$ because the affine connection of $F$ coincides with the Levi-Civita connection of $A$, by \cref{eq:ab_cc_pre}. For the derivation of the general form of geodesics, we may therefore assume that the geometry is given by $A$. Our point of departure will thus be the metric 
%
%
\begin{align}
&\D s^2  = \\
&-\D t^2 + (1+\varepsilon f_+(t-z)) \D x^2  + (1-\varepsilon f_+(t-z))\D y^2 + 2\varepsilon f_\times (t-z) \D x\,\D y + \D z^2,\nonumber
\end{align}
where $\varepsilon\ll 1$, and we can essentially follow the corresponding calculation in GR \cite{Rakhmanov_2009} until the moment at which the modified null condition becomes important. Using the coordinates $u= (t-z)/\sqrt{2}$ and $v = (t+z)/\sqrt{2}$ the geodesic equations to first order in $\varepsilon$ can be written as
\begin{align}
-\dot u \eqqcolon p_ v &= \text{const}, \label{eq:geod_1}\\
(1+\varepsilon f_+(u))\dot x +\varepsilon f_\times(u)\dot y\eqqcolon p_ x &= \text{const},\\
(1-\varepsilon f_+(u))\dot y +\varepsilon f_\times(u)\dot x\eqqcolon p_y &= \text{const}, \label{eq:geod_3}\\
\ddot v +\tfrac{1}{2} \varepsilon  \left(\dot{x}^2-\dot{y}^2\right) f_+'(u) + \varepsilon f_\times '(u)\dot x\dot y &= 0, \label{eq:geod_v}
\end{align}
where \eqref{eq:geod_1}-\eqref{eq:geod_3} are obtained as first integrals using the fact that $A$ is independent of the coordinates $v,x,y$. Here an overdot denotes a derivative with respect to the affine parameter $\sigma$. The first three equations can be rewritten to first order as $\dot u = -p_v$,
\begin{align}
\hspace{-7px}\dot x = (1-\varepsilon f_+(u))p_x - \varepsilon f_\times(u)p_y, \qquad \dot y = (1+\varepsilon f(u))p_y- \varepsilon f_\times(u)p_x,
\end{align}
and can be integrated (with respect to the affine parameter $\sigma$, chosen without loss of generality such that $\dot u = 1$) to
\begin{align}
u &= u_0 + \sigma, \\
x &= x_0 + \sigma\left[\left(1 - \varepsilon \bar f_+(\sigma)\right)p_x- \varepsilon \bar f_\times(\sigma)p_y\right],\\
y &= y_0 + \sigma\left[\left(1 + \varepsilon \bar f_+(\sigma)\right)p_y- \varepsilon \bar f_\times(\sigma)p_x\right],
\end{align}
where
\begin{align}
\bar f_{+,\times}(\sigma) \equiv \frac{1}{\sigma}\int_0^\sigma f_{+,\times}(u_0+\sigma)\D\sigma
\end{align}
is the averaged value of $f_{+,\times}$. The equation \eqref{eq:geod_v} for $v$ can be integrated to
\begin{align}
\dot v = -\tilde p_{u0}-\frac{1}{2} \varepsilon  \left(p_x^2-p_y^2\right) f_+(u_0+\sigma) - \varepsilon f_\times (u_0+\sigma)p_x p_y
\end{align}
where $\tilde p_{u0} = p_{u0}-\frac{\varepsilon}{2}(p_x^2-p_y^2)f(u_0)- \varepsilon f_\times (u_0)p_x p_y$ and where $p_u = -\dot v$ (not necessarily constant) and $p_{u0}$ is its initial value at $\sigma = 0$. Integrating once again, we obtain
\begin{align}
v = v_0 - \tilde p_{u0}\sigma-\frac{1}{2}\varepsilon (p_x^2-p_y^2)\sigma \bar f_+(\sigma)- \varepsilon \bar \sigma f_\times (\sigma)p_x p_y.
\end{align}
Any geodesic emanating from a given point $x_0^\mu$ can thus be described by the following parameterized path for certain values of the constants $p_x, p_y$ and $\tilde p_{u0}$:
\begin{align}
u(\sigma) &= u_0 + \sigma, \label{eq:geod_prim_u}\\
x(\sigma) &= x_0 + \sigma\left[\left(1 - \varepsilon \bar f_+(\sigma)\right)p_x- \varepsilon \bar f_\times(\sigma)p_y\right],\label{eq:geod_prim_x}\\
y(\sigma) &= y_0 + \sigma\left[\left(1 + \varepsilon \bar f_+(\sigma)\right)p_y- \varepsilon \bar f_\times(\sigma)p_x\right],\label{eq:geod_prim_y}\\
v(\sigma) &= v_0- \tilde p_{u0}\sigma-\tfrac{1}{2}\varepsilon (p_x^2-p_y^2)\sigma \bar f_+(\sigma)- \varepsilon \sigma\bar f_\times (\sigma)p_x p_y.\label{eq:geod_prim_v}
\end{align}
%
We need to know the specific expression for \textit{null} geodesics, however. The modified null condition or modified dispersion relation (MDR) for massless particles, $F=0$, is the first place where the Finslerian character of the gravitational wave enters. According to \cref{sec:Randers_causality}, the condition $F=0$ is equivalent to $A = -\beta^2$, i.e.
\begin{align}
    -2\dot u\dot v + (1+\varepsilon f_+(u)\!)\dot x^2 + (1-\varepsilon f_+(u)\!)\dot y^2 +2\varepsilon f_\times(u) \dot x\dot y=  -\lambda^2\dot u^2,
\end{align}
which, after substituting \eqref{eq:geod_prim_u}-\eqref{eq:geod_prim_v}, becomes $2\tilde p_{u0} + p_x^2+p_y^2=-\lambda^2$. We may therefore eliminate $\tilde p_{u0}$ by directly substituting this null condition into the expression \eqref{eq:geod_prim_v} for $v(\sigma)$. A generic null geodesic starting at $(u_0,x_0,y_0,v_0)$ at $\sigma=0$ can therefore be described by the following parameterized path,
\begin{align}
u &= u_0+\sigma, \\
x &= x_0 + \sigma\left[\left(1 - \varepsilon \bar f_+(\sigma)\right)p_x- \varepsilon \bar f_\times(u)p_y\right], \label{eq:middle1}\\
y &= y_0 + \sigma\left[\left(1 + \varepsilon \bar f_+(\sigma)\right)p_y- \varepsilon \bar f_\times(u)p_x\right], \label{eq:middle2}\\
v &= v_0 + \frac{\sigma}{2}(p_x^2+p_y^2+\lambda^2)-\frac{1}{2}\varepsilon (p_x^2-p_y^2)\sigma \bar f_+(\sigma)- \varepsilon \sigma \bar f_\times (\sigma)p_x p_y.
\end{align}
Here we can make two important observations:
\begin{enumerate}
\item the effect due to the MDR or modified null condition enters at order $\lambda^2$;
\item in the limit $\lambda\to 0$ we recover the null geodesics of a standard gravitational wave in GR \cite{Rakhmanov_2009}.
\end{enumerate}

\subsection{Radar distance}

Next, we plug in the boundary conditions at the receiving point $(u_0+\Delta u, x_0+\Delta x, y_0 + \Delta y, v_0+\Delta v)$. Note that $\sigma = \Delta u$ at that point, and hence from the middle two equations \eqref{eq:middle1} and \eqref{eq:middle2} we infer that
\begin{align}
p_x &= \frac{\Delta x}{\Delta u}\left(1 + \varepsilon \bar f_+(\Delta u)\right) +  \varepsilon \bar f_\times(\Delta u)\frac{\Delta y}{\Delta u}, \\
p_y &= \frac{\Delta y}{\Delta u}\left(1 - \varepsilon \bar f_+(\Delta u)\right)+  \varepsilon \bar f_\times(\Delta u)\frac{\Delta x}{\Delta u},
\end{align}
Plugging this into the $v$ equation yields
\begin{align}
\hspace{-4.5px}2\Delta u\Delta v = \Delta x^2 (1\!+\!\varepsilon \bar f_+(\Delta u)) + \Delta y^2 (1\!-\!\varepsilon \bar f_+(\Delta u)) + 2\varepsilon \bar f_\times \Delta x\Delta y +\! \lambda^2\Delta u^2,
\end{align}
or equivalently,
\begin{align}
\left(1-\frac{\lambda^2}{2}\right)\Delta t^2 = &\left(1 + \varepsilon \bar f(\Delta u)\right)\Delta x^2 + \left(1 - \varepsilon \bar f(\Delta u)\right)\Delta y^2 \nonumber\\
&+ 2\varepsilon \bar f_\times(\Delta u) \Delta x\Delta y +\left(1+\frac{\lambda^2}{2}\right)\Delta z^2 - \lambda^2\Delta z\Delta t,
\end{align}
where we have used that $-2\Delta u\Delta v = -\Delta t^2+\Delta z^2$.
This equation is solved to first order in $\varepsilon$ and $\lambda^2$ (neglecting $\varepsilon\lambda^2$ terms) by\footnote{In addition to this solution there is, formally, another solution to the equation. However, this other solution has the wrong zeroth order term, namely a negative one, which renders it physically irrelevant.}
\begin{align}\label{eq:delta_t_outgoing}
\Delta t = \Delta \ell &+ \left(\frac{\Delta x^2 - \Delta y^2 }{2\Delta\ell}\right)\varepsilon \bar f_+(\Delta u) + \left(\frac{\Delta x \Delta y }{\Delta\ell}\right)\varepsilon \bar f_\times(\Delta u)\nonumber\\
&+ \frac{1}{2}\left(\frac{\Delta x^2 + \Delta y^2 + 2 \Delta z^2}{2\Delta\ell} - \Delta z\right)\lambda^2,
\end{align}
where $\Delta \ell \equiv \sqrt{\Delta x^2 + \Delta y^2 + \Delta z^2}$ is the Euclidean spatial distance, which coincides with the radar distance in the case of flat Minkowski spacetime.

In principle, the RHS still depends on $t$ via $\bar f(\Delta u)$, so \eqref{eq:delta_t_outgoing} is not yet a closed formula for $\Delta t$. However, since $\bar f$ only appears multiplied with $\varepsilon$, and since we are only interested in the first order expression for $\Delta t$, any zeroth order expression for $\bar f$ suffices in this formula. We have

\begin{align}
\bar f(\Delta u)  &= \frac{1}{\Delta u}\int_0^{\Delta u} f(u_0+\sigma)\D\sigma = \frac{\sqrt{2}}{\Delta t - \Delta z}\int_0^{(\Delta t - \Delta z)/\sqrt{2}} f(u_0+\sigma)\D\sigma \nonumber\\
&= \frac{\sqrt{2}}{\Delta \ell - \Delta z}\int_0^{(\Delta \ell - \Delta z)/\sqrt{2}} f(u_0+\sigma)\D\sigma + \mathcal O(\varepsilon)\\
&= \frac{\sqrt{2}}{\Delta \ell - \Delta z}\int_0^{(\Delta \ell - \Delta z)/\sqrt{2}} f\left(\frac{1}{\sqrt{2}}(t_0-z_0)+\sigma\right)\D\sigma + \mathcal O(\varepsilon)\label{eq:avg_perturbation_zeroth_order}
\end{align}
since $\Delta t = \Delta \ell + \mathcal O(\varepsilon)$. We introduce another symbol for this expression, namely
\begin{align}\label{eq:avg_perturbation_zeroth_order_tz}
\bar f(\Delta \ell, \Delta z, t_0-z_0)\equiv \frac{\sqrt{2}}{\Delta \ell - \Delta z}\int_0^{(\Delta \ell - \Delta z)/\sqrt{2}} f\left(\frac{1}{\sqrt{2}}(t_0-z_0)+\sigma\right)\D\sigma,
\end{align} 
where the explicit display of the arguments serves to remind us that $\bar f$ depends only on $\Delta \ell, \Delta z$ and the initial value of $t-z$. Since $\varepsilon \bar f(\Delta u)  = \varepsilon \bar f(\Delta \ell, \Delta z, t_0-z_0) + \mathcal O(\varepsilon^2)$, it follows that we can rewrite \eqref{eq:delta_t_outgoing} to first order in $\varepsilon$ and $\lambda^2$ as
\begin{align}
\Delta t = \Delta \ell &+ \left(\frac{\Delta x^2 - \Delta y^2 }{2\Delta\ell}\right)\varepsilon \bar f_+(\Delta \ell, \Delta z, t_0-z_0)\nonumber\\
&+ \left(\frac{\Delta x \Delta y }{\Delta\ell}\right)\varepsilon \bar f_\times(\Delta \ell, \Delta z, t_0-z_0) \nonumber\\
&+ \frac{1}{2}\left(\frac{\Delta x^2 + \Delta y^2 + 2 \Delta z^2}{2\Delta\ell} - \Delta z\right)\lambda^2, \label{eq:Randers_time_elapsed_single_trip}
\end{align}
which is a closed expression for the elapsed coordinate time interval $\Delta t$ for a light ray traveling over a certain spatial coordinate distance, in terms of the spatial coordinate separations and the initial value of $t-z$. 

Now let us consider the complete trip, from $x^\mu_0$ to $x^\mu_0 + \Delta x^\mu$ and `back' to the original spatial location. The total coordinate time elapsed during this trip is the sum of the forward trip and the backward trip time intervals. Schematically:
\begin{align}
\Delta t_\text{tot} &= \Delta t(\Delta x,\Delta y,\Delta z,t_0-z_0) \nonumber\\
&+ \Delta t(-\Delta x, -\Delta y, -\Delta z,t_0+\Delta t-(z_0+\Delta z)),
%
\end{align}
since the spatial interval on the backward trip is simply minus the forward spatial interval, and since the `initial' value of $t-z$ for the backward trip is just the final value $t_0 -z_0 + \Delta t -\Delta z$ corresponding to the forward trip. Plugging in \eqref{eq:Randers_time_elapsed_single_trip} yields
\begin{align}\label{eq:Randers_time_elapsed_total_trip}
\Delta t_\text{tot} = 2\Delta \ell &+ \varepsilon\left(\frac{\Delta x^2 - \Delta y^2 }{2\Delta\ell}\right) \bar f_{+,\text{tot}} + \varepsilon \left(\frac{\Delta x \Delta y }{\Delta\ell}\right) \bar f_{\times,\text{tot}} \nonumber\\
&+ \frac{1}{2}\lambda^2\left(\frac{\Delta x^2 + \Delta y^2 + 2 \Delta z^2}{\Delta\ell} \right),
\end{align}
where $\bar f_{+,\text{tot}} = \bar f_{+,\text{forward}} + \bar f_{+,\text{backward}}$ and similarly for the $\times$-polarization, in terms of the forward and backward averaged amplitudes, respectively, given by
\begin{align}
\bar f_{+,\times,\text{forward}} &= \bar f_{+\times,}(\Delta \ell, \Delta z, t_0-z_0) \nonumber \\
&= \frac{\sqrt{2}}{\Delta \ell - \Delta z}\!\int_0^{(\Delta \ell - \Delta z)/\sqrt{2}}\! f_{+,\times}\!\left(\frac{1}{\sqrt{2}}(t_0-z_0)\!+\!\sigma\!\right)\D\sigma, \label{eq:bar_f_forward}\\
\bar f_{+,\times,\text{backward}} &= \bar f_{+,\times}(\Delta \ell, -\Delta z, t_0-z_0 + \Delta t - \Delta z) \label{eq:bar_f_backward}\\
& = \frac{\sqrt{2}}{\Delta \ell + \Delta z}\int_0^{(\Delta \ell + \Delta z)/\sqrt{2}} \!f_{+,\times}\!\left(\frac{1}{\sqrt{2}}(t_0+\Delta\ell-z_0 - \Delta z)\!+\!\sigma\!\right)\D\sigma. \nonumber
\end{align}
In the last expression we have replaced $\Delta t$ by $\Delta \ell$ in the argument of $f_{+,\times}$, because to zeroth order this makes no difference, and only the zeroth order expression for $\bar f_{+,\text{backward}}$ is relevant because $\bar f_{+,\text{backward}}$ always appears multiplied by $\varepsilon$ in the expressions we care about, like $\Delta t_\text{tot}$.


Equation \eqref{eq:Randers_time_elapsed_total_trip} gives the total coordinate time elapsed during the trip of the light ray forward and back. The next step in the calculation of the radar distance $R = \Delta \tau/2$ is to convert the coordinate time interval into the proper time interval measured by the stationary observer local to the emission and reception of the light ray.  This is where a second Finslerian effect enters. For such an observer we have $x=y=z=$ \textit{const} and hence the 4-velocity is given by $(\dot t,0,0,0)$, where we will assume without loss of generality that $\dot t>0$. The proper time measured by an observer is given by the Finslerian length along its worldline $\Delta \tau =-\int F\, \D \sigma$. If we use $\sigma = \tau$ as our curve parameter, differentiating with respect to it shows that the tangent vector to the curve should be normalized, satisfying $F=-1$. This is the Finsler equivalent of the fact that in GR the worldline of a particle parameterized by proper time should always satisfy $g_{\mu\nu}\dot x^\mu \dot x^\nu=-1$ (or $+1$, depending on the signature convention). In the case of our observer, the condition becomes
\begin{align}
F &= \text{sgn}(A)\alpha + |\beta| = \text{sgn}(-\dot t^2)\sqrt{|\dot t^2|} + \frac{|\lambda\dot t|}{\sqrt{2}} = -|\dot t| + \frac{|\lambda\dot t|}{\sqrt{2}} \\
&= \left(-1+\frac{\lambda }{\sqrt{2}}\right)\dot t \stackrel{!}{=} -1,
\end{align}
where we recall that $\lambda>0$. From this, it follows that
\begin{align}\label{eq:proper_time_vs_coord_time}
\Delta \tau = \left(1-\frac{\lambda }{\sqrt{2}}\right)\Delta t_\text{tot}
\end{align}
along the worldline of the stationary observer. Plugging in \eqref{eq:Randers_time_elapsed_total_trip} and \eqref{eq:proper_time_vs_coord_time} into $R = \Delta \tau/2$ we conclude that, to first order in $\varepsilon$ and second order in $\lambda$, the radar distance is given by
\begin{align}
%
%
R = \left(1 - \frac{\lambda }{\sqrt{2}}\right)\Delta \ell &+ \left(1 - \frac{\lambda }{\sqrt{2}}\right)\left(\frac{\Delta x^2 - \Delta y^2 }{4\Delta\ell}\right) \varepsilon\bar f_{+,\text{tot}} \label{eq:Randers_Radar_Distance}\\
&\hspace{0px}+\left(1 - \frac{\lambda }{\sqrt{2}}\right)\left(\frac{\Delta x \Delta y }{2\Delta\ell}\right)\varepsilon \bar f_{\times,\text{tot}}  + \frac{\lambda^2}{4}\left(\Delta\ell + \frac{\Delta z^2}{\Delta\ell} \right).\nonumber
\end{align}
Equation \eqref{eq:Randers_Radar_Distance} expresses the radar distance as a function of the spatial coordinate distances and the initial value of $t-z$ (the latter enters the expression via $\bar f_{+,\times,\text{tot}}$). In the limit $\lambda\to 0$ we recover the expression for the radar distance in the case of a standard gravitational wave in GR \cite{Rakhmanov_2009}:
\begin{align}
R = \Delta \ell + \varepsilon\left(\frac{\Delta x^2 - \Delta y^2 }{4\Delta\ell}\right) \bar f_{+,\text{tot}} + \varepsilon \left(\frac{\Delta x \Delta y }{2\Delta\ell}\right) \bar f_{\times,\text{tot}} + \mathcal O (\varepsilon^2).\label{eq:GR_radar_distance}
\end{align}
Before we move on, let us summarize in what ways the Finslerian parameter $\lambda$ has entered our analysis so far:
\begin{enumerate}
\item The null geodesics are altered due to the fact the Finsler metric induces a modified null condition or MDR. As a result, it takes a \textit{larger} coordinate time interval for a light ray to travel a given spatial coordinate distance. This effect works in all spatial directions, even the direction parallel to the propagation direction of the light ray (although the effect is somewhat different in this direction). This effect enters at order $\lambda^2$.
\item The ratio of proper time and coordinate time is altered with the result that \textit{less proper time is experienced per unit coordinate time}. This effect enters at order $\lambda$.
\end{enumerate}
There is, however, a third way in which the parameter enters. Namely in the relation between the coordinate distance and radar distance \textit{in the absence of the wave}. For a gravitational wave in GR, these conveniently coincide; in the case of our Randers waves, they do not. The expression \eqref{eq:Randers_Radar_Distance} for the radar distance derived above refers merely to coordinates. In order to make sense of the result, we need to express the RHS in terms of measurable quantities, like the radar distances in the various directions in the absence of the wave. Employing \eqref{eq:Randers_Radar_Distance}, we write
\begin{align}
\Delta X = \left(1 - \frac{\lambda }{\sqrt{2}}\right)\Delta x + \frac{\lambda^2}{4}\Delta x, \\
\Delta Y = \left(1 - \frac{\lambda }{\sqrt{2}}\right)\Delta y + \frac{\lambda^2}{4}\Delta y , \\
\Delta Z = \left(1 - \frac{\lambda }{\sqrt{2}}\right)\Delta z + \frac{\lambda^2}{2}\Delta z ,
\end{align}
for the radar distance in the $x,y$ and $z$ direction \textit{in the absence of the wave}, respectively, and 
\begin{align}
R_0 = \left(1 - \frac{\lambda }{\sqrt{2}}\right)\Delta \ell + \frac{\lambda^2}{4}\left(\Delta\ell + \frac{\Delta z^2}{\Delta\ell} \right),
\end{align}
for the radar distance \eqref{eq:Randers_Radar_Distance} in the relevant direction \textit{in the absence of the wave}. Eliminating the coordinate distances in favor of the physical radar distances by virtue of the inverse transformations
\begin{align}
\Delta x &= \Delta X\left(1 + \frac{\lambda }{\sqrt{2}} + \frac{\lambda^2}{4}\right),\\
\Delta y &= \Delta Y\left(1 + \frac{\lambda }{\sqrt{2}} + \frac{\lambda^2}{4}\right),\\
\Delta z &= \Delta Z\left(1 + \frac{\lambda }{\sqrt{2}}\right), \\
\Delta \ell &= R_0\left(1 + \frac{\lambda }{\sqrt{2}} + \frac{3}{4}\lambda^2\right) - \frac{\Delta z^2}{4 R_0}\lambda^2, \\
&= R_0\left(1 + \frac{\lambda }{\sqrt{2}} + \frac{\lambda^2}{4}\right) - \frac{\Delta Z^2}{4 R_0}\lambda^2,
\end{align}
valid to second order in $\lambda$,we obtain our final expression for the radar distance in the presence of the wave:
\begin{align}
R =  R_0 + \varepsilon\left(\frac{\Delta X^2 - \Delta Y^2 }{4R_0}\right)\bar f_{+,\text{tot}} + \varepsilon\left(\frac{\Delta X\Delta Y }{2R_0}\right)\bar f_{\times,\text{tot}}  + \mathcal O(\varepsilon^2, \lambda^3, \varepsilon\lambda^2).
\end{align}
This is a remarkable result. By expressing the radar distance in terms of the physical observables $\Delta X,\Delta Y$ and $R_0$ rather than merely coordinate separations, 
all dependence on $\lambda$ has disappeared to the desired order and the expression is identical to its GR counterpart \eqref{eq:GR_radar_distance}! We must conclude, therefore, that the effect of a Randers gravitational wave on an interferometer is virtually indistinguishable from that of a conventional GR gravitational wave. 

\section{Concluding remarks}

By obtaining an explicit expression for the radar distance of our Finslerian gravitational waves, we have shown that interferometer experiments are unable to distinguish such gravitational waves from general relativistic ones. This invites some discussion. On the one hand, the result is a bit disappointing since it indicates that such measurements cannot be used to test the Finslerian character of our universe. On the other hand, it also means that current gravitational wave measurements are all compatible with the idea that our spacetime has a Finslerian nature. Further research (theoretical as well as experimental) is necessary to obtain a definite answer to the question of whether Finslerian gravitational waves exist, or more generally, the question of whether our universe has a Finslerian nature. 

In this context, our radar distance result may be viewed not in the least as a proof of concept. It shows how clear physical predictions can be extracted from an \textit{a priori} abstract Finslerian model of spacetime. In this particular case, the predictions coincide with those of GR, but it would be highly worthwhile to repeat such an analysis in other cases of interest.

It is important to point out that Finslerian effects may also play a role in the \textit{generation} of gravitational waves during, say, a black hole merger event. This would lead to a Finslerian correction to the waveform and this \textit{could} be measured in interferometer experiments, at least in principle; especially since our results show how the waveform correlates with the experimental results---namely in exactly the same way in which the waveform of a general relativistic gravitational wave does. In order to be able to investigate this, however, Finslerian black hole solutions need to be better understood. A start in this direction has been made in \cite{cheraghchi2022fourdimensional,voicu2023birkhoff}, and further research (in part by the author) is actively ongoing.

Let us also point out some of the limitations of our investigation. First of all, it is by no means expected that the Finslerian gravitational waves discussed here should be the only possible ones. Although being much larger than even the complete class of \textit{all} Lorentzian (i.e. non-Finslerian) geometries, the class of \ab-metrics of Berwald type, to which we have restricted our analysis, is still quite restrictive in the large scheme of (Finsler geometric) things.

Moreover, even within the class of $(\alpha,\beta)$-metrics, our analysis is only valid for those metrics that can be regarded as `close' to a Lorentzian metric, such that they can be approximated by Randers metrics. So even though there is no observable difference between the Finslerian gravitational waves discussed here and their GR counterparts, there might very well be more general types of Finslerian gravitational waves that \textit{could} be distinguished observationally from general relativistic ones by means of interferometer experiments.

Furthermore, radar distance experiments are by no means the only way of probing our spacetime geometry. It might be possible to detect the Finslerian character of spacetime in some other way. We have not explored such possibilities here, but will certainly do so in the future.

Finally, we have assumed in our calculations that the amplitude of the gravitational waves as well as the Finslerian deviation from general relativity are sufficiently small such that a perturbative approach to first order in the former and second order in the latter is valid. It would be of interest to repeat the calculation to higher order in perturbation theory. We expect that this would in principle be a straightforward, yet possibly tedious, exercise.


%% file: ch_Unicorn_cosmology.tex
\chapter{An Exact Cosmological Unicorn Solution}\label{ch:unicorn_cosm}

Berwald and Landsberg spacetimes (see \Cref{ch:BerwaldLandsbergUnicorn}) may be thought of as incrementally non--pseudo-Riemannian. Or in physics terms: incrementally less reminiscent of GR. Every Berwald spacetime is also Landsberg, but provided one adheres to the most strict definition of a Finsler metric (more strict than ours) it has been a long-standing open question whether or not the opposite is true. In fact, Matsumoto stated in 2003 that this question represents the next frontier of Finsler geometry \cite{Bao_unicorns}, and as a token of their elusivity, Bao \cite{Bao_unicorns}  has called these non-Berwaldian Landsberg spaces \textit{`[\dots] unicorns, by analogy with those mythical single-horned horse-like creatures for which no confirmed sighting is available.'} Since 2006 some examples of unicorns have been obtained by Asanov \cite{asanov_unicorns}, Shen \cite{shen_unicorns} and  Elgendi \cite{Elgendi2021a} by relaxing the definition of a Finsler space slightly, resulting in a definition like ours. Even such examples of so-called $y$-local unicorns are still exceedingly rare. 

In this chapter, which is based on \citeref{ref6}{H6}, we present a new exact vacuum solution to Pfeifer and Wohlfarth's field equation, which is precisely such a unicorn. It falls into one of the classes introduced by Elgendi. Interestingly we find that these solutions have a physically viable light cone structure, even though in some cases the signature is not Lorentzian but positive definite. In fact, the Finslerian light cone turns out to be equivalent to that of the flat Minkowski metric, even in the cases with the signature anomaly! 


Furthermore, we find a natural cosmological interpretation of one of our solutions and a promising analogy with classical Friedmann-Lema\^itre-Robertson-Walker (FLRW) cosmology. In particular, this solution has cosmological symmetry, i.e. it is spatially homogeneous and isotropic, and it is additionally conformally flat, with conformal factor depending only on the timelike coordinate. We show that, just as in classical FLRW cosmology, this conformal factor can be interpreted as the scale factor of the universe; we compute this scale factor as a function of cosmological time, and we show that it corresponds to a linearly expanding (or contracting) Finslerian universe.

\section{A modification of Elgendi's unicorn metrics}
\subsection{Elgendi's class of unicorns}

Elgendi recently introduced a class of unicorns \cite{Elgendi2021a} with Finsler metric given by
\begin{equation}
    F = \left(a\beta+\sqrt{\alpha^2-\beta^2}\right)e^{\frac{a\beta}{a\beta+\sqrt{\alpha^2-\beta^2}}},
\end{equation}
in terms of a real, nonvanishing constant $a$ (not to be confused with the metric $a_{\mu\nu}$ this time) and 
\begin{align}
    \alpha = f(x^0)\sqrt{(y^0)^2+\phi(\hat y)}, \qquad \beta = f(x^0) y^0,
\end{align}
where $f$ is a positive real-valued function and $\phi(\hat y) = \phi_{ij} \hat y^i \hat y^j = \phi_{ij} y^i  y^j$ is a nondegenerate quadratic form on the space spanned by $\hat y = (y^1, y^2, y^3)$, with constant, symmetric coefficients $\phi_{ij}$. As before, Latin indices $i,j,\dots$ will run over $1,2,3$, whereas Greek induces $\mu,\nu,\dots$ will run over $0,1,2,3$. 
The geodesic spray of $F$ is given by\footnote{This was shown in \cite{Elgendi2021a}. We note, however, that all results from \cite{Elgendi2021a} used here and in what follows have been checked independently by the author. }
\begin{align}
    G^0 &= \left(\frac{2f(x^0)^2(y^0)^2-\alpha^2}{f(x^0)^2}+\frac{a^2-1}{a^2}\frac{\alpha^2-\beta^2}{f(x^0)^2}\right)\frac{f'(x^0)}{f(x^0)}\\
    G^{i} &= Py^i,
\end{align}
where 
\begin{align}
    P &= 2\left(y^0+\frac{1}{af(x^0)}\sqrt{\alpha^2-\beta^2}\right)\frac{f'(x^0)}{f(x^0)},
\end{align}
and the Landsberg tensor vanishes identically. This shows that these metrics are indeed Landsberg, but not Berwald, since the $i$-components of the spray are not
quadratic\footnote{Indeed, if each $G^i$ were quadratic this would imply that $\alpha^2-\beta^2 = (c_i y^i)^2$ for certain ($x$-dependent) coefficients $c_i$. Rewriting gives $\alpha^2 = \beta^2 + (c_i y^i)^2$ and taking two $y$-derivatives of this (in the open set $A>0$ for concreteness) shows that $a_{ij} = b_i b_j + c_i c_i$. That implies that $a_{ij}$ has at most rank 2, which is in contradiction with its definition as a 4-dimensional Lorentzian metric, which should have rank 4.} in $y$. Note that our $G^k$ is twice the $G^k$ in Elgendi's paper \cite{Elgendi2021a}, due to a difference in convention. Explicitly then, Elgendi's unicorns have the form
\begin{align}\label{eq:unipos}
     F = f(x^0)\left(y^0 + \sqrt{\phi(\hat y)}\right)e^{\frac{y^0}{y^0 + \sqrt{\phi(\hat y)}}}.
\end{align}
where we have absorbed the constant $a$ into a redefinition of $x^0$. (And so from here onward there is no possibility for confusion of the constant $a$ with the metric $a_{\mu\nu}$ anymore.) For our purposes, we will modify this expression slightly, though, for similar reasons as the ones that motivated our definition of the modified Randers metric in 
\cref{sec:Modified_Randers_Metrics}.

\subsection{The modified unicorn metric}

The expression \eqref{eq:unipos} defining the unicorn metric is only well-defined whenever $\phi(\hat y)\geq 0$. If $\phi$ is positive definite, this is automatically the case, but in other signatures this is not true in general. In order to extend the domain of definition of $F$, an obvious first approach would be to replace $\phi$ by its absolute value, $|\phi|$, leading to
\begin{align}\label{eq:Uni}
    F = f(x^0) \left(y^0 + \sqrt{|\phi(\hat y)|}\right) e^{\frac{y^0}{y^0 + \sqrt{|\phi(\hat y)|}}}\,.
\end{align}
%
%
From the physical point of view, this is still not completely satisfactory, however. This can be seen by considering the light cone corresponding to such a Finsler metric, given by $F=0$. Indeed, barring some potential problems with the exponent to which we will come back later, the light cone would be given by those vectors satisfying $y^0 = -\sqrt{|\phi|}$, which would imply that the light cone is contained entirely within the half-space $y^0<0$. Interpreting $y^0$ for the moment as a time direction, this would have the result that light rays can only propagate `backward in time' (with regards to their parameterization). 
If on the other hand, $y^0$ is a spacelike coordinate, the analogous statement would be that light cannot propagate in the spatial $y^0$-direction (in contrast to the $-y^0$-direction). One could perhaps argue that these distinctions only pertain to the light rays as \textit{parameterized} curves and hence are not necessarily physical, but even as a mathematical artifact, it seems odd.

This situation is reminiscent of the similar situation for Randers metrics discussed extensively in \cref{sec:Modified_Randers_Metrics}. The solution in that case was to replace the Randers metric with the modified Randers metric \eqref{eq:modified_randers_metric}. Here we will perform a similar modification of the unicorn metric based on the same ideas.

Our starting point will be the following Finsler metric:
\begin{align}\label{eq:uni_spacetime_pre}
    F_{0}
    = f(x^0) \left( |y^0| + \text{sgn}(\phi)\sqrt{|\phi|} \right) e^{\frac{|y^0|}{ |y^0| + \text{sgn}(\phi)\sqrt{|\phi|} }}\,,
\end{align} 
from which we define the \textit{modified unicorn metric} as:
\begin{align}\label{eq:uni_spacetime}
    F = \left\{\begin{matrix}
    F_{0} & \text{if} & |y^0| + \text{sgn}(\phi)\sqrt{|\phi|} \neq 0\\
    0 & \text{if} &  |y^0| + \text{sgn}(\phi)\sqrt{|\phi|} = 0
    \end{matrix}\right. \,.
\end{align}
It will be confirmed in \cref{sec:solvFgrav} that this still defines a non-Berwaldian Landsberg metric, justifying the term \textit{unicorn}.

The case distinction is necessary since the metric \eqref{eq:uni_spacetime_pre} is ill-defined at vectors satisfying $|y^0| + \text{sgn}(\phi)\sqrt{|\phi|}=0$, because of the division by the same number in the exponent. $F_0$ does not have a well-defined limit to such vectors either, because the exponent does not stay negative in such a limit. This issue was already present for Elgendi's unicorn \eqref{eq:unipos}, yet from a purely mathematical point of view, this is not necessarily a problem, as one can simply opt to exclude this set of vectors from the domain of $F$. From a physics perspective, however, we want to interpret the set $F=0$ as the possible propagation directions of light, so it should not be empty.

As we will see below, our definition of $F$ as in \eqref{eq:uni_spacetime} ensures the existence of a physically viable light cone while preserving also the unicorn property. This is 
why we prefer it in the physical context, even though it leads to a discontinuity at the light cone as discussed in more detail below in \cref{sec:detg}.


\subsection{Causal structure}\label{sec:unicorn_causal_structure}

First, we observe that regardless of the exact form or signature of the quadratic form $\phi$, our modified unicorn metrics have a light cone structure $F=0$ that is equivalent to that of a pseudo-Riemannian metric. 
\begin{prop}\label{prop:lightcone_unicorn}
    The light cone of the modified unicorn metric \eqref{eq:uni_spacetime} is given by 
    \begin{align}\label{eq:lightcone}
        \left(y^0\right)^2 + \phi = 0.
    \end{align}
\end{prop}
\begin{proof}
    The result follows from the following sequence of equivalences.
    \begin{align}
        F = 0 &\Leftrightarrow \text{sgn}(\phi)\sqrt{|\phi|}+|y^0|=0 \\
        &\Leftrightarrow  \text{sgn}(\phi)\sqrt{|\phi|}=-|y^0|\\
        &\Leftrightarrow  |\phi|=\left(y^0\right)^2 \quad\text{and}\quad \phi \leq 0\\
        &\Leftrightarrow  \phi=-\left(y^0\right)^2\\
        &\Leftrightarrow  \phi+\left(y^0\right)^2 = 0.
    \end{align}
\end{proof}
Depending on the signature of $\phi$, we can make a more precise statement. In the following corollary, the ordering of the plusses and minuses in expressions like $(+,+,+,-)$ and $(-,+,+,+)$ indicates which of the coordinates---in this case, $x^3$ and $x^0$, respectively---is the timelike one.
\begin{cor}\label{cor:lightcone_details}
    Let $F$ be the modified unicorn metric \eqref{eq:uni_spacetime} corresponding to some nondegenerate quadratic form $\phi_{ij}$. Then the following holds:
    \begin{itemize}
        \item For $\phi_{ij}$ of signature $(+,+,-)$ the light cone structure of $F$ is identical to that of the $(+,+,+,-)$ Minkowski metric.
        \item If $\phi_{ij}$ is negative definite, i.e. of signature $(-,-,-)$ then the light cone of $F$ is identical to that of the $(-,+,+,+)$ Minkowski metric.
        \item If $\phi_{ij}$ is positive definite, i.e. of signature $(+,+,+)$ then the light cone of $F$ is given by the zero vector, $y=0$.
        \item For $\phi_{ij}$ of signature $(-,-,+)$ the light cone structure of $F$ is identical to that of a pseudo-Riemannian metric of signature $(+,-,-,+)$.
    \end{itemize}
\end{cor}
\noindent This singles out the $(+,+,-)$ and $(-,-,-)$ signatures of $\phi_{ij}$ as the ones that are physically of interest, and we will mostly restrict our attention to these two cases in what follows. Given the fact that the light cone structure in these cases is so simple, it is natural---as discussed several times before---to interpret the interior of the future-- and past-pointing light cone as the cone of future-- and past-pointing timelike directions, respectively. In the $(+,+,-)$ case, this leads to the interpretation of the coordinate $x^3$ as the timelike coordinate, with the timelike cone given by $(y^3)^2>(y^0)^2+(y^1)^2+(y^2)^2$, while in the $(-,-,-)$ case, $x^0$ would be the timelike coordinate, with the timelike cone given by $(y^0)^2>(y^1)^2+(y^2)^2+(y^3)^2$.

In analogy to the discussion in \cref{sec:Randers_causality} it also follows, in the same way that it did for modified Randers metrics, that the sign of $F(x,y)$ is correlated with the causal character of $y$. In the current scenario, however, the correlation depends also on the signature of $\phi$. The details are provided by the following proposition.
\begin{prop}[Modified unicorn causal structure]\label{prop:mod_unicorn_causal_structure}
\hphantom{t}
    \begin{itemize}
        \item If $\phi$ has signature $(-,-,-)$ then 
        \begin{align*}
            y \text{ is timelike } \quad  \Leftrightarrow  \quad|y^0|+\sgn(\phi)\sqrt{|\phi|}> 0 \quad\Leftrightarrow\quad F>0\\
            y \text{ is spacelike } \quad\Leftrightarrow\quad |y^0|+\sgn(\phi)\sqrt{|\phi|}< 0 \quad\Leftrightarrow\quad F<0\\
            y \text{ is null }\quad \Leftrightarrow\quad  |y^0|+\sgn(\phi)\sqrt{|\phi|}= 0 \quad\Leftrightarrow\quad F=0
        \end{align*}
        \item If $\phi$ has signature $(+,+,-)$ then 
        \begin{align*}
            y \text{ is timelike } \quad\Leftrightarrow\quad  |y^0|+\sgn(\phi)\sqrt{|\phi|}< 0 \quad\Leftrightarrow\quad F< 0\\
            y \text{ is spacelike } \quad\Leftrightarrow\quad  |y^0|+\sgn(\phi)\sqrt{|\phi|}> 0 \quad\Leftrightarrow\quad F>0\\
            y \text{ is null } \quad\Leftrightarrow\quad  |y^0|+\sgn(\phi)\sqrt{|\phi|}= 0 \quad\Leftrightarrow\quad F=0
        \end{align*}
    \end{itemize}
\end{prop}
\begin{proof}
    We start with the $F<0$ equivalences. Since $f(x^0)$ is positive by assumption, we have
    \begin{align}
        F < 0 &\Leftrightarrow \text{sgn}(\phi)\sqrt{|\phi|}+|y^0|<0 \\
        &\Leftrightarrow  \text{sgn}(\phi)\sqrt{|\phi|}<-|y^0|\\
        &\Leftrightarrow  |\phi|>(y^0)^2 \quad\text{and}\quad \phi \leq 0\\
        &\Leftrightarrow  -\phi>(y^0)^2\\
        &\Leftrightarrow  (y^0)^2 + \phi< 0.
    \end{align}
     In the case with signature $(-,-,-)$ this inequality reads $(y^0)^2-(y^1)^2-(y^2)^2-(y^3)^2<0$, which says precisely that $y$ lies in the exterior of the light cone $(y^0)^2=(y^1)^2+(y^2)^2+(y^3)^2$, i.e. $y$ is spacelike. And in the $(+,+,-)$ case the inequality reads $(y^0)^2+(y^1)^2+(y^2)^2-(y^3)^2<0$, which says precisely that $y$ lies in the interior of the light cone $(y^0)^2+(y^1)^2+(y^2)^2=(y^3)^2$, i.e. $y$ is timelike. This completes the proof of the $F<0$ equivalences. The $F=0$ equivalences follow trivially from the definition of a null vector and \Cref{prop:lightcone_unicorn}. Since $F>0$ if and only if neither of the previous two cases $F<0$ or $F=0$ is realized, the remaining $F>0$ equivalences follow automatically.
\end{proof}


\subsection{Signature and regularity}\label{sec:detg}

Next, we investigate the signature of our modified unicorn metrics. When $\phi$ has Lorentzian signature, we denote by $\mathcal{S(\phi)}$ the set of all $\hat y = (y^1,y^2,y^3)$ that are $\phi$-spacelike and by $\mathcal{T(\phi)}$ the set of all $\hat y$ that are $\phi$-timelike.
\begin{prop}\label{prop:signature}
Consider a modified unicorn metric $F$ as in \eqref{eq:uni_spacetime}. 
\begin{itemize}
    \item If $\phi$ is positive definite or negative definite then $g_{\mu\nu}$ is positive definite on its entire domain of definition.
    \item If $\phi$ is Lorentzian then $g_{\mu\nu}$ has Lorentzian signature $(+,+,+,-)$ for all $y\in \R\times\mathcal S(\phi)$ at which it is defined and signature $(+,+,-,-)$ for all $y\in  \R\times \mathcal T(\phi)$ at which it is defined.
\end{itemize}
\end{prop}
\begin{proof}
We may choose coordinates such that $\phi = \phi(\hat y) = \varepsilon_1(y^1)^2+\varepsilon_2(y^2)^2+\varepsilon_3(y^3)^2$. Since the spacetime dimension is fixed to 4 the calculation of the determinant of the fundamental tensor is in principle a straightforward exercise whenever $F$ is sufficiently differentiable. It is given by
\begin{align}
    \det g 
    &=  \textrm{sgn}(\phi)\varepsilon_1\varepsilon_2\varepsilon_3 f(x^0)^8\exp\left(\frac{8|y^0|}{ |y^0| + \text{sgn}(\phi)\sqrt{|\phi|} }\right)\,.
\end{align}
The determinant already gives us a pretty good idea of what the possible signatures of $g_{\mu\nu}$ can be. In particular, since $g_{\mu\nu}$ is a four-dimensional matrix, it has Lorentzian signature if and only if its determinant is negative. 
\begin{enumerate}[a)]
    \item If $\phi_{ij}$ is positive definite then all $\varepsilon_i$ and $\textrm{sgn}(\phi(\hat y))$ are positive, and hence $\det g$ is positive.
    \item If $\phi_{ij}$ is negative definite then all $\varepsilon_i$ and $\textrm{sgn}(\phi(\hat y))$ are negative, and hence $\det g$ is positive. 
    \item If $\phi_{ij}$ has Lorentzian signature $(+,+,-)$ then $\det g$ is negative whenever $\textrm{sgn}(\phi(\hat y))>0$, i.e. on $\mathcal{S(\phi)}$, and $\det g$ is positive whenever $\textrm{sgn}(\phi(\hat y))<0$, i.e. on $\mathcal{T(\phi)}$.
    \item If $\phi_{ij}$ has Lorentzian signature $(-,-,+)$; then $\det g$ is negative whenever $\textrm{sgn}(\phi(\hat y))<0$, i.e. on $\mathcal{S(\phi)}$, and $\det g$ is positive whenever $\textrm{sgn}(\phi(\hat y))>0$, i.e. on $\mathcal{T(\phi)}$.
\end{enumerate}
This already shows that $g_{\mu\nu}$ is Lorentzian if and only if $\phi$ is Lorentzian and $y\in \R\times \mathcal S(\phi)$. But the sign of the determinant does not suffice to determine whether this signature is mostly plus or mostly minus. Similarly, it does not tell us much about the signature of  $g_{\mu\nu}$ when $\phi$ is positive or negative definite. In order to find out, we distinguish the following three cases.\\

\noindent\textbf{Case 1: $\phi$ Lorentzian and $y\in \R\times\mathcal S(\phi)$}\\
We first consider the case that $\phi$ is Lorentzian. Without loss of generality, we set $\phi(\hat y) = \varepsilon_1(y^1)^2+ \varepsilon_1(y^2)^2 - \varepsilon_1(y^3)^2$, where $\varepsilon_1=\pm 1$ selects if we are in case c) or d).

Now note that given a vector $y\in T_xM$ which is $\phi$-spacelike, it follows from the symmetries of the Finsler metric and in particular from the $3$-dimensional Lorentz symmetry of $\phi$ that we may always change coordinates, without changing the form of $\phi$ (and $F$), such that $y^2=y^3=0$.

For any choice of $\varepsilon_1$, we find by direct calculation, using that $\varepsilon_1^2=1$ and $|\varepsilon_1|=1$, that $g_{\mu\nu}$ is of the form
\begin{align}
g_{\mu\nu} = e^{\frac{2 |y^0|}{|y0|+\varepsilon_1|y1|}} f(x^0)^2
\begin{pmatrix}
M & 0 & 0\\
0 & 1 & 0 \\
0 & 0 & -1
\end{pmatrix}\,,
\end{align}
where $M$ is an ($\varepsilon_1$-dependent) $2\times 2$ matrix, whose explicit form we omit. Its relevant properties are that
\begin{align}\label{eq:det_and_trace_M}
     \det M = 1,\qquad \text{Tr}\,M = 2 + \frac{4(y^1)^2}{(|y^0|+\varepsilon_1|y^1|)^2}.
\end{align}
%
%
Since the determinant and trace are both positive it follows that $M$ is positive definite. Hence we conclude that $g_{\mu\nu}$ is of the mostly plus type $(+,+,+,-)$.\\

\noindent\textbf{Case 2: $\phi$ Lorentzian and $y\in \R\times\mathcal T(\phi)$}\\
In this case we may WLOG choose coordinates such that $\phi(\hat y) = \varepsilon_1(y^1)^2- \varepsilon_1(y^2)^2 - \varepsilon_1(y^3)^2$, where $\varepsilon_1=\pm 1$, and 
such that $y^2=y^3=0$. Again, by direct calculation we find that $g_{\mu\nu}$ is now of the form
\begin{align}
g_{\mu\nu} = e^{\frac{2 |y^0|}{|y0|+\varepsilon_1|y1|}} f(x^0)^2
\begin{pmatrix}
M & 0 & 0\\
0 & -1 & 0 \\
0 & 0 & -1
\end{pmatrix}\,,
\end{align}
where $M$ is again a $2\times 2$ matrix with the properties \eqref{eq:det_and_trace_M}, so that $M$ is again positive definite. Hence we conclude that in this case, $g_{\mu\nu}$ has signature $(+,+,-,-)$.\\

\noindent\textbf{Case 3: $\phi$ positive or negative definite}\\
In this case we may WLOG choose coordinates such that $\phi(\hat y) = \varepsilon_1 (y^1)^2+ \varepsilon_1 (y^2)^2 + \varepsilon_1 (y^3)^2$, where $\varepsilon_1=\pm 1$, and 
such that, for any given $y\in T_xM$, we have $y^2=y^3=0$. In this case $g_{\mu\nu}$ is of the form
\begin{align}
g_{\mu\nu} = e^{\frac{2 |y^0|}{|y0|+\varepsilon_1|y1|}} f(x^0)^2
\begin{pmatrix}
M & 0 & 0\\
0 & 1 & 0 \\
0 & 0 & 1
\end{pmatrix}\,,
\end{align}
where $M$ is again a $2\times 2$ matrix with the properties \eqref{eq:det_and_trace_M}, so that $M$ is again positive definite. Hence we conclude that $g_{\mu\nu}$ is positive definite.
\end{proof}
Let us point out some interesting features of the two physically reasonable scenarios we identified below \Cref{cor:lightcone_details} as a result of their satisfactory light cone structure, i.e. $\phi$ having $(-,-,-)$ or $(+,+,-)$  signature. In both cases the light cone is equivalent to that of Minkowski space, but surprisingly, in its interior---the timelike cone---the signature of $g$ is never Lorentzian. Indeed, if $\phi$ is negative definite then \Cref{prop:signature} shows that $g$ is positive definite everywhere, in particular inside the timelike cone. Similarly, if $\phi$ has signature $(+,+,-)$ then, with $x^3$ representing the timelike direction as pointed out right below \Cref{cor:lightcone_details}, the timelike cone is given by $(y^3)^2>(y^0)^2+(y^1)^2+(y^2)^2$. Any vector in this cone thus satisfies, in particular, the inequality $(y^3)^2>(y^1)^2+(y^2)$, which says precisely that $\hat y\in \mathcal T(\phi)$. It then follows from \Cref{prop:signature} that  $g$ has signature $(+,+,-,-)$ here and hence everywhere within the timelike cone.

In various alternative, more stringent definitions of Finsler spacetimes \cite{Javaloyes2019,Hohmann:2021zbt}, one requires the existence of a nonempty cone with certain properties on which the fundamental tensor has Lorentzian signature, in order to guarantee (among other things) the existence of a physically viable light cone structure. What we have just established here, however, is that there exist Finsler geometries that do have a perfectly viable light cone structure even while not having Lorentzian signature \textit{anywhere}. This is an interesting new observation about Finsler geometry in its own right: apparently, even in positive definite signature, a light cone structure may arise due to irregularities of the Finsler metric. The most notable irregularity in the present case is located at null vectors, i.e. vectors satisfying  $|y^0| + \text{sgn}(\phi)\sqrt{|\phi|}=0$. Indeed, symbolically, we have
\begin{align}
    \lim_{\substack{|y^0| + \text{sgn}(\phi)\sqrt{|\phi|}\to 0\\ |y^0| + \text{sgn}(\phi)\sqrt{|\phi|} < 0}}F = 0.
\end{align}
whereas
\begin{align}
    \lim_{\substack{|y^0| + \text{sgn}(\phi)\sqrt{|\phi|}\to 0\\ |y^0| + \text{sgn}(\phi)\sqrt{|\phi|} > 0}}F = \infty\,.
\end{align}
Taking into account \Cref{prop:mod_unicorn_causal_structure} this means that in the scenario where $\phi$ has signature $(+,+,-)$, the Finsler metric $F$ extends continuously to $0$ if one tends to the light cone from inside the timelike cone. The transition from the spacelike directions to the light cone is then discontinuous, however. In the $(-,-,-)$ scenario the situation is exactly the opposite. $F$ then extends continuously to $0$ if one tends to the light cone from the spacelike directions, whereas the transition from the timelike directions to the light cone is discontinuous.

\section{Exact solutions}\label{sec:solvFgrav}

Our next aim is to determine the form of the function $f(x^0)$---the only physical degree of freedom---in the modified unicorn metric \eqref{eq:uni_spacetime} by solving the Finsler gravity field equation. Since the latter simplifies tremendously for Landsberg spaces, let us start by confirming that the modified unicorn is still of this type. The geodesic spray of $F$ is given explicitly by
\begin{align}
    G^0 &= \left((y^0)^2 - |\phi|\right)\frac{f'(x^0)}{f(x^0)}\\
    G^{i} &= Py^i,\qquad i=1,2,3
\end{align}
where 
\begin{align}
    P = 2\left(|y^0|+\text{sgn}(\phi)\sqrt{|\phi|}\right)\text{sgn}(y^0)\frac{f'(x^0)}{f(x^0)}\,,
\end{align}
In analogy with Elgendi's original unicorn metrics \eqref{eq:unipos}, this geodesic spray is not quadratic in $y$, so the modified unicorn metrics \eqref{eq:uni_spacetime} are not Berwald. Let us now check that the Landsberg tensor still vanishes, justifying the name (modified) \textit{unicorn}. To this end we employ the definition \eqref{eq:Landsberg_def_intermsof_Spray} of the Landsberg tensor, $S_{\mu\nu\rho} = -\tfrac{1}{4}y_\sigma \bar\partial_\mu\bar\partial_\nu\bar\partial_\rho G^\sigma$, and note that only the $G^i$ terms ($i=1,2,3$) can give a nontrivial contribution, since $G^0$ is quadratic. For these terms, we compute that
\begin{align}
    \bar\partial_\mu\bar\partial_\nu\bar\partial_\sigma G^i = \bar\partial_\mu\bar\partial_\nu\bar\partial_\sigma\sqrt{|\phi|}y^i + 3\delta^ i{}_{(\mu}\bar\partial_\nu\bar\partial_{\sigma)}\sqrt{|\phi|} ,
\end{align}
where the round brackets around the indices denote symmetrization. To find the Landsberg tensor we need to contract this with $y_i = g_{i\mu}y^\mu = \tfrac{1}{2}\bar\partial_i F^2$, and it can be checked in a straightforward way that the latter can be written as some function times $\bar\partial_i\sqrt{|\phi|}$. It thus suffices to show that $\bar\partial_i\sqrt{|\phi|}\bar\partial_\mu\bar\partial_\nu\bar\partial_\sigma G^i=0$. Indeed, whenever $\mu\neq 0$, it follows from the homogeneity of $\sqrt{|\phi|}$ that the latter is equal to (for $\mu=0$ it vanishes immediately since $\phi$ does not depend on $y^0$)
\begin{align}
    \bar\partial_i\sqrt{|\phi|}&\left(\bar\partial_\mu\bar\partial_\nu\bar\partial_\sigma\sqrt{|\phi|}y^i + 3\delta^ i{}_{(\mu}\bar\partial_\nu\bar\partial_{\sigma)}\sqrt{|\phi|} \right) \\
    &\qquad\qquad= \sqrt{|\phi|}\bar\partial_\mu\bar\partial_\nu\bar\partial_\sigma\sqrt{|\phi|} + 3\bar\partial_{(\mu}\sqrt{|\phi|}\bar\partial_\nu\bar\partial_{\sigma)}\sqrt{|\phi|} \\
    &\qquad\qquad= \tfrac{1}{2}\bar\partial_\mu\bar\partial_\nu\bar\partial_\sigma \left(\sqrt{|\phi|}\right)^2 = \tfrac{1}{2}\bar\partial_\mu\bar\partial_\nu\bar\partial_\sigma |\phi| = 0,
\end{align}
which vanishes because $\phi$ is quadratic\footnote{Points satisfying $|\phi|=0$ can be safely ignored since these lie outside the domain $\mathcal A$ of the Finsler metric.}. Hence our modified unicorns are indeed non-Berwaldian Landsberg metrics, justifying their name, and ensuring that the field equation reduces to the (weakly) Landsberg field equation \eqref{eq:Pfeifer_Wohlfarth_eq_Landsberg}. The next proposition characterizes Ricci-flat modified unicorn metrics. Below we will see that any vacuum solution to the field equation must be Ricci-flat, hence this result even characterizes all vacuum solutions of modified unicorn type.
\begin{prop}\label{prop:ricci_flatness}
    $F$ is Ricci-flat if and only if $f$ has the form $f(x^0) = c_1 \exp\left(c_2 x^0\right)$, with $c_1,c_2=$\,const.
\end{prop}
\begin{proof}
    By definition, and using homogeneity and the fact that $N^\mu_\nu = \tfrac{1}{2}\bar\partial_\nu G^\mu$, we have
    \begin{align}
    %
    \text{Ric} &= R^\mu{}_{\mu\nu}y^\nu = (\delta_\mu N^\mu_\nu-\delta_\nu N^\mu_\mu) y^\nu \\
    &=  y^\nu((\partial_\mu - N^\rho_\mu\bar\partial_\rho )N^\mu_\nu-(\partial_\nu -N^\rho_\nu\bar\partial_\rho)N^\mu_\mu)\\
    &=  y^\nu \partial_\mu N^\mu_\nu-  y^\nu\partial_\nu N^\mu_\mu   -  y^\nu N^\rho_\mu\bar\partial_\rho N^\mu_\nu  + y^\nu N^\rho_\nu\bar\partial_\rho N^\mu_\mu  \\
    &=  \tfrac{1}{2}\left(y^\nu\partial_\mu \bar\partial_\nu G^\mu-  y^\nu\partial_\nu \bar\partial_\mu G^\mu\right) \\
    &\qquad  -  \tfrac{1}{4}\left(y^\nu\bar\partial_\mu G^\rho\bar\partial_\rho\bar\partial_\nu G^\mu  - y^\nu \bar\partial_\nu G^\rho\bar\partial_\rho\bar\partial_\mu G^\mu\right)\\
    &=  \partial_\mu  G^\mu-  \tfrac{1}{2}y^\nu\partial_\nu \bar\partial_\mu G^\mu  -  \tfrac{1}{4}\left(\bar\partial_\mu G^\rho\bar\partial_\rho G^\mu  - 2 G^\rho\bar\partial_\rho\bar\partial_\mu G^\mu\right).
    \end{align}
    Using the identities
    \begin{align}
        \begin{array}{ll}
            \bar\partial_0 P = 2 f'/f, &\qquad \bar\partial_0 G^0 = 2y^0f'/f,   \\        
            \bar\partial_0^2  G^0 = 2 f'/f, &\qquad \bar\partial_0\bar\partial_i G^0 = \bar\partial_0\bar\partial_i P = \bar\partial_0^2 P = 0,   \\
            y^i\bar\partial_i G^0= -2|\phi| f'/f, &\qquad  y^i\bar\partial_i P= 2\, \text{sgn}(\phi y^0)\sqrt{|\phi|} f'/f,
        \end{array}
    \end{align}
    one finds after some slightly tedious manipulations that the last two terms in the expression for the Ricci tensor can both be expressed as
    \begin{align}
        \bar\partial_\mu G^\rho\bar\partial_\rho G^\mu  = 2 G^\rho\bar\partial_\rho\bar\partial_\mu G^\mu = n P^2,
    \end{align}
    where $n = \dim M = 4$ in our case. Hence these terms cancel each other out precisely. Denoting $G^0 = \bar G^0 f'/f$ and $P = \bar P f'/f$, so that $\bar G^0$ and $\bar P$ do not depend on $x^\mu$, and using the fact that $\bar\partial_\mu G^\mu = n P$, one finds furthermore that 
    \begin{align}
    \partial_\mu G^\mu &= (\partial_0^2\log |f|)\bar G^0, \\
     y^\nu\partial_\nu \bar\partial_\mu G^\mu &=n y^0(\partial_0^2\log |f|)\bar P.
    \end{align}
    Consequently, we have
    \begin{align}
        \text{Ric} &=  \partial_\mu  G^\mu-  \tfrac{1}{2}y^\nu\partial_\nu \bar\partial_\mu G^\mu \\
        &= (\partial_0^2\log |f|)\left((1-n)(y^0)^2 -n |y^0| \text{sgn}(\phi)\sqrt{|\phi|} - |\phi| \right), \label{eq:Ricci_check_sgn}
    \end{align}
    which in dimension $n=4$ reduces to
    \begin{align}
        \text{Ric} &= -(\partial_0^2\log |f|)\left(3(y^0)^2 +4 |y^0| \text{sgn}(\phi)\sqrt{|\phi|} + |\phi| \right). \label{eq:Ricci_check_sgn_2}
    \end{align}
    If Ric $=0$ then we must, in particular, have 
    \begin{align}
        0=\bar\partial_0^2\text{Ric} = -6(\partial_0^2\log |f|).
    \end{align}
    It thus follows that Ric $=0$ if and only if $\partial_0^2\log |f| =0$, the general solution to which is given by the stated form of $f$.
\end{proof}
This shows that the following family of Finsler metrics are exact vacuum solutions to Pfeifer and Wohlfarth's field equation in Finsler gravity:
\begin{subequations}\label{prop:solution2and3}
\begin{align}\label{prop:solution2}
    F = \left\{\begin{matrix}
    F_{0} & \text{if} & |y^0| + \text{sgn}(\phi)\sqrt{|\phi|} \neq 0\\
    0 & \text{if} &  |y^0| + \text{sgn}(\phi)\sqrt{|\phi|} = 0
    \end{matrix}\right. \,, 
\end{align}
where
\begin{align}\label{prop:solution3}
    F_0 \!= \!c_1e^{c_2 x^0}\!\left( |y^0|\! +\! \text{sgn}(\phi)\sqrt{|\phi|} \right)\! \exp\!\left(\!\frac{|y^0|}{ |y^0| \!+ \!\text{sgn}(\phi)\sqrt{|\phi|} }\!\right)\!.
\end{align}
\end{subequations}
Furthermore, the following proposition shows that \textit{any} solution of the type \eqref{eq:uni_spacetime} must have this form. In other words, all vacuum solutions of modified unicorn type are in fact Ricci-flat.
\begin{prop}\label{prop:solution}
    A modified unicorn metric \eqref{eq:uni_spacetime} is a solution to the Finslerian field equations in vacuum if and only if $f(x^0) = c_1e^{c_2 x^0}$, i.e. if and only if it can be written as \eqref{prop:solution2and3}.
\end{prop}
%
%
%
%
\begin{proof}
    In four spacetime dimensions, the proof is straightforward and most easily performed in convenient coordinates in which $\phi$ is diagonal with all nonvanishing entries equal to $+1$ or $-1$. Due to the size of some of the expressions involved, however, we only give a sketch of the proof here. From \eqref{eq:uni_spacetime} one can directly compute $g_{\mu\nu}$ and then its inverse $g^{\mu\nu}$. From \eqref{eq:Ricci_check_sgn_2} together with the definition of the Finsler-Ricci tensor \eqref{eq:Ricci_defs} one can immediately compute $R_{\mu\nu}$. We omit the intermediate expressions because they are somewhat lengthy, but plugging all of this into the (weakly) Landsberg field equation \eqref{eq:Pfeifer_Wohlfarth_eq_Landsberg_pre}, we obtain
    \begin{align}
        \frac{-\partial_0^2\log |f|}{3\sqrt{|\phi|}}\left(-4\,\text{sgn}(\phi)|y^0|^3 -5y^0 \sqrt{|\phi|} + |\phi|^{3/2} \right) = 0.
    \end{align}
    This equation can only be satisfied for all $y^\mu$ for which it is defined if $\partial_0^2\log |f|=0$, in which case \eqref{eq:Ricci_check_sgn_2} shows that $F$ is in fact Ricci-flat and therefore, by \Cref{prop:ricci_flatness}, must have the form \eqref{prop:solution2and3}, as desired.
\end{proof}

\section{A linearly expanding universe}
We now turn to the physical interpretation of one of the unicorn solutions obtained above. We will focus on the case where $\phi$ has $(-,-,-)$ signature. In that case, we will see that the solution can be interpreted as a linearly expanding (or contracting) universe.

We start by noticing that for any time slice $x^0=$ const., $y^0=0$, the Finsler metric describing the spatial geometry is a function only of the 3-dimensional Riemannian Finsler Lagrangian $\omega^2 = (y^1)^2+(y^2)^2+(y^3)^2$. Since $\omega^2$ is homogeneous and isotropic, the spatial Finsler metric must be as well, for each time slice. This shows that the Finsler spacetime metric has cosmological symmetry, i.e. is spatially homogeneous and isotropic. Alternatively, it has been shown \cite{Hohmann:2020mgs,Pfeifer:2011xi} that a Finsler metric has cosmological symmetry if and only if it can be written in the form $F = F(x^0,y^0,\omega)$, where 
\begin{align}
    \omega^2 = \frac{(y^r)^2}{1-kr^2} + r^2 \left( (y^\theta)^2 + \sin^2\theta (y^\phi)^2\right), \qquad k=0,\pm 1
\end{align}
is the standard spatial cosmological line element in coordinates $(r,\theta,\phi)$, represented as a scalar function on $TM$. With $k=0$ this corresponds to our $\omega^2$ above, and hence we conclude also in this way that indeed, provided $\phi$ has $(-,-,-)$ signature, the modified unicorn metric has cosmological symmetry.

Second, it turns out that the modified unicorn metric is conformally flat. More precisely, we can write the Finsler metric
\begin{align}\label{eq:Finsler_cosm_analog}
    F = f(x^0) \bar F(y),
\end{align}
as a conformal factor $f(x^0)$ (which is a function only of the timelike coordinate) times the `reference' Finsler metric
\begin{align}
    \bar F = \left(|y^0| - \sqrt{|\phi(\hat y)|}\right)e^{\frac{|y^0|}{|y^0| - \sqrt{|\phi(\hat y)|}}},
\end{align}
which is flat. Indeed, since $\bar F$ has no $x$-dependence, it follows from \eqref{eq:spray_directly_from_L} that the spray vanishes identically. In other words, the geodesics of $\bar F$ are straight lines in these coordinates, and \eqref{eq:nonlinear_connection_and_spray_relations} implies that the curvature of $\bar F$ vanishes, $\bar N^\rho_\mu=0$.

This situation is reminiscent of FLRW cosmology, which also features cosmological symmetry as well as conformal flatness. Indeed, in the simplest (spatially flat) case, the FLRW metric can be written as $\D s^2 = -\D t^2 + a(t)^2(\D x^2 + \D y^2 + \D z^2)$, where $a(t)$ is the scale factor. Introducing the conformal time coordinate $\zeta = \int^t\tfrac{1}{a(t)}\D t$ the metric takes the form
\begin{align}
    \D s^2 = a(\zeta)^2\left(-\D \zeta^2 +\D x^2 + \D y^2 + \D z^2\right).
\end{align}
The Finsler metric corresponding to this line-element reads
\begin{align}
    F_\text{FLRW}(x,y) &= a(\zeta)\sqrt{\left|-(y^\zeta)^2 + (y^x)^2 + (y^y)^2 + (y^z)^2\right|} \\
    &= a(\zeta) \bar F_\text{FLRW}(y),\label{eq:FRLW_analog}
\end{align}
where $\bar F_\text{FLRW}$ is the flat `reference' metric in the FLRW case. Comparing \eqref{eq:FRLW_analog} to \eqref{eq:Finsler_cosm_analog} suggests that in the Finslerian case, the coordinate $x^0$ might have the interpretation of a kind of conformal time, while $f(x^0)$ might play a role analogous to the scale factor $a(\zeta)$. Since the transformation $t\to \zeta$ is characterized by $\partial t/\partial \zeta = a$ and $\partial t/\partial x^i = 0$, it makes sense in this analogy to perform a similar (inverse) coordinate change $x^0\to \tilde t$ according to $\partial \tilde t/\partial x^0 = f$ and $\partial\tilde t / \partial x^i = 0$ in the Finsler metric. Here $\tilde t$ is a new coordinate that, if possible, we would like to interpret as \textit{cosmological time}, analogous to $t$ in the FLRW metric. In fact, we will omit the tilde and just write $\tilde t = t$ from here onwards. Then the relation between the coordinates implies that $y^t = f y^0$ and hence the modified unicorn metric attains the form
\begin{align}
     F  &=  \left( |y^t| - f(t) \sqrt{|\phi|} \right) e^{\frac{|y^t|}{ |y^t|  - f(t) \sqrt{|\phi|} }}.
\end{align}
where $f(t)$ is shorthand for $f(x^0(t))$. Indeed, it is now clear that:
\begin{itemize}
    \item $f(t)$ has the interpretation of a scale factor, for the spatial geometry at a time slice $t=t_0$ is given by\footnote{We have removed the irrelevant overall minus sign in order to render the spatial geometry positive definite.} 
    \begin{align}
        F_\text{sp} = F\big|_{y^t=0, t=t_0} = f(t_0)\sqrt{(y^1)^2+(y^2)^2+(y^3)^2},
    \end{align}
    which is just the Finsler metric of flat three-dimensional Euclidean space multiplied by $f(t_0)$; equivalently,
    \begin{align}
        \D s^2 = f(t_0)^2\left((\D x^1)^2+(\D x^2)^2+(\D x^3)^2\right).
    \end{align} 
    \item The coordinate $t$ has the interpretation of cosmological time since it corresponds (up to a constant\footnote{Along the worldline of a stationary observer, which we may parameterize as $x^\mu(t) = (t,x^i_0)$, $\dot x^\mu = (1,0,0,0)$, we have $F(x,\dot x) = |\dot x^t|e^{|\dot x^t|/|\dot x^t|} = e$, so that $\D\tau/\D t = e$. Hence, strictly speaking, $t=\tau/e$ is a constant multiple of the proper time.}) to the proper time of a stationary observer with $\dot x^i = 0$. Note that the latter describes a geodesic.
\end{itemize}
If we additionally require the metric to be a vacuum solution then we know from \Cref{prop:solution} that $f(x^0) = c_1e^{c_2 x^0}$. In that case it follows from $\partial t/\partial x^0 = f(x^0) = c_1e^{c_2 x^0}$ that
\begin{align}
    x^0(t) = \frac{1}{c_2} \ln\left(\frac{c_2}{c_1}(t-c_3)\right)\,,
\end{align}
where $c_3$ is another integration constant, and hence we find that the scale factor $f(t)$ as a function of cosmological time $t$ is given by
\begin{align}
    f(t)  \equiv f(x^0(t)) = c_2 (t - c_3)\,.
\end{align}
Thus the modified unicorn vacuum solutions describe a linearly expanding or contracting universe.

As an additional mathematical curiosity, it turns out that these solutions are not only Ricci-flat and conformally flat, as already noted, but flat, in the sense that the nonlinear curvature tensor $R^\rho{}_{\mu\nu} = \delta_\mu N^\rho_\nu - \delta_\nu N^\rho_\mu$ vanishes identically. This follows by explicit computation (which we omit here). Even though $R^\rho{}_{\mu\nu} = 0$, the geometry is nontrivial. This would be impossible in pseudo-Riemannian geometry. Indeed, in pseudo-Riemannian geometry a vanishing curvature tensor would imply that there exist coordinates in which geodesics are straight lines, or, equivalently, in which the spray vanishes identically, $G^\rho=0$. In the case at hand, though, we are dealing with a Finsler metric that is non-Berwaldian, meaning that, whatever coordinates we use, the spray $G^\rho$ will never be quadratic, and hence, in particular, it will never vanish identically. This proves the nonexistence of a coordinate system in which the geodesics of the unicorn metric are straight lines, showing that the geometry can justly be characterized as nontrivial, even though it has zero curvature.\\

\section{Concluding remarks}

The results obtained in this chapter motivate a systematic search for cosmological Landsberg spacetimes that solve the field equations. Indeed, while the Berwald property has been shown to be too restrictive to be able to model nontrivial cosmological dynamics \cite{Hohmann:2020mgs}, we have shown here that---going up one step in generality---Landsberg spaces do have this capacity. With the recent results characterizing cosmological symmetry in Finsler spacetimes \cite{Hohmann:2020mgs} and Elgendi's machinery for constructing unicorns using conformal transformations \cite{elgendi_2020, Elgendi2021a}, we have a lot of tools at our disposal to study such cosmological unicorns. This is work in progress.



%% file: Discussion.tex
\chapter*{Discussion and Outlook}\label{Discussion}
\addcontentsline{toc}{chapter}{Discussion and Outlook}

\fancyhead{}
\fancyhead[LE]{\textit{Discussion and Outlook}}
\fancyhead[RO]{\textit{Discussion and Outlook}}

The results obtained in this dissertation can be divided into two main (interrelated) categories: 
\begin{itemize}
    \item the characterization of Berwald spaces and their metrizability (\cref{part:Berwald_metrizability});
    \item vacuum solutions in Finsler gravity (\cref{part:vacuum_sols}).
\end{itemize}


Berwald spaces are relevant not only from an intrinsic mathematical point of view but also in the context of Finsler gravity. 
They represent the first step up in complexity relative to \psR manifolds, and they are relatively easy to handle, computationally. In \cref{ch:Characterization_of_Berwald} we introduced a new characterization of Berwald spaces and we explored some of its consequences. In particular, we obtained a general necessary and sufficient condition for \ab-metrics to be of Berwald type and we applied this Berwald condition to several specific \ab-metrics: Randers metrics (for which the result is well known), exponential \ab-metrics, $m$-Kropina metrics (for which the result is also known) and generalized $m$-Kropina metrics. 
While based mostly on our published work \citeref{ref1}{H1}, we have presented these results here with a substantially heightened focus on mathematical rigor relative to the original article. Apart from the applications discussed here, these results have also found application in the classification of cosmological Berwald spaces \cite{Hohmann:2020mgs,VoicuHabilitationThesis} and it seems worthwhile to investigate in the future whether there are more possible applications.


While all positive definite Berwald spaces defined on $\mathcal A = TM_0$ are pseudo-Riemann metrizable according to Szab\'o's well-known metrization theorem, it was demonstrated in \cref{ch:Metrizability} that this is not true any longer in the general setting of Berwald spaces of arbitrary signature. We provided a general argument for this along with a simple explicit counterexample, based on \citeref{ref3}{H2}. Next, we investigated the situation more closely in the case of generalized \mkrop spaces with closed null 1-form, and (standard) \mkrop spaces with arbitrary 1-form, and we obtained the precise necessary and sufficient conditions for local metrizability. These results are based in part on, but considerably generalize the findings of \citeref{ref2}{H3}. 


Interestingly, in all of the cases considered, we concluded that local metrizability is equivalent to the symmetry of the Finsler-Ricci tensor. One of these implications---that a locally metrizable affine connection requires a symmetric Ricci tensor---is trivial, but the opposite implication---that a symmetric Finsler-Ricci tensor implies local metrizability---is quite wonderfully surprising. Of course, this equivalence could be purely incidental and specific to the types of metrics considered here, but one is naturally led to wonder whether such a result might hold more generally. It is thus very much of interest to study the metrizability of other types of Berwald spaces and see whether the equivalence holds up. The ultimate goal would be to prove in complete generality that a Berwald space is locally metrizable if and only if it has a symmetric Finsler-Ricci tensor, or find a counterexample to this conjecture.

\cref{ch:Metrizability} concluded with the local classification of locally metrizable, Ricci-flat \mkrop spaces in $1+3$D whose 1-forms have constant causal character. These findings have not been published yet and will appear in a forthcoming article. Two cases can be distinguished: either the 1-form is null or it is not. In the first case, we found that locally, only a single family of such spaces exists, namely with $\alpha$ given by a classical pp-wave metric, and $\beta$ its defining covariantly constant null 1-form. In the latter case, we found that for each (admissible) value of $m$ there is locally, a unique (necessarily locally metrizable) Ricci-flat \mkrop space of Berwald type, namely with $\alpha$ given by the flat Minkowski metric and $\beta$ having constant components in `flat' coordinates. We have restricted to the scenario that $\alpha$ is of Lorentzian signature, but analogous results can be proven easily in other signatures.  

Since Ricci-flatness is a sufficient (and in some cases even necessary) condition for a Berwald space to be a vacuum solution to the Finsler gravity field equations, these results have direct consequences for the classification of vacuum solutions of \mkrop type. Specifically, apart from the above-mentioned `trivial' solutions, a Berwald \mkrop space can only be a vacuum solution if one of the following two conditions is met:
\begin{enumerate}[(i)]
    \item the 1-form is null and $F$ is not locally metrizable;
    \item the 1-form is not null and $F$ is not Ricci flat.
    \item the causal character of $b$ is not constant.
\end{enumerate}
All that is currently known about such scenarios is that some examples of Ricci-flat spaces (and hence vacuum solutions) satisfying (i) exist; whether spaces satisfying (ii) or (iii) exist is unknown. To complete the classification of \mkrop vacuum solutions of Berwald type, a thorough investigation of each of the cases (i)-(iii) is required.\\

And that brings us to \cref{part:vacuum_sols}, in which vacuum solutions to Pfeifer and Wohlfarth's field equation in Finsler gravity were the central topic. We introduced several classes of exact vacuum solutions, classified them, and discussed their physical interpretation.

To begin with, we proved in \cref{ch:Berwald_solutions} that any \ab-metric constructed from a vacuum solution to Einstein's equation $\alpha$ and a covariantly constant 1-form $\beta$ is a Ricci-flat vacuum solution in Finsler gravity, allowing us to deform solutions in general relativity (GR) to solutions in Finsler gravity. We then classified such solutions and showed that all nontrivial ones are generalizations of the well-known general relativistic pp-wave solutions. For Randers metrics as well as modified Randers metrics---a slightly different version of the Randers metric that we introduced and studied in depth because of its preferable causal properties---we even demonstrated that all vacuum solutions of Berwald type must be of this form, classifying (locally) all Berwald-Randers vacuum solutions. These results are based on \citeref{ref4}{H4} and \citeref{ref5}{H5}.

While such classification results are intrinsically valuable from a mathematical point of view, it is essential from a physics perspective to understand what kind of physics such solutions represent. That is why, in \cref{ch:lin}, we investigated the observational signature of our pp-wave type Finsler spacetimes, based on \citeref{ref5}{H5}. After a physically motivated twofold linearization procedure---linearizing both in the departure from flatness as well as in the departure from GR---we found that these solutions can be interpreted as Finslerian gravitational waves and we were able to obtain an explicit expression for the radar distance, the main observable in gravitational wave interferometry. The result was quite surprising: if such a Finslerian gravitational wave were to pass the Earth, its effect on an interferometer would be indistinguishable from a similar general relativistic gravitational wave!

This is on the one hand somewhat disappointing, as it suggests that interferometer measurements may not have the power to reveal possible Finslerian characteristics of our universe. On the other hand, it also means that current gravitational wave measurements are all compatible with the idea that spacetime has a Finslerian nature. Moreover, while research in Finsler gravity is generally quite abstract, our radar distance result---which may be viewed not in the least as a proof of concept---shows how clear physical predictions can be extracted from an \textit{a priori} abstract Finslerian model of spacetime. Repeating such an analysis in other scenarios (e.g. more generic types of Finslerian gravitational waves, higher-order analyses, or completely different Finsler spacetimes) may lead to additional insight into the observational signature of Finsler gravity.

Another highly interesting direction for future research in this context would be to investigate whether Finslerian effects also play a role in the generation of gravitational waves during, say, a black hole merger event. It is to be expected that such effects would lead to a Finslerian correction to the waveform and this \textit{could}, in principle, be measured in interferometer experiments,  especially since our results show how the waveform correlates with the experimental results---namely in exactly the same way in which the waveform of a general relativistic gravitational wave does. To investigate this, Finslerian black hole solutions first need to be better understood, and research in this direction is actively ongoing, in part by the author. All 4-dimensional spherically symmetric Finsler metrics of Berwald type have recently been classified \cite{cheraghchi2022fourdimensional} and, while a Birkhoff-type theorem has been proven for Ricci-flat spherically symmetric Berwald spacetimes \cite{voicu2023birkhoff}, it will be demonstrated in a forthcoming article that non--Ricci-flat spherically symmetric vacuum solutions more general than the Schwarzschild geometry do exist, suggesting that the gravitational field around a black hole might indeed feature Finslerian effects.

While all solutions discussed up to this point have been of Berwald type, we introduced and studied a (non-Berwaldian Landsberg) unicorn solution to the vacuum field equation in \cref{ch:unicorn_cosm}, which is based on \citeref{ref6}{H6}. Its mathematical interest lies in the fact that this is the first and only non-Berwaldian solution to Pfeifer and Wohlfarth's equation currently known in the literature and that it is of unicorn type, but in addition to that, we have seen that it can be interpreted very naturally in a cosmological setting as well. Indeed, we have shown that the solution has cosmological symmetry, i.e. is spatially homogeneous and isotropic, that it has a well-behaved light cone structure, and that its (flat) spatial slices evolve in cosmological time by means of a scale factor---reminiscent of the Friedmann-Lema\^itre-Robertson-Walker (FLRW) metric---which has a linear dependence on cosmological time. This leads to the natural interpretation of our unicorn solution as describing a linearly expanding (or contracting) Finslerian universe.

As an additional curiosity, we found that the requirement of a physically viable light cone structure does not, strictly speaking, necessitate Lorentzian signature, as is often assumed. This is illustrated by the signature anomaly of our above-mentioned cosmological unicorn solution, which indeed has positive definite signature everywhere on its domain, and yet has a light cone structure that is equivalent to that of flat Minkowski space. This surprising observation raises the question of how strictly one should adhere to the standard requirement of Lorentzian signature in the context of Finsler spacetime geometry. Of course, there are other reasons why Lorentzian signature might still be a desirable property---extremization properties of geodesics, existence, and uniqueness (in the right context) of solutions to the field equations, the propagation speed of the gravitational interaction, and more---but nevertheless it is worth thinking about this question in depth.


Our results also motivate a systematic search for cosmological Landsberg spacetimes that solve the field equations. Since (properly Finslerian) cosmological solutions of Berwald type are necessarily static \cite{Hohmann:2020mgs} and hence not particularly interesting, any nontrivial such Landsberg spacetime must necessarily be a unicorn. With the recent results characterizing cosmological symmetry in Finsler spacetimes \cite{Hohmann:2020mgs}, Elgendi's machinery for constructing unicorns using conformal transformations \cite{elgendi_2020, Elgendi2021a}, and the discovery of the first unicorn solution discussed here, we are very optimistic about this endeavor. This is work in progress.


Another important next step in the study of Finsler gravity, not only in the context of cosmology but in general, is the investigation of solutions to the field equation in the presence of matter. Promising ideas about the Finslerian description of matter have arisen recently. For instance, taking into account the individual motion of each gas particle in a kinetic gas, one obtains an energy-momentum tensor that is direction-dependent \cite{Hohmann:2019sni,Hohmann:2020yia}. Such an energy-momentum tensor cannot be accommodated in GR for obvious reasons, and indeed, in the usual Einstein-Vlasov treatment \cite{Andreasson:2011ng,10.1063/1.4817035} the individual motion of the gas particles is averaged over. Finsler gravity, however, is perfectly well capable of incorporating this direction-dependence and hence it could very well be that Finsler gravity is able to provide a more accurate description of the gravitational field of a kinetic gas. Currently, matter solutions in Finsler gravity are still completely unexplored, but with the before-mentioned ideas in mind, it seems a very promising avenue.\\

Finally, let us zoom out and return to the broader perspective. Fundamental physics stands before an enormous challenge: the unification of gravity with quantum mechanics. 
Ultimately, the research covered in this dissertation stands in service of this goal. 
Indeed, the possibility, suggested by quantum gravity research, that local Lorentz invariance---which lies at the heart of GR---may not be fundamental leads directly to the notion of Finsler spacetime geometry, since the latter beautifully accommodates departures from Lorentz invariance already at the classical level. This has been the primary motivation for our work. Of course, the topics covered here 
represent only a tiny fraction of the larger picture. Yet, we are hopeful that our results, though modest in scope, will prove instrumental in the understanding of gravity and that, 
eventually, they may even shine light on the reconciliation of gravity with quantum mechanics. With this grand prospect, we have come to the end of this dissertation.

\newpage


%% file: Appendix.tex
\renewcommand{\thechapter}{A}
\appendix

\chapter{Special Pseudo-Riemannian Metrics}

In this appendix we discuss several special types of pseudo-Riemannian metrics and their standard expression in coordinates, starting with metrics that admit a closed null 1-form. But first, a remark about the notation is in order. In several places below, $b$ will \textit{a priori} denote a 1-form. But we will occasionally use the same symbol $b$ to denote its vector equivalent via the metric-induced isomorphism. For example, $b = \D u$ means that $b_i = \partial_i u$ and $b = \partial_v$ means that $b^i=\partial x^i/\partial v$. No confusion should be possible as to what is meant.

\section{Metrics admitting a closed null 1-form}\label{sec:Kundt_sort_of}

\begin{prop}\label{prop:metric_w_closed_null_1form}
Any \psR metric admitting a closed null 1-form $b$ is locally (whenever $b\neq 0$) given by
\begin{align}
    \D s^2 = -2\D u\D v + H\D u^2 + 2 W_b\D u\D x^b + h_{bc}\D x^b\D x^c,
\end{align}
in coordinates $(u,v,x^a)$, $a=3,\dots,n$, with $b = \D u$, where $H,W_a$ are smooth functions, and where, for fixed $u$ and $v$, $h_{ab}$ is a \mbox{pseudo-Riemannian} metric of dimension $n-2$.
\end{prop}
\begin{proof}
First, we may pick coordinates $(v,x^2,\dots,x^n)$ around $p$ adapted to $b$ in the sense that $b =\partial_v$, i.e. $b^i = \delta^i_1$. At this point, the metric has the general form $a = a_{ij}\D x^i\otimes\D x^j$. 
The null character of $b$ manifests as the fact that $a_{11}=a_{vv}=0$ in these coordinates. Because $b$ is closed and hence locally exact, we may write, locally, $b= \D u$ for some function $u(v,x^2,\dots,x^n)$.  Equivalently, $b_i = \partial_i u$. Note also that $\partial_i u = b_i = a_{ij}b^j = a_{ij}\delta^j_1 = a_{i 1}$. Since $a_{11}=0$, it follows that $\partial_v u=\partial_1 u=0$. As $b\neq 0$ by assumption, there must be some $i\geq 2$ such that $\partial_i u = a_{i1}\neq 0$ in a neighborhood of $p$. Order the coordinates $x^2,\dots,x^n$ such that this is true for $i=2$, i.e. assume without loss of generality that $a_{21}\neq 0$. Next, define the map
\begin{align}
x = (v,x^2,\dots,x^n)\mapsto \tilde x = (v,u(x^2,\dots,x^n),x^3,\dots,x^n).
\end{align}
Its Jacobian matrix and its inverse are given by 
\begin{align}
J^i{}_j &= \frac{\partial\tilde x^i}{\partial x^j} = \begin{pmatrix}
1 & 0 &  0&\hdots & 0 \\
0 & a_{21} & a_{31} & \hdots & a_{n1} \\
0 & 0 & 1 &  \hdots & 0 \\
\vdots & \vdots &  \vdots & \ddots & \vdots \\
0 & 0 & 0 & \hdots & 1
\end{pmatrix}, \\
(J^{-1})^i{}_j &= \frac{\partial x^i}{\partial \tilde x^j} = \begin{pmatrix}
1 & 0 & 0 & \hdots & 0 \\
0 & 1/a_{21} & -a_{31}/a_{21} &  \hdots & a_{n1}/a_{21} \\
0 & 0 & 1 &  \hdots & 0 \\
\vdots & \vdots & \vdots & \ddots & \vdots \\
0 & 0 & 0 &  \hdots & 1
\end{pmatrix},
\end{align}
and since $\det J = a_{21}\neq 0$ this matrix is invertible, so $x\mapsto \tilde x$ is a local diffeomorphism at $p$. It remains to find the form of the metric in the new coordinates. We have
\begin{align}
\tilde a_{ij} = \frac{\partial x^k}{\partial \tilde x^i} \frac{\partial x^\ell}{\partial \tilde x^j}a_{k\ell},\qquad \text{i.e.} \qquad \tilde a = J^{-1T}aJ^{-1}.
\end{align}
Therefore we have
\begin{align}
\tilde a_{11} &= (J^{-1}{}^T)_1{}^i a_{ij}(J^{-1}{})^j{}_1 = a_{11} = 0,\\
\tilde a_{12} &= (J^{-1}{}^T)_1{}^i a_{ij}(J^{-1}{})^j{}_2 = 1, \\
\tilde a_{1b} &= (J^{-1}{}^T)_1{}^i a_{ij}(J^{-1}{})^j{}_b = a_{12}(-a_{b1}/a_{21}) + a_{1b} = 0,\quad b = 3,\dots, n.
\end{align}
This shows that $a = \tilde a_{ij}\D \tilde x^i\D \tilde x^j = 2\D u\D v + H\D u^2 + 2 W_b\D u\D x^b + h_{bc}\D x^b\D x^c$ for certain functions $H$, $W_a$, $h_{ab}$, and hence after a redefinition $v \to -v$ we may write the metric in the form
\begin{align}
a = -2\D u\D v + H\D u^2 + 2W_b\D u\D x^b + h_{bc}\D x^b\D x^c.
\end{align}
It follows from the easily checked fact that $\det h = -\det a \neq 0$ that $h_{ab}$ is itself a \mbox{pseudo-Riemannian} metric of dimension $n-2$.
\end{proof}
In matrix form, such a metric reads, schematically, 
\begin{align}
    g = \begin{pmatrix}
        H   & -1 & W_a \\
        -1  & 0  & 0 \\
        W_a & 0  & h_{ab}
    \end{pmatrix}
\end{align}
with inverse is given by
\begin{align}
    g^{-1} &= \begin{pmatrix}
        0   & -1 & 0\\
        -1  & -H +  h^{ab}W_aW_b  & h^{ab}W_b  \\
        0 & h^{ab}W_b   & h^{ab}
    \end{pmatrix} \\
    &=  \begin{pmatrix}
        0   & -1 & 0\\
        -1  & -H +  W^2  & W^a  \\
        0 & W^a  & h^{ab}
    \end{pmatrix},
\end{align}
where $h^{ab}$ is the inverse of $h_{ab}$, indices have been raised with $h$, and $W^2 = h^{ab}W_aW_b$. We note that the Christoffel symbols with upper index $u$ have a particularly simple form, 
\begin{align}\label{eq:christu}
    \Gamma^u_{ij} &= \tfrac{1}{2}\partial_v g_{ij}.
\end{align}

\section{Metrics admitting a parallel vector field}
\label{app:metrics_w_parallel_vector_field}

Next, we discuss \psR metrics that admit a parallel (i.e. covariantly constant) vector field. Note that the existence of a parallel vector field is equivalent to the existence of a parallel 1-form of the same causal character. We distinguish the two cases that the 1-form is null, or that the 1-form is not null, starting with the former. In this case, the resulting metric is called a CCNV\footnote{CCNV stands for \textit{covariantly constant null vector}.} metric or a pp-wave metric.  
\begin{prop}\label{prop:coords_cc_null_1form}
Any \psR metric admitting a parallel null 1-form $b_i$ is locally (whenever $b\neq 0$) given by
\begin{align}\label{eq:CCNV_metric}
    \D s^2 = -2\D u\D v + H(u,x^a)\D u^2 + 2W_b(u,x^a)\D u\D x^b + h_{bc}(u,x^a)\D x^b\D x^c,
\end{align}
in coordinates $(u,v,x^a)$, $a=3,\dots,n$, with $b = \D u$, and where, for each fixed $u$, $h_{bc}$ is a \psR metric of dimension $n-2$. Conversely, $\D u$ is null and parallel with respect to any metric of the form \eqref{eq:CCNV_metric}.
\end{prop}
\begin{proof}
    It is easy to see that a 1-form is parallel if and only if it is closed and satisfies the Killing equation. Hence $b_i$ is in particular closed and it follows from \Cref{prop:metric_w_closed_null_1form} that locally we may write
    \begin{align}
        \D s^2 = -2\D u\D v + H\D u^2 + 2W_b\D u\D x^b + h_{bc}\D x^b\D x^c,  \qquad b = \D u.
    \end{align}
    Computing $\nabla_i b_j$ explicitly in these coordinates, using the fact that $b_i = \delta^u_i$ and $g^{ui} = -\delta^i_v$ and $g_{iv} = 0$, yields
\begin{align}\label{2}
\nabla_ib_j  = -\frac{1}{2}\frac{\partial a_{ij}}{\partial v}.
\end{align}
Since $b_i$ is parallel, all these components must vanish, and consequently all metric functions must be independent of $v$. The converse statement that $\D u$ is always null and parallel is straightforward to check explicitly.
\end{proof}
If, on the other hand, the 1-form is not null, the metric decomposes locally as a direct product of a flat 1-dimensional metric and an arbitrary $(n-1)$-dimensional metric. In order to prove this we employ the Frobenius theorem.

Recall that a distribution is called involutive if it is closed under the Lie bracket, and it is called completely integrable (we follow the terminology of e.g. \cite{lee2012smoothmanifolds}) if it is spanned by a set of coordinate vector fields in some cubic chart. The Frobenius theorem states that any involutive distribution is completely integrable. We will take this result for granted.
%
%
%
%
%
%

The following is a consequence of a more general result (see e.g. Appendix C of  \cite{straumann2012generalrelativity}), but here we give a self-contained proof directly from the Frobenius theorem. To ease the notation, coordinate labels will momentarily run from $0$ to $n-1$ rather than $1$ to $n$, even though the signature is not necessarily Lorentzian but in principle arbitrary. Greek indices $\mu,\nu,\rho,\dots$ run from $0$ to $n-1$, whereas Latin indices $i,j,k,\dots$ run from $1$ to $n-1$.
\begin{prop}\label{prop:metric_w_non_null_parallel_vector}
    Any \psR metric admitting a nonnull (i.e. $|b|^2\neq 0$) parallel vector field $b^\mu$ can locally be written as a product metric
    \begin{align}
        \D s^2 = \pm(\D x^0)^2 + h_{ij}(x^k)\D x^i\D x^j.
    \end{align}
    where $b = b_0\,\D  x^0$ with $b_0=$ constant, and where $h_{ij}$ is an $(n-1)$-dimensional pseudo-Riemannian metric.
\end{prop}
\begin{proof}  
    We apply the Frobenius theorem twice. First to the distribution spanned by $b^\mu$, and then to the distribution orthogonal to $b^\mu$. Since any vector commutes with itself, it follows immediately that span$\{b\}$ is involutive and hence completely integrable, so there exist coordinates $y^\mu$ such that $b = \partial/\partial y^0$. We keep this coordinate system in mind.
    
    Next, consider the distribution $D$ consisting of all vectors orthogonal to $b^\mu$. Note that since $|b|^2\neq 0$, by assumption, the orthogonal complement $b^\perp$ of $b$ in $T_xM$ is an $(n-1)$-dimensional linear subspace, and symbolically $T_xM = \mathrm{span}\{b\}\oplus b^\perp$. Hence $D$ is an $(n-1)$-dimensional distribution. Now let $v,w\in D$. Then, by torsion-freeness, we have $[v,w] = \nabla_v w - \nabla_w v$, and by metric-compatibility, $\nabla g = 0$, we have $g(\nabla_v w, b) = -g(w, \nabla_v b)$ and $g(\nabla_w v, b) = -g(v, \nabla_w b)$. Hence, since $b$ is parallel we see that
   \begin{align}
       g([v,w], b) &= g(\nabla_v w, b) - g(\nabla_w v,b) \\
       &= -g(w, \nabla_v b) + g(v, \nabla_w b) \\
       &= 0.
   \end{align}
   This shows that $[v,w]\in D$ and hence $D$ is involutive. By the Forbenius theorem, the $(n-1)$-dimensional distribution $D$ is therefore completely integrable, that is, there exist local coordinates $z^0,\dots, z^{n-1}$ such that $D$ is spanned by $\partial/\partial z^1,\dots, \partial/\partial z^{n-1}$. 

   Now we combine the two coordinate systems $y^\mu$ and $z^\mu$. Namely, we define new coordinates $x^\mu$ by setting $x^0 = z^0$ and $x^i = y^i$ for $i=1,\dots,n-1$. This is a valid coordinate transformation as long as the $\D x^\mu$ are linearly independent. To see that the latter is the case, suppose that at some point in $M$ we have $c_\mu\D x^\mu =0$, i.e.
   \begin{align}
       c_0\D z^0 = -c_i\D y^i.
   \end{align}
   Then, by construction, we see that the LHS annihilates $D = b^\perp$, whereas the RHS annihilates $b$. Hence the vector corresponding to this 1-form would be contained in both span$(b)$ and $b^\perp$. But, since $|b|^2\neq 0$, we have span$(b)\cap b^\perp = \{0\}$, so it follows that we must have $c_0=c_i=0$. Hence $x^\mu$ are valid coordinates on $M$.

   By construction, $b$ is a multiple of $\partial/\partial x^0$. This is because $b = \partial/\partial y^0$, meaning that $\D y^i(b) =0$ for each $i$. Therefore $\D x^i(b) = \D y^i(b) = 0$.
   
   Next, we will prove that in the $x^\mu$ coordinates, $g_{0i}=0$. First notice that if a 1-form $\omega$ annihilates the orthogonal complement of $b$, i.e. if $\omega(v)=0$ for all $v$ satisfying $g_{\mu\nu}v^\mu b^\nu=0$, then we must have $\omega = \omega_0\D z^0$, which is clear from the form of $D$ in $z^\mu$-coordinates. But we have $\D x^0 = \D z^0$ and hence it follows as well that if a 1-form $\omega$ annihilates the orthogonal complement of $b$ then $\omega = \omega_0\D x^0$, or equivalently $\omega_i=0$. We know, however, that $\omega_\mu = g_{\mu\nu}b^\nu$ trivially annihilates the orthogonal complement of $b$. This therefore implies that $\omega_i = g_{i\nu}b^\nu=0$. Now recall that $b^\mu = b^0\partial_0$. It follows that $\omega_i = g_{i0}=0$, as claimed.

    Since the cross terms in the metric vanish and $b$ is a multiple of $\partial/\partial x^0$, we also have $b = b_0\D x^0$. It then follows from the fact that $b$ is closed that $b_0 = b_0(x^0)$.
    
   So far what we have shown is that the metric and 1-form have the form
   \begin{align}
       \D s^2 = g_{00}(x^\mu) (\D x^0)^2 + g_{ij}(x^\mu)\D x^i\D x^j,\qquad b = b_0(x^0)\D x^0
   \end{align}
   Next, we explicitly evaluate the equation $\nabla_\mu b_\nu=0$ in the new coordinates. The $ij$-component shows that $\nabla_i b_j =-b_0\Gamma ^0_{ij} = b_0 g^{00}\partial_0g_{ij} = 0$, hence $\partial_0 g_{ij}=0$. And the $0i$-component shows that $\nabla_0 b_i = -b_0\Gamma^0_{0i} = -b_0g^{00} \partial_ig_{00} =0$, hence  $\partial_ig_{00}=0$. From the latter it follows that we can do a final coordinate transformation $x^0\mapsto \bar x^0$ such that $\D\bar x^0 = \sqrt{|g_{00}(x^0)|} \D x^0$. Then, after dropping the bar over $x^0$, the metric attains the form 
    \begin{align}
       \D s^2 = \pm (\D x^0)^2 + g_{ij}(x^k)\D x^i\D x^j.
   \end{align}
   Then $\Gamma^0_{00}=0$ and hence $\nabla_0 b_0=\partial_0 b_0=0$ implies that $b_0 = $ const. Finally, since $\det g_{\mu\nu} = \pm \det g_{ij}$, the $(n-1)$-dimensional metric $g_{ij}$ is nondegenerate and hence pseudo-Riemannian. This completes the proof.
\end{proof}
\begin{cor}\label{cor:EFE_sol_with_parallel_field}
    In 4 spacetime dimensions, the only vacuum solution to Einstein's field equations admitting a nonnull parallel vector field or 1-form is (locally) flat Minkowski spacetime.
\end{cor}
\begin{proof}
    Let $g_{\mu\nu}$ be such a metric. By \Cref{prop:metric_w_non_null_parallel_vector}, it can be written locally as 
    \begin{align}
        \D s^2 = \pm(\D x^0)^2 + h_{ij}(x^k)\D x^i\D x^j.
    \end{align}
    It is a simple exercise to show that the Ricci curvature decomposes as 
    \begin{align}
        R_{00} = R_{0i}=0,\qquad R_{ij} = \tilde R_{ij},
    \end{align}
    where $\tilde R_{ij}$ is the Ricci tensor of $h_{ij}$.  If $g_{\mu\nu}$ is a vacuum to Einstein's field equations, $R_{\mu\nu}=0$, it therefore follows that $\tilde R_{ij}=0$ and $h_{ij}$ is Ricci-flat. However, in three dimensions it is a standard result that any Ricci-flat pseudo-Riemannian metric is flat, which follows from the Ricci-decomposition of the Riemann tensor because the Weyl tensor vanishes identically. Hence $h_{ij}$ is flat. As the direct product of two flat pseudo-Riemannian manifolds, $g_{\mu\nu}$ is itself flat.
\end{proof}
See also \cite[\S 35.1.1]{stephani_kramer_maccallum_hoenselaers_herlt_2003} for an alternative argument and related results.

\section{Some properties of the pp-wave metric}

Consider the pp-wave metric \eqref{eq:CCNV_metric} in 4 spacetime dimensions,
\begin{align}
    \D s^2 = -2\D u\D v + H(u,x^a)\D u^2 + 2W_b(u,x^a)\D u\D x^b + h_{bc}(u,x^a)\D x^b\D x^c,
\end{align}
$a,b,c,\dots = 1,2$. Under the assumption that this is a vacuum solution to Einstein's field equations, i.e. that the metric is Ricci-flat, it follows from the results in \cite{ehlers1962exact} (see also section 24.5 in \cite{stephani_kramer_maccallum_hoenselaers_herlt_2003}) that the functions $W_a$ can be transformed away and $h_{ab}$ can be transformed into $\delta_{ab}$ by means of an appropriate coordinate transformation\footnote{In \cite{ehlers1962exact}, or rather its English translation \cite{Jordan2009_repub}, the term \textit{normal hyperbolic} is used, which is meant to indicate that the metric is \textit{of Lorentzian signature}.}. Such a vacuum pp-wave metric can thus be written in standard form
\begin{align}\label{eq:pp_waves_final_form}
    \D s^2 = -2\D u\D v + H(u,x,y)\D u^2 + \D x^2 + \D y^2.
\end{align}
Then the only nonvanishing component of the Ricci tensor is $R_{uu} = -\tfrac{1}{2}\Delta H$, where $\Delta = \partial_x^2 + \partial_y^2$, which yields the following.
\begin{prop}\label{eq:metric_cc_null_1form_vac_sol}
    A \psR metric in $(1+3)$D admitting a parallel null 1-form is Ricci-flat, i.e. is a vacuum solution to Einstein's field equation, if and only if it can be written locally as 
    \begin{align}\label{eq:pp_waves_final_form2}
    \D s^2 = -2\D u\D v + H(u,x,y)\D u^2 + \D x^2 + \D y^2,\qquad \Delta H=0.
\end{align}
\end{prop}

\chapter{A Basic Linearity Result}

The aim of this appendix is to prove a basic result, \Cref{prop:lemapp}, concerning linearity and constancy of certain functions on $\R^n$, used in \Cref{theor:generalized_mKrop} and \Cref{theor:generalized_mKrop2}. In what follows, $f$ and $g$ will be real-valued functions an open subset of $\R^n$. Moreover, to prevent confusion, we stress that indices $i,j,k,\dots$ in expressions like $x^i$ will always be indices, not powers, i.e. $x^2$ denotes the second component of the vector $x\in\R^n$.
\begin{lem}\label{lem:lem1}
If the function $x\mapsto f(x)x^i$ is affine for each 
$i$, i.e. $f(x)x^i =c^i_kx^k+d^i$, on an open subset $U\subset \R^n$ with $n\geq 2$ such that $0\notin U$, then $f(x)$ is constant on $U$.
\end{lem}
\begin{proof}
Multiplying $f(x)x^i =c^i_kx^k+d^i$ with $x^j$ and $f(x)x^j =c^j_kx^k+d^j$ with $x^i$ shows that for all $i,j$ we must have
\begin{align}
    c^i_kx^kx^j +d^ix^j = c^j_kx^kx^i +d^jx^i,
\end{align}
which implies that 
\begin{align}
    d^ix^j=d^jx^i,\qquad c^i_kx^kx^j = c^j_kx^kx^i.
\end{align}
The first equation implies (e.g. by differentiation) that $d^i = 0$ for each $i$. The second equation can be written as 
\begin{align}
    c^i_k\delta^j_l x^kx^l =  c^j_k\delta^i_l x^kx^l,
\end{align}
and hence, by differentiating twice, we obtain that
\begin{align}
    c^i_k\delta^j_l + c^i_l\delta^j_k = c^j_k\delta^i_l + c^j_l\delta^i_k.
\end{align}
Taking any $i\neq j=k=l$ (this is where the assumption $n\geq 2$ is needed) this reduces to $2c^i_k=0$. Hence the matrix of coefficients $c^i_j$ must be diagonal. Next, taking any $i=k\neq l=j$, the equation reduces to $c^i_i = c^j_j$ (no summation). Hence $c^1_1=\dots = c^n_n \eqqcolon c$. Substituting this in the original assumption that $f(x)x^i =c^i_kx^k+d^i$ yields $f(x)x^i =c x^i$ and hence, since $x\neq 0$ on $U$, this implies that $f(x) = c$ is constant.
\end{proof}
%
%
%
\begin{lem}\label{prop:lemapp}
If $g$ is differentiable on an open subset $U\subset \R^n$ with $n\geq 3$ 
and if the function $x\mapsto f(x)b^i + g(x)x^i$ is linear for each $i$, i.e. $f(x)b^i + g(x)x^i = c^i_j x^j$, then $f(x)$ is linear and $g(x)$ is constant.
\end{lem}
\begin{proof}
Consider $b^i$ and $x^i$ as the components of two vectors $\vec b$ and $\vec x$ in $\R^n$ and pick a basis such that $\vec b = (1,0\dots,0)$ in that basis. The assumption that $f(x)b^i + g(x)x^i$ is linear then splits into two, namely: 
\begin{enumerate}[(a)]
    \item $f(x) + g(x)x^1$ is linear;
    \item $g(x)x^i$ is linear for all $i>1$.
\end{enumerate}
Writing $x = (x^1,\bar x)$, condition (b) implies that, for any $x^1$ and any $i=1,\dots,n-1$, the map 
\begin{align}
    \R^{n-1}\ni\bar x\mapsto g(x^1,\bar x)\bar x^i &= g(x)x^{i+1} = c^{i+1}_j x^j = c^{i+1}_1 x^1 + \sum_{j=1}^{n-1} c^{i+1}_{j+1} \bar x^{j}
\end{align}
is affine on the open set $U_{x^1} = \{\bar x\in\R^{n-1}\,:(x^1,\bar x_0)\in U\,\}$. 
 Hence, by \Cref{lem:lem1}, this map is constant on $U_{x^1}\setminus\{0\}$, but this implies, by continuity, that the map is constant on $U_{x^1}$. This means that $g(x)$ can only depend on $x^1$, i.e. we may write $g(x) = h(x^1)$. Now consider the specific case of $i=2$. According to (b) and the result just obtained, $h(x^1)x^2$ is linear, that is, we can write $h(x^1)x^2= c^2_j x^j$. Differentiation with respect to $x^2$ then shows that $h(x^1) = c^2_2= $ constant, and hence $g(x)=$ constant. Plugging this into (a) immediately implies that $f(x)$ is linear, completing the proof. 
\end{proof}
%
%

\chapter{Determinant of an Indefinite \texorpdfstring{$(\alpha,\beta)$}{(\textalpha,\textbeta)}-Metric}
\label{sec:ab_determinant}

\fancyhead[LE]{\small\textit{\nouppercase{\leftmark}}}
\fancyhead[RO]{\small\textit{\nouppercase{\leftmark}}}

Here we derive the formula (see \eqref{eq:det_ab} below) for the determinant of a not necessarily positive definite \ab-metric. This formula, which appeared originally (with a typo) in \citeref{ref5}{H5}, is a generalization of a well-known result for positive definite \ab-metrics, and the proof is not substantially different from the positive definite analog. We consider Finsler metrics of the form $F=\alpha\phi(s)$, where $s = \beta/\alpha$, $\alpha = \sqrt{|A|}=\sqrt{|a_{ij}y^i y^j|}$, $A = a_{ij}y^i y^j =  \text{sgn}(A)\alpha^2$, and where $a_{ij}$ is assumed to be a pseudo-Riemannian metric, i.e. not necessarily Riemannian/positive definite.

In complete analogy with the positive definite case, it can be shown by direct calculation that the fundamental tensor $g_{ij}\equiv \tfrac{1}{2}\bar\partial_i\bar\partial_j F^2$ is given by
\begin{align}\label{eq:alpha_beta_metric_fundamental_tensor}
g_{ij} = \sgn(A)\rho a_{ij} + \rho_0 b_i b_j + \rho_1(b_i\alpha_j + \alpha_i b_j) + \rho_2\alpha_i\alpha_j,
\end{align}
where we have defined $\alpha_i= a_{ij}y^j/\alpha$, and with coefficients given by
\begin{align}
\rho &= \phi(\phi-s\phi'),\\
\rho_0&= \phi \phi'' + \phi'\phi',\\
\rho_1 &=  -(s\rho_0 - \phi \phi') = -\left[s(\phi \phi'' + \phi'\phi') - \phi\phi '\right],\\
\rho_2 &=  -s\rho_1 = s\left[s(\phi \phi'' + \phi'\phi') - \phi\phi '\right].
\end{align}
The only difference here with the positive definite case is the factor sign$(A)$ appearing in the first term in  \eqref{eq:alpha_beta_metric_fundamental_tensor}. Denoting $\dim M = n$, we can write this in matrix notation as
\begin{align}\label{eq:eqeq}
g = \sgn(A)\rho \left(a + UWV^T\right),
\end{align}
in terms of the three matrices
\begin{align}
W = \frac{\sgn(A)}{\rho}\mathbb I_{4\times 4},\qquad U = (\vec b, \vec b, \vec \alpha, \vec \alpha), \qquad V = (\rho_0 \vec b, \rho_1 \vec\alpha, \rho_1 \vec b, \rho_2 \vec\alpha).
\end{align}
Note that $U$ and $V$ are both $n\times 4$ matrices and $W$ is a $4\times 4$ matrix. It is a well-known result (one of the matrix determinant lemmas, see e.g. \cite{harville2008matrix}) that if $a$ is an invertible matrix (which it is in our case) then the determinant of the expression in brackets in \eqref{eq:eqeq} is equal to
\begin{align}
\det \left(a + UWV^T\right) = \det \left(\mathbb I_{4\times 4} + WV^Ta^{-1}U\right)\det a.
\end{align}
It follows that
\begin{align}
\det g = \sgn(A)^n\rho^n\det \left(\mathbb I_{4\times 4} + WV^Ta^{-1}U\right)\det a.
\end{align}
The matrix product $WV^Ta^{-1}U = \tfrac{\sgn(A)}{\rho}V^Ta^{-1}U$ can be evaluated by explicit computation and reads
\begin{align}
WV^Ta^{-1}U = \frac{\epsilon}{\rho}\left(
\begin{array}{cccc}
 |b|^2 \text{$\rho_0 $} & |b|^2 \text{$\rho_0$} & \epsilon s \text{$\rho_0$} & \epsilon s \text{$\rho_0$} \\
 \epsilon s \rho  & \epsilon s \rho  & \epsilon \rho  & \epsilon \rho  \\
 |b|^2 \text{$\rho_1$} & |b|^2 \text{$\rho_1$} & \epsilon s \text{$\rho_1$} & \epsilon s \text{$\rho_1$} \\
 \epsilon s \text{$\rho_2$} & \epsilon s \text{$\rho_2$} & \epsilon \text{$\rho_2$} & \epsilon \text{$\rho_2$} \\
\end{array}
\right),
\end{align}
where we have written $\epsilon = \sgn(A)$. Hence we obtain
\begin{align}
\det g&= \epsilon^n\rho^n \det a  \det\left(\mathbb I_{4\times 4}+\frac{\epsilon}{\rho}\left(
\begin{array}{cccc}
 |b|^2 \text{$\rho_0 $} & |b|^2 \text{$\rho_0$} & \epsilon s \text{$\rho_0$} & \epsilon s \text{$\rho_0$} \\
 \epsilon s \rho  & \epsilon s \rho  & \epsilon \rho  & \epsilon \rho  \\
 |b|^2 \text{$\rho_1$} & |b|^2 \text{$\rho_1$} & \epsilon s \text{$\rho_1$} & \epsilon s \text{$\rho_1$} \\
 \epsilon s \text{$\rho_2$} & \epsilon s \text{$\rho_2$} & \epsilon \text{$\rho_2$} & \epsilon \text{$\rho_2$} \\
\end{array}
\right)\right) \\
&= \epsilon^n\rho^n \det a  \det\left(\mathbb I_{4\times 4}+\frac{1}{\rho}\left(
\begin{array}{cccc}
 \epsilon |b|^2 \text{$\rho_0 $} & \epsilon |b|^2 \text{$\rho_0$} &  s \text{$\rho_0$} & s \text{$\rho_0$} \\
  s \rho  &  s \rho  &  \rho  & \rho  \\
 \epsilon |b|^2 \text{$\rho_1$} & \epsilon |b|^2 \text{$\rho_1$} & s \text{$\rho_1$} & s \text{$\rho_1$} \\
s \text{$\rho_2$} & s \text{$\rho_2$} &  \text{$\rho_2$} &  \text{$\rho_2$} \\
\end{array}
\right)\right)\\
&= \phi^{n+1}(\phi-s\phi')^{n-2}(\phi-s\phi' +  (\epsilon |b|^2-s^2)\phi'')\det a_{ij}.
\end{align}
Some useful identities that we have used are: $\alpha_i = \sgn(A) y_i/\alpha$ so that $\alpha_i\alpha^i = \sgn(A)$ and $\alpha_i b^i = \sgn(A) s$. We conclude that
\begin{align}\label{eq:det_ab}
\boxed{
\det g = \sgn(A)^n\phi^{n+1}(\phi-s\phi')^{n-2}(\phi-s\phi' +  (\sgn(A) |b|^2-s^2)\phi'')\det a.}
\end{align}
In the case that $\alpha$ is positive definite, sign$(A)= 1$ everywhere, so the formula reduces to the standard result (see e.g. \cite{ChernShen_RiemannFinsler}). We remark that \eqref{eq:det_ab} can also be obtained directly from the positive definite version, by noticing that $F$ and hence $g$ is invariant under $a\to -a$. If $A<0$ then $-A>0$, hence the positive definite formula applies if we replace $a$ by $-a$. However, under this sign flip, $|b|^2$ changes by a sign and $\det a$ changes by $(-1)^n$, whereas everything else in the formula for the determinant is invariant. This reasoning immediately leads to the formula above.

\newpage

\chapter{Additional Proofs}\label{app:add_proofs}

\fancyhead{} 
\fancyhead[LE]{\small\textit{\nouppercase{\leftmark}}}
\fancyhead[RO]{\small\textit{\nouppercase{\rightmark}}}

\section{Metric-compatibility of a nonlinear connection}\label{app:metric_compatibility}

\begin{prop}
Let $(M,L)$ be a Finsler space and $N$ a torsion-free, homogeneous nonlinear connection on $M$. TFAE:
\begin{enumerate}[(i)]
\item $\delta_i L =0$,
\item $\nablad g_{ij} = 0$.
\end{enumerate}
\end{prop}
\begin{proof}
The implication $(i)\Rightarrow(ii)$ can be checked explicitly using the expression \eqref{eq:nonlinear_connection_explicit} for the canonical connection. In terms of the spray
\begin{align}
    G^k = g^{kj}G_j,\qquad G_j = \tfrac{1}{2}\left(y^m\partial_m\bar\partial_j L - \partial_j L\right),
\end{align}
we simply have $N^k_i = \frac{1}{2}\bar\partial_i G^k$, by the fundamental lemma of Finsler geometry. We need to show that
\begin{align}
    \nablad g_{ij} = y^k\delta_k g_{ij} - N^k_ig_{kj} - N^k_jg_{ik} = 0.
\end{align}
To that end, consider the fact that we can write
\begin{align}
    N^k_ig_{kj} \!= \!\tfrac{1}{2} \bar\partial_i \left(g^{\ell k}G_\ell\right)g_{kj} \!=\! \tfrac{1}{2} \!\left(\bar\partial_i G_j \!+\!   g_{kj}\bar\partial_i g^{\ell k}G_\ell\right) \!=\! \tfrac{1}{2} \!\left(\bar\partial_i G_j \!-\!  \bar\partial_i g_{kj}G^k\right).
\end{align}
By noting that the two individual terms can be written as
\begin{align}
    \bar\partial_i G_j = \partial_{[i}\bar\partial_{j]}L + y^m\partial_m g_{ij}, \qquad G^k = N^k_my^m,
\end{align}
where the second one follows from Euler's theorem, it follows that
\begin{align}
    N^k_ig_{kj} &= \tfrac{1}{2} \left(\partial_{[i}\bar\partial_{j]}L + y^m\partial_m g_{ij} - y^mN^k_m\bar\partial_i g_{kj}\right)\\ 
    &= \tfrac{1}{2} \left(\partial_{[i}\bar\partial_{j]}L + y^m\partial_m g_{ij} - y^mN^k_m\bar\partial_k g_{ij}\right)\\ 
    &= \tfrac{1}{2} \left(\partial_{[i}\bar\partial_{j]}L + y^m\delta_m g_{ij}\right).
\end{align}
It thus follows that 
\begin{align}
    N^k_ig_{kj}+N^k_jg_{ik} = y^k\delta_k g_{ij},
\end{align}
which gives the desired result, proving the first implication.

Conversely, to prove that $(ii)\Rightarrow(i)$, note that we have
\begin{align}
 y^i\delta_i L = \nablad L = \nablad \left(g_{jk}  y^j y^k\right) = \nablad g_{jk} y^j y^k,
\end{align}
since $\nablad y^i = 0$.  So $\nablad g_{ij}=0$ implies $0 = y^i\delta_i L$, which means also that
\begin{align}
0 = \bar\partial_j\left(y^i\delta_i L\right) = \delta_j L + y^i\bar\partial_j \delta_i L.
\end{align}
Thus it suffices to show that $y^i\bar\partial_j \delta_i L=0$. Since $[\bar\partial_j,\delta_i] = -\bar\partial_j N^k_i\bar\partial_k$ we have
\begin{align}
y^i\bar\partial_j \delta_i L &= y^i\delta_i\bar\partial_j  L - y^i\bar\partial_j N^k_i\bar\partial_kL \\
&= y^i\delta_i\bar\partial_j  L - y^i\bar\partial_i N^k_j\bar\partial_kL - y^iT^k{}_{ij}\bar\partial_kL \\
&= y^i\delta_i\bar\partial_j  L - N^k_j\bar\partial_kL - y^iT^k{}_{ij}\bar\partial_kL \\
&= \nablad\left(\bar\partial_j L\right) -y^iT^k{}_{ij}\bar\partial_kL \\
&= \nablad\left(2 g_{jk}y^k\right) -y^iT^k{}_{ij}\bar\partial_kL \\
&= -y^iT^k{}_{ij}\bar\partial_kL \\
&= -2T_{kij}y^iy^k
\end{align}
where in the second line we used the definition of the torsion tensor, in the third line we used homogeneity of $N^k_{ij}$,  in the fourth line we used the definition of $\nablad $, then we expressed $\bar\partial_j L$ in terms of $g_{ij}$, and finally in the last line we used the Leibniz rule for $\nablad$ and the fact that $\nablad y^k=0$ and $\nablad g_{jk}=0$. Combining this with what we found above, we conclude that if $\nablad g=0$ then 
\begin{align}
\delta_j L = y^iT^k{}_{ij}\bar\partial_kL
\end{align}
and hence if additionally the torsion of $\nablad$ vanishes then we must indeed have $\delta_iL=0$.
\end{proof}

\section{Generality of generalized \texorpdfstring{$m$}{m}-Kropina metrics}\label{sec:mKropGenerality}

\Cref{theor:generalized_mKrop} assumes that $\phi'\neq 0$ on $\mathcal A$. Here we generalize the theorem to the case where no such condition is presupposed. In what follows, we will denote by $s$ both the function $s:(x,y)\mapsto \beta(x,y)/\alpha(x,y)$ as well as the function values $s\in \R$. So we will write, for instance, that $s\in s(U)$ if the \textit{value} $s$ lies in the image of the set $U$ under the \textit{function} $s$. We start with a stronger version of \Cref{lem:top}, where the assumption that $U$ is connected is omitted.
\begin{lem}\label{lem:top2}
    Suppose that $b_i$ is nowhere vanishing. If $U\subset\mathcal A$ is open then the image $s(U)$ is a union of intervals that each have nonempty interior, and 
    $s(U)\subset \overline{s(U)^0\setminus C}$ for any finite set $C$. 
\end{lem}
\begin{proof}
    Since $U$ is an open subset of the manifold $TM$, it is itself a manifold and hence it is locally connected. As a result, we may write $U = \cup_i U_i$ where each $U_i$ is open and connected. By \Cref{lem:top}, each $s(U_i)$ is an interval with nonempty interior, and we have $s(U_i)\subset \overline{s(U_i)^0\setminus C}$ for each $i$. Hence 
    \begin{align}
        s(U)&=\cup_i s(U_i)\subset \cup_i\overline{s(U_i)^0\setminus C}\subset \overline{\cup_is(U_i)^0\setminus C} \subset \overline{\left(\cup_is(U_i)\right)^0\setminus C}\\
        & = \overline{\left(s(\cup_iU_i)\right)^0\setminus C} = \overline{s(U)^0\setminus C},
    \end{align}
    as desired.
\end{proof}
As a result, if some ODE for $\phi(s)$ is satisfied for all $s\in s(U)\setminus C$, where $U\subset \mathcal A$ is some open set and $C$ a finite set, then the ODE may be solved (explicitly) on the open set $s(U)^0\setminus C$ using standard methods, and this uniquely determines $\phi$ on all of $s(U)$ by continuous extension. The following result is the generalization \Cref{theor:generalized_mKrop} that we alluded to.
\begin{theor}\label{theor:generalized_mKrop2}
     Let $F = \alpha\phi(\beta/\alpha)$ be a properly Finslerian $(\alpha,\beta)$-metric with $\dim M>2$. Suppose furthermore that $\mathcal A$ is connected. Then TFAE:
     \begin{enumerate}[(i)]
         \item F is Berwald and there exist nowhere vanishing $\lambda,\rho,\sigma\in C^\infty(M)$ such that
         \begin{align}\label{eq:TtensorTheorem2}
             T^k{}_{ij} = \lambda b^k b_i b_j + \rho\left(b_ i\delta^k_j+ b_ j\delta^k_i \right) + \sigma b^k a_{ij};
         \end{align}
         \item  $\beta$ is closed and there are nonvanishing constants $c,d,m$ and a nowhere vanishing $p\in C^\infty(M)$ such that         
         \begin{align}
         \phi(s) &= \pm s^{-m}(c + d s^2)^{(m+1)/2}\\
        \nabla_i b_j &= p\left\{cm|b|^2 a_{ij} + \left[c(1-m) + \epsilon d |b|^2\right]b_i b_j\right\} \label{eq:generalized_mKrop_Berwald_cond_theor_closed2}
    \end{align}
     \end{enumerate}
     In that case $\rho = -\sigma = -cmp$ and $\lambda=\epsilon d p$.
 \end{theor}
 \begin{proof}
     Assuming $(ii)$ then, according to \Cref{prop:gen_mKrop_Berwald_closed}, $F$ is Berwald and the affine connection is given by \eqref{eq:generalized_mKrop_Berwald_affine_connection_closed} and from this we infer that $T^k_{ij}$ is clearly of the form \eqref{eq:TtensorTheorem2} with $\lambda = \epsilon d p$ and $\rho = -\sigma = -c m p$. Hence $(ii)$ implies $(i)$. We will split the proof that $(i)$ implies $(ii)$ over the two lemmas below. Hence, with these lemmas, the proof of the theorem is complete.
\end{proof}
Denote $B = \{\phi'=0\}\subset\mathcal A$ and $\tilde{\mathcal A}=\mathcal A\setminus \overline{B^0}$. There are two cases that we will distinguish. Either $B^0=\emptyset$ or $B^0\neq\emptyset$. Each of the two cases gets its own lemma.
\begin{lem}
    \Cref{theor:generalized_mKrop2} holds under the additional assumption that $B^0 = \emptyset$.
\end{lem}
\begin{proof}   
     It suffices to show that $(i)$ implies $(ii)$. The Berwald condition \eqref{ab_Berwald_condition} must be satisfied on $\tilde{\mathcal A}$ with $T^k_{ij}$ given by \eqref{eq:TtensorTheorem2}, in which case this condition reduces to
     \begin{align}
         y^j\nabla_i b_j \!=\! \lambda |b|^2 \beta b_i \!+\! 2\rho \beta b_i \!+\! \sigma |b|^2 y_i \!+\! \frac{\epsilon \psi}{\alpha}\!\left( \lambda\beta^2 b_i \!+\! \rho(\beta y_i \!+\! A b_i) \!+\!\sigma \beta y_i \right)\!,\!
     \end{align}
     where $\psi = \phi/\phi' - s$. We can write this also as
     \begin{align}\label{eq:to_subst_into2}
         y^j\nabla_i b_j \!-\! \left( \lambda |b|^2 \beta b_i \!+\! 2\rho \beta b_i \!+\! \sigma |b|^2 y_i\right) \!=\! \frac{\epsilon \psi}{\alpha}\!\left( \lambda\beta^2 \!+\! \rho A \right) b_i \!+\!  \frac{\epsilon \psi}{\alpha}\left( \rho \!+\! \sigma\right)\beta y_i,\!
     \end{align}
     where we have collected all manifestly linear terms on the LHS and grouped the RHS by a $b_i$ term and a $y_i$ term. Since the LHS is linear, the RHS must be so as well, and in fact, by \Cref{prop:lemapp} this can only be achieved if both terms on the RHS are linear, i.e. if  $\frac{\epsilon \psi}{\alpha}\left( \rho + \sigma\right)\beta$ is independent of $y$ and if $\frac{\epsilon \psi}{\alpha}\left( \lambda\beta^2 + \rho A \right)$ is linear. Since $A=0$ is excluded, by definition, from $\mathcal A$ and since $\mathcal A$ has only a single connected component, by assumption, $\mathcal A$ must be contained in one of the open sets $A>0$ or $A<0$. And that, in particular, implies that $\epsilon=\sgn(A)$ is just a constant on $\mathcal A$. Thus we need $\frac{\psi}{\alpha}\left( \rho + \sigma\right)\beta$ to be independent of $y$ and $\frac{ \psi}{\alpha}\left( \lambda\beta^2 + \rho A \right)$ to be linear. \\
     
     \noindent\textbf{Step 1) Claim:} $\rho = -\sigma$\\
     First we consider the $y_i$ term, which is linear in $y$ iff $\psi\left( \rho + \sigma\right)\beta/\alpha=$ is independent of $y$. The only way this can be true, assuming $F$ is properly Finslerian, is if $\rho = -\sigma$. To see this, suppose that $\rho\neq -\sigma$. Then we must have $ \psi\beta/\alpha = \psi s\eqqcolon\ell$ where $\ell$ is a priori a function of $x$, but since the LHS depends only on $s$, $\ell$ must actually be constant. Hence $\psi = \phi/\phi ' - s = \ell/s$ for all $(x,y)\in\tilde{\mathcal A}$ with $s\neq 0$. As a differential equation in the variable $s$ this must hold for all $s\in s(\tilde{\mathcal A})\setminus\{0\}$. Since $\mathcal A$ is open and connected, we can apply \Cref{lem:top2} and its consequences: we can integrate the differential equation for $\phi(s)$ on the interior of $s(\tilde{\mathcal A})\setminus\{0\}$ and then extend the solution by continuity to all of $s(\tilde{\mathcal A})$, i.e. for all relevant values of $s$. Moreover, since $B$ is assumed to have empty interior, $\tilde{\mathcal A}$ is dense in $\mathcal A$. Hence, by continuity, $s(\tilde{\mathcal A})$ is dense in $s(\mathcal A)$. This means that $\phi$ can be obtained for \textit{all} values of $s$ by continuous extension.
     
     We will now apply this scheme. Assuming wlog that $\ell\neq 0$ (for otherwise $\phi -s \phi '=0$ and $\det g_{ij}=0$, by \eqref{eq:det_ab}) we find $\ln |\phi| = \int \left(s + \ell/s\right)^{-1}\D s= \ln(|c(\ell + s^2)|)/2$, where $c$ is an integration constant (that may, at this stage, be different on each connected component). Then $\phi = \sqrt{c(\ell + s^2)}$ for $s$ in the interior of $s(\tilde{\mathcal A})\setminus\{0\}$, and by the argument given above this must be the form of $\phi$ for all values of $s$. Hence $F^2 = \alpha^2 \phi^2 = c(\ell\alpha^2 + \beta^2)$ is pseudo-Riemannian. This is excluded in the premise of the theorem and hence it is a contradiction, so we must indeed have $\rho = -\sigma$.\\

    \noindent\textbf{Step 2) Deriving an ODE in $s$}\\
     Next, we consider the $b_i$ term, setting $\sigma=-\rho$. This term is linear iff the coefficient of $b_i$ is linear, which is the case iff 
     \begin{align}\label{eq:second_der_for_linearity2}
         \bar\partial_i\bar\partial_j\left[\frac{\psi}{\alpha}\left( \lambda\beta^2 + \rho A \right)\right] =0.
     \end{align}
     For the first derivative, we have
     \begin{align}
         \bar\partial_i\left[\frac{\psi}{\alpha}\left( \lambda\beta^2 + \rho A \right)\right] &= \frac{\psi'\bar\partial_i s}{\alpha}\left( \lambda\beta^2 + \rho A \right) - \frac{\psi}{\alpha^2}\bar\partial_i\alpha\left( \lambda\beta^2 + \rho A \right) \nonumber\\
         &\qquad+ \frac{\psi}{\alpha}\left( 2\lambda\beta b_i + 2\rho y_i \right) \\
         &\hspace{-60px}=  \frac{\psi'}{\alpha^2}\left(b_i - \frac{\beta y_i}{A}\right)\left( \lambda\beta^2 + \rho A \right) - \frac{\epsilon \psi y_i}{\alpha^3}\left( \lambda\beta^2 + \rho A \right)  \nonumber\\
         &\qquad+ \frac{\psi}{\alpha}\left( 2\lambda\beta b_i+ 2\rho y_i \right) \\
         &\hspace{-60px}=  \frac{1}{\alpha^2}\left[-\left(\frac{\beta \psi'}{A} + \frac{\epsilon \psi}{\alpha}\right)\left( \lambda\beta^2 + \rho A \right)  + 2\alpha \rho\psi\right] y_i \nonumber\\
         &\qquad+ \{b_i\text{ terms}\},\label{eq:first_der_for_linearity2}
     \end{align}
     where we have used the identities
     \begin{align}
        \bar\partial_i \alpha = \frac{\epsilon y_i}{\alpha} ,\qquad \bar\partial_i s = \frac{1}{\alpha}\left(b_i - \frac{\beta y_i}{A}\right),
     \end{align}
     and separated the $y_i$ and the $b_i$ parts, the latter of which are irrelevant. The reason for this is the following. By inspecting the situation (or working it out exactly) it is clear that the second derivative in \eqref{eq:second_der_for_linearity2} will be of the form $f a_{ij} + g h_{ij}$, where $f,g$ are functions and $h_{ij}$ is a linear combination of $b_i b_j, b_i y_j, y_i b_j$ and $y_i y_j$. Hence $h_{ij}$ has at most rank 2, which is strictly less than the rank of $a_{ij}$. That means that \eqref{eq:second_der_for_linearity2}, i.e. the equation $f a_{ij} + g h_{ij}=0$, can only hold if $f=0$, for otherwise we could write $a_{ij} =(g/f)h_{ij}$ and $a_{ij}$ and $h_{ij}$ would have the same rank. It turns out that it suffices for what we aim to prove to consider this coefficient $f$, as we will see shortly. Differentiation of the $b_i$ terms in \eqref{eq:first_der_for_linearity2}, however, will never yield terms proportional to $a_{ij}$. In fact, the only way we get terms proportional to $a_{ij}$ is by directly differentiating $y_i$ in \eqref{eq:first_der_for_linearity2}. No other terms resulting from the product rule are proportional to $a_{ij}$ either. This  simplifies this calculation enormously because it means that in order to get the $a_{ij}$-coefficient $f$ from \eqref{eq:first_der_for_linearity2} we effectively need only replace $y_i$ by its derivative $\bar\partial_j y_i = a_{ij}$ to obtain
     \begin{align}
         \bar\partial_i\bar\partial_j\left[\frac{\psi}{\alpha}\left( \lambda\beta^2 + \rho A \right)\right] =\qquad\qquad\qquad\qquad\qquad\qquad\qquad\qquad\qquad\\
          \frac{1}{\alpha^2}\left[-\left(\frac{\beta \psi'}{A} + \frac{\epsilon \psi}{\alpha}\right)\left( \lambda\beta^2 + \rho A \right)  + 2\alpha \rho\psi\right]a_{ij} + g h_{ij}.
     \end{align}
     As we argued above, the vanishing of this implies that we must have
     \begin{align}
          0&=\frac{1}{\alpha}\left[-\left(\frac{\beta \psi'}{A} + \frac{\epsilon \psi}{\alpha}\right)\left( \lambda\beta^2 + \rho A \right)  + 2\alpha \rho\psi\right]\\
          &= -\left(\frac{s \psi'}{\epsilon \alpha^2} + \frac{\epsilon \psi}{\alpha^2}\right)\left( \lambda\beta^2 + \rho A \right)  + 2 \rho\psi\\
          &= -\epsilon\left(s \psi'+ \psi\right)\left( \lambda s^2 + \rho \epsilon \right)  + 2 \rho\psi, 
     \end{align}
     or, after rewriting,
     \begin{align}\label{eq:ODE_for_psi2}
         \frac{\psi'}{\psi} = \frac{ \eta- s^2}{ s( s^2 + \eta )}, \qquad \eta \equiv \epsilon\rho/\lambda.
     \end{align}
     As before, since the LHS depends only on $s$, the RHS should also not depend on $x$ explicitly, which implies that $\eta$, which is a priori a function on $M$, should actually be a constant. Hence \eqref{eq:ODE_for_psi2} is an ODE for $\psi(s)$. \\
     
     \noindent\textbf{Step 3) Solving the ODE to obtain $\phi(s)$}\\
     In analogy with how we solved the differential equation for $\phi$ earlier in this proof, this one is solved uniquely by 
     \begin{align}\label{eq:psi_unique_sol2}
         \psi = \frac{\tilde c s}{s^2 + \eta},
     \end{align}
     for all $s\in s(\tilde {\mathcal A})\setminus\{\sqrt{-\eta},-\sqrt{-\eta}\}$, where $\tilde c$ is an integration constant. We find $\phi$ on $K^0$ via $\psi = \phi/\phi' - s$. This can now be written as
     \begin{align}
         \frac{\phi'}{\phi} = \left(\frac{\tilde cs}{s^2 + \eta}+s\right)^{-1} = \frac{d s^2 -  c m}{d s^3 + c s},
     \end{align}
     where we have introduced an arbitrary constant $d$ and defined $ c = d(\tilde c+\eta)$ and $m = - d \eta/ c = -\eta/(\eta+\tilde  c)$. This is uniquely solved by
     \begin{align}
         \phi(s) = \tilde d s^{-m}( c+d s^2)^{(m+1)/2},
     \end{align}
     for $s\in s(\tilde {\mathcal A})\setminus\{\sqrt{-\eta},-\sqrt{-\eta}\}$, where we may absorb the integration constant $\tilde d$ (up to sign) into $c$ and $d$. Hence $\phi$ attains the desired form
     \begin{align}\label{eq:pghut2}
         \phi(s) = \pm s^{-m}( c+d s^2)^{(m+1)/2}
     \end{align}
     for all $s\in s(\tilde {\mathcal A})\setminus\{\sqrt{-\eta},-\sqrt{-\eta}\}$. Once again, by \Cref{lem:top2} and the fact that $\tilde{\mathcal A}$ is dense in $\mathcal A$, we can again extend the solution by continuity to all relevant values of $s$. In other words, $\phi$ has the form \eqref{eq:pghut2} for all relevant values of $s$, as desired. In principle, the constants $c,d,m$ and the sign $\pm$ could have different values on different connected components of $s(\mathcal A$), but since $\mathcal A$ is assumed to be connected, $s(\mathcal A)$ is also connected and hence the constants and the sign are actually fixed over all of $s(\mathcal A)$.\\
     %
     %
     %
     
     \noindent\textbf{Step 4) The condition on $\nabla_i b_j$}\\
     Finally we substitute the form \eqref{eq:psi_unique_sol2} of $\psi$ and the relation between $\rho,\sigma$ and $\lambda$ into the Berwald condition \eqref{eq:to_subst_into2}, leading to
     \begin{align}\label{eq:consistency_check2}
         y^j\nabla_i b_j &= \left( \lambda |b|^2 \beta b_i + 2\rho \beta b_i + \sigma |b|^2 y_i\right) + \epsilon \lambda \tilde c\beta b_i,
     \end{align}
     from which we can infer by differentiation that for $\nabla_i b_j$ is symmetric under $i\leftrightarrow j$. This implies that the 1-form $\beta$ is closed and hence, by \Cref{prop:gen_mKrop_Berwald_closed}, that the desired Berwald condition holds. This completes the proof. 
 \end{proof}
 \begin{rem}
      As a consistency check for step 4 of the proof, we may also derive the desired Berwald condition directly from \eqref{eq:consistency_check2}.  
      For this, we note that the definitions of $c$ and $m$ in terms of $\tilde c$ and $\eta$ above may be inverted as $\tilde c = c(1+m)/d$ and $\eta  = -mc/d$. Using also the relations between $\lambda,\rho,\sigma$, we find that \eqref{eq:to_subst_into2} turns into
     \begin{align}
         y^j\nabla_i b_j &= \left( \lambda |b|^2 \beta b_i + 2\beta b_i + \sigma |b|^2 y_i\right) + \epsilon \lambda \tilde c\beta b_i \\
         &= \left( \lambda |b|^2 \beta b_i + 2\epsilon\eta\lambda \beta b_i - \epsilon\eta\lambda  |b|^2 y_i\right) + \epsilon \lambda \frac{c(1+m)}{d}\beta b_i\\
         &=  \lambda\left[\left( |b|^2   + 2\epsilon\eta    + \epsilon  \frac{c(1+m)}{d} \right) \beta b_i -\epsilon\eta |b|^2 y_i\right]\\
         &=  \lambda\left[\left( |b|^2   - \frac{2\epsilon mc}{d}    +   \frac{\epsilon c(1+m)}{d} \right) \beta b_i +\frac{\epsilon mc}{d} |b|^2 y_i\right]\\
         %
         &=  \lambda\left[\left( |b|^2   + \frac{\epsilon c}{d}\left(1-m\right) \right) \beta b_i -\epsilon\frac{-mc}{d} |b|^2 y_i\right],
     \end{align}
     from which we can infer by differentiation that
     \begin{align}
         \nabla_i b_j &= \frac{\epsilon \lambda}{d}\left\{\left[ \epsilon d |b|^2   +  c\left(1-m\right) \right]  b_ib_j + mc |b|^2 a_{ij}\right\},
     \end{align}
     which is precisely \eqref{eq:generalized_mKrop_Berwald_cond_theor_closed2} if we identify $p=\epsilon \lambda/d$.
 \end{rem}
 \begin{lem}
     \Cref{theor:generalized_mKrop2} holds under the additional assumption that $B^0 \neq \emptyset$.
 \end{lem}
 \begin{proof}
     Again, it suffices to prove that $(i)$ implies $(ii)$. Most of the proof of the previous lemma is still valid. In particular, the conclusion that $\phi$ must be of the form 
     \begin{align}\label{eq:ppdsd}
         \phi(s) = \pm s^{-m}( c+d s^2)^{(m+1)/2}
     \end{align}
     still holds on $s(\tilde{\mathcal A})$. The difference is that $\tilde{\mathcal A}$ is currently not dense in $\mathcal A$ and so we cannot extend this by continuity to all of $s(\mathcal A)$. However, since we defined $B = \{\phi'=0\}\subset\mathcal A$ and $\tilde{\mathcal A}=\mathcal A\setminus \overline{B^0}$, where we will agree that the closure is taken, by definition, in $A$, we have
     \begin{align}
         \mathcal A = \tilde{\mathcal A} \cup \overline{B^0}, \qquad s(\mathcal A) = s(\tilde{\mathcal A}) \cup s(\overline{B^0}),
     \end{align}
     so it remains to see what happens for $s\in s(\overline{B^0})$. First, on $s(B^0)$, we have $\phi'=0$, by definition. By \Cref{lem:top2} and the remark right below it, this implies that 
     \begin{align}\label{eq:temdfjk}
         \phi(s) = k
     \end{align}
     for all $s\in s(\overline{B^0})$. Thus what we have shown is that for all $s$ in  $s(\mathcal A)$, $\phi$ is given by either by \eqref{eq:ppdsd} or by \eqref{eq:temdfjk}. Note that $s(\mathcal A)$ is an interval because $\mathcal A$ is connected and $\phi$ assumed continuous, even smooth. Moreover, by \Cref{lem:top2}, both $s(\overline{B^0})$ and $s(\tilde{\mathcal A})$ are a union of intervals with nonempty interior. We want to prove that \eqref{eq:ppdsd} holds \textit{everywhere}. Suppose not. Consider the interface between an interval on which \eqref{eq:ppdsd} is satisfied and an interval on which \eqref{eq:temdfjk} is satisfied. $\phi$ is assumed to be smooth on this interface. That means that all derivatives of \eqref{eq:ppdsd} must tend to zero (which are the derivatives of \eqref{eq:temdfjk}) if $s$ tends to this interface. However, one can check that there are no values of $s$ for which all derivatives of \eqref{eq:ppdsd} tend to zero, except if $cm = 0$, in which case $F$ is either degenerate (which would be a contradiction) or pseudo-Riemannian. But that means that on every interval on which \eqref{eq:ppdsd} is satisfied, $F$ must in fact be pseudo-Riemannian. And that simply means that $F$ is \psR as a whole. This is a contradiction with our assumptions and therefore this completes the proof.
 \end{proof}

\newpage

%% file: ListOfPublications.tex
\cleardoublepage
\phantomsection
\addcontentsline{toc}{chapter}{List of Publications}
\chapter*{List of Publications}
\fancyhead{}
\fancyhead[RE]{\textit{List of Publications}}
\fancyhead[LO]{\textit{List of Publications}}

\large
\textbf{This dissertation is based on:}
\normalsize
\vspace{5px}

\begin{enumerate}[{[H1]}]
\item \label{ref1} C. Pfeifer, \textbf{S. Heefer}, A. Fuster. \textit{Identifying Berwald Finsler Geometries}, \href{https://doi.org/10.1016/j.difgeo.2021.101817}{Differential Geom. Appl. 79, 101817 (2021)}, \href{https://arxiv.org/abs/1909.05284}{\tt arXiv:1909.05284}
\item \label{ref3} A. Fuster, \textbf{S. Heefer}, C. Pfeifer, N. Voicu.  \textit{On the Non Metrizability of Berwald Finsler Spacetimes}, \href{https://doi.org/10.3390/universe6050064}{Universe 6, 64 (2020)}, \href{https://arxiv.org/abs/2003.02300}{\tt arXiv:2003.02300}
\item \label{ref2}\textbf{S. Heefer}, C. Pfeifer, J. van Voorthuizen, A. Fuster. \textit{On the Metrizability of m-Kropina Spaces with Closed Null 1-form}, \href{https://doi.org/10.1063/5.0130523}{J. Math. Phys. 64, 022502 (2023)}, \href{https://arxiv.org/abs/2210.02718}{\tt arXiv:2210.02718}
\item \label{ref4}\textbf{S. Heefer}, C. Pfeifer, A. Fuster. \textit{Randers pp-Waves}, \href{https://doi.org/10.1103/PhysRevD.104.024007}{Phys. Rev. D 104, 024007 (2021)}, \href{https://arxiv.org/abs/2011.12969}{\tt arXiv:2011.12969}
\item \label{ref5}\textbf{S. Heefer}, A. Fuster. \textit{Finsler Gravitational Waves of $(\alpha,\beta)$-type and their Observational Signature},  \href{https://doi.org/10.1088/1361-6382/acecce}{Class. Quantum Grav. 40 184002 (2023)}, \href{https://arxiv.org/abs/2302.08334}{\tt arXiv: 2302.08334}

\item \label{ref6}\textbf{S. Heefer}, C. Pfeifer, A. Reggio, A. Fuster. \textit{A Cosmological Unicorn Solution to Finsler Gravity}, \href{https://doi.org/10.1103/PhysRevD.108.064051}{Phys. Rev. D 108, 064051 (2023)}, \href{https://arxiv.org/abs/2306.00722}{\tt arXiv: 2306.00722}
\end{enumerate}

\vspace{10px}

\newpage







%% file: Acknowledgments.tex

\cleardoublepage
\phantomsection
\addcontentsline{toc}{chapter}{Acknowledgments}

\fancyhead{}

\chapter*{Acknowledgments}
\fancyhead[RE]{\textit{Acknowledgments}}
\fancyhead[LO]{\textit{Acknowledgments}}


The past couple of years have been an incredible time for me. It has been a blessing to have had the opportunity to completely immerse myself in the world of academia, work on such fascinating topics, and go to numerous inspiring conferences all around Europe. There are quite some people who have been instrumental in this journey and I would like to express my sincere gratitude to all of them.\\[-10px]

First of all, my co-promotor and daily supervisor, Andrea, thanks to whom all of this has been possible. Your research ideas and 
advice provided a solid foundation from the outset. You also gave me the freedom to pursue my own ideas without too much pressure and never forced me to do anything I was not comfortable with (except maybe take some days off from time to time---to my own benefit, of course). This has resulted in a very comfortable and productive way of working and doing research. \\[-10px]

Second, my promotor, Luc, who would always ask good and difficult questions about my research that would incentivize me to think more deeply about certain aspects of it. Thanks for the support and the great discussions.\\[-10px]

All other people who I collaborated and coauthored with over the last couple of years, most notably Christian and Nicoleta. I thoroughly enjoyed our joint projects and I hope we will continue our collaboration in the future. \\[-10px]

All people---too many to name---with whom I had inspiring conversations and discussions. I'm particularly grateful for having been part of the COST action CA18108: \textit{Quantum gravity phenomenology in the multi-messenger approach (QG-MM)}, which considerably broadened my perspective not just on my own research but on science in general.\\[-10px]

My committee members. Thanks, first of all, for taking the time to read my dissertation and join the defense ceremony. But I'm especially thankful for the valuable feedback that I received on the draft version, which has certainly improved the quality of the dissertation.\\[-10px]


My office mates and fellow ADG group members, Andrea, Bart, Finn, Gautam, Gijs, Georgios, Lars, Luc, Nicky, Remco, Rick, Stephan. I greatly enjoy our discussions (about math or otherwise) over lunch, coffee or just when we really ought to be working. I'm sure we'll have many more. Special thanks to Nicky and Rick for being my paranymphs. And special thanks to Gijs for taking the time to proofread the dissertation---in quite some technical detail. Thanks also to all my other colleagues at TU/e (CASA and beyond) who have made for a very pleasant atmosphere at the university.\\[-10px]

I warmly thank all my family and friends for their interest in my endeavors, their support, and the welcome distractions. In particular, a big thanks goes out to the notorious `Vrijdagavond' group, Dirk, Frank, Ivo, Jasper, Joran, Kjell, Loek, Niek, Roeland, Thomas, Tim, Tom, and Thijs, with whom I have so many fantastic memories. I count myself very fortunate to have such a great group of friends. Special thanks to Loek for taking the time to proofread a part of the dissertation.\\[-10px]

Finally and perhaps most importantly, I'm deeply thankful to my parents and my brother, who have always supported me unconditionally and believed in me. Thank you for always being there for me. 

\begin{flushright}
Sjors Heefer\\
Eindhoven, April 2024
\end{flushright}

\newpage